\documentclass[a4paper,10pt,final]{report}

\usepackage[round]{natbib}
\usepackage{amsmath,amssymb,graphicx,setspace,fancyhdr}
\usepackage[ps2pdf=true,pdfborder=0,colorlinks=true,linkcolor=black,citecolor=black]{hyperref}
\usepackage[all]{hypcap}

\newcommand{\be}{\begin{equation}}
\newcommand{\ee}{\end{equation}}
\newcommand{\figref}[1]{Figure \ref{#1}}
\newcommand{\secref}[1]{Section \ref{#1}}
\newcommand{\chapref}[1]{Chapter \ref{#1}}
\def\link{\prec\!\!*\,}
\def\CS{\mathcal{C}}
\def\mink{\mathbb{M}}
\def\p{\hat{\phi}}
\def\lv{\langle 0 |}
\def\rv{| 0 \rangle}
\def\adag{\hat{a}^\dagger}
\def\a{\hat{a}}

\def\Ale{\bar{A}}
\def\cint{\mathbb{A}}
\def\X{X}
\def\Y{Y}
\def\GR{(G_R)} %Notation for retarded propagator
\def\GA{(G_A)}
\def\GF{(G_F)}
\def\ve{\vec{e}}
\def\E{\hat{E}}
\def\ID{i \Delta}
\def\q{\hat{q}}
\def\th{^\textrm{th}}
\def\sh{\vec{\sigma}}
\def\vp{\vec{p}}
\def\vq{\vec{q}}
\def\vk{\vec{k}}
\def\F{F}
\def\chap{Chapter~}

\begin{document}
\thispagestyle{plain}
\begin{titlepage}
\thispagestyle{plain}
\setcounter{page}{1}
\begin{center}
\textsc{\LARGE Quantum Fields on Causal Sets}\\[1cm]
by \\[1cm]
{\large Steven Paul Johnston}\\
MSci Mathematics, Imperial College London, 2006
\\ [3cm]
Submitted to the Department of Physics at Imperial College London
in partial fulfillment of the requirements for the degree of\\
[1cm]
Doctor of Philosophy \\
(Theoretical Physics, University of London)\\
[1cm]
and\\[1cm]Diploma of Imperial College London\\[1cm]
\vfill
September 2010\\[1cm]

\end{center}

\end{titlepage}

\section*{Declaration}

I declare that this work is entirely my own, except where otherwise stated. Parts of \chapref{chap:PathIntegrals} appear in 
\citet{Johnston:2008, Johnston:2009:DICE}. Parts of \chapref{chap:FreeQFT} appear in \citet{Johnston:2009}.
\\[1cm]
Signed: \line(1,0){200}\hfill September 2010.
\vfill
\section*{Acknowledgements}
\setcounter{page}{2}
It is a pleasure to thank the numerous people that have helped me over the last four years, during which time it has been a privilege to work on causal set theory. I thank my supervisor, Fay Dowker, for her support and advice. Her knowledgeable and balanced guidance has been invaluable.

I also thank Rafael Sorkin for repeated helpful exchanges, David Rideout for help with simulations, Joe Henson and Graham Brightwell for their willingness to discuss my work. Also Sumati Surya for her friendly hospitality and interest. I am grateful to Johan Noldus who provided a crucial insight for the work in Chapter 4. I am also grateful to Noel Hustler and Bernhard Schmitzer for their interest and discussions.

I thank the theoretical physics group at Imperial College for their help over the past four years. Amongst my fellow students I thank Leron Borsten for his encouragement and friendship, Dionigi Benincasa for being a good travelling companion,
Duminda Dahanayake, Lydia Philpott, Will Rubens, Jamie Vicary, Ben Withers and James Yearsley for their friendship and varied  perspectives on current physics.

Of course I thank my friends and family, in particular my parents for their support. Special mention belongs to my wife Ashley---I thank her for her love, constant support and encouragement.

\newpage

\phantomsection
\addcontentsline{toc}{chapter}{Abstract}
\begin{abstract}
\setcounter{page}{3}
\thispagestyle{plain}
Causal set theory provides a model of discrete spacetime in which spacetime events are represented by elements of a causal set---a locally finite, partially ordered set in which the partial order represents the causal relationships between events. The work presented here describes a model for matter on a causal set, specifically a theory of quantum scalar fields on a causal set spacetime background.

The work starts with a discrete path integral model for particles on a causal set. Here quantum mechanical amplitudes are assigned to trajectories within the causal set. By summing these over all trajectories between two spacetime events we obtain a causal set particle propagator. With a suitable choice of amplitudes this is shown to agree (in an appropriate sense) with the retarded propagator for the Klein-Gordon equation in Minkowski spacetime.

This causal set propagator is then used to define a causal set analogue of the Pauli-Jordan function that appears in continuum quantum field theories. A quantum scalar field is then modelled by an algebra of operators which satisfy three simple conditions (including a bosonic commutation rule). Defining time-ordering through a linear extension of the causal set these field operators are used to define a causal set Feynman propagator. Evidence is presented which shows agreement (in a suitable sense) between the causal set Feynman propagator and the continuum Feynman propagator for the Klein-Gordon equation in Minkowski spacetime. The Feynman propagator is obtained using the eigendecomposition of the Pauli-Jordan function, a method which can also be applied in continuum-based theories.

The free field theory is extended to include interacting scalar fields. This leads to a suggestion for a non-perturbative S-matrix on a causal set. Models for continuum-based phenomenology and spin-half particles on a causal set are also presented.
\end{abstract}
\setcounter{page}{4}
\begin{spacing}{0.9}
\begin{small}
\cleardoublepage\phantomsection\pdfbookmark{Contents}{contents}
\tableofcontents
\cleardoublepage\phantomsection\pdfbookmark{List of Figures}{figures}
\listoffigures
\end{small}
\end{spacing}
\newpage

\pagestyle{fancy}
\fancyhf{}
\rhead{\textit{\small{\thesection\ \rightmark}}}
\cfoot{\thepage}

\renewcommand{\chaptermark}[1]{\markboth{#1}{}}
\renewcommand{\sectionmark}[1]{\markright{#1}{}}

\chapter{Introduction} \label{chap:Intro}

\section{Quantum gravity}

\begin{quote}
\ldots Pauli asked me what I was working on. I said I was trying
to quantize the gravitational field. For many seconds he sat silent, alternately shaking
and nodding his head.
He finally said ``That is a very important problem--but it will take someone really smart!''

\flushright{Bryce DeWitt, in \citet{DeWitt:2009}.}
\end{quote}

The two great pillars of $20\th$ century physics are general relativity and quantum
mechanics. General relativity is our best theory of gravity and describes the behaviour of extremely large objects---planets, stars, galaxies, etc. Quantum mechanics (and its successors quantum field theory and the Standard Model) is our best theory of the small-scale behaviour of matter---elementary particles and the forces between them. 

Both these pillars are well supported by experimental results and observations but they each use different ideas and concepts to describe the world. One of the major tasks for $21^\textrm{st}$ century physics is to unify these two pillars into one physical theory. This, as yet unknown, theory is given the name \emph{quantum gravity}.

Since general relativity and quantum mechanics are supposed to describe the same universe, we assume it is possible to combine them into \emph{one} theory with one physical language and one mathematical description. This is the challenge---we need to keep enough of each theory so we still agree with experiment while dropping assumptions that are not supported by experiment. The hope is that by tweaking one or both of the theories they can be married together more naturally.

The work presented here deals with one particular approach to the problem of quantum gravity called \emph{causal set theory}. The central idea is to drop the assumption that spacetime is continuous. By assuming that there is a fundamental discreteness to spacetime we're learning from the discreteness inherent in quantum mechanics and, it is hoped, are closer to the theory of quantum gravity.

\section{Spacetime and causality}

Our best description of space and time has undergone many revisions up to the current day. Newton envisaged space and time as separate absolute, rigid entities. Einstein's theory of special relativity showed that instead it is more natural to think of space and time as different aspects of the same thing---spacetime. This unified description of spacetime was vindicated with the theory of general relativity in which gravity was successfully described as the curvature of spacetime.

Since then the accepted description of spacetime has remained essentially unchanged. It has become the background arena for the development of other physical theories. For example quantum mechanics was originally conceived in a non-relativistic spacetime and then in relativistic flat and ultimately curved spacetimes. Each development offered unexpected insights into how quantum theory and gravity can co-exist.

For the current work a noteworthy feature of the spacetime of general relativity is that it is continuous. Loosely speaking this means that spacetime can be arbitrarily sub-divided into smaller and smaller pieces without ever reaching a ``smallest piece of spacetime''. In causal set theory, by contrast, spacetime is modelled by a discrete structure---a causal set. In this model there \emph{are} ``smallest pieces'' of spacetime. We don't notice the discreteness in our day-to-day lives, so the idea goes, because these pieces are so extremely tiny. This is similar to water in a bathtub---although the water appears to be a smooth continuous fluid it's really made of many tiny water molecules.

Other theories of quantum gravity have modelled spacetime in different ways---it is a 10 dimensional manifold in string theory, 11 dimensional manifold in M-theory, a spacetime foam, a spin network etc. It is safe to say there is, as yet, no commonly accepted successor to the description of spacetime given by general relativity.

\subsection{Spacetime in general relativity}

We now review how general relativity models spacetime. In particular we focus on the notion of relativistic causality---good references include \citet{Penrose:1972}, \citet{Hawking:1973}.

In general relativity spacetime is modelled by a four-dimensional Lorentzian manifold\footnote{This is usually assumed to be a connected, Hausdorff, smooth ($C^\infty$) manifold. These extra conditions of are both physically reasonable and mathematically convenient.} $(M,g)$. The manifold $M$ represents the collection of all spacetime events and the metric $g$ is a symmetric non-degenerate tensor on $M$ of signature\footnote{We shall consistently use the $(+,-,\ldots,-)$ signature for our Lorentzian manifolds.} $(+,-,-,-)$.
Points in $M$ represent idealisations of spacetime events in the limit of the event happening in smaller and smaller regions of spacetime.

At each point $p \in M$ tangent vectors $X \in T_pM$ can be classified as either \emph{timelike}, \emph{null} or \emph{spacelike} depending on the whether $g(X,X)$ is positive, zero or negative respectively.

Timelike tangent vectors at a point $p \in M$ can be divided into two types. For timelike tangent vectors $X, Y \in T_pM$ we can define an equivalence relation $X \sim Y \iff g(X,Y) > 0$ and find there are two equivalence classes\footnote{See, for example, \citet[p82-83]{Geroch:1985}.}. We arbitrarily label one ``future-directed'' and one ``past-directed'' and think of the labels as defining a local arrow of time at the point. Null vectors can be given a time-orientation depending on whether they are limits of future or past-directed timelike vectors.

A Lorentzian manifold is \emph{time-orientable} if we can make a consistent continuous choice of future-directed and past-directed timelike or null vectors everywhere in the manifold.

Smooth curves in $M$ (at least $C^1$, i.e. those for which tangent vectors are everywhere defined) can be classified according to their tangent vectors:
\begin{itemize}
\item Chronological (or timelike): The tangent vector is always timelike,
\item Null : The tangent vector is always null,
\item Spacelike : The tangent vector is always spacelike,
\item Causal (or non-spacelike): The tangent vector is always timelike or null.
\end{itemize}

In a time-orientable Lorentzian manifold a timelike or causal curve is \emph{future} (resp. \emph{past}) \emph{directed} depending on whether its tangent vector is everywhere future (resp. past) directed.

\subsection{Causal structure} \label{sec:CausalStructure}

We now give a brief introduction to the causal structure of a time-orientable Lorentzian manifold. For introductions to causality theory \citet{Penrose:1972}, \citet[\chap6]{Hawking:1973} and \citet{Minguzzi:2006} are good references. The important theorems from our point of view are contained in  \citet{Hawking:1976,Malament:1977} and \citet{Levichev:1987}.

The causal structure of a Lorentzian manifold $(M,g)$ is defined in terms of smooth curves within $M$. For two points $x,y \in M$ we write $x \ll y$ (read ``$x$ chronologically precedes $y$'') if and only if there exists a future-directed timelike curve\footnote{There are other equivalent ways to define $\ll$ and $\preceq$ which may be technically more convenient---for example using ``trips'' and ``causal trips'' \citep{Penrose:1972}.} from $x$ to $y$. We write $x \preceq y$ (read ``$x$ causally precedes $y$'') if and only if there exists a future-directed causal curve from $x$ to $y$. The knowledge of which pairs of points are causally (chronologically) related defines the manifold's causal (chronological) structure.

If there are no closed causal curves in the spacetime (i.e. no distinct $x$ and $y$ such that $x \preceq y$ and $y \preceq x$) then we say the Lorentzian manifold is \emph{causal}. This is just one of a number of causality conditions that can be imposed on a Lorentzian manifold. 

For a causal Lorentzian manifold the chronological relation is:
\begin{itemize}
\item Irreflexive: For all $x \in M$, we have $x \not\ll x$,
\item Transitive: For all $x,y,z \in M$, we have\footnote{We shall use the notation $xRyRz$ as shorthand for $xRy$ and $yRx$ for any binary relation $R$.} $x \ll y \ll z \implies x \ll z$.
\end{itemize}
The causal relation is:
\begin{itemize}
\item Reflexive\footnote{Essentially because the curve consisting of a single point is a causal curve.}: For all $x \in M$, we have $x \preceq x$,
\item Antisymmetric: For all $x, y \in M$, we have $x \preceq y \preceq x \implies x = y$.
\item Transitive: For all $x,y,z \in M$, we have $x \preceq y \preceq z \implies x \preceq z$,
\end{itemize}
These conditions mean that the pair $(M,\ll)$ is an irreflexive partial order and that $(M,\preceq)$ is a (reflexive) partial order (\secref{sec:PartialOrders} contains a brief introduction to partial orders).

It is useful to define the \emph{chronological future} (resp. \emph{past}) of $x \in M$ as $I^+(x) := \left\{ y \in M : x \ll y \right\}$ (resp. $I^-(x) := \left\{ y \in M : y \ll x \right\}$). Similarly the \emph{causal future} (resp. \emph{past}) of $x$ is $J^+(x) := \left\{ y \in M: x \preceq y \right\}$ (resp. $J^-(x) := \left\{ y \in M: y \preceq x \right\}$).

The causal structure contains a lot of information about the Lorentzian manifold. Before elaborating on this idea we mention another causality condition that a Lorentzian manifold can satisfy.

A Lorentzian manifold is \emph{past} (resp. \emph{future}) \emph{distinguishing} if $ I^-(x) = I^-(y)$ implies $x = y$ (resp.  $I^+(x) = I^+(y)$ implies $x = y$). A Lorentzian manifold is \emph{distinguishing} if it is both past and future distinguishing.

We can now present two important theorems comparing the causal structures of different Lorentzian manifolds. To help us we say that for two Lorentzian manifolds $(M,g)$ and $(M',g')$ a bijection $f : M \to M'$ is a \emph{chronological isomorphism} if it preserves the chronological structure: i.e. for all $x,y \in M$ we have $x \ll y \iff f(x) \ll f(y)$. It is a \emph{causal isomorphism} if it preserves the causal structure: i.e. for all $x,y \in M$ we have $x \preceq y \iff f(x) \preceq f(y)$. 

We have the following theorems:

\begin{description}
\item[Malament's Theorem] Suppose that $(M,g)$ and $(M',g')$ are two distinguishing Lorentzian manifolds and $f: M \to M'$ is a chronological isomorphism then $f$ is a smooth conformal isometry (Meaning $f$ is a smooth map and that $f_* g = \Omega^2 g'$ for some conformal factor $\Omega : M' \to \mathbb{R}$).
\end{description}

This result appears as Theorem 2 in \citet{Malament:1977}\footnote{In \citet{Malament:1977} the term ``causal isomorphism'' is used to mean what we've called ``chronological isomorphism''.}. It relies heavily on previous results by \citet{Hawking:1976} relating the causal structure and topology of a Lorentzian manifold.
\begin{description}
\item[Levichev's Theorem] Suppose that $(M,g)$ and $(M',g')$ are two distinguishing Lorentzian manifolds and $f: M \to M'$ is a causal isomorphism then $f$ is a smooth conformal isometry.
\end{description}

This result is essentially Theorem 2 in \citet{Levichev:1987} (here extended trivially to a bijection between two different Lorentzian manifolds). It it an extension of the Malament theorem to the case of a \emph{causal} isomorphism. This is achieved by characterising the chronological relation of a distinguishing Lorentzian manifold in terms of the causal relation.

The theorems just presented mean that the conformal geometry of a distinguishing Lorentzian manifold is entirely determined by its causal structure. For a four-dimensional Lorentzian manifold the metric $g$ has 10 independent components. Fixing the conformal factor fixes $\det(g)$ which is equivalent to fixing one of these ten components. We can thus say that the causal structure determines ``${ 9/10}\textrm{ths}$'' of the metric.

This wealth of information contained in the causal structure is the reason that causal set theory takes the causal partial order as fundamental. While causality is one of the main ingredients in causal set theory the other is spacetime discreteness.

\section{Discreteness}

One of the central lessons from quantum mechanics is that nature is discrete---that at a fundamental level matter is made of small indivisible \emph{quanta}. When trying to combine general relativity and quantum mechanics one obvious modification of general relativity is to somehow make spacetime discrete. This is the approach taken by causal set theory and we now discuss the idea.

\subsection{The Planck scale} \label{sec:PlanckScale}

If spacetime really is discrete then what size is the discreteness scale? Presumably it is so small the its effects have gone unnoticed so far. A common choice for the size of the discreteness scale is the Planck scale. This is a ``natural scale'' determined by three dimensionful physical constants.

General relativity relies on two dimensionful quantities $G$, Newton's constant and $c$, the speed of light in a vacuum. Quantum mechanics relies on $\hbar$, the (reduced) Planck's constant. These physical constants take the values:
\begin{align}
G &= 6.674 28 \times 10^{-11} \textrm{ m}^3\, \textrm{kg}^{-1}\, \textrm{s}^{-2},\\
c &= 299,792,458 \textrm{ ms}^{-1},\\
\hbar &= 1.054571628(53) \times 10^{-34} \textrm{ kg m}^2\, \textrm{s}^{-1}.
\end{align}
They can be combined to form a length $P_l$, a time $P_t$ and a mass $P_m$:
\begin{align}
P_l&= \sqrt{\frac{G \hbar}{c^3}} = 1.616 252(81) \times 10^{-35} \textrm{ m},\\
P_t&= \sqrt{\frac{G \hbar}{c^5}}= 5.391 24(27) \times 10^{-44} \textrm{ s},\\
P_m&= \sqrt{\frac{\hbar c}{G}}= 2.176 44(11) \times 10^{-8} \textrm{ kg}.
\end{align}
Within the quantum gravity community the expectation is that any form of spacetime discreteness will become manifest at the Planck scale. To put it another way, it is expected that the smallest pieces of spacetime will have a spacetime volume around, say,  $P_t P_l^3$. This scale is \emph{extremely} small. As an example, suppose we imagine a world which contains one spacetime event in every $P_t P_l^3$ volume of spacetime. In this world the number of events in one cubic-metre-second (i.e. $1 \textrm{ s m}^3$) is around $4.4 \times 10^{147}$, an extremely large number.

\subsection{Technical problems with the continuum}

One of the main motivations for considering that spacetime might be discrete is that the continuum model has a number of technical deficiencies. These take the form of infinities which appear in quantum field theory and general relativity which we now briefly review.

\subsubsection{Quantum field theory}
\begin{quote}What we need and shall strive after is a change in the
fundamental concepts, analogous to the change in 1925 from Bohr to Heisenberg and
Schr\"odinger, which will sweep away the present difficulties automatically.

\flushright{Paul Dirac (1949) in \citet[p183]{Kragh:1992}}
\end{quote}

\begin{quote} I must say that I am very dissatisfied with the situation, because this so-called ``good theory'' does involve neglecting infinities which appear in its equations, neglecting them in an arbitrary way. This is just not sensible mathematics. Sensible mathematics involves neglecting a quantity when it is small -- not neglecting it just because it is infinitely great and you do not want it!
\flushright{Paul Dirac (1975) in \citet[p184]{Kragh:1992}}
\end{quote}

Many calculations in quantum field theory lead to divergent answers. These are typically in the form of divergent integrals or sums and their appearance held up the progress of particle physics for decades. The difficulties caused by these infinities were overcome in the 1940s and 50s with the development of renormalisation. This procedure recognises that the physical constants (such as mass and charge) present in the theoretical model (say quantum electrodynamics) are not the same as those that are measured experimentally. This is because the experimentally measured mass and charge differ from the parameters that enter the theory due to ongoing, ever-present particle interactions.

It was discovered that by re-expressing the theory in terms of the experimentally measured parameters finite results could be obtained. At the same time the bare parameters had be adjusted so that, when their values were renormalised, the measured values were obtained. The cost to this procedure is that, to obtain agreement in a continuum theory, the bare values must become infinite themselves.

The origin of the divergences (before they are renormalised away) can be traced back to the small-scale behaviour of the theory. This, in turn, depends on the model for spacetime that is being used at very small length-scales. We follow Dirac (as quoted above) in seeking a change in the fundamental concepts that will remove these divergences. It is possible that introducing discrete spacetime will achieve this.

\subsubsection{General relativity}

The gravitational equations of general relativity ensure that under physically realistic conditions the spacetime geometry will form a \emph{singularity}. This is the name used when a physical quantity (such as spacetime curvature) becomes infinite as well as for other, more subtle difficulties (see \citet[\chap 8]{Hawking:1973}). While the exact nature of the singularity may vary its existence indicates that the equations of general relativity have broken down---they no long provide physically sensible answers.

The singularities of general relativity may be a mathematical artifact of the theory or they may actually occur in spacetime. Either way the description of spacetime at a singularity is not accounted for by general relativity. If spacetime is modelled by a discrete structure then we can expect that the discreteness will tame the singular behaviour and either (i) ensure that no singularity occurs or (ii) provide a description of spacetime at the singularity.

\subsection{Conceptual problems with the continuum}

\begin{quote}
However painful its loss may be, by losing it [the continuum] we probably lose something that is very well worth losing.
\flushright{Erwin Schr\"odinger in \citet[p29-30]{Schrodinger:1951}}
\end{quote}

On a more philosophical note we briefly review some of the conceptual problems with a continuum spacetime.

\subsubsection{Volumes}
\begin{quotation}
If one is to accept the physical reality of the continuum, then one must accept that there are as many points in a volume of diameter $10^{-13}$cm, or $10^{-33}$cm, or $10^{-1000}$cm, as there are in the entire universe.
\flushright{Roger Penrose in \citet[p334]{Klauder:1972}}
\end{quotation}

If spacetime is a continuum then every spacetime region contains the same number of points. In a continuum spacetime we cannot count the number of points in a spacetime region to find its volume---every such count gives infinity. To get a notion of spatial or spacetime volume one must use a volume measure which assigns a real number to regions to represent their volume.
While perfectly well-defined mathematically, some measure theoretic results are physically absurd. One famous example is the Banach-Tarski paradox in which a sphere can be cut up into different pieces which can then be reassembled to have \emph{twice} the volume of the initial sphere!

\subsubsection{Construction of the continuum}
\begin{quote}
If the history of mathematics had developed differently, then we might, by now, have formed a very different view from the one now prevalent of the nature of space and time, and of many other physical concepts.

\flushright{Roger Penrose in \citet[p334]{Klauder:1972}}
\end{quote}

Another difficulty with the continuum is the non-uniqueness of its mathematical construction. The prototypical example of a continuum is the real numbers $\mathbb{R}$. This is constructed from the integers through the usual route of equivalence classes of Cauchy sequences of rational numbers. This route, however, leaves a number of jumping-off points from which we could construct something else. If our equivalence relation between the Cauchy sequences was different, for example, we could arrive at non-standard analysis (a branch of mathematics which rigorously allows infinitesimally small and infinitely great quantities).

To some extent the particular continuum that we have arrived at is the result of a historical accident. The physical motivation for the construction of the continuum certainly comes in large part from the appearance of the continuity of space. The mathematical process of working this into a rigorous framework has led to choices being made which are not directly physically motivated and, as such, may be the wrong ones if we are to model the actual space(time) that makes up our universe.

\chapter{Causal Sets} \label{chap:CausalSets}

\begin{quote}
So far this continuity [of spacetime] has been established for distances down to about $10^{-15}$cm by experiments on pion scattering.
Thus it may be that a manifold model for space-time is inappropriate for distances less than $10^{-15}$cm and that we should use theories in which space-time has some other structure on this scale.
\flushright{Stephen Hawking and George Ellis, (\citeyear[p57]{Hawking:1973}})
\end{quote}

Having discussed quantum gravity generally we now look in detail at causal set theory.
Non-technical introductions include \citet{Dowker:2005,Dowker:2006}. The term ``causal set'' was coined in \citet{Bombelli:1987} and good reviews include \citet{Sorkin:2003,Henson:2006}.

\section{Partial orders} \label{sec:PartialOrders} 

Central to the causal set program is the notion of a partial order. We have already glimpsed these objects in \secref{sec:CausalStructure} so it is worthwhile to once and for all define what we are talking about (for a comprehensive introduction see, for example, \citet[\chap 3]{Stanley:1986}).

Let $S$ be a set. A \emph{relation} $R$ on $S$ is a subset of $S \times S$, i.e. a relation is a set of ordered pairs of elements of $S$. If an ordered pair $(s,t)$ is in $R$ we write $sRt$. We will write $sRtRu$ to mean $sRt$ and $tRu$.

A \emph{partially ordered set} (or \emph{poset}) is a set $S$ together with a relation $R$ which is 
(i) reflexive\footnote{It is sometimes useful to define an \emph{irreflexive partially ordered set} to be a set $S$ with a relation $R$ which is transitive and irreflexive. $R$ is irreflexive if it is not reflexive: for all $s \in S$, we have $(s,s) \not\in R$.}: for all $s \in S$, $s Rs$; (ii) antisymmetric: for all $s,t \in S$,  $s Rt Rs$ implies $s = t$; and (iii) transitive: for all $s,t,u \in S$, $sRtRu$ implies $s R u$.

A typical example of a partially ordered set is the integers, ordered by the relation $\leq$ meaning ``is less than order equal to''. For example we have $1\leq1, -10\leq7, 1\leq4, 5\leq10$ etc.

Another example is the set of all subsets of a set $X$, ordered by the relation $\subseteq$ meaning ``is a subset of''. For example if $X = \{a,b,c\}$ then we have $\{ a \} \subseteq \{ a,b\}$, $\{a,c\}\subseteq\{ a,b,c\}$ etc.

\section{What is a causal set?} \label{sec:WhatIsACausalSet}

A \emph{causal set} (or \emph{causet}) is a locally finite partially ordered set. This means it is a pair
$(\CS,\preceq)$ with a set $\CS$ and a partial order relation $\preceq$ defined on $\CS$ that is
\begin{itemize}
\item Reflexive\footnote{We choose the \emph{reflexive} convention to most closely agree with the \emph{causal} (rather than chronological) relation in a causal Lorentzian manifold. In addition using a reflexive relation simplifies the definition of the incidence algebra (see \secref{Sec:IncidenceAlgebra}) since a single element of $\CS$ is regarded as a causal interval:  $\{u\} = [u,u]$.}: For all $u \in \CS$, $u \preceq u$,
\item Antisymmetric: For all $u,v \in \CS$,  $u \preceq v \preceq u$ implies $u = v$,
\item Transitive: For all $u,v,w \in \CS$, $u \preceq v \preceq w$ implies $u \preceq w$,
\item Locally finite: For all $u,v \in \CS$, $\left| \left[ u,v \right] \right| < \infty$.
\end{itemize}
The set $[u,v]:=\{w \in \CS | u \preceq w \preceq v\}$ is a \emph{causal interval} and $\left| A \right|$ denotes the cardinality of a set $A$. We write $u \prec v$ to mean $u \preceq v$ and $u \neq v$.

The set $\CS$ represents the set of spacetime events and the partial order $\preceq$ represents the causal relationship between pairs of events. If $x \preceq y$ we say ``$x$ precedes $y$'' or ``$x$ is to the causal past of $y$''. As can be seen from \secref{sec:CausalStructure} the first three conditions for $(\CS,\preceq)$ are in complete analogy with the causal relation on a causal Lorentzian manifold. The last condition---local finiteness---is where spacetime discreteness enters. 

Local finiteness ensures that any causal interval contains only a finite number of spacetime events. This, in turn, allows for a natural notion of volume for a causal set spacetime, the volume of a spacetime region (i.e. a subset of $\CS$) being simply the number of elements it contains. For causal intervals, at least, this volume is always finite and, up to a proportionality factor dependent on which units are used, can be identified as the physical volume of the region. In ``fundamental'' units\footnote{These are usually assumed to be the Planck units from \secref{sec:PlanckScale}.} the proportionality constant would be equal to one. In this case volume and number are measuring the same thing: ``Number = Volume''.

This situation is in stark contrast to a Lorentzian manifold in which every causal interval which is not a single point contains an uncountable infinity of points. To assign volumes to regions in such a manifold extra structure is required, e.g. a volume measure derived from a metric.

As we saw in \secref{sec:CausalStructure} the causal structure of a Lorentzian manifold determines the manifold up to a conformal factor in the metric. The only thing we were missing was a volume measure which would fix the conformal factor. On a causal set we have a causal ordering together with an implicit notion of volume. This strongly motivates the claim that these two pieces of complementary information are enough to specify the large-scale structure of spacetime: ``Order + Number = Geometry''.

In the most na\"ive framework we claim that spacetime is a single \emph{enormously} large causal set. In principle this causal set would have a ``manifold-like'' causal structure and contain so many elements that it could be well-approximated by a Lorentzian manifold. This would explain why physics based on Lorentzian manifolds works so well on large scales. 
More realistic frameworks (which give less prominence to a \emph{single} causal set) suggest that spacetime could be a classical ensemble of causal sets or even a large quantum superposition of many different causal sets (see \secref{sec:Dynamics}).

If spacetime is a single fixed causal set then how do we do physics on it? That question will form the basis for the majority of this thesis. Perhaps we can ask the question another way: If Newton had began his Principia with the phrase ``Let the universe be a causal set\ldots'' then what might physics look like now?

\subsection{Similar approaches}

We briefly outline a few similar models for spacetime that incorporate discreteness and causality at a fundamental level.

\subsubsection{Myrheim}

In \citet{Myrheim:1978} the causal structure of a Lorentzian manifold is given centre stage. Starting with a causal ordering and volume measure Myrheim constructs timelike geodesic lengths, coordinates and gravitational field equations (Sec 2). Applying these constructions to a discrete spacetime is then discussed (Sec 3) and it is emphasised that the constructions are only meaningful when applied to regions large enough that their volumes contain a statistically significant number of points. He concludes with a speculation that the dimension of spacetime might have a statistical basis.

\subsubsection{'t Hooft}

In \citet[Sec B]{'tHooft:1978} there is speculation that spacetime might be modelled by a lattice. Imposing requirements of ``general invariance'' lead to the use of chronological ordering to give structure to the lattice. This ordering is implicitly taken to be locally finite which allows discussion of timelike and spacelike distances in the lattice (Sec 10). To conclude he speculates that the dynamics of such a theory should be governed by a non-local action (Sec 12).

\subsubsection{Hemion} \label{Sec:HemionModel}

In \citet{Hemion:1980} a model for discrete spacetime is proposed based on a partially ordered set $W$. A form of discreteness is imposed by assuming that, for all $x,y\in W$, $| \{ z \in W | z \preceq x \textrm{ and } z \not\preceq y\} | < \infty$, a condition which he refers to as ``locally finite'' (this differs significantly from the locally finite condition of the current work). He proceeds to model particles in $W$ (see \secref{Sec:HemionParticleModel} of the current work) as well as discussing embedding $W$ into Minkowski spacetime. In Secs 7 and 8 attempts are made to model gravitational forces.

In \citet{Hemion:1988} the model is modified and now $W$ is locally finite in our sense (a condition which he calls ``discrete''). He focuses on modelling classical electrodynamics on the discrete spacetime using the action-at-a-distance formulation (see \secref{Sec:ActionAtADistance}). A key part of both his works is a notion of ``position'' in a partially ordered set \citep[Sec 3.6]{Hemion:1988}. Every element in $W$ defines a position but in general there are more positions.

\section{Correspondence with the continuum}

If spacetime is a causal set then we must explain why spacetime looks like a four-dimensional Lorentzian manifold on large-scales\footnote{Where ``large'' really means $10^{-15}$cm and up.}. Ultimately this task requires a theory of dynamics for causal sets---some physical rule or model which would explain why spacetime looks large and four-dimensional (and even a solution of Einsteins equations). 

Here we discuss how causal sets and continuum manifolds can be compared before reviewing ideas for causal set dynamics.

\subsection{Embeddings}

To compare causal sets to continuum Lorentzian manifolds we use the notion of an embedding.

An \emph{embedding} of a causal set $(\CS,\preceq)$ into a Lorentzian manifold $(M,g)$ is a map $f : \CS \to M$ which preserves the causal relations:
\be x \preceq y \textrm{ in $\CS$ } \, \iff \, f(x) \preceq f(y) \textrm{ in $M$}. \ee
This captures the idea that a causal set can be embedded into a Lorentzian manifold if their causal structures can be matched up.

A \emph{faithful embedding} of a causal set $(\CS,\preceq)$ into a Lorentzian manifold $(M,g)$ is an embedding such that the images of the causal set elements are uniformly distributed in $M$ according to the volume measure on $M$. Further we require that the characteristic scale over which the manifold's geometry varies is much larger than the embedding scale.

A fundamental conjecture of causal set theory is that if a causal set can be faithfully embedded into two Lorentzian manifolds $(M,g)$ and $(M',g')$ then the two manifolds are similar on large scales. This is called the \emph{Hauptvermutung}\footnote{This means ``main conjecture'' and originally referred to a 1908 conjecture in geometric topology. It may be a bad choice of name because the 1908 conjecture later turned out to be false!}. We restrict to ``similar on large scales'' rather than ``identical'' because a faithful embedding is blind to the structure of the manifold at scales smaller than the discreteness scale.

\subsection{Sprinklings}

Determining whether a causal set can faithfully embed into a given Lorentzian manifold is a difficult task. We can take a short-cut by constructing causal sets which automatically faithfully embed into Lorentzian manifolds. This is done through \emph{sprinkling}.

This involves generating a causal set by randomly placing points into a causal Lorentzian manifold. We place points according to a Poisson process with a sprinkling density $\rho$ such that the probability of placing $n$ points in a region of spacetime volume $V$ is 
\be \textrm{Prob($n$ points in volume $V$)} = \frac{(\rho V)^n}{n!} e^{-\rho V}.\ee
Here $\rho$ is a dimensionful parameter called the \emph{sprinkling density} which is a measure of the number of points placed in a unit volume. This process ensures that the expected number of points in a region of volume $V$ is $\rho V$. Having sprinkled the points we generate a causal set in which the elements are the sprinkled points and the causal relation is ``read-off'' from the manifold's causal relation restricted to the sprinkled points. We restrict to a \emph{causal} Lorentzian manifold because if the manifold had closed causal curves then the read-off causal relations might not be antisymmetric.

If the causal set has been generated by sprinkling into a manifold then, by definition, it can be faithfully embedded into the manifold.

\begin{figure}[h!]
\begin{center}
\includegraphics[width = 0.8\textwidth]{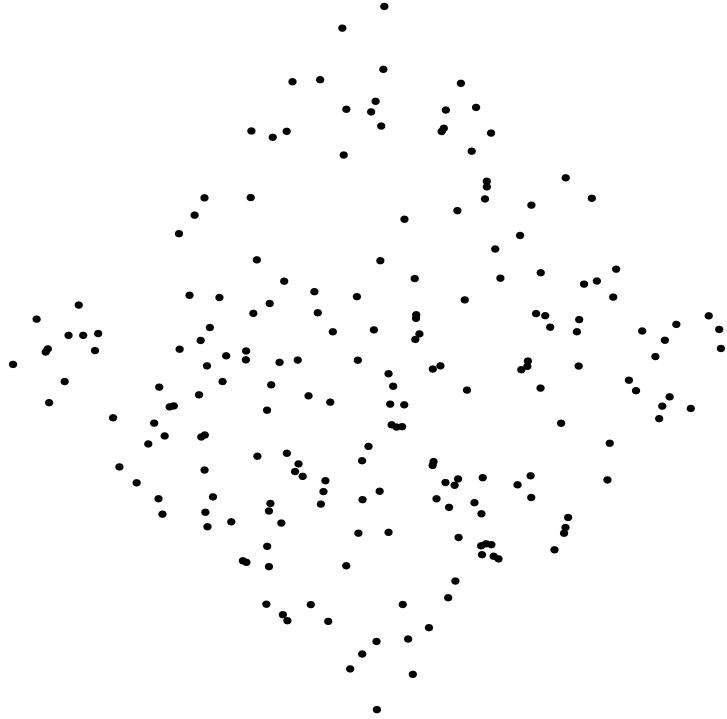}
\caption[A 100 element sprinkling]{An example distribution of 100 points sprinkled into a causal interval in 2-dimensional Minkowski spacetime.}
\label{fig:SprinkleExample}
\end{center}
\end{figure}

\subsubsection{Lorentz invariance} \label{sec:LorentzInvariance}

At first sight the sprinkling process seems unnecessarily complicated---why place points randomly? why use a Poisson distribution? The underlying reason for these choices is that we don't want the generated causal set to pick out a particular direction in spacetime.

To see this idea in action it is simplest to consider a sprinkling into Minkowski spacetime. Suppose we fix a coordinate frame and sprinkle our points with density $\rho$. If we now perform a Lorentz transformation on all the sprinkled points their coordinates will change but the \emph{distribution} of the transformed points will be statistically identical---it will still be a Poisson distribution. Essentially this is because the spacetime volume of a region is a Lorentz-invariant quantity and the sprinkling process depends only on such volumes. In particular the expected number of points in a region of volume $V$ is $\rho V$ both before and after we perform the Lorentz transformation.

Now suppose, on the other hand, we had fixed a frame and then placed points on a regular, say hyper-cubic, lattice. Applying a Lorentz transformation to these points gives a statistically \emph{different} distribution of points. If we applied a high boost to the lattice there will be regions with many points and regions with very few points---the points will no long be regularly distributed.

For this reason we use a random Poisson process when sprinkling. The statistical distribution of points is the same in all frames---we have ``statistical Lorentz invariance''. If we define a fundamental physical theory on such a sprinkled causal set and then look at the large-scale continuum theory that it gives rise to then the continuum theory will be Lorentz invariant.

Unfortunately this Lorentz invariance comes at a cost. For a sprinkling into infinite Minkowski spacetime each causal set element will have an infinite number of ``nearest neighbours''\footnote{In a causal set two elements $u$ and $v$ are ``nearest neighbours'', or \emph{linked}, if there is no other element causally between them---see \secref{sec:Definitions}.}. This in stark contrast to a hyper-cubic lattice in which every point has only a finite number of nearest neighbours.

The appearance of this infinity was one of the first objections to the causal set proposal \citep{Moore:1988}. It was suggested that any physical theory on a causal set would ``have the nasty property that every point is influenced by an infinity of `nearest neighbours' which, in a given frame, are arbitrarily far back in time.''
This may have relevance to the current work because, as we shall see, the work of \chapref{chap:PathIntegrals} is defined for causal sets of any cardinality whereas the work of \chapref{chap:FreeQFT} and \chapref{chap:InteractingQFT} holds only for finite causal sets.

\subsection{Dynamics} \label{sec:Dynamics}

Having settled on a kinematical model spacetime---i.e. ``spacetime is a causal set''---there remains the difficult problem of defining a dynamics for causal sets. We now review some ideas in this area.

\subsubsection{Sequential Growth Models}

One of the main ideas being explored for a dynamics of causal sets is a sequential growth model in which causal sets are built one element at a time. This has been realised concretely in a classical growth model based upon probabilities \citep{Rideout:1999,Rideout:2000}.

In these models probabilities are assigned to different ways a new element can be added to the existing causal set. By imposing physically motivated requirements on the probability assignments all possible probability rules can be parametrised. By building causal sets in this way probabilities are assigned to partial finite causal sets or, by running the process to infinity, to countably infinite causal sets.

These classical sequential growth models were envisaged as a ``classical warm-up'' to be replaced one day by a quantum mechanical model. The most na\"ive attempt to do so would replace the real probabilities with \emph{complex} probability amplitudes. This would allow interference between the amplitudes which, it is hoped, would allow large manifold-like causal sets to emerge (this is considered in \citet[\S4]{Dowker:2010a}).

Surprisingly, given its interim status, there is evidence that one type of classical sequential growth model generates causal sets which are similar to 4-dimensional de Sitter spacetime \citep{Ahmed:2009}.

\subsubsection{Einstein-Hilbert action and Ricci Scalar} \label{sec:EinsteinHilbert}

More recent work has attempted to find an action principle for causal sets. The idea is to find a causal set analog of the 4-dimensional Einstein-Hilbert action. This has been achieved by \citet{Benincasa:2010}. The causal set action is a combinatorial sum of weights assigned to sub-structures in the causal set.

The expression for the action for a causal set $\CS$ is \citep[eq (14)]{Benincasa:2010}:
\be S^{(4)}[\CS] = N - N_1 + 9 N_2 - 16 N_3 + 8 N_4, \ee
where $N$ is the number of elements in $\CS$ and $N_i$ is the number of $(i+1)$-element causal intervals in $\CS$.

While further work is needed on this approach it is hoped that this action (presumably used in a quantum mechanical sum-over-histories formalism) would enforce the causal sets to be large, four-dimensional and even solutions to Einstein's equations.

\section{Definitions} \label{sec:Definitions}

It is worthwhile to gather together some definitions and notation which will be used extensively later on in this work (we have already seen some of these definitions, e.g. that of a causal interval). Good references include \citet[\chap 3]{Stanley:1986}.

If $u \preceq v$ and $u \neq v$ we write $u \prec v$ (we shall refer to this irreflexive relation $\prec$ as a \emph{strict} causal relation). If $u, v \in \CS$ are unrelated (meaning $u \not\preceq v$ and $v \not\preceq u$) we write $u || v$.

A pair $(Q,\preceq')$ is a \emph{sub-poset} of a poset $(P,\preceq)$ if $Q$ is a subset of $P$ and the order relation $\preceq'$ is equal to $\preceq$ when restricted to elements in $Q$: i.e. $u \preceq' v \iff u \preceq v$ for all $u,v \in Q$.

An \emph{interval} (or \emph{causal interval} or \emph{Alexandrov set}) is the set $[u,v]:=\{w \in \CS : u \preceq w \preceq v\}$ whenever $u \prec v$. Note that with this convention the ``end-points'' of the interval are included: $u, v \in [u,v]$. Also $[u,u] = \{ u \}$.

A \emph{labelling} assigns an integer subscript to the elements of $\CS$: labelling them as $v_x$ for $x=1,\ldots,|\CS|$. A \emph{natural labelling} is a labelling such that $v_x \preceq v_y \implies x \leq y$. 

A \emph{total order} (or \emph{totally ordered set} or \emph{chain}) is a partial order in which any two elements are related---that is, there are no unrelated pairs of elements. One familiar example of a total order is ($\mathbb{Z},\le$), i.e. the integers under the usual ``less than or equal to'' order relation.

A subset of a causal set $(\CS,\preceq)$ is a \emph{chain} if it is totally ordered when regarded as a sub-poset of $(\CS,\preceq)$. A finite chain of length $n$ is a sequence of distinct elements $u_0 \prec u_1 \prec u_2 \prec \ldots \prec u_n$.
A \emph{multichain} is a chain with repeated elements. A multichain of length $n$ is a sequence of (not necessarily distinct) elements $u_0 \preceq u_1 \preceq u_2 \preceq \ldots \preceq u_n$.
An \emph{antichain} is a set of elements which are mutually unrelated.

A \emph{link} (or \emph{covering relation} or \emph{nearest neighbour relation}) is a relation $u \prec v$ such that there
exists no $w \in \CS$ with $u \prec w \prec v$. We say $u$ and $v$ are \emph{nearest neighbours} (or $v$
\emph{covers} $u$) and write $u \link v$.
A \emph{path} is a subset $P \subset \CS$ which is a maximal (or saturated) chain. This means $P$ is a chain with
no element $w \in C - P$ such that $u \prec w \prec v$ for some $u, v \in P$. A finite path of length $n$ is a sequence of distinct elements $u_0 \link u_1 \link u_2 \link \ldots \link u_n$.

A \emph{linear extension} of a causal set $(\CS,\preceq)$ is a total order $(\CS, \leq)$ which is consistent with the partial order. This means $u \preceq v \implies u \leq v$ for all $u,v \in \CS$.

The \emph{direct} (or \emph{Cartesian}) \emph{product} of two partial orders $(\CS_1,\preceq_1)$ and $(\CS_2,\preceq_2)$ is the partial order $(\CS_1 \times \CS_2,\preceq)$ based on the Cartesian product of the sets $\CS_1$ and $\CS_2$ with order relation $(u_1,u_2) \preceq (v_1,v_2) \iff u_1 \preceq_1 v_1 \textrm{ and } u_2 \preceq_2 v_2$.

A set of elements $\{v_1, \ldots, v_n\}$ will be called \emph{non-Hegelian} if they are mutually unrelated and have identical (strict) causal relations to the rest of the causal set. That is $(u \prec v_x \prec w) \iff (u \prec v_y \prec w)$ for all $x, y = 1,\ldots, n$.

\section{Representing causal sets}

A finite causal set $(\CS, \preceq)$ can be represented in a number of equivalent ways. One way it to simply list the elements of $\CS$ together with the order relations. An example causal set we shall use to illustrate the different representations is:
\begin{equation} \label{eq:CausetExample1} \mathcal{C} = \{v_1,v_2,v_3,v_4,v_5,v_6\},\end{equation}
\begin{equation} \label{eq:CausetExample2} \begin{split}
& v_1 \preceq v_1,v_1 \preceq v_2,v_1 \preceq v_3,v_1 \preceq v_4, v_1 \preceq v_5, v_1 \preceq v_6 \\
& v_2 \preceq v_2, v_2 \preceq v_3,v_2 \preceq v_5, v_2 \preceq v_6 \\
& v_3 \preceq v_3, v_3 \preceq v_5, v_3 \preceq v_6\\
& v_4 \preceq v_4, v_4 \preceq v_5, \\
& v_5 \preceq v_5,\\
& v_6 \preceq v_6.
\end{split} \end{equation}
One can verify almost mechanically that this pair $(\CS,\preceq)$ satisfy the four conditions of a causal set.

\subsection{Hasse diagrams and directed graphs}

\begin{figure}[h!]
\begin{center}
\includegraphics[width = 0.5\textwidth]{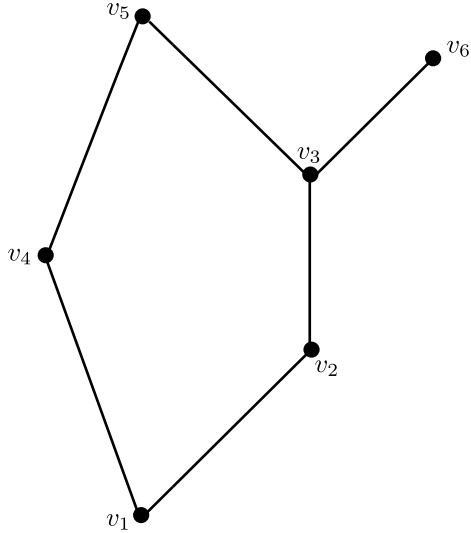}
\caption{The example causal set drawn as a Hasse diagram.}
\label{fig:ExampleHasse}
\end{center}
\end{figure}

Another way to represent a finite causal set is by a \emph{Hasse diagram}. Here the elements of $\CS$ are drawn as points in the page and if two elements are linked $u \link v$ then $u$ is positioned lower than $v$ with a line drawn connecting them (see \figref{fig:ExampleHasse}).

\begin{figure}[h!]
\begin{center}
\includegraphics[width = 0.5\textwidth]{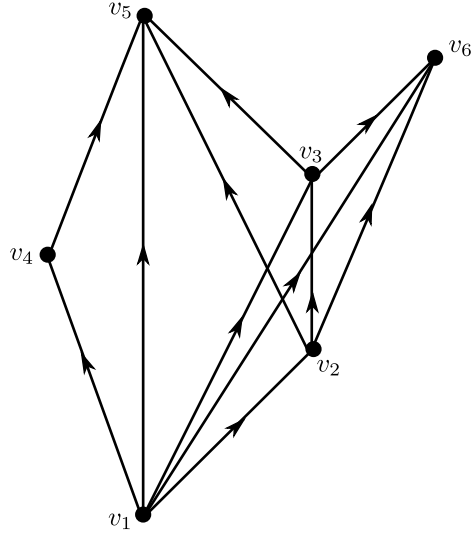}
\caption{The example causal set drawn as a directed graph.}
\label{fig:ExampleDigraph}
\end{center}
\end{figure}

A similar, but ultimately more cluttered, approach is to draw the causal set as a \emph{directed graph}. Here elements are again drawn as points in the page but all relations (not just links) between distinct elements are drawn in. A relation $u \prec v$ is drawn as a directed line from $u$ to $v$ (see \figref{fig:ExampleDigraph}).

We can illustrate some of the definitions from the previous section within $(\CS,\preceq)$. As an example, the relation $v_1 \prec v_4$ is a link (i.e. $v_1 \link v_4$) but the relation $v_1 \prec v_5$ is not a link (because there are elements causally between $v_1$ and $v_5$, e.g. $v_1 \prec v_4 \prec v_5$). The sequence $v_1 \prec v_3 \prec v_5$ is a chain but not a path (because the relation $v_1 \prec v_3$ is not a link). The causal interval $[v_1, v_6]$ is equal to $[v_1,v_6] = \{v_1,v_2, v_3, v_6 \}$.

\subsection{Adjacency matrices}

As far as this work is concerned the most useful way to represent a finite causal set is by its adjacency matrix.
If $\CS$ contains $p$ elements we can label them $v_1, v_2, \ldots v_p$ and define two $p\times p$ matrices.

The \emph{causal matrix} is defined by 
\be C_{xy} := \left\{ \begin{array}{ll} 1 & \textrm{ if } v_x \prec v_y \\ 0 & \textrm{otherwise.} \end{array} \right. \ee 

The \emph{link} matrix is defined by 
\be L_{xy} := \left\{ \begin{array}{ll} 1 & \textrm{ if } v_x \link v_y \\ 0 & \textrm{otherwise.} \end{array} \right. \ee
Here $x,y = 1,2,\ldots,p$ index positions in the matrices. $C$ and $L$ are both zero on the main diagonal and, from the definition of a causal set, the labelling can always be chosen to ensure they are both strictly upper triangular matrices (in which case the labelling is a \emph{natural labelling}).

One way to calculate $L$ from $C$ is to compute
\be L = C - (C^2 > 0),\ee
where, for a real matrix $A$, $(A>0)_{xy} = 1$ if $A_{xy} > 0$ and $0$ otherwise.

If $(\CS_1,\preceq_1)$ and $(\CS_2,\preceq_2)$ are two finite causal sets with causal matrices $C_1$ and $C_2$, the direct product $(\CS_1 \times \CS_2,\preceq)$ (see \secref{sec:Definitions}) has causal matrix 
\be (I+C_1) \otimes (I+C_2) - I \otimes I = C_1 \otimes C_2 + C_1 \otimes I+ I \otimes C_2.\ee
where $\otimes$ denotes the Kronecker product.

\subsubsection{Example}

For our example causal set $(\CS,\preceq)$ (defined in \eqref{eq:CausetExample1} and \eqref{eq:CausetExample2}) we have:
\be C = \left(
\begin{array}{cccccc}
  0 & 1 & 1 & 1 & 1 & 1\\
  0 & 0 & 1 & 0 & 1 & 1 \\
  0 & 0 & 0 & 0 & 1 & 1\\
  0 & 0 & 0 & 0 & 1 & 0\\
  0 & 0 & 0 & 0 & 0 & 0\\
  0 & 0 & 0 & 0 & 0 & 0\\
\end{array}
\right),
\qquad
L = \left(
\begin{array}{cccccc}
  0 & 1 & 0 & 1 & 0 & 0\\
  0 & 0 & 1 & 0 & 0 & 0\\
  0 & 0 & 0 & 0 & 1 & 1\\
  0 & 0 & 0 & 0 & 1 & 0 \\
  0 & 0 & 0 & 0 & 0 & 0\\
  0 & 0 & 0 & 0 & 0 & 0\\
\end{array}
\right).\ee

\subsubsection{Chain lengths}

Powers of these matrices have the following useful properties \citep[p115]{Stanley:1986}: 
\begin{align}
\left(C^n\right)_{xy} &= \textrm{The number of chains of length $n$ from $v_x$ to $v_y$,}\\ 
\left(L^n\right)_{xy} &= \textrm{The number of paths of length $n$ from $v_x$ to $v_y$,}\\
\left((I + C)^n\right)_{xy} &= \textrm{The number of multichains of length $n$ from $v_x$ to $v_y$,}
\end{align}
(where $I$ is the $p\times p$ identity matrix).

We can can appreciate why these formulae hold by considering the $C^2$ example. Here we have 
\be (C^2)_{xy}:=\sum_{a=1}^p C_{xa} C_{ay}. \ee
For each $a=1,\ldots,p$ the summand is only non-zero if there exists a chain $v_x \prec v_a \prec v_y$. When such a chain exists the summand is equal to 1. Thus the sum is equal to the number of chains of length 2 from $v_x$ to $v_y$. A similar argument applies for other powers of $C$ and $L$.

For finite causal sets both $C$ and $L$ are nilpotent matrices (meaning that raising them to a high enough power gives the zero matrix). This means that power series in $C$ and $L$ truncate. For example for a complex number $z$ we have
\be D(z) := (I - z C)^{-1} = I + z C + (z C)^2 + \ldots = \sum_{n=0}^\infty (z C)^n, \ee
\be E(z) := \exp(z C) = I + z C + \frac{(zC)^2}{2} + \ldots = \sum_{n=0}^\infty \frac{(z C)^n}{n!}. \ee
where $\exp(zC)$ is the matrix exponential of $zC$ and both power series truncate (eventually).

We see that $D(z)_{xy}$ and $E(z)_{xy}$ are polynomials in $z$ with degree equal to the length of the longest chain from $v_x$ to $v_y$. One way to calculate this degree is to make use the following easily verified formula for the degree of a polynomial $P(z)$:
\be \deg(P) = \lim_{z \to \infty} \frac{z P'(z)}{P(z)}, \ee
where $P'$ is the derivative of $P$.

Using this the length of the longest chain from $v_x$ to $v_y$ is equal to
\be \lim_{z \to \infty} \frac{z D(z)'_{xy}}{D(z)_{xy}} = \lim_{z \to \infty} \frac{z (C D(z)^2)_{xy}}{D(z)_{xy}} = \lim_{z \to \infty} \frac{z E(z)'_{xy}}{E(z)_{xy}} = \lim_{z \to \infty} \frac{z (C E(z))_{xy}}{E(z)_{xy}}, \ee
where we've used $D(z)' = C D(z)^2$ and $E(z)' = C E(z)$.

In numerical simulations substituting a large, but finite, value of $z$ gives a good approximation to the length of the longest chain.

We mention an interesting result for calculating the \emph{total} number of chains (or paths) of different lengths in a finite causal set. We have \citep{Stanley:1996}, for a finite causal set $(\CS,\preceq)$ with causal matrix $C$ and link matrix $L$:
\begin{quote}The coefficient of $z^n$ in $\det(I + z(J-C))$ (resp. $\det(I + z(J-L))$) is the total number of chains (resp. paths) of length $n$ in $\CS$.
\end{quote}
Here $J$ is the ``all ones'' matrix: $J_{xy} = 1$ for $x,y=1,\ldots,p$.

\subsubsection{Volumes}

Squaring $I+C$ or $C$ can be used to compute the cardinality of the causal intervals in the causal set. We have:
\begin{align}
|[v_x,v_y] | = \left| \left\{ w \in \CS | v_x \preceq w \preceq v_y \right\} \right|&= ((I+C)^2)_{xy},\\
\left| \left\{ w \in \CS | v_x \prec w \prec v_y \right\} \right| &= (C^2)_{xy}.
\end{align}

\def\V0{V_0}

These cardinalities are just dimensionless numbers. If we assign a fundamental spacetime volume\footnote{Which, presumably, should have a mass-dimension of $[\V0] = M^{-d}$ if the causal set corresponds to a $d$-dimensional spacetime.} $\V0$ to all causal set elements then the spacetime volume of a region with $N$ elements is equal to $N \V0$.

If the volume of the causal set elements are not all equal\footnote{We mention this for completeness although it goes against the spirit of the causal set approach in which volume simply \emph{is} number.} (e.g. if one element is regarded as having twice the spacetime volume of another, say) then we can define a diagonal matrix $V$ such that $V_{xx}$ is the volume of element $v_x$. We then have 
\be \textrm{Vol}([v_x,v_y]) = ((I+C)V(I+C))_{xy}. \ee

\chapter{Path Integrals} \label{chap:PathIntegrals}

\begin{quote}
An electron has an amplitude to go from point to point in space-time, which I will call ``$E(A$ to $B)$.'' Although I will represent $E(A$ to $B)$ as a straight line between two points, we can think of it as the sum of many amplitudes --- among them, the amplitude for the electron to change direction at points $C$ or $C'$ on a ``two-hop'' path, and the amplitude to change direction at $D$ and $E$ on a ``three-hop'' path---in addition to the direct path from $A$ to $B$. The number of times an electron can change direction is anywhere from zero to infinity, and the points at which the electron can change direction on its way from $A$ to $B$ in space-time are infinite. All are included in $E(A$ to $B)$.
\flushright{Richard Feynman, (\citeyear[p92]{Feynman:1985}})
\end{quote}

In this chapter we describe an approach to modelling particles on a causal set. The main results are a collection of models for defining discrete path integrals on a causal set. These lead to propagators for particles on a causal set which, in a suitable sense, match the continuum propagators when the causal set is generated by sprinkling into Minkowski spacetime.

The main results of this chapter appear in \citet{Johnston:2008,Johnston:2009:DICE}. One of the models also appears in a Wolfram demonstration \citep{JohnstonDemo}.

\section{Particle models}

The question of how to model matter on a causal set has been addressed by a number of people. As a result there are a number of different models, each using different physical and mathematical ideas. Here we review the models which treat matter as particles rather than fields (fields will be addressed in \chapref{chap:FreeQFT}).

\subsection{Swerves}

In general relativity free classical point particles follow geodesics in spacetime. It is tempting to use this geometrical rule to model free point particles on a causal set. The most studied models of this type are the \emph{swerves models}. The first of these was proposed by \citet{Dowker:2003} to model a massive point particle.

The idea behind this approach is to find a Markovian propagation rule that determines the worldline trajectory of a point particle.  The trajectory is built up iteratively, one element at a time, with the next element being chosen by applying the rule. The rule depends on the structure of the causal set as well as the past trajectory of the particle (either the entire past trajectory or the past trajectory only up to some finite ``forgetting time'').

The aim is to find a rule such that the large-scale behaviour of a massive point particle when propagating on a sprinkled causal set is to follow a timelike geodesic. Of course, for a particular sprinkled causal set, this will not be possible exactly due to the random distribution of sprinkled points. In general the particle will be forced to \emph{swerve} slightly as it attempts to hug a geodesic as closely as possible. It is this swerving behaviour that gives the approach its name and would provide a clear signal of underlying spacetime discreteness.

A number of swerves models have been developed which give the correct large-scale geodesic behaviour (see \citet{Philpott:2009a,Philpott:2009b,Philpott:2010} for full details). 

The phenomenology of these models has focussed on their effect on particle propagation when the continuum limit is taken. It turns out that, so long as the underlying propagation rule is Lorentz-invariant, the continuum behaviour is described by a Lorentz-invariant diffusion equation depending on the mass of the particle and a ``diffusion parameter''. By comparing this continuum description to experiments and observations the size of the diffusion parameter can be constrained.

This continuum description can be modified to describe massless particles \citep[Sec IV]{Philpott:2009a} as well as massless particles with polarisation \citep{Contaldi:2010} (e.g. photons from the cosmic microwave background). In these massless cases, however, there is, at yet, no underlying causal set propagation rule that leads to the continuum description.

\subsection{Hemion classical model} \label{Sec:HemionParticleModel}

In the Hemion model for discrete spacetime (\citet[Sec 2]{Hemion:1980}, \citet[Sec 3.4]{Hemion:1988}) particles are modelled as infinite chains---infinite totally ordered sets of spacetime points which represent the particle's worldline. 
The particles in this model are classical and cannot be created or destroyed. Nevertheless Hemion acknowledges that future developments could lead to a model that includes quantum behaviour such as ``the phenomena of creation and annihilation, vacuum loops, and in general all the particle structures normally considered in Feynman diagrams'' (p1181).

To accommodate these classical particles it is assumed that the poset $W$ has ``particle structure'', meaning it is the disjoint union of totally ordered subsets. The elements of $W$ then make up the particle worldlines and points in empty space are ``positions'' in $W$ \citep[Sec 3.6]{Hemion:1988}.

In \citet{Hemion:1988} he focuses on defining an interacting point particle model for classical electrodynamics. This is done within the Fokker action-at-a-distance framework for classical electrodynamics (which we shall return to in \secref{Sec:ActionAtADistance}).

\subsection{Discrete path integrals}

The two models just described suffer the serious draw-back that they are entirely classical. This is problematic when seeking a fundamental model for matter at ultra-small length scales because on such scales the effects of quantum mechanics are very important. On small scales it is simple incorrect to model matter as a collection of classical point particles with precisely defined, unique spacetime worldlines\footnote{It remains possible, however, that quantum mechanics emerges from a deeper deterministic theory (see, e.g. \citet{'tHooft:2006}). In this case, however, the deterministic objects in the deeper theory would not correspond to the particles in the emergent quantum mechanics.}.

A better approach is to base the particle model on quantum mechanics from the outset. The work presented here does just this by basing the model on the path-integral formulation of quantum mechanics.

The path integral approach was initiated by \citet{Dirac:1933} and ultimately brought to completion by \citet{Feynman:1948:NonRelQM}. A full introduction is given in \citet{FeynmanHibbs} and a very readable introduction for the lay-person is given in \citet{Feynman:1985}.

In this formulation the ``propagator'' for a point particle is given centre stage. This is a complex valued function of two spacetime points which describes the quantum mechanical propagation of the particle. In the path integral formalism it is obtained as a quantum mechanical path integral. Complex probability amplitudes are assigned to all possible trajectories that the particle can take and, by summing these amplitudes up over all trajectories, the propagator is obtained.

That, in principle, is the idea. When it comes to calculating this path integral for, say, a non-relativistic particle in the continuum, there are unwelcome mathematical difficulties which arise. Since the amplitudes are complex numbers the path integral cannot be defined as a bone fid\'e integral over a space of paths. Instead the path ``integral'' is really a prescription for 1) discretising the paths, 2) performing the path integral over skeletonised paths and 3) the taking the continuum limit.

If spacetime is not a continuum then perhaps these mathematical difficulties can be overcome, perhaps the sum over particle paths can be defined directly and rigorously. Indeed, if spacetime is modelled as a causal set then, as we shall see, this is precisely what happens.

We mention that path integrals on discrete spacetime have been considered before. In \citet{Gudder:1988}, for example, path integrals on a hyper-cubic lattice were described. The idea of path integrals on causal sets was considered by \citet{Meyer:1997} but unfortunately never developed. Some work with propagators on a causal set has been done by \citet{Daughton:1993, Salgado:2008, Sorkin:2009} (see \secref{sec:InvertingGreens} for a summary).
Discrete path integral models have also been inspired by the Feynman checkerboard model (see \secref{sec:Checkerboard} for details).

As it turns out, a very similar approach to the one described in this chapter was developed independently as a summer project at the University of San Jos\'e (the results of which appeared in \citet{Scargle:2009}). Their aim was to develop a model for the effect of spacetime discreteness on photon energy. The result was a path integral model for photons which took the ``sum over \emph{all} paths'' spirit of the formulation literally---the paths summed over in their model include future-directed, past-directed and spacelike paths.

\section{Causal set path integrals} \label{Sec:PathIntegralModels}

To define path integrals on a causal set we have to make two choices: which trajectories to sum over and what
amplitudes to assign to each trajectory. The two most obvious choices for trajectories are all chains between two elements or all paths between two elements.

Since the causal set is locally finite there are only a finite number of chains or paths between any two elements. Assigning each one an appropriate amplitude we can then simply sum the amplitudes to obtain the propagator. For every pair of elements this sum will exist since we are just summing a finite collection of complex numbers. We must then attempt to choose amplitudes which give us the correct propagator.

We shall present a number of different models, each suitable for obtaining a different continuum propagator when the causal set is generated by sprinkling into Minkowski spacetime of different dimensions. In all the models the amplitude assigned to each trajectory depends on the length of the trajectory.

\subsection{The models}

We can picture a point particle travelling along a chain or path sequentially from one element to another. The progress along the trajectory can be broken down into `hops' from one element to the next as well as `stops' at each element of the trajectory (the initial and final elements are not regarded as stops). For a chain or path of length $n$ there are $n$ hops and $n-1$ stops. The amplitude for the whole trajectory is then the product of the amplitudes for each hop and each stop it contains. This talk of hops and stops is an echo of \citet[footnote 3, p91]{Feynman:1985} as well as the quote at the start of this chapter.

We first consider the simplest case where the hop and stop amplitudes are \emph{constant}. If $a$ is the amplitude for the particle to hop once along the trajectory and $b$ is the amplitude for the particle to stop once then the amplitude for a chain or path of length $n$ (so there are $n$ hops and $n-1$ intermediate stops) is the product $a^n b^{n-1}$.

We begin by working with a fixed finite causal set $(\CS,\preceq)$ with $p$ elements. How the ideas extend to arbitrary causal sets is presented in \secref{Sec:GeneralCausalSets}.

To start we define a $p \times p$ matrix $\Phi$.
If we sum over chains we define 
\be \Phi:= a C. \ee 
where $C$ is the causal matrix for $\CS$.
If we sum over paths we define 
\be \Phi:= a L. \ee 
where $L$ is the link matrix for $\CS$.
The total amplitude to go from $v_x$ to $v_y$ along a trajectory of any length is then $K_{xy}$ where $K$ is the $p
\times p$ matrix 
\be \label{eq:CausetPropagator} K:= \Phi + b\Phi^2 + b^2\Phi^3 + \ldots = \sum_{n=1}^\infty b^{n-1} \Phi^n,\ee 
where matrix multiplication is used to compute each $n\th$ power of $\Phi$.

\begin{figure}[!h]
\begin{center}
\includegraphics[width = \textwidth]{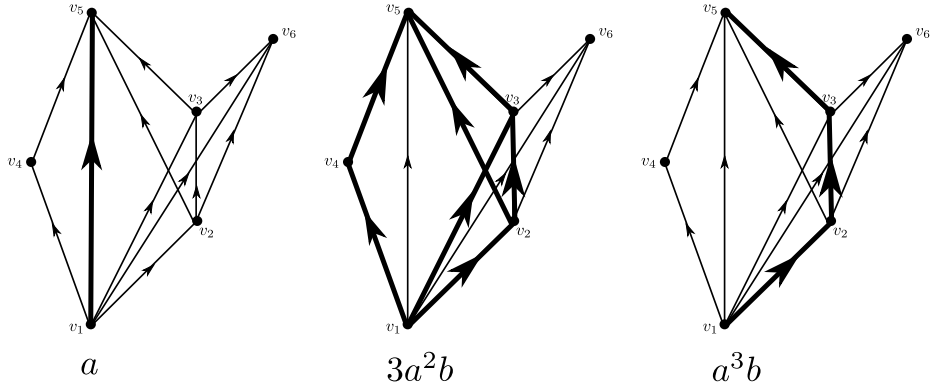}
\caption[Path integral for the example causal set]{The amplitude $K_{15} = a + 3 a^2 b + a^3 b^2$ when summing over chains in the example causal set $(\CS,\preceq)$.}
\label{fig:ExamplePathInt}
\end{center}
\end{figure}

Each term in this sum is the contribution to the total amplitude from chains or paths of a particular length (see \figref{fig:ExamplePathInt}). As
an example, \be b(\Phi^2)_{xy} = \sum_{z=1}^p b \Phi_{xz} \Phi_{zy}, \ee is the sum (over all intermediate
positions $v_z$) of the amplitudes for all length two trajectories $v_x$ to $v_z$ to $v_y$ (compare this to \citet[eq 5]{Feynman:1948:NonRelQM}).

Since the causal set is finite and the trajectories move only forwards in time the sum in
\eqref{eq:CausetPropagator} terminates and we have \be K = \Phi(I - b \Phi)^{-1}, \ee where $I$ is the $p\times p$ identity matrix. This matrix inverse is simple to perform if the causal set has been labelled to ensure $\Phi$ is strictly upper triangular. In this case the upper triangular matrix $I - b\Phi$ can be inverted using elementary row operations.

The question we now face is whether there are values of $a$ and $b$ such that $K_{xy}$, for a causal set
generated by a sprinkling into Minkowski spacetime, is approximately equal to an appropriate continuum propagator.

We first observe that the particles we are modelling do not have any internal properties. Therefore we expect the continuum propagator will be a propagator for a scalar particle. It's also clear that the propagator value $K_{xy}$ is zero unless $v_x \preceq v_y$ (later we'll denote the $K$ matrix as $K_R$ to emphasise this). The requirement that the propagator be zero outside the future lightcone suggests we try to compare the causal set propagator with the \emph{retarded propagator} for the Klein-Gordon equation.

\section{Propagators in the continuum} \label{sec:KGPropagators}

The propagators in $d$-dimensional Minkowski spacetime scalar quantum field theory are Green's functions of the
Klein-Gordon equation:
\be \label{eq:GreenKleinGordon} (\Box + m^2)G^{(d)}_m(x) = \delta^d(x).\ee 
Here $x = (x^0,\vec{x})$, $m$ is the mass of the particle, $\delta^d$ is the $d$-dimensional Dirac delta function and we choose units with $\hbar = c =1$. The d'Alembertian is given by \be \Box:= \frac{\partial^2}{\partial {x^0}^2} - \frac{\partial^2}{\partial
{x^1}^2} - \frac{\partial^2}{\partial {x^2}^2} - \ldots - \frac{\partial^2}{\partial {x^{d-1}}^2}. \ee 
We can think of $G_m^{(d)}(x)$ as the propagation amplitude for the particle to travel from the origin to $x \in \mink^d$. More generally, using the translation-invariance of $\mink^d$, $G_m^{(d)}(y-x)$ is the propagation amplitude for the particle to travel \emph{from} $x$ \emph{to} $y$ (in curved spacetimes the propagator depends on $x$ and $y$ separately, see \secref{sec:CurvedSpacetimes}). 

To obtain the propagator $G^{(d)}_m(x)$ explicitly we define the Fourier transform by \be \widetilde{f}(p) := \int d^dx
f(x) e^{i p x}, \qquad f(x) = \frac{1}{(2 \pi)^d}\int d^d p \widetilde{f}(p) e^{-ip x}, \ee where $p x := p_0
x^0 - \vec{p}\cdot \vec{x}$. Using this we can solve equation \eqref{eq:GreenKleinGordon} to obtain 
\be \label{eq:PropMomSpace}
\widetilde{G}^{(d)}_m(p) := -\frac{1}{p_0^2 - \vec{p}^2 - m^2},\ee so \be G^{(d)}_m(x) := -\frac{1}{(2 \pi)^d}
\int d^d p \frac{e^{-ipx}}{p_0^2 - \vec{p}^2 - m^2}. \ee The mass-dimension of this propagator is
$[G^{(d)}_m] = M^{d-2}$.

The integrand in these expressions contains poles so the Minkowski spacetime propagator is only uniquely
defined if we specify a contour of integration or, equivalently, boundary conditions for the solution of
\eqref{eq:GreenKleinGordon}. 
The \emph{retarded propagator}, which we'll denote $\GR_m^{(d)}(x)$, is the unique Green's function which is only non-zero in the future lightcone---that is, $\GR_m^{(d)}(y-x)$ is zero unless $x \preceq y$.

This boundary condition can be imposed by avoiding the poles in \eqref{eq:PropMomSpace} at $p_0 = \pm \sqrt{\vec{p}^2 + m^2}$ by two small semi-circles in the upper-half $p_0$ complex plane. This is equivalent to \be \label{eq:GRDef} \GR^{(d)}_m(x) := \lim_{\epsilon \to 0^+} -\frac{1}{(2 \pi)^d} \int d^d p
\frac{e^{-ipx}}{(p_0+i \epsilon)^2 - \vec{p}^2 - m^2}. \ee 

This integral can be evaluated in various dimensions. The rigorous calculations require the use of the theory of distributions (See \citet{Gelfand:1964} for the general theory and \citet{DeJager:1967} for the theory applied in detail to the $d=4$ Klein-Gordon equation).

We list the expressions\footnote{$(G_R)_m^{(1)}$ is obtained by a straightforward calculation from \eqref{eq:GRDef}. The expressions for $(G_R)_m^{(d)}$ with $d=2,3,4$ are given in \citet[Example 2.3, p144-145]{PDE:1994}. The physical 4-dimensional case is covered in great detail in \citet[\S 15]{Bogoliubov:1959} and \citet{DeJager:1967}.} for dimensions $d=1,2,3,4$:
\begin{align}
\label{eq:KR1d}\GR_m^{(1)}(x)&=\theta(x^0) \frac{\sin(m x^0)}{m} &\\
\label{eq:KR2d}\GR_m^{(2)}(x)&=\theta(x^0) \theta(\tau^2) \frac{1}{2} J_0(m \tau) &\\
\GR_m^{(3)}(x)&=\theta(x^0)\theta(\tau^2) \frac{1}{2 \pi}\frac{\cos\left(m \tau \right)}{\tau} &\\
\label{eq:KR4d}\GR_m^{(4)}(x)&= \theta(x^0)\theta(\tau^2)\left( \frac{1}{2 \pi} \delta(\tau^2) - \frac{m}{4\pi} \frac{J_1(m \tau)}{\tau}\right)&
\end{align}

Here $\tau:= \sqrt{(x^0)^2-\vec{x}^2}$ is the proper length of the vector $x$, $J_\alpha$ is a Bessel function of the first kind of order $\alpha$,
$\delta$ is the Dirac delta function and
\be \theta(\alpha) = \left\{ \begin{array}{ll} 1
& \textrm{if $\alpha \geq 0$} \\ 0 & \textrm{if $\alpha < 0$}. \end{array}
\right. \ee 
For a comprehensive introduction to Bessel functions the reader is directed to the classic \citet{Watson:1958}.

For ease of reference we list the massless $\GR_0^{(d)}$ propagators\footnote{These are also given in \citet[Example 2.2, p144]{PDE:1994}.} (which are simply obtained by taking the $m\to 0$ limit of each $\GR_m^{(d)}$):
\begin{align}
\GR_0^{(1)}(x)&=\theta(x^0) x^0 \\
\label{eq:KR2dMassless}\GR_0^{(2)}(x)&=\theta(x^0) \theta(\tau^2) \frac{1}{2} \\
\GR_0^{(3)}(x)&=\theta(x^0)\theta(\tau^2) \frac{1}{2 \pi \tau} \\
\label{eq:KR4dMassless}\GR_0^{(4)}(x)&= \theta(x^0)\theta(\tau^2) \frac{1}{2 \pi} \delta(\tau^2) 
\end{align}

The \emph{advanced propagator} $\GA_m^{(d)}(x)$ is obtained from the boundary conditions that it be non-zero only in the past lightcone. It is related to the retarded propagator by $\GA_m^{(d)}(x) = \GR_m^{(d)}(-x)$. The Feynman propagator is obtained by more complicated boundary conditions---see \secref{sec:FeynmanPropagator} for details.

\section{Dimensional analysis}

If we wish the causal set propagators to match the Klein-Gordon propagators then, of course, their mass-dimensions should be the same. This simple condition can help to constrain the form of the $a$ and $b$ amplitudes.

Firstly we observe that the amplitudes $a^n b^{n-1}$ must all have equal mass-dimension (since we are adding them together) for $n=1,2,\ldots$. This implies that $[ab]=1$, that is, the product $ab$ is dimensionless.

Now, in $d$-dimensional Minkowski spacetime the mass-dimension of the Klein-Gordon propagators are $[G^{(d)}_m] = M^{d-2}$. Matching this to $[a^n b^{n-1}]$ for all $n$ gives:
\be [a] = M^{d-2} \qquad [b] = M^{2-d} \qquad  [ab] = 1. \ee
The only dimensionful quantities available to use on a sprinkled causal set are the sprinkling density $\rho$ and the mass of the particle $m$. These satisfy
\be [\rho] = M^d \qquad [m] = M. \ee
This is as far as we can go without additional assumptions---there are a number of ways of combining $\rho$ and $m$ to create amplitudes with dimensions $M^{d-2}$ and $M^{2-d}$.

If, however, we assume that $a$ is independent of the particle mass then $a$ must only depend on $\rho$. This fixes the form of $a$ to be
\be \label{eq:aampdimensions} a = A \rho^{1-2/d},\ee
where $A$ is a dimensionless constant. As it turns out our $a$ amplitudes will have this form.

\section{Expected values}

Causal sets generated by sprinklings into a Lorentzian manifold will not be identical because the particular
points that are sprinkled are chosen randomly. The value of the propagator calculated from one sprinkled causal
set therefore depends in detail on the particular causal set that is generated. To compare the causal set and continuum propagators we shall be interested in the \emph{expected value} (and variance) for the causal set propagator calculated
by averaging over all possible sprinkled causal sets.

To understand why, suppose we model spacetime by a causal set generated by sprinkling into $\mink^d$ with density $\rho$. For all pairs of sprinkled points we can calculate the causal set propagator by summing over trajectories in the manner described in \secref{Sec:PathIntegralModels}. We can also compute the continuum propagator for all pairs of points (using their coordinates in $\mink^d$ and the formulae in \secref{sec:KGPropagators}). We then wish to compare these two sets of propagator values. In addition we wish to compare the values for a \emph{typical} sprinkling, not just the particular causal set we sprinkled. To do this we will compute the expected value of the causal set propagator and compare it to the continuum propagator.

To compute the expected values for sprinklings into $d$-dimensional Minkowski spacetime we first fix two points $x, y \in \mink^d$. We then sprinkle a causal set. There is zero probability that the sprinkled causal set will contain $x$ and $y$ so we then add $x$ and $y$ to the sprinkled causal set. We then count the number of chains and paths and calculate the value of the propagator between the two points. Averaging these values over all sprinkled causal sets (with a fixed sprinkling density $\rho$), we obtain the expected number of chains and paths from $x$ to $y$ and the expected value of the propagator between $x$ and $y$.

\subsection{Summing over chains} \label{sec:SummingOverChains}

To calculate the expected number of chains between two points $x$ and $y$ in $d$-dimensional Minkowski spacetime
we define\footnote{This function on $\mink^d$ should not be confused with the causal matrix $C_{xy}$.} \be C(y-x) := \left\{ \begin{array}{ll} 1 & \textrm{ if } x \preceq y \\ 0 & \textrm{otherwise.}
\end{array} \right. \ee
Translation invariance of Minkowski spacetime ensures $C(y-x)$ is only a function of the separation $y-x$. The
expected number of chains of length one from $x$ to $y$ is given by \be C_1(y-x) = C(y-x). \ee The expected
number of chains $x \prec z_1 \prec \ldots \prec z_{n-1} \prec y$ of length $n > 1$ is given by \citet[p49-50]{Meyer:1988}:
\be C_n (y-x) := \rho^{n-1}\int \cdots \int d^d z_1 \cdots d^dz_{n-1} C(y-z_{n-1}) C(z_{n-1} - z_{n-2}) \cdots
C(z_1-x). \ee 
This integral can be evaluated in closed form \citep[Theorem III.2, p50]{Meyer:1988}. For $n > 1$ we have:
\be \label{eq:ChainExpectedValue} C_n(y-x) = C(y-x) \frac{(\rho V(x-y))^{n-1}}{n-1} \left( \frac{\Gamma(d+1)}{2}\right)^{(n-2)} \frac{\Gamma(\omega) \Gamma(2 \omega)}
{\Gamma((n-1) \omega) \Gamma(n \omega)}, \ee
where\footnote{The notation in \citet{Meyer:1988} is different to that used here. His $d$ refers to the \emph{spatial} dimension (i.e. our $d$ minus 1), he sets $\rho = 1$ and he writes $\langle C_k \rangle$ to denote the expected number of chains of length $k+1$ (i.e. $\langle C_{n-1} \rangle = C_n(x-y)$).} $\omega:=d/2$ and $V(x-y) = V(y-x)$ denotes the $d$-dimensional volume of the causal interval between $x$ and $y$ and $\Gamma(z)$ is the Gamma function.

In general we have, \citep[eq (44)]{Rideout:2006}:
\be \label{eq:CausalVolume} V(x-y) = \frac{\pi^{\frac{d-1}{2}}}{2^{d-1} d \,\Gamma((d+1)/2)} \tau^d,\ee
where $\tau$ is the proper time from $x$ to $y$.

The expected value for a propagator which sums amplitudes assigned to chains is given by\be
\label{eq:ChainInfiniteSum} K_C(y-x) := \sum_{n=1}^\infty a^n b^{n-1}  C_n (y-x). \ee This satisfies the
integral equation \be \label{eq:ChainSumExpectedValue} K_C(y-x) = a C(y-x) + ab \rho \int d^dz\, C(y-z)
K_C(z-x). \ee

\subsection{Summing over paths} \label{sec:SummingOverPaths}

To calculate the expected number of paths between two sprinkled points $x$ and $y$ in $d$-dimensional Minkowski
spacetime we define \be P(y-x) := \left\{
\begin{array}{ll} e^{- \rho V(x-y)} & \textrm{ if } x \preceq y \\ 0 & \textrm{otherwise.} \end{array} \right.
\ee where $V(x-y)$ is the $d$-dimensional Minkowski spacetime volume of the causal interval between $x$ and $y$.
The expected number of paths of length one from $x$ to $y$ (i.e. the probability that $x \link y$) is given by \be P_1(y-x) = P(y-x).\ee The expected number of paths $x \link z_1 \link \ldots \link z_{n-1} \link y$ of length $n > 1$ is given
by \citep[eq 2.5.5, p75]{Bombelli:1987:PhD}: \be P_n (y-x) = \rho^{n-1}\int \cdots \int d^d z_1 \cdots d^dz_{n-1} P(y-z_{n-1})
P(z_{n-1} - z_{n-2}) \cdots P(z_1-x). \ee 
Unfortunately no closed-form expression is known for $P_n(y-x)$ for general $n$ and $d$ (although an attempt is made to find  an asymptotic form for the expectation of the \emph{total} number of paths for sprinklings into $\mink^2$ in \citet[p76-77]{Bombelli:1987:PhD}. In Appendix \ref{chap:ExpectedNumberOfPaths} we attempt to determine $P_n$ as fully as possible).

The expected value for the propagator which sums over paths
is given by\be \label{eq:PathInfiniteSum} K_P(y-x) := \sum_{n=1}^\infty a^n b^{n-1}  P_n (y-x). \ee This satisfies
the integral equation \be \label{eq:PathSumExpectedValue} K_P(y-x) = a P(y-x) + ab\rho \int d^dz\,
P(y-z) K_P(z-x). \ee

\section{1+1 dimensional Minkowski spacetime} \label{sec:1+1pathintegral}

It turns out (in the sense that we get the right answer) that in 1+1 dimensions the propagator requires us to sum over chains. 

\subsection{Fourier transform calculation}

Fourier transforming \eqref{eq:ChainSumExpectedValue} the integral, being a convolution, becomes a product and we have \be
\widetilde{K}_C(p) = a \widetilde{C}(p) + ab\rho \,\widetilde{C}(p) \widetilde{K}_C(p), \ee or \be
\label{eq:ChainSumMomSpace} \widetilde{K}_C(p) = \frac{a \widetilde{C}(p)}{1 - ab\rho \widetilde{C}(p)}. \ee
In 1+1 dimensions the function $C(x)$ has Fourier transform 
\be \widetilde{C}(p) = -\frac{2}{(p_0 + i \epsilon)^2 - p_1^2}, \ee 
since $C(x) = 2 \GR^{(2)}_0(x)$ (and setting $m=0$ in \eqref{eq:PropMomSpace}). Substituting this into
\eqref{eq:ChainSumMomSpace} we have \be \widetilde{K}_C(p) = \frac{-\frac{2a}{(p_0+i\epsilon)^2 - p_1^2}}{1 +
\frac{2 ab \rho}{(p_0+i\epsilon)^2 - p_1^2}} = -\frac{2a}{(p_0+i\epsilon)^2 - p_1^2 + 2ab\rho}.\ee Equating this
to the $d=2$ version of equation \eqref{eq:PropMomSpace} we find \be \label{eq:1+1Amplitudes} a = \frac{1}{2}, \qquad b =
-\frac{m^2}{\rho}, \ee are the correct amplitudes.

In 1+1 dimensions $[\rho] = M^2$ so the amplitudes assigned to the chains are dimensionless: $[a] = [b] = [a^n
b^{n-1}] = 1$ (for $n=1,2,\ldots$).

\subsection{Direct calculation}

An alternative approach is to use the explicit form for $C_n$ given in \eqref{eq:ChainExpectedValue} to calculate $K_C$ for sprinklings into 1+1 dimensions. 

Using $V(x-y) = \tau^2/2$, $d=2$ and substituting \eqref{eq:ChainExpectedValue} into \eqref{eq:ChainInfiniteSum} we have:
\begin{align} K_C(y-x) :=&C(y-x)\left( a + \sum_{n=2}^\infty \frac{a^n b^{n-1} \rho^{n-1} \tau^{2n-2}}{(n-1) 2^{n-1}} \left( \frac{\Gamma(3)}{2}\right)^{\!(n-2)} \!\!\frac{\Gamma(1) \Gamma(2)}
{\Gamma(n-1) \Gamma(n )}\right), \\
 =& C(y-x)\left(a+ \sum_{n=2}^\infty a^n b^{n-1} \rho^{n-1} \frac{1}
{2^{n-1} \Gamma(n) \Gamma(n )} \tau^{2n-2} \right),\\=& C(y-x) a I_0\left(\sqrt{2 a b \rho} \, \tau\right),\end{align}
where $I_0$ is a modified Bessel function of the first kind. This satisfies $I_0(i z) = J_0(z)$.

We thus see that with $a = 1/2$, $b=-m^2/\rho$ (so $2 ab \rho = -m^2$) we have:
\be K_C(y-x) = C(y-x)\frac{1}{2} I_0(i m \tau) = C(y-x)\frac{1}{2} J_0(m \tau) = \GR^{(2)}_m(y-x). \ee

\section{3 + 1 dimensional Minkowski spacetime} \label{sec:3+1pathintegral}

\def\GRT{(\widetilde{G}_R)}

It turns out that in 3+1 dimensions the propagator requires us to sum over paths. Fourier transforming
\eqref{eq:PathSumExpectedValue} we get \be \label{eq:PathSumMomSpace} \widetilde{K}_P(p) = \frac{a
\widetilde{P}(p)}{1 - ab\rho \widetilde{P}(p)}. \ee In 3+1 dimensional Minkowski spacetime $V(x-y)
:= \frac{\pi}{24} \tau_{xy}^4$ is the volume of the causal interval between $x$ and $y$. We therefore have
\be P(y-x) = \left\{ \begin{array}{ll} e^{-\rho V(x-y)} = e^{-\frac{\pi}{24} \rho \tau_{xy}^4} & \textrm{if } x \preceq y \\
0 & \textrm{otherwise.} \end{array} \right. \ee The function
\be f_\rho(z,c) := \left\{ \begin{array}{ll} \sqrt{\rho} e^{-\pi c \rho z^2} & \textrm{if } z \geq 0 \\
0 & \textrm{if } z < 0,\end{array} \right. \ee (with a real constant $c > 0$) satisfies \be \lim_{\rho \to
\infty} f_\rho(z,c) = \frac{1}{2 \sqrt{c}} \delta(z), \ee where $\delta(z)$ is the Dirac delta function\footnote{This limit is in the sense of distributions (see \citet[Example 2, p36-37]{Gelfand:1964}).}. We
therefore see
\be \label{eq:3+1Limit} \lim_{\rho \to \infty} \sqrt{\rho} \,P(y-x) = \lim_{\rho \to \infty} \left\{ \begin{array}{ll} f_\rho(\tau_{xy}^2, \frac{1}{24}) & \textrm{if } x^0 \leq y^0 \\
0 & \textrm{otherwise} \end{array} \right. = \left\{ \begin{array}{ll} \frac{\sqrt{24}}{2} \delta(\tau_{xy}^2) & \textrm{if } x \preceq y \\
0 & \textrm{otherwise.} \end{array} \right. \ee Setting $m=0$ in \eqref{eq:PropMomSpace} and using \eqref{eq:KR4dMassless}
we see that in 3+1 dimensions we have
\be \GR^{(4)}_0(y-x) = \left\{\begin{array}{ll} \frac{1}{2 \pi} \delta(\tau^2_{xy})& \textrm{ if } x \preceq y \\ 0 & \textrm{otherwise,}\end{array} \right. \ee
\be \GRT^{(4)}_0(p) = -\frac{1}{(p_0+i\epsilon)^2 - \vec{p}^2}.\ee
These imply, taking Fourier transforms and using \eqref{eq:3+1Limit}, that\footnote{\label{foot:Fourier} We make use of the fact that \[ \lim_{\rho \to \infty} f_\rho = f \iff \lim_{\rho \to \infty} \tilde{f}_\rho = \tilde{f}\]
for suitable functions $f$ \citep[Thm 13.2, p135]{Champeney:1989}.}
\be \lim_{\rho \to \infty} \sqrt{\rho}\; \widetilde{P}(p) = 2 \pi \frac{\sqrt{24}}{2} \GRT^{(4)}_0(p) = -2 \pi\sqrt{6} \frac{1}{(p_0+i\epsilon)^2 - \vec{p}^2}.\ee 
Setting \be a = A \sqrt{\rho}, \qquad b = \frac{B}{\rho}, \ee where $A$ and $B$ are (possibly dimensionful) constants independent of $\rho$ we substitute into \eqref{eq:PathSumMomSpace} to get
\be \widetilde{K}_P(p) = \frac{A \sqrt{\rho}\widetilde{P}(p)}{1 - AB\sqrt{\rho}\widetilde{P}(p)}.\ee As $\rho$ tends to infinity this becomes
\be \lim_{\rho \to \infty} \widetilde{K}_P(p)=\frac{-\frac{A C}{(p_0+i\epsilon)^2 - \vec{p}^2}}{1 + \frac{AB C}{(p_0+i\epsilon)^2 -
\vec{p}^2}}  = -\frac{AC}{(p_0+i\epsilon)^2 - \vec{p}^2 + ABC},\ee where $C := 2\pi\sqrt{6}$.

Equating this with the $d=4$ version of equation \eqref{eq:PropMomSpace} we have \be AC = 1, \qquad B = -m^2,\ee so \be \label{eq:3+1Amplitudes}
a = \frac{\sqrt{\rho}}{2 \pi \sqrt{6}}, \qquad b = -\frac{m^2}{\rho},\ee are the correct amplitudes.

In 3+1 dimensions $[\rho] = M^4$ so the amplitudes assigned to the paths have mass-dimension $M^2$: $[a] = M^2$,
$[b] = M^{-2}$, $[a^n b^{n-1}] = M^2$ (for $n=1,2,\ldots)$.

\subsection{Summing over chains}

In 3+1 dimensions we have to sum over paths. We briefly mention what happens if we sum over chains.

Setting $d=4$ and substituting \eqref{eq:ChainExpectedValue} into \eqref{eq:ChainInfiniteSum} we have:
\begin{align} K_C(y-x) :=& C(y-x) \left(a + \sum_{n=2}^\infty a^n b^{n-1} \left( \rho V \right)^{n-1} \frac{\left(\frac{\Gamma(5)}{2}\right)^{n-2} \Gamma(2) \Gamma(4)}{(n-1) \Gamma(2n-2) \Gamma(2n)}\right),\\
 =& C(y-x) \left(a + a \sum_{n=2}^\infty (a b \rho V)^{n-1} \left(12\right)^{n-2}\frac{6}{(n-1) \Gamma(2n-2) \Gamma(2n)}\right),\\
\label{eq:3+1SumOverChains} =& C(y-x)\frac{a}{c} \left( I_1(c) + J_1(c)\right),\end{align}
where $c = 2 (12 a b \rho V)^{1/4}$

 While it may be of interest to know this result it does not seem possible to find $a$ and $b$ such that \eqref{eq:3+1SumOverChains} is equal to $\GR_m^{(4)}$ (even in the infinite $\rho$ limit). It is possible that summing over chains with different amplitudes (not of the form $a^n b^{n-1}$) may reproduce the continuum propagator (in a sense, this is what we are doing when we sum over paths because every path is a chain).

\section{Comparison with the continuum} \label{sec:PathIntegralComparison}

The causal set propagators just described can be calculated for a particular sprinkling using a computer. We can then compare the causal set and continuum propagators with the following steps:
\begin{enumerate}
 \item Fix a sprinkling density $\rho$ and a particle mass $m$.
 \item Sprinkle a causal set $\CS$ with density $\rho$ into a causal interval\footnote{We use a causal interval to ensure that the causal structure of the causal set matches that of $\mink^d$---i.e. there are no edge effects such as sub-causal-intervals being ``chopped off'' by the edges of the region. Another way to say this is that a causal interval is causally convex (a region $R \subset \mink^d$ is causally convex if $x, y \in R \implies [x,y] \subseteq R$.)} in $\mink^d$.
 \item Compute $C$ and $L$, the causal and link matrices for $\CS$.
 \item Compute $K_R$, the propagator matrix.
 \item For each pair of causally related points $v_x \preceq v_y$ compute their proper time separation $\tau_{xy}$ using their coordinates in $\mink^d$.
 \item Plot $(K_R)_{xy}$ against $\tau_{xy}$ for all pairs of related points and compare it with $\GR^{(d)}_m$ as a function of proper time.
\end{enumerate}

We now perform this comparison for sprinklings into 1+1 and 3+1 dimensional Minkowski spacetime.

\subsection{1+1 dimensions} \label{sec:1+1Comparison}

In \secref{sec:1+1pathintegral} we showed that the causal set propagator for sprinklings into $\mink^2$ is given by:
\be K_R := \frac{1}{2} C\left(I + \frac{m^2}{2 \rho} C\right)^{-1},\ee
(where we sum over chains and use the $a$ and $b$ amplitudes from \eqref{eq:1+1Amplitudes}).

\begin{figure}[!hp]
\begin{center}
\includegraphics[width = \textwidth]{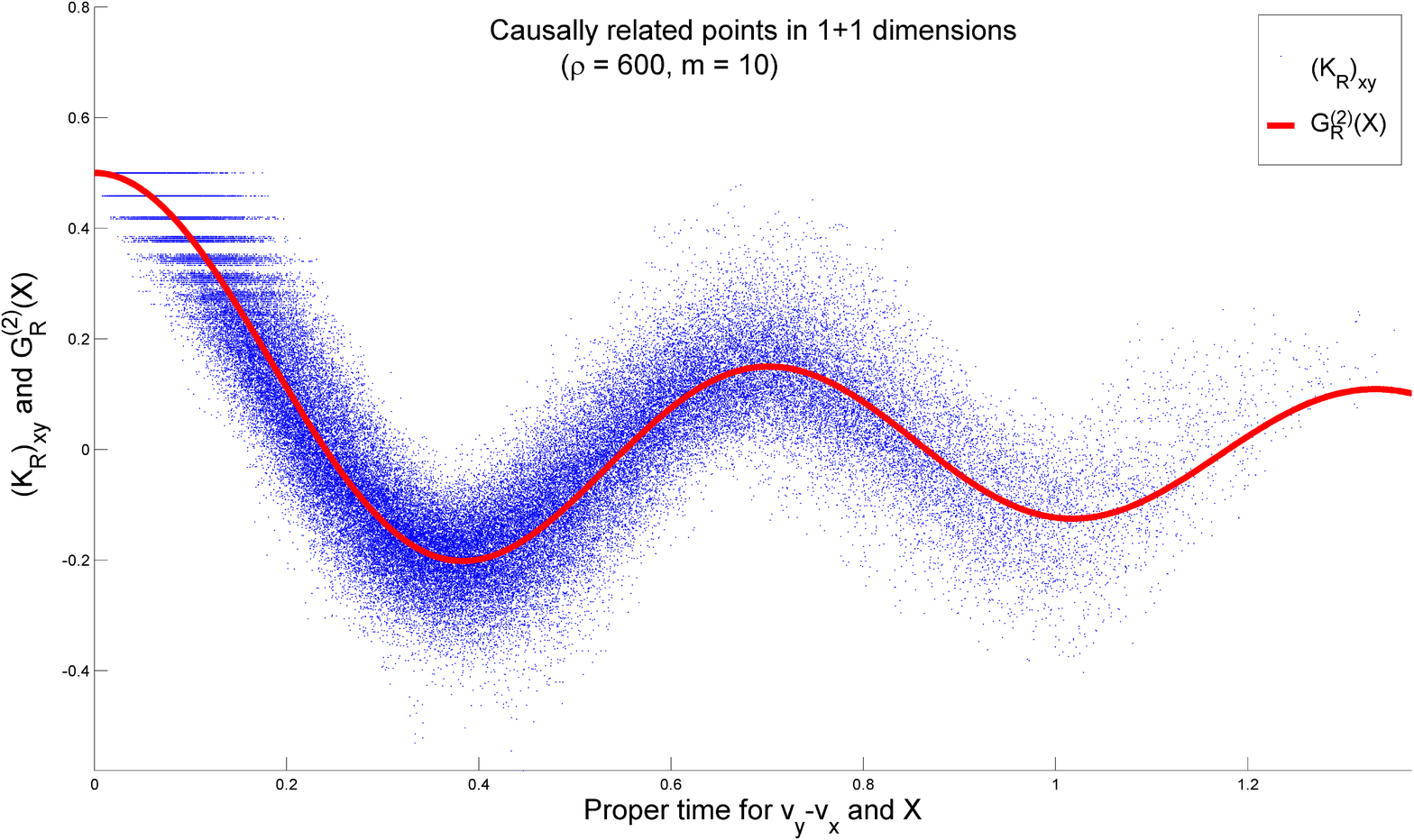}
\caption[Retarded propagator for 1+1 dimensional sprinkling]{The causal set retarded propagator for a sprinkling in 1+1 dimensions with $\rho=600, m = 10$. Each dot corresponds to a pair of causally related sprinkled points.}
\label{fig:1+1PathInt}
\vspace{2cm}
\includegraphics[width = \textwidth]{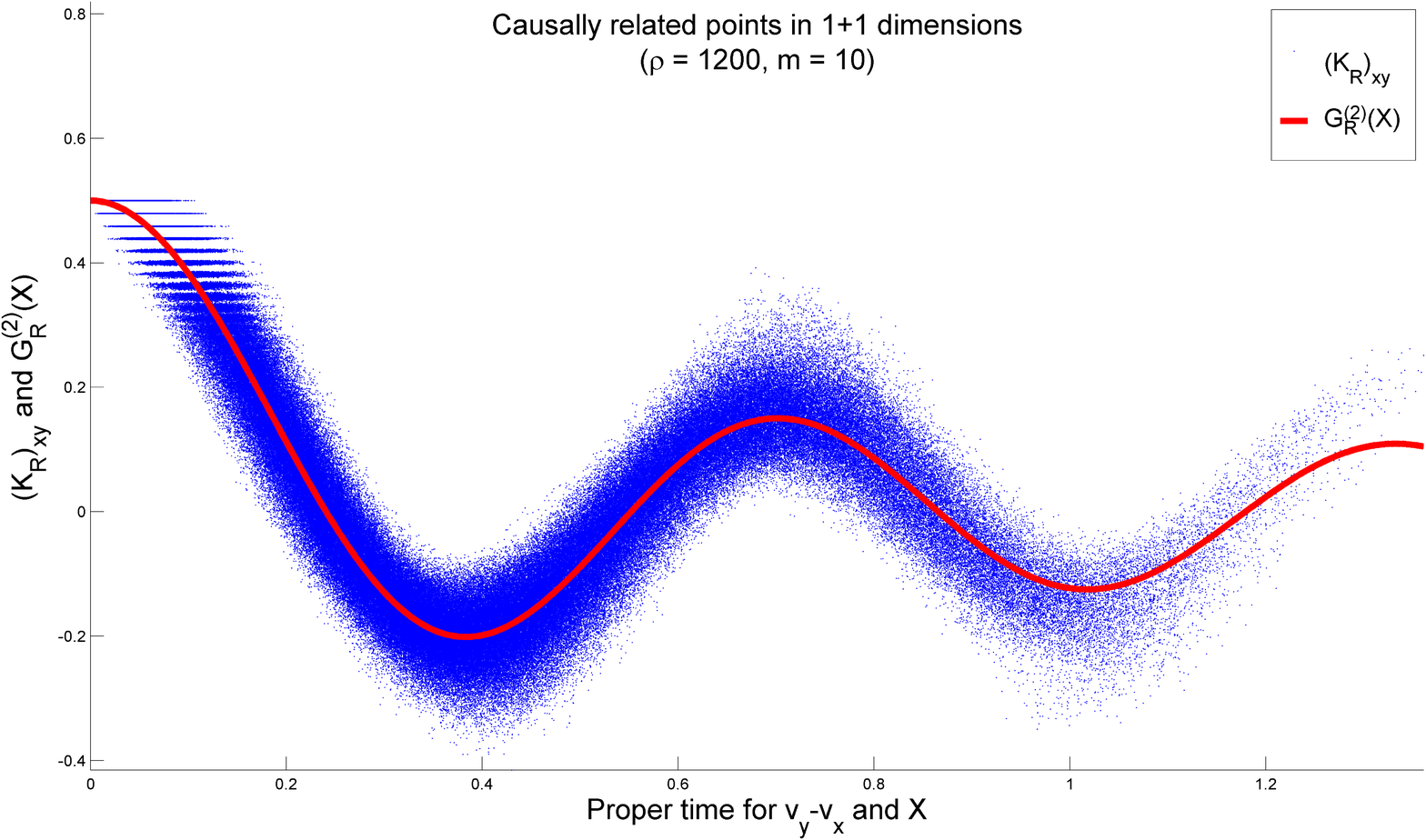}
\caption[Retarded propagator for 1+1 dimensional sprinkling]{The causal set retarded propagator for a sprinkling in 1+1 dimensions with $\rho=1200, m = 10$.}
\label{fig:1+1PathInt1200}
\end{center}
\end{figure}

In \figref{fig:1+1PathInt} and \figref{fig:1+1PathInt1200} we plot this propagator for two sprinklings with $m=10$ and for $\rho = 600$ and $\rho = 1200$. Every dot that is plotted corresponds to a pair of causally related sprinkled points. For that pair their proper time separation $\tau_{xy}$ is the horizontal coordinate and the value of the propagator $(K_R)_{xy}$ is their vertical coordinate.

\begin{figure}[!h]
\begin{center}
\includegraphics[width = \textwidth]{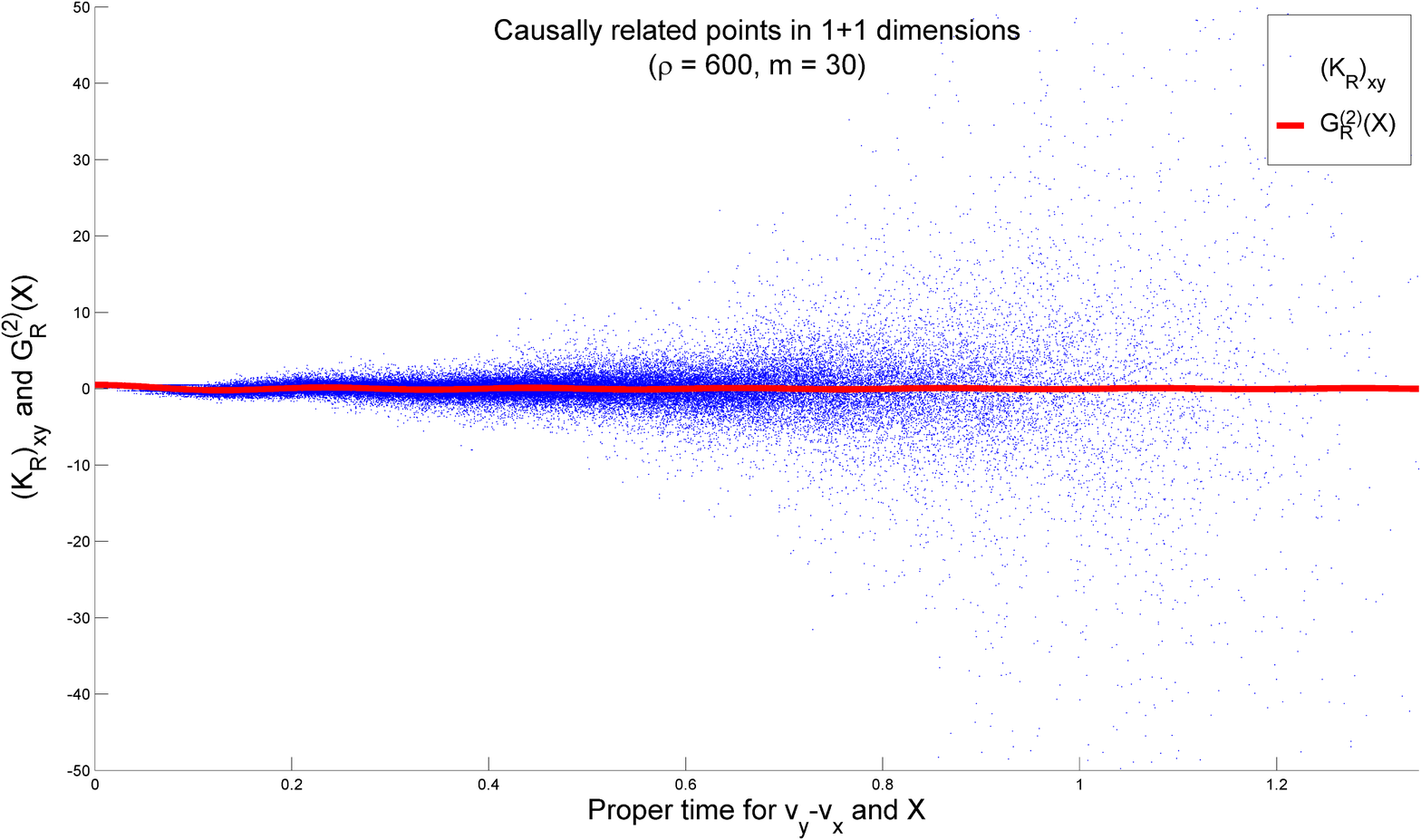}
\caption[Retarded propagator for 1+1 dimensional sprinkling]{The causal set retarded propagator for a sprinkling in 1+1 dimensions with $\rho=600, m = 30$. There is poor agreement with the continuum for $m^2 > \rho$.}
\label{fig:1+1PathIntLargeMass}
\end{center}
\end{figure}

As we can see there is good agreement between the causal set propagator and the continuum propagator. In general this agreement holds provided $m^2 \ll \rho$. If $m^2 > \rho$ the plotted points are spread more widely away from the continuum curve. In \figref{fig:1+1PathIntLargeMass} we plot the results for $\rho = 600, m = 30$. This has $m^2/\rho = 1.5$ and shows poor agreement with the continuum. This behaviour suggests that the variance of the propagator is small provided $m^2 \ll \rho$.

Since we are not plotting the expected value of the propagator, but rather the propagator calculated for a single causal set, there are fluctuations away from the continuum propagator. These appear as a spread in the dots away from the $\GR^{(2)}_m$ curve. These fluctuations are due to the particular random distribution of points used in the sprinkling. If we fix the mass and the volume of the causal interval we find that these fluctuations decrease as the density is increased. As an example of this we can compare the plots for $\rho=600$ and $\rho = 1200$---the spread of the plotted points in \figref{fig:1+1PathInt1200} is less than in \figref{fig:1+1PathInt}.

For small proper times the causal set propagator becomes ``discretised''---the values clump into a few horizontal layers. This is because points separated by a small proper time will only contain a few other points in the causal interval between them. If this number is small (say 1,2,3 or 4) then there's only a few values the propagator can take (e.g. if the $n$ points in between are mutually unrelated the propagator is $a + n a^2 b$). When the proper time is larger there are many more points contained in the causal interval and subsequently a larger range of values for the propagator. This leads to the values becoming more ``smeared out'' and not confined to a few layers. Another way to think about this is that when we sum over a large number of paths there is more interference between the amplitudes assigned to paths of different lengths.

\subsection{3+1 dimensions} \label{sec:3+1Comparison}

In \secref{sec:3+1pathintegral} we showed that the causal set propagator for sprinklings into $\mink^4$ is given by:
\be K_R := \frac{\sqrt{\rho}}{2\pi \sqrt{6}} L\left(I + \frac{m^2}{2 \pi \sqrt{(6 \rho)}} L\right)^{-1},\ee
(where we sum over paths and use the $a$ and $b$ amplitudes from \eqref{eq:3+1Amplitudes}).

When comparing this with the continuum, however, there are difficulties which did not arise in the 1+1 dimensional case. Firstly, for sprinklings into $\mink^4$ the expected value of the propagator only equals the continuum propagator in the \emph{infinite} density limit. A realistic sprinkling density (e.g. at Planckian density) would be \emph{large but finite}. We therefore expect to get good agreement only for very large sprinkling densities\footnote{The sprinkling process is only defined for finite densities. It makes no sense to sprinkle an ``infinite density'' causal set.}.

Investigating the behaviour of the propagator for large sprinkling densities is difficult to do through numerical simulations. This is because current simulations cannot cope with enough sprinkled points to ensure a large density over a large spacetime volume.

High density sprinklings \emph{can} be achieved, however, if we sprinkle a moderate number of points into a small volume. Unfortunately in this case the behaviour of the propagator is only investigated within such small volumes---i.e. for small proper times. For the 3+1 dimensional case the small proper time behaviour is dominated by the delta-function term in the propagator (which, for finite densities, is smeared out away from the lightcone).

\begin{figure}[!h]
\begin{center}
\includegraphics[width = \textwidth]{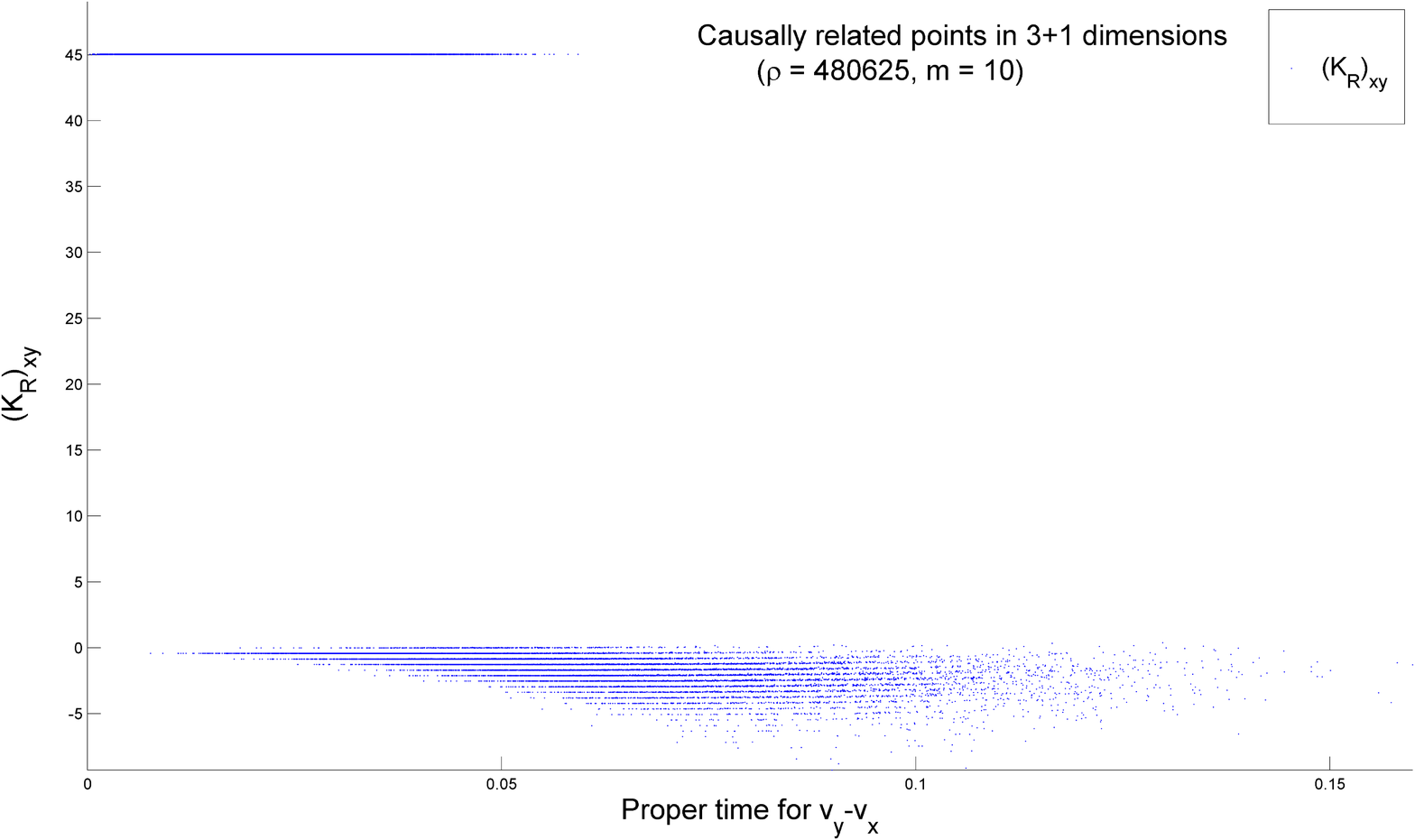}
\caption[Retarded propagator for 3+1 dimensional sprinkling]{The causal set retarded propagator for a sprinkling in 3+1 dimensions.}
\label{fig:3+1PathInt}
\end{center}
\end{figure}

In \figref{fig:3+1PathInt} we plot the propagator for a sprinkling into a small causal interval with $\rho = 480625, m = 10$. The horizontal line present for small proper times corresponds to the propagator value for pairs of linked elements. Its value is $a=\sqrt{\rho}/(2 \pi \sqrt{6}) \approx 45$. The other values present for larger proper times are contributions from elements joined by paths of length greater than 1 (as in the 1+1 dimensional case these values are ``discretised'' into horizontal layers).

We do not yet see the Bessel function behaviour within the future lightcone that is present in the continuum propagator $\GR_m^{(4)}$. Presumably this behaviour would be apparent if a simulation could be performed with a large sprinkling density over a large spacetime volume. This is a task for future work.

A second difficulty is that we cannot compare the two propagators directly. The continuum propagator contains a delta-function term on the forward lightcone. This is a distributional object which we cannot plot explicitly. One solution to this would be to apply the continuum and causal set propagators to a family of test functions and compare the results. We have not done this comparison since the current simulations suggest a much higher density is needed to obtain good results.

We note that, extrapolating from the 1+1 dimensional case, we would expect good agreement with the continuum propagator (for very large $\rho$) provided $m^2 \ll \sqrt{\rho}$. This constraint on the mass of the particle is easily satisfied for realistic sprinkling densities. Assuming a Planckian sprinkling density we let $\rho$ be the inverse of the Planck
4-volume: $\rho =c^7/(G \hbar)^2$. The heaviest known elementary particle is the top quark with a mass $m =
171.3 \pm 1.2$GeV \citep{Amsler:2008}. In Planck units, so $\rho = 1$, this is $m \approx 1.4 \times 10^{-17}$. This
certainly satisfies $m^2/\sqrt{\rho} \approx 2 \times 10^{-34} \ll 1$. If the masses of particles modelled on the causal set are of order the masses of known elementary particles then the causal set propagator (for a Planckian density sprinkling) should well-approximate the continuum propagator.

\section{Mass scatterings} \label{sec:MassScatterings}

Having established suitable amplitudes for causal sets generated by sprinkling into 1+1 and 3+1 dimensional Minkowski spacetime we can now better understand why they take the values they do.

The mass dependence for both propagators is entirely contained in the amplitude $b = -m^2/\rho$. By setting the mass to zero we see that the series for the propagator truncates to give $K = \Phi$ where $\Phi_{xy}$ is just the amplitude for the particle to `hop' from $v_x$ to $v_y$. We thus see that the hop amplitude for a massive particle is equal to total propagation amplitude for a massless particle.

This suggests that a free massive particle is equivalent to a massless particle undergoing repeated ``self-interactions'' at each stop. The strength of these interactions is governed by $b$. Through these ``mass scatterings'' the massless propagator is ``dressed up'' to give the massive propagator. We have kept the quotation marks since we have not yet developed a theory of interacting particles.

The same idea of a massless particle undergoing self-interaction mass scatterings can be dealt with in the continuum. To demonstrate this we suppose that $G_0(x)$ is the massless propagator for the Klein-Gordon equation:
\be \Box G_0(x) = \delta^d(x).\ee
Defining the $d$-dimensional convolution of two functions $f$ and $g$ on $\mink^d$ as
\be (f * g)(x):=\int d^dy f(x-y) g(y),\ee
we now consider the function defined by
\be \label{eq:MassiveSeries} G_m(x):=G_0(x) - m^2 (G_0 * G_0)(x) + m^4(G_0 * G_0 * G_0)(x) + \ldots. \ee
The $n\th$ term in this infinite sum involves the convolution of $n$ copies of $G_0$ with 	a coefficient of $(-m^2)^{n-1}$.
Now we observe that
\be \label{eq:BoxConvolution} \Box (G_0 * f)(x) = ((\Box G_0) * f)(x) = (\delta^d * f)(x) = f(x), \ee
for suitable functions $f$. Therefore we see that
\begin{align}\Box G_m(x) &= \delta^d(x) -m^2 G_0(x) + m^4 (G_0 * G_0)(x) + \ldots \\
&= \delta^d(x) -m^2 G_m(x). \end{align}
or
\be (\Box +m^2) G_m(x) = \delta^d(x). \ee
This shows that $G_m$ is a propagator for the \emph{massive} Klein-Gordon equation.

This series can be simply expressed in Fourier space where
\be \tilde{G}_0(p) = -\frac{1}{p^2}, \ee
\be \tilde{G}_m(p) = -\frac{1}{p^2} - m^2 \frac{1}{(p^2)^2} - m^4 \frac{1}{(p^2)^3} - \ldots = -\frac{1}{p^2} \left(\sum_{n=0}^\infty \left(\frac{m^2}{p^2}\right)^n \right) = -\frac{1}{p^2 - m^2},\ee
since the Fourier transform of $f * g$ is equal to $\tilde{f} \tilde{g}$.

These \emph{formal} calculations are given just to suggest that the massive propagator can be defined in terms of the massless propagator. We have made no attempt to rigorously define the convergence of the series or determine the validity of operations like \eqref{eq:BoxConvolution}. One attempt at a rigorous analysis is given by \citet{Aste:2007}.

A model for the massive retarded propagator based on a random walk in the continuum using the series \eqref{eq:MassiveSeries} is given by \citet{Dugne:2003}.

\section{Non-constant `hop' and `stop' amplitudes} \label{sec:Nonconstant}

We now consider a generalisation of the model in which the amplitude for the hops and stops are allowed to depend on position.

On a finite $p$-element causal set we now have a whole $p \times p$ matrix of possible hop amplitudes. As before we denote this by $\Phi$ with the amplitude for a single hop from $v_x$ to $v_y$ equal to $\Phi_{xy}$. We will place the possible stop amplitudes into a diagonal $p \times p$ matrix $\Psi$ with the amplitude to stop at $v_x$ equal to $\Psi_{xx}$.

The total amplitude, summed over all trajectories from $v_x$ to $v_y$ is then $K_{xy}$ where $K$ is the matrix
\be \label{eq:CausetPropagatorNonConstant} K := \Phi + \Phi \Psi \Phi + \Phi \Psi \Phi \Psi \Phi +\ldots= \Phi \sum_{n=0}^\infty \left(\Psi \Phi\right)^n. \ee
In this sum $(\Phi (\Psi \Phi)^{n-1})_{xy}$ is the contribution from trajectories with $n$ hops (and $n-1$ stops) from $v_x$ to $v_y$.

Without a restriction on the allowed amplitudes we cannot guarantee that this sum converges. As it stands, we have not ruled out the possibility of hops between spacelike elements or hops from one element into its causal past.
We can even allow ``hops'' from one element to itself (e.g. if $\Phi_{xx} \neq 0$)! 

If the hops are restricted to be suitably causal then we can guarantee the sum will converge. If we ensure
\be \label{eq:HopRestriction} \Phi_{xy} \neq 0 \implies v_x \prec v_y, \ee
then the trajectories being summed over move only forwards in time and the sum terminates. In this case we have
\be \label{eq:GeneralRetarded} K = \Phi(I - \Psi \Phi)^{-1}. \ee

If we lift the restriction \eqref{eq:HopRestriction} then it's possible the series will not terminate. In this case the matrix geometric series will still converge if the eigenvalues of the matrix $\Psi \Phi$ are all less than one in absolute value.

A model which allowed arbitrary hops was considered by \citet{Scargle:2009}. In their notation $\Phi = \mathbb{A}$ and $\Psi = I$ where $I$ is the $p\times p$ identity matrix.

\subsection{Sprinklings into other dimensions}

For sprinklings into $\mink^1$ and $\mink^3$ we have to allow non-constant hop amplitudes. Following the discussion in \secref{sec:MassScatterings} we see that the massive propagators $\GR_m^{(1)}$ and $\GR_m^{(3)}$ can be obtained from the massless retarded propagators $\GR_0^{(1)}$ and $\GR_0^{(3)}$ through a ``mass scattering'' series. We shall therefore reproduce the massless propagator on a causal set and use that to define the massive propagator.

\subsubsection{0+1 dimensions} \label{sec:0+1dim}

In $\mink^1$ we have $x = (x^0)$ and 
\be \GR_0^{(1)}(x) = x^0 \theta(x^0). \ee

We choose the causal set analogue of this function to be $\Phi = \frac{1}{\rho} (C^2 + C)$ (although there exist other possible choices). This is a matrix of non-constant hop amplitudes with the correct mass-dimension $\left[\Phi\right] = M^{-1}$. Following \secref{sec:MassScatterings} we shall continue to use $b = -m^2/\rho$. The propagator is then
\be K = \Phi(I - b \Phi)^{-1}.\ee
Since a sprinkling into $\mink^1$ is a total order we can compute $K$ explicitly. If we use the unique natural labelling for the elements of $\CS$ we have $v_1 \prec v_2 \prec v_3 \ldots \prec v_p$ and
\be C = \left(\begin{array}{ccccc}
 0 & 1 & 1 & 1 & 1 \,\cdots \\
 0 & 0 & 1 & 1 & 1 \,\cdots \\
 0 & 0 & 0 & 1 & 1 \, \cdots \\
 0 & 0 & 0 & 0 & 1 \,\cdots \\
 \vdots & \vdots & \vdots & \vdots & \ddots 
\end{array}\right)
\ee

We can use combinatorics to compute the number of chains of length $n=1,2,3,\ldots$ from $v_x$ to $v_y$ ($x, y = 1,\ldots,p$). To specify a chain of length $n$ from $v_x$ to $v_y$ (assuming $v_x \prec v_y$) we have to choose $n-1$ elements from the set of $y-x-1$ elements which lie between $v_x$ and $v_y$ in the total order. The number of ways of doing this (i.e. the number of chains) is given by a binomial coefficient:
\be (C^n)_{xy} = \binom{y - x - 1}{n-1} = \binom{y-x-1}{y-x-n}, \ee
for $x < y$. If $v_y \preceq v_x$ (so $y-x-1 < 0$) the number of chains is zero but this the first of these binomial coefficients is non-zero\footnote{Since for $n<0$ and $m > 0$, we have \[\binom{n}{m} = (-1)^{m} \binom{-n+m-1}{m}.\]}. 

We can express the number of chains between any two elements by a \emph{single} binomial coefficient, valid for all $x,y=1,\ldots,p$ if we use the convention that $\binom{n}{m} = 0$ if  $m$ is negative. In this case we have
\be \label{eq:TotalOrderPowers} (C^n)_{xy} = \binom{y - x - 1}{y-x-n},\ee
valid for all $x,y=1,\ldots,p$.

 To compute the propagator we need to calculate $((C^2 + C)^n)_{xy}$. By expanding this using the binomial theorem we have
 \be\nonumber ((C^2 + C)^n)_{xy} = \sum_{k=0}^n \binom{n}{k} (C^{2k} C^{n-k})_{xy} =  \sum_{k=0}^n \binom{n}{k} (C^{n+k})_{xy} \ee
  \be \label{eq:TotalOrderPowers2}= \sum_{k=0}^n \binom{n}{k} \binom{y - x - 1}{y-x-n-k} = \binom{y-x+n-1}{y-x-n}, \ee
  where in the last step we used Vandermonde's identity\footnote{Vandermonde's identity states that \[\sum_{k=0}^n \binom{n}{k} \binom{m}{r-k} = \binom{n+m}{r}.\]}.
 We therefore have that $\Phi^n$ is given by
\be (\Phi^n)_{xy} = \frac{1}{\rho^n}\binom{y-x+n-1}{y-x-n}.\ee
The propagator $K$ is therefore
\be K_{xy} = \sum_{n=1}^\infty (\Phi^n)_{xy}\left(\frac{-m^2}{\rho}\right)^{n-1}= \sum_{n=1}^\infty \binom{y-x+n-1}{y-x-n} \frac{1}{\rho^n} \left(\frac{-m^2}{\rho}\right)^{n-1}.\ee
This sum can be evaluated explicitly. We have
\everymath{\displaystyle}
\be \label{eq:0+1prop} K_{xy} = \left\{\begin{array}{cc} \frac{2 \sin \left(2 (y-x) \arcsin\left(\frac{m}{2 \rho}\right)\right)}{m \sqrt{4-\frac{m^2}{\rho^2}}} & \textrm{if $x < y$} \\ \\
0 & \textrm{if $x \geq y$.} \end{array} \right. \ee\everymath{}

This is an exact expression which has not been averaged over sprinklings. We can ask how it compares to the continuum propagator in the following way.

Fix two points $X,Y \in \mink^1$ and suppose that $X \preceq Y$ and $Y-X = \tau$ is their proper time separation. We now lay down a totally ordered causal set which contains $X$ and $Y$ and suppose the causal interval between them contains $N$ elements. The density of this causal set is then $\rho = N/\tau$.

Using this causal set we then calculate the propagator from $X$ to $Y$ to give:
\be \frac{2 \sin \left(2 (N-1) \arcsin\left(\frac{m \tau}{2 N}\right)\right)}{m \sqrt{4-\frac{(\tau m)^2}{N^2}}}. \ee
The $N \to \infty$ limit of this can be evaluated and is equal to $\GR_m^{(1)}(Y-X)$.

\subsubsection{2+1 dimensions}

In $\mink^3$ we have $x = (x^0,x^1,x^2)$ and 
\be \GR_0^{(3)}(x) = \theta(x^2) \theta(x^0)\frac{1}{2 \pi \tau} =\theta(x^2) \theta(x^0)\frac{1}{2 \pi \sqrt{(x^0)^2 - \vec{x}^2}}. \ee
To reproduce this $\tau$-dependence on a causal set we shall use the relationship between $\tau$ and the volume of a causal interval $V$ given by
\be V(x-y) = \frac{\pi}{12} \tau^3. \ee
where $\tau$ is the proper time from $x$ to $y$. This gives
\be \GR_0^{(3)} = \theta(x^2) \theta(x^0) \frac{1}{2\pi} \left(\frac{12 V}{\pi}\right)^{-\frac{1}{3}}.\ee

We want the mass dimensions of the massless propagator $K$ to be $[K] = M$. In $\mink^3$ we have $[\rho] = M^3$ so we expect that $\Phi$ will contain a factor of $\sqrt[3]{\rho}$.

The dimensionful volume of the causal interval between $v_x$ and $v_y$ is equal to $\frac{1}{\rho}(C+I)^2_{xy}$. With this in mind we define the proper time dependence in terms of the cube root of the volume of a causal interval (with appropriate constant factors):
\everymath{\displaystyle}
\be \Phi_{xy} :=  \left\{\begin{array}{cc}\frac{1}{2\pi}\left(\frac{\pi \rho}{12}\right)^\frac{1}{3}( (C+I)^2)_{xy}^{-\frac{1}{3}}& \textrm{if $v_x \prec v_y$} \\ 0 & \textrm{ otherwise.} \end{array} \right. \ee
\everymath{}
which is a matrix of non-constant hop amplitudes.

We then define the massive propagator as $K = \Phi(I - b \Phi)^{-1}$ with $b=-m^2/\rho$.

\subsubsection{Higher dimensions}

We have seen that to obtain the massive retarded propagator it is sufficient to determine a causal set analogue of the massless retarded propagator. There are expressions for the massless retarded propagators in higher dimensions given in \citet{Soodak:1993}. They consist of retarded combinations of step functions, delta functions and derivatives of delta functions\footnote{Different expressions for the massless retarded propagator appear in \citep[p155, Exercise a]{Vladimirov:1971}. It seems likely that mistakes may have slipped in during translation of this work so the \citet{Soodak:1993} expressions seem more likely to be correct.} which it might be possible to reproduce on a causal set.

\section{Summary of the models}

We have found path integral models for causal sets generated by sprinkling with density $\rho$ into $\mink^d$ with $d=1,2,3,4$. The retarded propagator for a particle of mass $m$ is a matrix $K_R$ defined by
\be K_R = \Phi\left(I + \frac{m^2}{\rho}\Phi\right)^{-1}, \ee
where $\Phi$ is a $p\times p$ matrix. If $v_x \prec v_y$ then $\Phi_{xy}$ is given by:
\everymath{\displaystyle}
\[\begin{array}{| c | c | c | c | c |}
d & 1& 2 &3&4 \\ \hline & & & & \\
\Phi_{xy} & \frac{1}{\rho}(C^2+C)_{xy} & \frac{1}{2} C_{xy} & \frac{1}{2\pi}\left(\frac{\pi \rho}{12 ((C+I)^2)_{xy}}\right)^{1/3} & \frac{\sqrt{\rho}}{2 \pi \sqrt{6}}L_{xy} \\& & & & \\
\end{array}\]\everymath{}
\!otherwise $\Phi_{xy}=0$. Here $C$ is the causal matrix for the causal set and $L$ is the link matrix for the causal set.

\subsection{Model philosophy}

We have presented four causal set propagators whose expected values (for large sprinkling densities) agree with the retarded propagators in dimensions $d=1,2,3,4$. This apparent success is actually somewhat troubling.
We would have preferred a \emph{single} model which agreed with the retarded propagators in \emph{different} dimensions. As it stands, to obtain agreement one needs to know in advance the dimension of the background spacetime that is being sprinkled into.

The most direct solution to this difficulty is to recognise that the $d=4$ case is the only physically relevant one\footnote{Setting aside the possibility that spacetime has extra dimensions.}. If we use the $d=4$ model (i.e. summing over paths) for \emph{all} causal sets then when the causal set happens to be a sprinkling into $\mink^4$ the propagators will agree. When the causal set is a sprinkling into $\mink^2$, say, the propagators will disagree but this may be no great loss.

Indeed, perhaps it is too optimistic to hope for a single causal model which will agree with the propagators in different dimensions. This is because the mass-dimension of the retarded propagators is $\lbrack \GR_m^{(d)} \rbrack = M^{d-2}$, a dimension-dependent quantity. The only dimensionful quantity available on a general causal set is the fundamental spacetime volume $V_0$. Therefore, if we were ingenious enough to find a model which agreed in all dimensions, it must depend on $V_0$ in an appropriate way to ensures the propagator's mass-dimension is correct. The trouble is that the mass-dimension of $V_0$ is not determined a priori. The dimensionality of $V_0$ has to be set by hand---presumably being set to $M^{-4}$ for any causal set.

To sum up, we suggest that the $d=4$ sum over paths model should be used for all causal sets (but this won't stop us from using the other models to test our ideas).

\section{Generalisations of the models}

\subsection{Infinite causal sets} \label{Sec:GeneralCausalSets}

We now consider generalising the path integral model to infinite causal sets.

\subsubsection{Incidence algebra} \label{Sec:IncidenceAlgebra}

Since a causal set is locally finite the propagator $K(u,v)$ for any two elements $u \preceq v$ can be calculated by
applying the methods of \secref{Sec:PathIntegralModels} to the finite interval $[u,v]$. There is another way,
however, to view the path integral framework which uses the \emph{incidence algebra} of a causal set.

For a causal set $\CS$ (not necessarily finite) we denote the set of all intervals by 
\be\textrm{Int}(\CS):= \{[u,v] |\, u, v \in \CS ,\, u \preceq v \}. \ee
For a function $f : \textrm{Int}(\CS) \to \mathbb{C}$ we write $f(u,v)$ for $f([u,v])$.

The incidence algebra \citep[Sec 3.6, p113]{Stanley:1986} of $\CS$ over $\mathbb{C}$, denoted
$I(\CS,\mathbb{C})$, is then the associative algebra of all functions \be f: \textrm{Int}(\CS) \to \mathbb{C}, \ee
with multiplication defined by \be (f * g) (u,v):= \sum_{u \preceq w \preceq v} f(u,w)g(w,v).\ee 
The sum here is
finite (so $f * g$ is defined) because the causal set is locally finite. $I(\CS,\mathbb{C})$ is an associative
algebra with two-sided identity \be \delta(u,v):= \left\{\begin{array}{ll} 1 & \textrm{ if } u = v
\\ 0 & \textrm{ otherwise. } \end{array}\right. \ee We note that we could use an algebraic field other
than $\mathbb{C}$ in defining the algebra. The causal and link matrices, which were defined only for finite causal
sets, generalize to the algebra elements \be C(u,v):= \left\{ \begin{array}{ll} 1 & \textrm{ if } u \prec v \\
0 & \textrm{ otherwise, }  \end{array}\right. \ee 
\be L(u,v):= \left\{ \begin{array}{ll} 1 & \textrm{ if } u \link v \\ 0 & \textrm{ otherwise. } \end{array}\right. \ee Powers of these functions, under $*$, satisfy: 
\be C^n (u,v) = \textrm{The number of chains of length $n$ from $u$ to $v$,}\ee 
\be L^n (u,v) = \textrm{The number of paths of length $n$ from $u$ to $v$,}\ee
\be (\delta + C)^n (u,v) = \textrm{The number of multichains of length $n$ from $u$ to $v$.}\ee 

The path integral work in \secref{Sec:PathIntegralModels} can be done using the incidence algebra. Phrasing the
method this way rather than using adjacency matrices allows us to work with infinite causal sets without
restricting to a finite sub-causal-set. For finite causal sets the incidence algebra and adjacency matrix
methods are entirely equivalent.

If we sum over chains we define an element $\Phi$ of $I(\CS,\mathbb{C})$ by \be \Phi(u,v):= a C(u,v).\ee If we
sum over paths we define \be \Phi(u,v):=  a L(u,v). \ee We then have, in a manner similar to the finite case,
that \be K := \delta + \Phi + b (\Phi * \Phi) + b^2 (\Phi * \Phi * \Phi) + \ldots, \ee 
is the algebra element for the propagator. Here $K(u,v)$ is the quantum mechanical
amplitude that the particle will travel from $u$ to $v$ along a trajectory of any length.

Formally we have \be K = \delta + \Phi * (\delta - b \Phi)^{-1},\ee but it is not immediately clear if an
inverse of $\delta - b \Phi$ exists. Applying Proposition 3.6.2 in \citet{Stanley:1986} shows,
however, that $(\delta - b \Phi)^{-1}$ exists so $K$ can be written in this form. The proposition also shows
that $K(u,v)$ depends only on the values of $\Phi$ for intervals contained within $[u,v]$.

It is pleasing to note that the choice of modelling spacetime as a causal set enables us to define its incidence
algebra which can then be used to define path integrals. The choice of causal set spacetime fits naturally with
the rules governing quantum mechanical amplitudes.

\subsection{Non-sprinkled causal sets}

The basic amplitudes for sprinkled causal sets depend on the sprinkling density $\rho$. For causal sets not
generated by a sprinkling, however, we must make sense of the $\rho$ that appears in
$a$ and $b$. To do this we assume that an arbitrary causal set element is assigned a fundamental spacetime
volume $V_0$. When the causal set is generated by a sprinkling into Minkowski spacetime with density $\rho$ we
have $V_0 = 1/\rho$. When the causal set is not generated by a sprinkling we simply replace the $\rho$ that appears in our amplitudes by $1/V_0$.

As an example, if the causal set represents a macroscopically 1+1 dimensional spacetime we sum over chains and assume $[V_0] =
M^{-2}$. This gives \be a = \frac{1}{2}, \qquad b = -m^2 V_0. \ee
If the causal set represents a macroscopically 3+1 dimensional spacetime we sum over paths and assume $[V_0] =
M^{-4}$. This gives \be a = \frac{1}{2 \pi \sqrt{6 V_0}}, \qquad b = -m^2 V_0. \ee

\subsection{Multichain path integral}

We have presented models for summing over chains and paths within a causal set. We now describe a model which sums amplitudes assigned to \emph{multichains}. For two related elements $u \preceq v$ in a causal set there are infinitely many multichains from $u$ to $v$. This is because multichains can contain arbitrarily long sequences of identical elements. Consider, for example, the sequence of multichains $u \preceq v$, $u \preceq u \preceq v$, $u \preceq u \preceq u \preceq v$ of lengths 1, 2 and 3 respectively. This infinitude of multichains is in stark contrast to the finite number of chains and paths between any two elements. Summing amplitudes assigned to multichains raises the issue of convergence of the sum---something which was not an issue for the chain and path sums.

\subsubsection{The model}

\def\aa{\bar{a}}
\def\bb{\bar{b}}

The amplitude we assign to a multichain of length $n$ is $\aa^n \bb^{n-1}$ for two complex number $\aa$ and $\bb$. For a finite causal set $(\CS,\preceq)$ with $p$ elements we then define a $p\times p$ matrix
\be \Phi:= \aa (I+C). \ee 
where $C$ is the causal matrix for $\CS$.

The propagator, summing amplitudes assigned to multichains, is then
\be  K:= \Phi + \bb\Phi^2 + \bb^2\Phi^3 + \ldots = \sum_{n=1}^\infty \bb^{n-1} \Phi^n,\ee 
just as in the chain and path models.

The convergence of this matrix geometric series, however, is only assured if the eigenvalues of $\bb\Phi$ are less than one in absolute values. Choosing a natural labelling for the causal set we can ensure that $C$ is strictly upper-triangular so that the diagonal entries of $\Phi$ are $\aa$. We thus see that the eigenvalues of $\bb \Phi$ are all equal to $\aa \bb$.

Assuming $|\aa\bb| < 1$ we then have that the geometric series converges and that 
\be K = \Phi(I - \bb \Phi)^{-1},\ee

\subsubsection{Relation to chain path integral}

We now show that there are values of $\aa$ and $\bb$ which ensure that the propagator calculated by summing over multichains is equal to the propagator $K$ calculated by summing over chains.

The result is obtained by straightforward calculation. We have:
\begin{align} K &= a C(I-abC)^{-1} = \frac{1}{b}(I - abC)^{-1} - \frac{I}{b} \\&= \frac{1}{b}\left( I+ ab I - ab(I + C)\right)^{-1} - \frac{I}{b}\\
&= \frac{1}{b(1+ab)}\left(I - \frac{ab}{1+ab} (I+C)\right)^{-1}-\frac{I}{b} \\
&= \frac{-a}{1+ab} I + \sum_{n=1}^\infty \frac{a^n b^{n-1}}{(1+ab)^{n+1}} (I+C)^n \end{align}

Choosing $\aa = a(1+ab)^{-2}$, $\bb = b(1+ab)$ then gives us the correct correspondence. Technically we have only found $\aa$ and $\bb$ that give the correct correspondence for distinct elements  $v_x \neq v_y$. When looking at the propagator to go from one element to itself the chain sum path integral always satisfies $K_{xx} = 0$ whereas the multichain path integral satisfies $K_{xx} = \aa/(1-\aa \bb)$.

\subsection{Path integral on a lightcone lattice}

A 1+1 dimensional lightcone lattice is an example of a causal set, albeit a particularly \emph{regular} causal set. Because of its regularity we can explicitly calculate the propagator when summing over chains in the lattice.

First of all we must determine the causal matrix for the lightcone lattice. We start by observing that the causal order relation in 1+1 Minkowski spacetime is a direct product of two total orders. This is best seen if we use lightcone coordinates $u=(x^0+x^1)/\sqrt{2}$ and $v=(x^0-x^1)/\sqrt{2}$. For two points $x = (u_1,v_2)$ and $y=(u_2,v_2)$ we then have $x \preceq y$ if and only if $u_1 \leq u_2$ and $v_1 \leq v_2$.

\begin{figure}[!h]
\begin{center}
\includegraphics[width = \textwidth]{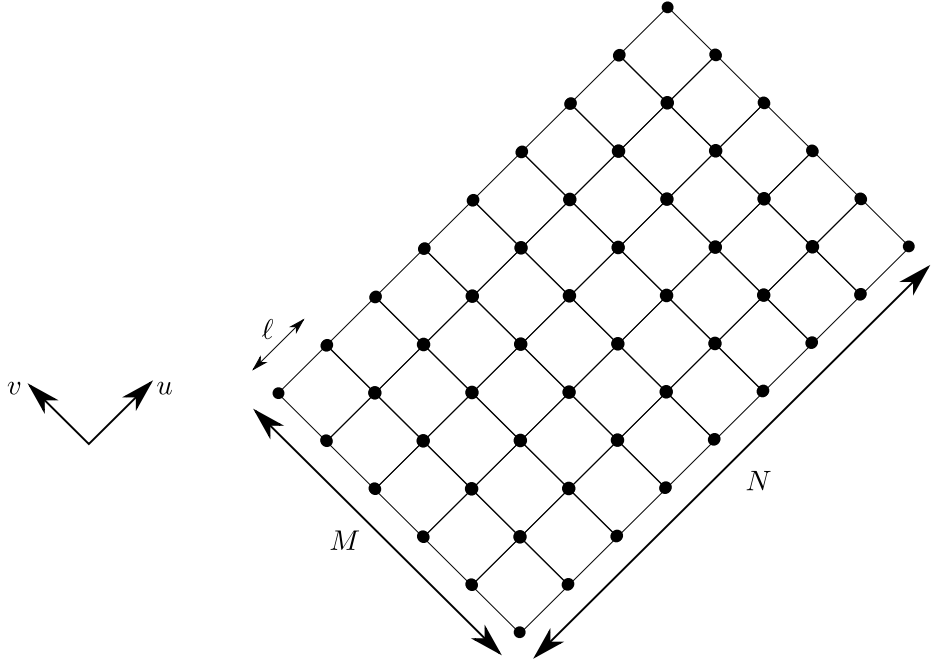}
\caption[A 1+1 lightcone lattice]{A 1+1 lightcone lattice with lattice spacing $\ell$.}
\label{fig:Lightconelattice}
\end{center}
\end{figure}

Consider the total ordered set of integers $(\mathbb{Z}_n, \leq)$ where $\mathbb{Z}_n := \{ 1,2,\ldots,n\}$ and $\leq$ is the usual ``less than or equal to'' order relation. If our lightcone lattice is a $N \times M$ rectangular array of points (see \figref{fig:Lightconelattice}) with lightcone coordinates $(i, j)$ (with $i=1,\ldots,N$, $j=1,\ldots,M$) then the causal set is just the product order $(\mathbb{Z}_N, \leq) \times (\mathbb{Z}_M, \leq) = (\mathbb{Z}_N \times \mathbb{Z}_M,\preceq)$.

It's helpful to define the $n\times n$ matrix $A_n$ given by
\be (A_n)_{ij} = \left\{ \begin{array}{ll} 1 & \textrm{if $i \leq j$} \\ 0 & \textrm{ otherwise.} \end{array}\right.\ee
for $x,y=1,\ldots,n$ (note that this is non-zero on the diagonal).

The causal matrix for $(\mathbb{Z}_N \times \mathbb{Z}_M,\preceq)$ is then the $(NM)\times (NM)$ matrix
\be C = A_N \otimes A_M - I_N \otimes I_M, \ee
where $I_n$ is the $n\times n$ identity matrix.

We then have that \be (i,j) \prec (i',j') \iff (e_i \otimes e_j)^T C (e_{i'} \otimes e_{j'}) = 1. \ee
Powers of $A_n$ are given by a binomial coefficient (with the convention that $\binom{x}{y} = 0$ if $y$ is negative):
\be((A_n)^r)_{ij} = \binom{r+j-i-1}{j-i}, \ee
(this can be shown using \eqref{eq:TotalOrderPowers}).

We can now calculate the propagator when summing over chains in the usual way:
\be K = a C + a^2 b\, C^2 + \ldots = \sum_{n=1}^\infty a^n b^{n-1} \left(A_N \otimes A_M - I_N \otimes I_M \right)^n.\ee
If $b=0 $  then $K = aC$. If $b\neq 0$ we have
\begin{align} K &= \frac{1}{b} \left(I_N \otimes I_M - ab(A_N \otimes A_M - I_N \otimes I_M)\right)^{-1} - \frac{1}{b} I_N \otimes I_M \\
&= \frac{1}{b} \left((1+ab) I_N \otimes I_M - ab(A_N \otimes A_M)\right)^{-1} - \frac{1}{b} I_N \otimes I_M \\
&=  \frac{1}{b(1+ab)} \left(I_N \otimes I_M - \frac{ab}{1+ab}(A_N \otimes A_M)\right)^{-1} - \frac{1}{b} I_N \otimes I_M \\
&=  \frac{1}{b(1+ab)} \sum_{n=0}^\infty \left(\frac{ab}{1+ab}\right)^n (A_N \otimes A_M)^n - \frac{1}{b} I_N \otimes I_M \end{align}
The amplitude to propagate from $x=(i,j)$ to $y=(i',j')$ (with $i' > i$ and $j' > j$) is then $K_{xy}$ which is given by
\begin{align} &= (e_i \otimes e_j)^T K (e_{i'} \otimes e_{j'}), \\&=  \frac{1}{b(1+ab)} \sum_{n=0}^\infty \left(\frac{ab}{1+ab}\right)^n \binom{n+i'-i-1}{i'-i} \binom{n+j'-j-1}{j'-j}. \end{align}
This sum can be evaluated explicitly. For $i' > i$ and $j' > j$ it is given by a hypergeometric function:
\begin{align}
&= \frac{a}{(1+ab)^2} \phantom{1}_2F_1 \left(\left. \begin{array}{cc} i'-i+1, & j'-j+1 \\ \multicolumn{2}{c}{$1$} \end{array} \right\vert \frac{ab}{1+ab} \right) \\
\label{eq:LatticePropagator} & = \frac{a}{(1+ab)^{1-i'+i}}\phantom{1}_2F_1 \left(\left. \begin{array}{cc} i'-i+1, & j-j' \\ \multicolumn{2}{c}{$1$} \end{array} \right\vert - ab \right) 
\end{align}
where the second line was obtained from the first by a hypergeometric identity\footnote{\citet[eq 15.3.4]{Abramowitz:1965}:\[ \phantom{1}_2F_1 \left(\left. \begin{array}{cc} a,& b \\ \multicolumn{2}{c}{$c$} \end{array} \right\vert z \right) = (1-z)^{-a} \phantom{1}_2F_1 \left(\left. \begin{array}{cc} a,& c-b \\ \multicolumn{2}{c}{$c$} \end{array} \right\vert \frac{z}{z-1} \right).\]}. For $x=(i,j), y=(i',j')$ with $i' \leq i$ or $j' \leq j$ we have $K_{xy} = 0$.

Having obtained an explicit form for the retarded propagator on a $1+1$ lightcone lattice we can ask how it compares to the continuum propagator. Fix two points $x,y$ in 1+1 dimensional Minkowski spacetime. Without loss of generality we choose $x = (0,0)$, $y=(U,V)$. We now lay down a lightcone lattice which includes $x$ and $y$ and calculate the propagator on the lattice.

We suppose the lattice between $x$ and $y$ is an $N \times M$ rectangular array (see \figref{fig:Lightconelattice}). 
The total number of lattice points is $N M$ and the total spacetime volume of the region is $UV$. This gives a density of lattice points of 
\be \rho = \frac{NM}{UV} = \frac{2 N M}{\tau^2},\ee
where $\tau$ is the proper time from $x$ to $y$.

We can plug in the appropriate $a$ and $b$ amplitudes:
\be a = \frac{1}{2} \qquad b = -\frac{m^2}{\rho} = -\frac{(m \tau)^2}{2 NM},\ee 
Substituting these into \eqref{eq:LatticePropagator} we have the amplitude to propagate from $x$ to $y$
\be \label{eq:1+1:LatticProp}\frac{1}{2}\left(1-\frac{(m\tau)^2}{4 NM}\right)^{1-N} \phantom{1}_2F_1 \left(\left. \begin{array}{cc} N+1, & -M \\ \multicolumn{2}{c}{$1$} \end{array} \right\vert \frac{(m \tau)^2}{4 NM}\right). \ee

Perhaps surprisingly the limit as $N, M \to \infty$ of this hypergeometric function can be calculated. Using \citep[Sec 5.7, p154]{Watson:1958} we find the limit of \eqref{eq:1+1:LatticProp} is equal to $\frac{1}{2} J_0(m \tau)$. Thus the continuum limit of the lattice propagator is indeed $\GR_m^{(2)}(y-x)$.

\section{General spacetimes}

So far in this chapter we have dealt with path integral models which, when computed on causal sets sprinkled into $\mink^d$, correctly reproduce the flat-spacetime retarded propagators for various dimensions. In defining these models our aim was to choose parameters so that the causal set propagator would match with the flat spacetime propagator.

Here we examine to what extent the models agree with retarded propagators in Lorentzian manifolds other than Minkowski and to what extent they need modification.

\subsection{Non-trivial topology}

If we sprinkle into a Lorentzian manifold with a different topology to $\mink^d$ the path integral model requires modification when dealing with regions that are large enough to ``see the topology''.

\subsubsection{Spatial circle}

\begin{figure}[!h]
\begin{center}
\includegraphics[width = 0.4\textwidth]{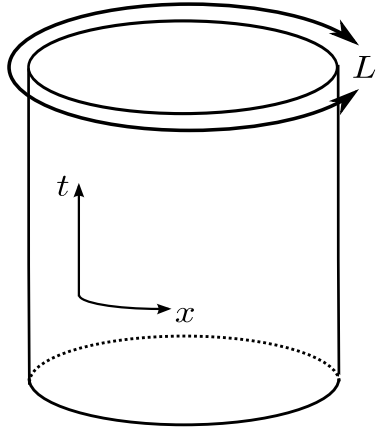}
\caption{The cylinder spacetime.}
\label{fig:Cyl}
\end{center}
\end{figure}

Consider the 1+1 dimensional cylinder spacetime $M = \mathbb{R} \times S^1$ obtained by identifying the spatial part of $\mink^2$ periodically (see \figref{fig:Cyl}). Trying to match the causal set propagator to the continuum propagator here is revealing.

The continuum retarded propagator on $M$ can be computed by the method of images. One imagines unwrapping the cylinder to give multiple copies of a strip of Minkowski spacetime. Points in this strip have coordinates $(t,x) \in \mathbb{R} \times [0,L]$ (where $L$ is the circumference of the cylinder). The propagator $G(x,y)$ for a pair of points $x, y \in M$ is found by summing the Minkowski spacetime propagator for the image points of $x$ and $y$ under the periodic identification.

Explicitly we have
\be G(x,y) = \sum_{n=-\infty}^\infty \GR_m^{(2)}(y-x+nv), \ee
where $v = (0,L)$ is the vector that translates from one strip to the next. For fixed $x$ and $y$ the contribution to $G(x,y)$ actually involves only a finite sum because $\GR_m^{(2)}(y-x+nv) \neq 0$ for only a finite number of $n$.

\begin{figure}[!hp]
\begin{center}
\includegraphics[width = 0.8\textwidth]{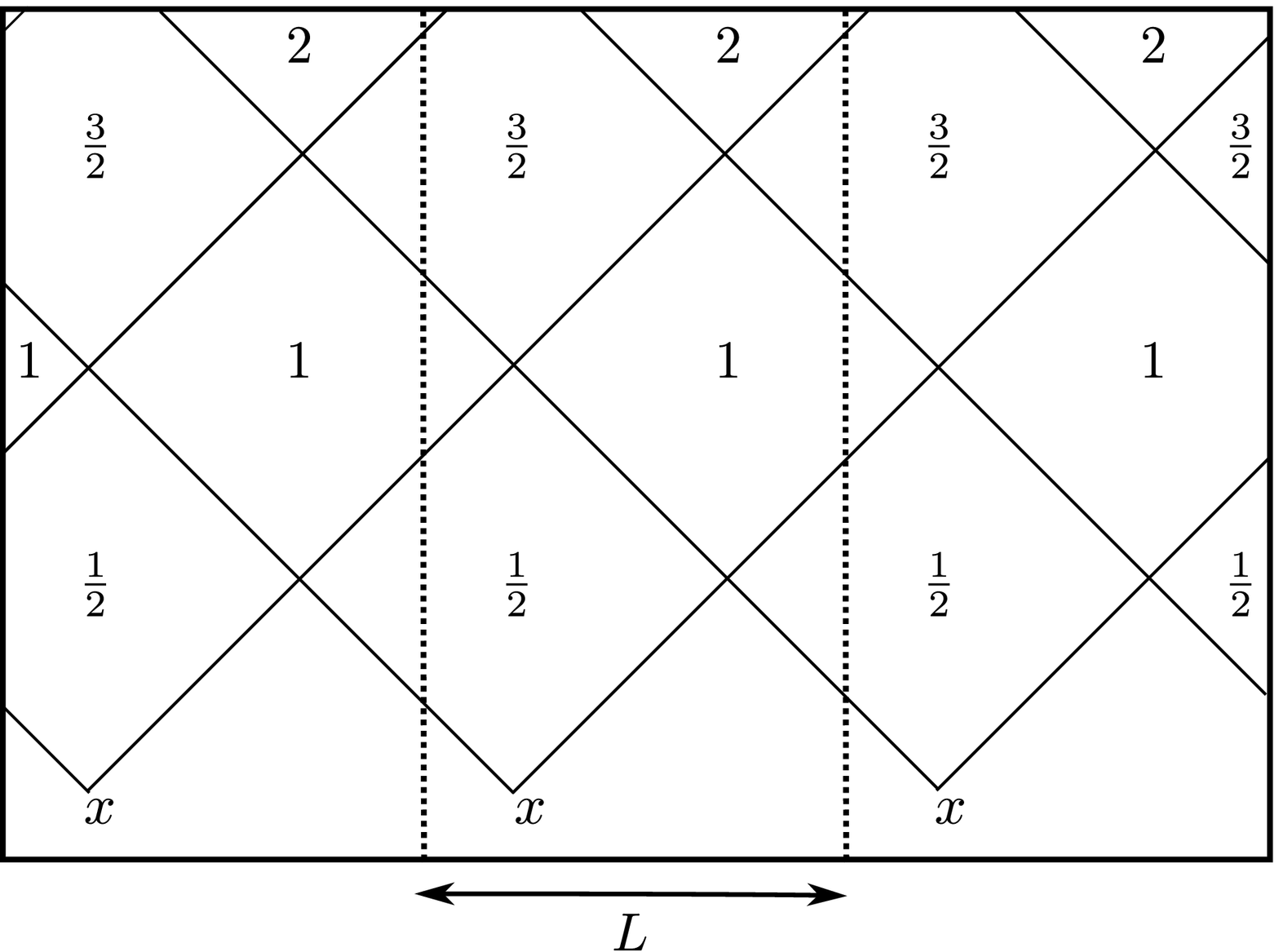}\\
\caption[Massless continuum propagator on the cylinder]{Values of the massless propagator for the cylinder spacetime. We have drawn multiple copies of the unwrapped the cylinder.}
\label{fig:Cylinder}
\vspace{2cm} 
\includegraphics[width = 0.8\textwidth]{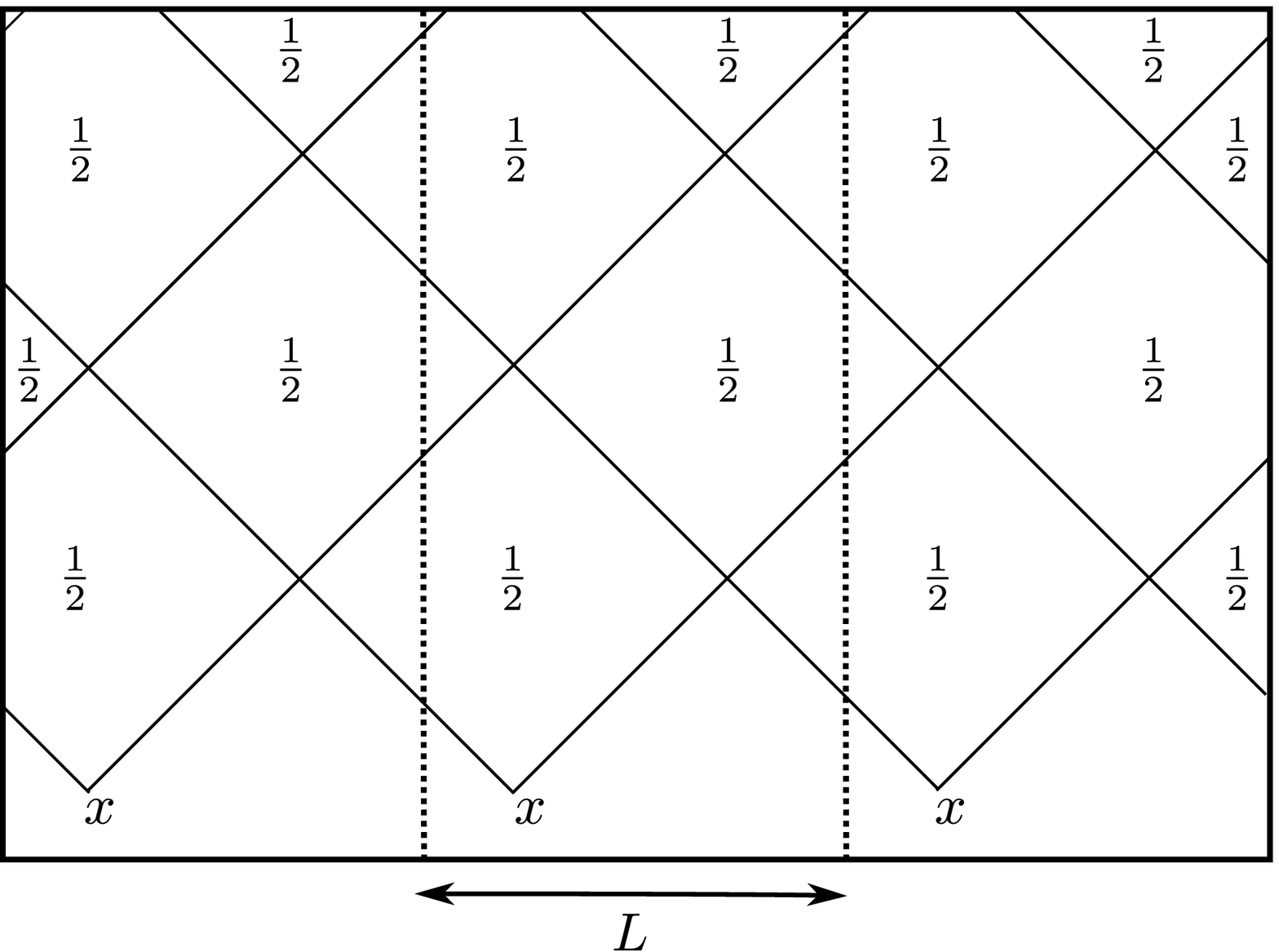}\\
\caption[Massless causal set propagator on the cylinder]{The expected value of the massless causal set propagator for sprinklings into the cylinder spacetime.}
\label{fig:CylinderCauset}
\end{center}
\end{figure}

The massless case illustrates the issues involved most clearly. In \figref{fig:Cylinder} we show the multiple copies of the unwrapped cylinder. For a fixed point $x \in M$ the values of the massless propagator $G(x,y)$ are shown when $y$ lies within different regions in $M$.

In the causal set path integrals suitable for sprinklings into 2-dimensional spacetimes the massless propagator is $K_R := 1/2 C$ where $C$ is the causal matrix for the sprinkling. In \figref{fig:CylinderCauset} we show, for the same point $x$, the expected value of the causal set propagator for the same regions (i.e. the expected value of $1/2 C$ for sprinklings into $M$.)

If the region being sprinkled into does not wrap around the cylinder, i.e. it's too small to ``see the topology'', then the expected value of the causal set propagator equals the continuum propagator. However, for \emph{general} regions on the cylinder, it is clear from the figures that the expected value of the causal set propagator does not equal the continuum propagator---the causal set path integral does not take account of the cylinder topology.

The difficulty is that the causal relations only encode the spatial topology \emph{implicitly}. As an example, consider two causally related sprinkled points $x, y \in M$. There may be a causal curve that goes from $x$ to $y$ without wrapping around the cylinder. Then again, there may be a causal curve that wraps once around (in either direction) or twice around etc. In short, there may be many classes of causal curves from $x$ to $y$, each distinguished by the number of times the curves wrap around the cylinder (i.e. each class is a homotopy class of future-directed causal curves\footnote{See \citet[Chpt 34]{Geroch:1985} for an introduction to homotopy. The massless $G(x,y)$ is essentially just one half the number of homotopy classes of future-directed causal curves joining $x$ and $y$.}). When we sprinkle a causal set into $M$ this wealth of information (i.e. the number of times causal curves can wind around the cylinder) is simply recorded as ``$x \preceq y$''.

If we only focussed on the pair $x$ and $y$ then it appears the causal relations cannot tell us this topological information. Nevertheless the causality theorems discussed in \secref{sec:CausalStructure} ensure that the causal relations \emph{do} encode the topology of the manifold. The information, however, is encoded in the whole causal structure, not just a single relation.

Within the causal set program a crucial idea is that the topology of the manifold can be deduced from a sprinkled causal set if the sprinkling density is sufficiently high. Recovering this information, however, is difficult (some approaches to this include \citet{Major:2007,Surya:2008}). If we were able to determine the topology of the cylinder spacetime in terms of the sprinkled causal set then it might be possible to modify the path integral to ensure it agrees with the continuum.

\subsubsection{Spatial 2-torus}

The causal set path integral is in an even worse situation in 4 dimensions. Consider the spacetime $M = \mathbb{R} \times T^3$, where the 3-dimensional space is periodically identified to make a spatial 2-torus. Suppose we sprinkled into this spacetime and computed the massless causal set propagator. Since this is a 4-dimensional spacetime the massless propagator is just $\sqrt{\rho}/(2\pi\sqrt{6}) L$ where $L$ is the link matrix for the sprinkling.

In this spacetime, however, if you go far enough into the future of any sprinkled point eventually there will be \emph{no more links}. The links cease (or more correctly, are exponentially suppressed) because eventually, due to the spatial identifications, the causal interval between two events separated far enough in time is so large that it is very unlikely to contain no sprinkled points.

The continuum massless propagator is not exponentially suppressed in the far future which indicates that the causal set path integral does not agree for sprinklings into this manifold.

\subsubsection{Modifications to the model}

One can invent ad hoc modifications of the path integral model to fix its deficiencies for particular spacetimes (for a modification that ensures agreement for the 1+1 dimensional cylinder see \citet{Schmitzer:2010}). Since this depends on knowing the particular background spacetime, however, this is unsatisfactory. Instead we would like a framework that would work for sprinklings into any causal Lorentzian manifold, of any topology.

It's likely that to correctly modify the model we would need a theory of ``causal set homotopy''. This would allow us to assign to each causal relation information about the homotopy class of causal curves that connect the two sprinkled points. Ideally this would be deduced from the causal set itself without recourse to the background manifold. The usual homotopy theory is heavily based on continuum ideas: i.e. continuous deformations of continuous curves. It might be enough to find a suitable topology for ``curves'' (for example, chains or paths) in the causal set such that continuous deformations can be defined.

\subsection{Curved spacetimes} \label{sec:CurvedSpacetimes}

Perhaps the most interesting question is how well the causal set path integral model extends to sprinklings into \emph{curved} spacetimes. Now we briefly outline the formalism for scalar field theory in general Lorentzian manifolds (we follow \citet{Fulling:1989} and \citet[\S3.2]{Birrell:1982}).

If $(M,g)$ is a Lorentzian manifold then the generalisation of the Klein-Gordon equation for a scalar field $\phi$ is \citep[eq 6.2, p117]{Fulling:1989}: 
\be (\Box + m^2 + \zeta R(x)) \phi(x) = 0. \ee
Here $\zeta$ is a constant, $R(x)$ is the Ricci curvature of the manifold and
\be\Box \phi = \frac{1}{\sqrt{g}} \partial_\mu \lbrack g^{\mu \nu} \sqrt{g} \partial_\nu \phi\rbrack, \ee
is the generalisation of the d'Alembertian to curved spacetimes (with $g := |\det(g^{\mu \nu})|$).

Green's functions for this equation are distributions $G(x,y)$ satisfying \citep[p125]{Fulling:1989}:
\be (\Box_x + m^2 + \zeta R(x)) G(x,y) = \frac{\delta(x-y)}{\sqrt{g}}. \ee
Solutions to this equation depend on the particular Lorentzian manifold that is being used and calculating $G(x,y)$ explicitly is very difficult for all but the simplest cases. We note, however, that for any globally hyperbolic spacetime this equation is guaranteed to possess unique retarded and advanced solutions (characterised by their support properties in spacetime \citep[p79, p125]{Fulling:1989}).

For sprinklings into a causal Lorentzian manifold the expectation value of the causal set propagator, averaged over sprinklings, is a retarded function. It is possible that for certain Lorentzian manifolds this function reproduces the continuum retarded Green's function.

One can investigate this by looking at simple examples. One of the simplest are sprinklings into 1+1 dimensional Lorentzian manifolds. Such manifolds are conformally flat so their causal relations can be calculated from the causal relations in $\mink^2$.

Another place where progress might be made is for sprinklings into spacetimes with constant curvature: i.e. de Sitter or anti-de Sitter (for definitions see \citet[\S 5.2, p124]{Hawking:1973}). Again these are conformally flat so their causal relations can be computed using those of $\mink^d$. In addition \citet[Thm III.4, p68]{Meyer:1988} gives expressions for the expected number of chains for sprinklings into de Sitter and anti-de Sitter. While the expressions are complicated it is possible that they could be used to find an exact expression for the expected value of the sum-over-chains propagator in these spacetimes.

We mention that the path integral has the potential to be cope with position-dependent curvature. This is because curvature of the background manifold affects the distribution of sprinkled points. This in turn affects the distribution of chains and paths in the causal set which then affects the combinatorics in the sum over trajectories. It remains possible that without changing the hop and stop amplitudes the model could correctly reproduce the curvature dependence in the continuum Green's functions.

\subsubsection{Huygens Principle and Tails}

One unusual phenomenon that occurs in curved spacetime is the presence of \emph{tails}. This is the name used when the Green's function for the d'Alembertian is non-zero for chronologically related points rather than just for null-related points. In effect these non-zero ``tail-terms'' are due to ``back-scattering'' of the field off the spacetime curvature.
The \emph{absence} of tails in the propagation of fields is one way to characterise Huygens' principle: that radiation should propagate only along the lightcone (\citet[p7182]{Bombelli:1994}, \citet{Sonego:1991}).

The massless propagator for sprinklings into 3+1 dimensional spacetimes is proportional to the link matrix of the causal set. This means that in general its expectation value will be strongly peaked on the lightcone and exponentially small away from the lightcone. We see therefore that we don't expect the causal set propagator to reproduce a Green's function with a tail-term. Physically this says that massless particles propagate along the lightcone\footnote{or at least the causal set analogue of the lightcone.}, i.e. Huygens' principle holds for the causal set model.

It thus seems that the 3+1 dimensional model would require modification if it were to reproduce a Green's function with a tail. Since these tails have not been observed experimentally it is intriguing to speculate that the causal set propagator might be the physically correct one!

\section{Uses for the retarded propagator}

We've concentrated so far on ways to derive the retarded propagator on a causal set. In the next chapter we shall use this work to obtain a quantum scalar field theory and the Feynman propagator. For now we briefly mention some possible uses that the retarded propagator could be put to on its own.

\subsection{Yang-Feldman formalism}

The Yang-Feldman formalism \citep{Yang:1950} is an approach to quantum field theory that uses the retarded and advanced propagators. Their paper deals with quantum electrodynamics but the ideas apply to other theories. We briefly outline the method applied to a self-interacting scalar field.

For such a theory the equation of motion is
\be (\Box + m^2) \p(x) = \hat{f}(x), \ee
where $\p$ is a field operator (acting on some Fock space) and $\hat{f}$ is an operator representing an interaction (possibly depending on $\p$, e.g. $\hat{f}(x) = \lambda \p^3(x)$). This equation can be integrated using the retarded and advanced propagators to give
\be \label{eq:YFIn} \p(x) = \p_{in}(x) + \int d^dy \GR^{(d)}_m(x-y) \hat{f}(y), \ee
\be \label{eq:YFOut} \p(x) = \p_{out}(x) + \int d^dy \GA^{(d)}_m(x-y) \hat{f}(y), \ee
where $\p_{in}$ and $\p_{out}$ are solutions to the free field equation:
\be (\Box + m^2) \p_{in}(x) = (\Box + m^2) \p_{out}(x) = 0. \ee
The incoming (outgoing) field coincides with $\p$ in far past (future) and represent what the field would have been if the interaction was absent. Both the incoming and outgoing fields satisfy the same free field commutation relations:
\be \lbrack \p_{in}(x), \p_{in}(x) \rbrack = \lbrack \p_{out}(x), \p_{out}(x) \rbrack = i \Delta(y-x), \ee
where $\Delta(x) := \GR_m^{(d)}(x) - \GA_m^{(d)}(x)$ is the Pauli-Jordan function\footnote{This will play a central role in the next chapter.}. Since the incoming and outgoing fields satisfy identical commutation relations they must be related by a unitary transformation in the following way:
\be \label{eq:YFCommutation} \p_{out} = \hat{S}^{-1} \p_{in} \hat{S}. \ee
This unitary operator $\hat{S}$ defines the S-matrix for the theory.

To evaluate the matrix elements of the $\hat{S}$ operator the equations \eqref{eq:YFIn} and \eqref{eq:YFOut} are solved iteratively by successive approximations. These approximations are then combined with \eqref{eq:YFCommutation} to compute the matrix elements to a particular order (in a suitable coupling constant). This procedure is described in general terms in \citet[\S8-7, p167]{Jauch:1955}. Apparently this procedure is entirely equivalent to the usual Feynman diagram methods \citep[p169]{Jauch:1955}.

This approach gives a central role to the retarded and advanced propagators. If we only have these propagators available on a causal set then perhaps they could be used to define a causal set quantum field theory. In the next chapter, however, we show how to define quantum field theory on a causal set in a simpler manner.

\subsection{Feynman tree theorem}

In quantum field theory the Feynman propagator plays an important role. When evaluating Feynman diagrams, for example, it is the \emph{Feynman} propagator that is associated to each leg of the diagram. It is interesting to wonder if a similar use can be found for the \emph{retarded} propagator which reproduces the same results---can we reformulate quantum field theory using the retarded propagator?

A tantalising step in this direction is suggested in \citet{Feynman:2000} (also available in \citet[p355-375]{Klauder:1972}). There a method is described which relates scattering amplitudes calculated with the Feynman propagator to scattering amplitudes calculated with the retarded propagator.

The method relies on the fact that the Feynman propagator $G_F(x)$ and retarded propagator $G_R(x)$ are related by
\be G_F(x) = G_R(x) + G_s(x), \ee
where $G_s(x)$ is a solution to the Klein-Gordon equation.

By substituting this equation into the amplitude for a Feynman diagram containing a closed loop we can re-express the amplitude as sums of ``tree diagram'' amplitudes (see \citet{Feynman:2000} for full details). These tree diagrams contain no closed loops and it is the \emph{retarded} propagator that is assigned to each leg. A single loop diagram splits in this way into a sum over a number of different tree diagrams, each containing additional on-shell incoming and outgoing particles. This result is known as the Feynman tree theorem.

In the paper it is subsequently claimed that the amplitude for complete processes can be obtained by calculating with the \emph{retarded propagator only}, but making sure to include suitable additional on-shell particles---see \citet[eq $(22)^\prime$]{Feynman:2000}.

This method potentially offers a way to re-express quantum field theory using only the retarded propagator. It requires, however, knowing how to sum over incoming and outgoing particles which have definite on-shell momentum. Carrying this over to define a field theory on a causal set might be possible but requires the notion of on-shell momentum on a causal set (see \secref{sec:NormalModes} for ideas in this direction).

\subsection{Action-at-a-distance} \label{Sec:ActionAtADistance}

It is possible to reformulate classical (and quantum) electrodynamics as a particle theory involving direct inter-particle interactions. In such a theory there are no fields, only particles subject to certain relativistic interactions. Such a reformulation is known as an ``action-at-a-distance'' theory.

Perhaps the best known such reformulation is the Wheeler-Feynman absorber theory \citep{Wheeler:1945,Wheeler:1949}. This is a reformulation of classical electrodynamics in which there is no radiation field. Instead the theory describes the motion of charged particles subject to direct interactions. These interactions are governed by both the retarded \emph{and advanced} propagators. The advanced effects, which are usually thought to lead to paradoxical ``effect precedes cause'' behaviour, turn out to cancel completely if the effect of the rest of the universe is included. Matter elsewhere in the universe plays an essential role as a giant absorber whose advanced effects are needed to ensure that the theory appears to use only the retarded propagator.

The physical picture in this theory is very different to the usual electrodynamics but, perhaps surprisingly, it is possible to show that the theory reproduces the same results as conventional electrodynamics.

Given that such a familiar theory as classical electrodynamics can be so radically reformulated yet still reproduce the same physical results, it is interesting to speculate on whether a reformulation of standard physical theories (in such action-at-a-distance terms) could help when trying to define them on a causal set. Certainly the non-locality of a causal set (in which each element is linked to a large number of other elements) fits more closely with an action-at-a-distance formulation rather than a local differential description.

One attempt to formulate action-at-a-distance electrodynamics on a discrete spacetime was given in \citet{Hemion:1988}. There he formulates a theory of particles which exist within a locally finite partially ordered set. By identifying analogues of proper distances in the discrete spacetime he aims to assign an action to particle configurations. The hope is that by extremising this action it is possible to identify the dynamically allowed particle configurations. Unfortunately the continuum action depends on the velocities of particles and Hemion side-steps the difficulties present in defining such velocities on a discrete spacetime \citep[p1193]{Hemion:1988}.

The action-at-a-distance formulation of physics has also been advocated by \citet{Hoyle:1974}. In this set of lectures they discuss reformulating quantum electrodynamics as an action-at-a-distance theory. Interestingly in their Section 8.1 they discuss mass scatterings very similar to those in our \secref{sec:MassScatterings}. They discuss a path integral in which the particle can, in our language, ``hop'' into the future \emph{or past} lightcone at each step (i.e they use half the sum of the retarded and advanced massless propagators as their massless propagator).

It is them emphasised that, since such zig-zag paths can stray arbitrarily far from the initial or final points, to assign them the correct amplitude we must know how the ``mass field'' $m$ behaves everywhere in the universe (p144). If we allow ignorance of this behaviour and restrict the path integral to a finite region then, they claim, we can derive the Feynman propagator as the sum of the free particle propagator plus the ``response of the universe'' (p144-145). If this is correct then it presents an intriguing way to derive the Feynman propagator from the retarded and advanced propagators.

It's possible that action-at-a-distance ideas could lead to a new way to define physical theories on a causal set. Unfortunately in theories of this type the analysis of even a small-scale physical system requires that the behaviour of the entire universe be included. To implement this on a causal set it seems clear that knowledge of the entire causal set is needed, as well as a way to ``screen out'' the background behaviour of the rest of the universe. While this sounds like a  formidable task it's possible that future developments could lead in this direction.

\chapter{Free Quantum Field Theory} \label{chap:FreeQFT}

\begin{quote}
I had asked Salam ``what IS a quantised field'', and received the answer ``Good; I was afraid you would ask me something I did not know. A quantised field, phi(x) at the point x of space-time, is that operator assigned by the physicist using the correspondence principle, to the classical field phi at the point x''. I went away thinking about this; then I realised that what I needed was a statement of WHICH operator is assigned by the physicist.

\flushright{Ray Streater, 1957, in \citet{Streater:2000}.}
\end{quote}

\section{Motivation}

The best theory to describe matter at a fundamental level is currently quantum field theory. This theoretical framework grew out of efforts to combine quantum mechanics with special relativity in the 1920s-1940s. The most successful quantum field theory is the Standard Model which was developed in the 1960s-1970s. This successfully describes the electromagnetic, weak and strong fundamental forces. Despite the \emph{physical} successes of the quantum field theory framework its \emph{mathematical} underpinnings have remained poorly-defined.

One of the major obstacles to a mathematically well-defined quantum field theory is the presence of divergences. These are typically divergent integrals or sums which must be regularised before finite answers can be calculated. The removal of the divergences presented a serious obstacle in the development of the theoretical framework. In the 1940s renormalisation techniques were developed which avoided the divergences and enabled finite answers for physical predictions to be obtained. These predictions matched experiments very well and renormalisable quantum field theories became established as a central part of modern physics.

The modern perspective on renormalisation is that quantum field theories provide only an \emph{effective description} of matter. The idea is that we simply do not know the precise theory of physics at very small length scales (or equivalently very high energies). If we admit our ignorance about this and ``integrate out'' the small-scale behaviour of the theory we are left with an effective description that gives us the correct answer at the length-scales we're probing.

An alternative approach is to try to formulate a model for matter which avoids the divergences from the beginning. This is one of the motivations for considering discrete spacetime and is the approach we are aiming at.

\subsection{Discrete spacetime and field theory}

An obvious way to try to curb the divergences is to discretise spacetime. The most direct way, and one that it frequently used, is to ignore relativistic invariance and simply replace the continuum spacetime with a hyper-cubic lattice.

One early approach along these lines was the 1930 suggestion by Heisenberg that the world is a ``lattice world''---a \emph{Gitterwelt}. In this sketch of an idea the differential equations of physics were to be replaced by finite-difference equations. By this route it was hoped that the divergences could be avoided.

Discrete spacetime also appears in a more pragmatic, less foundational approach known as lattice field theory. Since the continuum theory is difficult to solve analytically one can try to numerically solve the equations iteratively on a computer. To make the problem well-posed for numerical simulations this usually involves Wick rotating the theory to Euclidean space and then laying down a hyper-cubic lattice. This approach is particularly used for strongly-interacting quantum field theory where the usual perturbative expansion is not useful. Nevertheless the use of a Wick rotation and a hyper-cubic lattice mean the lattice that is used shares few of the properties of physical spacetime (e.g. spacetime's Lorentzian signature and Lorentz-invariance). 

\subsection{Causal set field models}

Quantum field theory on a causal set has been previously considered by a number of people. There are two main approaches to the problem---modelling matter \emph{as} the causal set or modelling matter \emph{on} the causal set.

\section{Scalar fields as geometry}

One approach to modelling fields on a causal set follows the spirit of Wheeler's geometrodynamics in which matter and spacetime are unified---indeed matter \emph{is} spacetime. Inspired by this approach one can imagine that individual particles or field configurations are really particular patterns of causal relations or clusters of causal set elements which propagate (or perhaps, repeat) through the causal set.

We mention a few ways in which the causal set itself could be used to encode a matter field (methods 1 and 2 have been considered by David Rideout).

\subsection{Method 1}

For a causal set $(\CS,\preceq)$ we define an equivalence relation $u \sim v$ if $u, v \in \CS$ share identical (strict) causal relations (i.e. $u$ and $v$ are a ``non-Hegelian pair''):
\be u \sim v \textrm{ if, for all $w, w' \in \CS$, we have } w \prec u \prec w' \iff w \prec v \prec w'.\ee
Each equivalence class under $\sim$ is an antichain of some cardinality (which may be equal to one). We can create a new causal set $(\CS',\preceq')$ by starting with the set of equivalence classes: $\CS' := \{[u] : u \in \CS \}$. This set inherits an order relation from $\CS$ in a natural way:
\be u \preceq v \implies [u] \preceq' [v], \ee
for all $u,v \in \CS$.

\begin{figure}[h]
\begin{center}
\includegraphics[width = 0.6\textwidth]{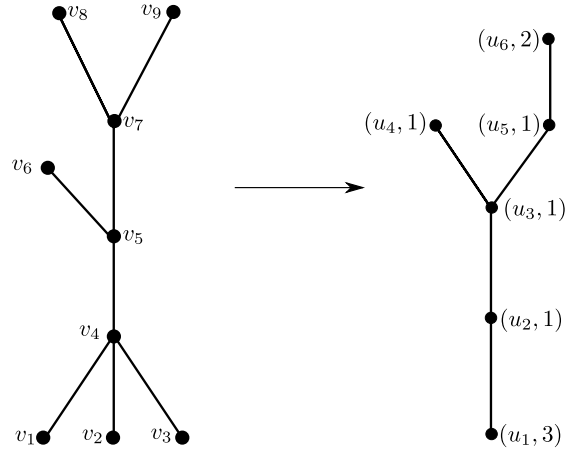}
\caption[Encoding a scalar field in terms of the causal set---method 1]{Two Hasse diagrams illustrating method 1. We have $u_1 = \{v_1,v_2,v_3\},\, u_2=\{v_4\},\, u_3 = \{v_5\},\, u_4 = \{v_6\},\, u_5 = \{v_7\},\,u_6=\{v_8,v_9\}$.}
\label{fig:ScalarField1}
\end{center}
\end{figure}

Our original causal set $(\CS,\preceq)$ is equivalent to the new causal set $(\CS',\preceq')$ if every element of $\CS'$ is assigned a natural number equal to the cardinality of the equivalence class (see \figref{fig:ScalarField1}).

In this way we can encode a positive integer-valued scalar field in terms of the causal set itself.

\subsection{Method 2}

Another approach involves dropping the requirement that $\preceq$ is antisymmetric. This leaves us with a \emph{preposet} (or \emph{quasi-ordered set}) $(\CS, \preceq)$ in which $\preceq$ is reflexive and transitive \citep[p153, Ex. 1a]{Stanley:1986}. If $\preceq$ is not antisymmetric then causal loops are allowed---these are $n$ distinct elements related as: $v_1 \prec v_2 \prec \ldots v_{n-1} \prec v_n \prec v_1$.

\begin{figure}[h]
\begin{center}
\includegraphics[width = 0.6\textwidth]{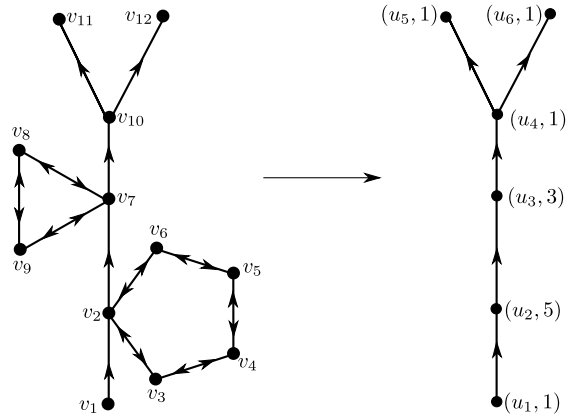}
\caption[Encoding a scalar field in terms of the causal set---method 2]{Directed graphs illustrating method 2. We have $u_1 = \{v_1\},\, u_2=\{v_2,v_3,v_4,v_5,v_6\},\, u_3 = \{v_7,v_8,v_9\},\, u_4 = \{v_{10}\},\, u_5 = \{v_{11}\},\,u_6=\{v_{12}\}$.}
\label{fig:ScalarField2}
\end{center}
\end{figure}

We can define an equivalence relation $u \sim v$ if $u,v \in \CS$ are in the same causal loop:
\be \textrm{$u\sim v$ if $u \preceq v$ and $v \preceq u$}. \ee
If we denote the set of equivalence classes under $\sim$ as $\CS'$ then we can define a causal set $(\CS',\preceq')$ if we define
\citep[p153, Ex. 1b]{Stanley:1986}:
\be [u] \preceq' [v] \textrm{ if there exists $u' \in [u], v' \in [v]$ such that $u' \preceq v'$}.\ee
We can thus view each loop as representing a single spacetime event endowed with a positive integer $n$ equal to the cardinality of the equivalence class (see \figref{fig:ScalarField2}).

Using methods 1 or 2 a positive-integer-valued scalar field can be encoded directly in terms of the causal set itself (albeit by perhaps slightly modifying the definition of a causal set in method 2).

\subsection{Kaluza-Klein theory on a causal set}

Another approach to modelling matter on a causal set is to attempt to define Kaluza-Klein theory on a causal set\footnote{This was suggested to the author by Sumati Surya.}. In standard continuum Kaluza-Klein theory the dynamics of pure gravity in a higher dimensional Lorentzian manifold give rise to coupled gravity-matter fields in a lower dimensional Lorentzian manifold.

To implement this idea for causal set theory one could imagine sprinkling into $\mink^4 \times S$ where $S$ is some compact Riemannian manifold (e.g. a circle or torus). One could then write down the gravitation dynamics for this causal set (e.g. the Einstein-Hilbert causal set action discussed in \secref{sec:EinsteinHilbert}). The idea is that this expression would split easily into two parts: (i) the 4-dimensional causal set action and (ii) an action for matter fields. By correctly identifying the action for matter fields it may be possible to define suitable quantities intrinsic to the causal set to serve as the matter fields themselves. So far, however, this remains only a sketch of an idea.

\section{Scalar fields on a causal set}

An alternative to modelling matter \emph{as} the causal set itself is to model matter \emph{on} the causal set. This approach is similar to the usual theories of matter described on a background Lorentzian manifold.

\subsection{The d'Alembertian}

A free classical scalar field on a Lorentzian manifold satisfies the Klein-Gordon equation. A natural step towards defining field theory on a causal set is to somehow discretise this equation. This involves finding a causal set analogue of the d'Alembertian differential operator.

There have been two main approaches to this problem. Both of them aim to find an expression intrinsic to the causal set that plays the role of the d'Alembertian operator.

Once we have such an expression we can check its validity as follows. Fix a finite region $R \subset \mink^d$ with a real scalar field $\phi : R \to \mathbb{R}$ defined on it. Fix a point $x \in R$. If we sprinkle a causal set into $R$ there is zero probability that $x$ will be a sprinkled point. We therefore add $x$ to the sprinkled causal set (and label it as element $v_1$, say). The generated causal set inherits the scalar field from $R$ by assigning the value of $\phi$ at each sprinkled point as the value of the causal set scalar field.

We can now apply the causal set d'Alembertian to the causal set field to obtain the d'Alembertian of the field at the element $v_1$ (i.e. the element corresponding to the point $x \in R$). By averaging this value over all sprinklings into $R$ (with the same density) we obtain the expected value of the causal set d'Alembertian of $\phi$ at $x$. We can compare this to $\Box \phi (x)$ to see if the causal set expression gives the right answer.

\subsubsection{Method 1: Inverting the Green's function} \label{sec:InvertingGreens}

The first attempt to obtain the d'Alembertian on a causal set was pursued by \citet{Daughton:1993} and later by \citet{Salgado:2008}. They start with the continuum equation satisfied by a Green's function $G(x)$ for the d'Alembertian in $\mink^d$:
\be \label{eq:DAlembertian} \Box G(x) = \delta^d(x).\ee
The idea is that on a finite $p$-element causal set this equation should be replaced by a matrix equation:
\be B G = I, \ee
where $B$ and $G$ are now $p \times p$ matrices and $I$ is the identity matrix\footnote{Strictly speaking, on dimensional grounds, the identity matrix should be multiplied by a factor of $1/V_0$ where $V_0$ is the dimensionful fundamental volume assigned to the causal set elements. This is to ensure that the causal set equation has the same mass-dimension as the continuum equation it seeks to emulate. In \citet{Daughton:1993} and \citet{Salgado:2008} this issue is ignored as they set $V_0 = 1$.}. $B$ is the candidate d'Alembertian operator and $G$ is a causal set analogue of a continuum Green's function.

As we have seen in the previous chapter it is possible to obtain causal set analogues for the retarded (and advanced) d'Alembertian Green's functions for sprinklings into $\mink^d$ for $d=1,2,3,4$. The idea is to invert these matrices to obtain a candidate for the d'Alembertian operator: $B := G^{-1}$.

Unfortunately the matrix for the retarded Green's function $K_R$ is non--invertible (we can see this because $K_R$ is zero on the diagonal and a labelling can be chosen to ensure it is strictly upper triangular). Two options have been explored to side-step this difficulty. Firstly one can include a non-zero diagonal value for the retarded propagator (in which case the matrix is guaranteed to be invertible). Secondly one can work with the symmetric half-advanced, half-retarded Green's function equal to $(K_R + K_A)/2 = (K_R + K_R^T)/2$ (which, depending on the causal set, may or may not be invertible).

For sprinklings into $\mink^2$, for example, $K_R = 1/2 C$ (where $C$ is the causal matrix for the causal set) and the two candidate d'Alembertians are either
\be B = \left(c I + \frac{1}{2} C\right)^{-1}, \ee
for some constant $c$ (usually take to be $1/2$) or
\be B = \left(\frac{1}{4}(C + C^T)\right)^{-1}. \ee

Both \citeauthor{Daughton:1993} and \citeauthor{Salgado:2008} have performed extensive numerical simulations which attempt to check the validity of these d'Alembertian operators for sprinklings into $\mink^2$. Their results suggest that the expected value of the causal set d'Alembertian is a good approximation to the continuum d'Alembertian for fields which vary slowly on the discreteness scale and are zero on the boundary of the sprinkled region.

\subsubsection{Method 2: Summing over layers}

A second approach to the d'Alembertian has been pursued by \citet{Sorkin:2007, Sorkin:2009, Benincasa:2010}. Here the approach is more direct and starts with an attempt to finite-difference the d'Alembertian differential operator. The requirement that this be done in a Lorentz-invariant way leads to the idea that the d'Alembertian of a field at an element $v_x$ in the causal set should be a sum over the values of the field on ``layers'' of causal set elements to the past of $v_x$.

These layers are defined in terms of the cardinality of the inclusive causal interval between an element and $v_x$. Layer $n$ is the set of all elements $u$ such that $|[u,v_x]| = n+1$. So, for example, the first layer is the set of all past links of $v_x$.

The d'Alembertian of the field at $v_x$ is then a sum of the field values over different layers, each layer weighted by a suitable coefficient. By looking at the expected value of such expressions it is possible to choose combinations of layers such that the causal set d'Alembertian is equal to the continuum d'Alembertian in the infinite sprinkling density limit. This method has a surprising connection with the retarded propagators derived in the previous chapter (see \secref{sec:3+1PathPDE}).

Unfortunately the fluctuations of the causal set d'Alembertian are large and grow with the sprinkling density. One way around this is to introduce a meso-scale which lies somewhere between the discreteness scale and the nuclear scale. The layers being summed over are then fattened out to the size of the meso-scale and in this way the fluctuations are dampened down (see \citet{Sorkin:2007} for details).

It remains to be seen how these d'Alembertians expressions can be used to define a field theory on a causal set.

\subsection{Field actions}

An alternative approach to defining scalar fields on a causal set has been pursued by \citet{Sverdlov:2009a, Sverdlov:2009b}. This approach starts with a classical field defined on a Lorentzian manifold. The Lagrangian for such a field is them re-expressed in terms of the causal structure, the volume element and the proper time, all of which have causal set analogs.

This expression can be simply discretised by making the appropriate substitutions. In this way it's possible to define a Lagrangian for a scalar field on the causal set. The resulting expressions, however, are very complicated and one has to work hard to ensure they don't pick out a particular reference frame.

Further work along these lines has been pursued by \citet{Sverdlov:2008a, Sverdlov:2008b, Sverdlov:2008c, Sverdlov:2009:PhD}. The same method, of re-expressing continuum quantities and then discretising, is applied to gauge fields and spinor fields. This top-down approach, which starts with continuum quantities and then finds a suitable discretisation is a good first-step but leads to causal set expressions that are complicated and somewhat artificial. An approach which instead starts with a theory defined on a causal set and then shows that it reproduces the continuum theory is more appealing.

\section{Quantum scalar field theory in Minkowski spacetime}

Before describing the model for free scalar quantum field theory on a causal set we first review scalar quantum field theory in $d$-dimensional Minkowski spacetime $\mink^d$.

\subsection{Feynman propagator} \label{sec:FeynmanPropagator}

In \secref{sec:KGPropagators} the retarded, $\GR_m^{(d)}(x)$, and advanced, $\GA_m^{(d)}(x)$, propagators for the Klein-Gordon in spacetimes of dimension $d=1,2,3,4$ were given.

Here we describe another propagator for the Klein-Gordon equation---the Feynman propagator\footnote{This is also known as the \emph{causal propagator}. It is somewhat mis-named, apparently having been first introduced by Stueckelberg and Rivier \citep[Footnote 30, p141]{Bogoliubov:1959}.}. This is a Green's function for the Klein-Gordon equation:
\be  (\Box + m^2)\GF^{(d)}_m(x) = \delta^d(x),\ee 
but with a different boundary condition. This boundary condition is usually described in terms of the way positive and negative frequency modes are propagated but here, instead, we shall simply use the usual Fourier transform contour prescription to define:
\be \label{eq:GFDef} \GF^{(d)}_m(x) := \lim_{\epsilon \to 0^+} -\frac{1}{(2 \pi)^d} \int d^d p \frac{e^{-ipx}}{p^2 - m^2 + i \epsilon}. \ee

As with the retarded and advanced propagators a rigorous treatment of this relies on the theory of distributions. This can be done in different dimensions (\citet[\S2.8, p287]{Gelfand:1964} evaluate this Fourier transform explicitly). In \citet[eq 2.77, p23]{Birrell:1982} it is stated that the general form of $\GF_m^{(d)}$ in $d$-dimensions is given by\footnote{We have replaced their $n$ by $d$, their $\sigma$ by $\frac{1}{2} \tau^2$, multiplied their equation by $-1$ and rearranged factors of $2$ and $i$ etc.}
\be \GF_m^{(d)}(x) = \lim_{\epsilon \to 0^+} \frac{\pi}{2} \frac{(-1)^d}{(2 \pi)^{d/2}} \left(\frac{m}{\sqrt{\tau^2 - i \epsilon}}\right)^{d/2-1} H^{(2)}_{d/2-1} \left( m \sqrt{\tau^2 - i \epsilon} \right). \ee
This agrees with the expression given in \citet[eq 9, p289]{Gelfand:1964} when appropriate factors of $i$, $2\pi$ etc are taken into account.

The $\epsilon \to 0$ limit must be taken within the framework of distribution theory. In the $d=4$ case, for example, a delta-function term is present which is not readily apparent unless care is taken with the limit. The Feynman propagator can be expressed in $d=1,2,3,4$ dimensions as\footnote{The $d=1$ expression can be calculated directly. The $d=2$ and $d=3$ expressions appear in \citet[eqs 13 and 16]{Zhang:2008}. The $d=4$ expression is derived in detail in \citet[\S15, p 147]{Bogoliubov:1959} and \citet{DeJager:1967}.}:
\begin{align}
\GF_m^{(1)}(x)&= -\frac{\exp(-i m s)}{2 i m}\\
\GF_m^{(2)}(x)&=\frac{1}{4} H_0^{(2)}(m s) \\
\GF_m^{(3)}(x)&=\frac{-i}{4 \pi} \frac{\exp(-i m s)}{s} \\
\label{eq:KF4d}\GF_m^{(4)}(x)&= \frac{1}{4 \pi} \delta(s^2) - \frac{m}{8\pi} \frac{H_1^{(2)}(m s)}{s}
\end{align}
where $s = \sqrt{(x^0)^2 - \vec{x}^2}$ for $(x^0)^2 \geq \vec{x}^2$ and
$s=-i\sqrt{\vec{x}^2-(x^0)^2}$ for $(x^0)^2 < \vec{x}^2$. Note that $s = \lim_{\epsilon \to 0^+} \sqrt{\tau^2-i\epsilon}$. $H_\alpha^{(2)}$ is a Hankel function of the second kind of order $\alpha$ and $\delta(s^2)$ is the Dirac delta-function.

We note that, from the definition of $G_F$, its real part is given by:
\be \label{eq:RealFeynman} \Re[\GF_m^{(d)}(x)] = \frac{\GR_m^{(d)}(x)+\GA_m^{(d)}(x)}{2}. \ee

We also note that the in $d=2$ case the massless limit of the imaginary part of the Feynman propagator diverges (i.e. the massless limit of $\GF_m^{(2)}$ does not exist).

\subsection{Quantum fields}

A free real bosonic scalar field is represented by an algebra of field operators $\p(x)$ (which act on a Fock space $F$) satisfying the following conditions:
\begin{align}
1.\;\; & \label{eq:KleinGordon} (\Box + m^2)\p(x) = 0,\\
2.\;\; & \label{eq:Hermitian} \p(x) = \p(x)^\dagger,\\
3.\;\; & \label{eq:Commutator} [\p(x),\p(y)] = i \Delta(y-x),
\end{align}
Here $\lbrack \hat{A}, \hat{B} \rbrack := \hat{A} \hat{B} - \hat{B} \hat{A}$ is the commutator of two operators and $\Delta(x)$ is the \emph{Pauli-Jordan function} (or \emph{commutator function}) defined as the difference between the retarded and advanced propagators \citep{Bogoliubov:1959}:
\be \label{eq:PauliJordan} \Delta(x) := G_R(x) - G_A(x).\ee

In addition there exists a Poincar\'e invariant vacuum state $\rv \in F$. With the fields so defined 
the Feynman propagator is given by the vacuum expectation value of the time-ordered product of two field operators:
\be \label{eq:GEDef} G_F(y-x) = i \lv T \p(x) \p(y) \rv. \ee
The time-ordering has time increasing from right to left. The vacuum expectation value $\lv \p(x) \p(y) \rv$ (without time-ordering) is called the \emph{two-point function} (or \emph{Wightman function}).

We note that applying $(\Box_x + m^2)$ to the commutator in \eqref{eq:Commutator} gives 
\be \label{eq:CommutatorKleinGordon} [(\Box_x + m^2)\p(x),\p(y)] = (\Box_x + m^2)i\Delta(y-x) = 0. \ee
That is, even if only \eqref{eq:Hermitian} and \eqref{eq:Commutator} hold, we have that $(\Box_x +
m^2)\p(x)$ commutes with all the $\p(y)$\footnote{This was emphasised to the author by Johan Noldus.}.

\section{Quantum scalar field theory on a causal set}

Our aim now is to define scalar quantum field theory on a causal set. In particular we are interested in defining a candidate Feynman propagator on the causal set. These results appear in \citet{Johnston:2009}. 

\subsection{The Pauli-Jordan function} \label{sec:PauliJordan}

The starting point for our model is the causal set analogue of the Pauli-Jordan function.

Let $(\CS, \preceq)$ be a finite causal set with $p$ elements generated by a sprinkling into a finite causal interval $\cint \subset \mink^d$. Using the results of the previous chapter we define a $p\times p$ matrix $K_R$ as the appropriate definition of the retarded propagator. It then follows that the transpose of this matrix $K_A := K_R^T$ is the appropriate definition of the advanced propagator. The real skew-symmetric matrix defined by
\be \Delta := K_R - K_A, \ee
is then the causal set analogue of the Pauli-Jordan function (compare with \eqref{eq:PauliJordan}).

The matrix $i\Delta$ is skew-symmetric\footnote{A matrix $M_{xy}$ is skew-symmetric if $M_{xy} = -M_{yx}$, i.e. $M^T = -M$.} and Hermitian\footnote{The Hermitian conjugate of a matrix $M_{xy}$ is the matrix $(M^\dagger)_{xy}:= M_{yx}^*$. A matrix is Hermitian if $M = M^\dagger$.}. These two properties ensure its
rank\footnote{The rank of a matrix is the number of non-zero eigenvalues it possesses.} is even \citep{Perlis:1958} and its non-zero eigenvalues appear in real positive and negative pairs. We label its eigenvalues and normalised eigenvectors as follows:
\be i \Delta u_i =\lambda_i u_i, \qquad \qquad i \Delta v_i = -\lambda_i v_i, \qquad \qquad i\Delta w_k = 0,\ee
where $\lambda_i > 0$, $2s$ is the rank of $i\Delta$ and $i=1,\ldots,s$; $k=1,\ldots,p-2s$.

These eigenvectors form an orthonormal basis for $\mathbb{C}^p$ and can be chosen such that 
\be u_i = v_i^*, \quad w_k = w_k^*,\ee
\be \label{eq:EigenOrtho} u_i^\dagger u_j = v_i^\dagger v_j = \delta_{ij}, \quad w_k^\dagger w_l = \delta_{kl}, \quad u_i^\dagger v_j = w_k^\dagger u_i = w_k^\dagger v_i = 0, \ee
for all $i,j=1,\ldots,s$; $k,l=1,\ldots,p-2s$ (where $z^*$ denotes complex conjugate of $z$ and $u^\dagger:=(u^*)^T$ is the Hermitian conjugate of a column vector $u$).

In terms of these eigenvectors the Pauli-Jordan function is given by
\be \ID = \sum_{i=1}^s \lambda_i u_i u_i^\dagger - \sum_{i=1}^s \lambda_i v_i v_i^\dagger.\ee

\subsubsection{The $Q$ matrix} \label{sec:QMatrix}

It's useful to define the Hermitian, positive semi-definite\footnote{A $p\times p$ matrix $M_{xy}$ is positive semi-definite if it is Hermitian and $z^\dagger M z \geq 0 $ for all $z \in \mathbb{C}^p$.} $p\times p$ matrix 
\be \label{eq:Qdef} Q:= \sum_{i=1}^s \lambda_i u_i u_i^\dagger,\ee
such that $i \Delta = Q - Q^* = Q - Q^T$. This $Q$ is obtained from $i\Delta$ by restricting $i\Delta$ to its ``positive eigenspace'' (i.e. the subspace spanned by eigenvectors with positive eigenvalues). This matrix will turn out to play the role of the two-point function in the causal set quantum field theory.

We can express $Q$ in another way using the eigen-decomposition of $i\Delta$. Since $i\Delta$ is Hermitian there exists a unitary matrix $U$ (whose columns are the eigenvectors of $i\Delta$) and a diagonal matrix $D$ (whose diagonal entries are the eigenvalues of $i\Delta)$ such that 
\be\label{eq:Eigendecomp} i\Delta = U D U^\dagger.\ee 

If we define two diagonal matrices containing the positive and negative eigenvalues:
\be (D^+)_{xy} := \left\{ \begin{array}{ll} D_{xy} & \textrm{ if } D_{xy} > 0 \\ 0 & \textrm{ otherwise} \end{array}\right.\quad (D^-)_{xy} := \left\{ \begin{array}{ll} D_{xy} & \textrm{ if } D_{xy} < 0 \\ 0 & \textrm{ otherwise} \end{array}\right.\ee
such that $D = D^+ + D^-$ then we have $Q = U (D^+) U^\dagger$. In practice this is a simple way to compute $Q$ on a computer.

Another way to characterise $Q$ is as
\be \label{eq:QSqrt} Q = \frac{1}{2} \left(\sqrt{(i\Delta)^2} + i\Delta\right). \ee
The matrix $(i\Delta)^2$ is positive semi-definite and, as such, possesses a unique positive semi-definite square root \citep[Thm VI.9]{Reed:1972}. It is this matrix that we denote by $\sqrt{(i\Delta)^2}$.
It follows from $Q - Q^* = i\Delta$ that the imaginary part of $Q$ is $\Im[Q] = \Delta/2$. Combining this with \eqref{eq:QSqrt}, we have
\be \Re[Q] = \frac{\sqrt{(i\Delta)^2}}{2}, \quad \Im[Q] = \frac{\Delta}{2}.\ee 

Yet another way to characterise the $Q$ matrix is as the unique Hermitian positive semi-definite matrix that is closest to $i\Delta$ as measured in the Frobenius norm \citep[p324]{Dorf:1997}.

\subsection{Fields on a causal set} \label{sec:FieldsOnCausalSet}

To model a free real scalar field on the causal set $(\CS,\preceq)$ we  suppose that to every element $v_x \in \CS$ we assign a field operator $\p_x$ (which acts on some Hilbert space $H$). These field operators are then defined to satisfy the following three conditions:
\begin{align}
1.\;\; & \label{eq:HermitianCondition} \p_x = \p_x^\dagger,\\
2.\;\; & \label{eq:CommutationCondition} \lbrack \p_x,\p_y \rbrack = i \Delta_{xy},\\
3.\;\; & \label{eq:ZeroEigenvalueCondition} i\Delta w = 0 \implies \sum_{x'=1}^{p} w_{x'} \p_{x'} = 0,
\end{align}
for $x,y = 1,\ldots,p$. In condition 3 $w$ is a $p$-component column vector of complex numbers.
The first two conditions are natural generalisations of the continuum case (compare \eqref{eq:Hermitian} and \eqref{eq:KleinGordon}). Condition 3 ensures that any linear combination of field operators
that commutes with all the field operators must be zero. By \eqref{eq:CommutatorKleinGordon} this is the analogue
of imposing the Klein-Gordon equation on the field operators (for further discussion of this point, see \secref{sec:Condition3}).

From these field operators we can define new operators\footnote{These are normalised differently to the operators appearing in \citet{Johnston:2009}.}
\be \label{eq:CreationAnnihilationDef} \a_i := \sum_{x=1}^{p} \frac{1}{\sqrt{\lambda_i}} (v_i)_x \p_x, \qquad \adag_i := \sum_{x=1}^{p} \frac{1}{\sqrt{\lambda_i}} (u_i)_x \p_x.\ee
for $i=1,\ldots,s$. They satisfy $(\adag_i)^\dagger = \a_i$ and
\begin{align}
\label{eq:AAC}&\lbrack \a_i,\a_j\rbrack = \frac{v_i^T i\Delta v_j}{\sqrt{\lambda_i \lambda_j}}  = \frac{-\lambda_j u_i^\dagger v_j}{\sqrt{\lambda_i \lambda_j}} = 0,\\
\label{eq:CCC}&\lbrack \adag_i, \adag_j\rbrack = \frac{u_i^T i\Delta u_j}{\sqrt{\lambda_i \lambda_j}} = \frac{\lambda_j v_i^\dagger u_j}{\sqrt{\lambda_i \lambda_j}} = 0,\\
\label{eq:ACC}&\lbrack \a_i,\adag_j\rbrack = \frac{v_i^T i\Delta u_j}{\sqrt{\lambda_i \lambda_j}} = \frac{\lambda_j u_i^\dagger u_j}{\sqrt{\lambda_i \lambda_j}} = \delta_{ij}.
\end{align}
which are readily recognised as the commutation relations for $s$ creation operators (the $\adag_i$) and $s$ annihilation operators (the $\a_i$).

The transformation from the $\p_x$ operators to the $\a_i$ and $\adag_i$ operators can be inverted to give
\be \label{eq:FieldDef} \p_x = \sum_{i=1}^s \sqrt{\lambda_i}(u_i)_x \a_i + \sqrt{\lambda_i}(v_i)_x \adag_i, \ee
(here we use the orthogonality of the eigenvectors \eqref{eq:EigenOrtho} as well as \eqref{eq:ZeroEigenvalueCondition}).

We now define a vacuum state vector $\rv \in H$ by the conditions that $\a_i \rv = 0$ for all
$i=1,\ldots,s$ and $\lv0\rangle = 1$. This allows us to recognise that $H$ is the Fock space spanned by basis vectors
$(\adag_1)^{n_1}(\adag_2)^{n_2}\cdots (\adag_s)^{n_s} \rv$ for integers $n_i \geq0$, $i=1,\ldots,s$.

\subsection{Uniqueness and condition 3} \label{sec:Condition3}

As we have just shown, any algebra of causal set field operators $\p_x$ that satisfies the three conditions:
\begin{align}
1.\;\; & \tag{\ref{eq:HermitianCondition}} \p_x = \p_x^\dagger,\\
2.\;\; & \tag{\ref{eq:CommutationCondition}} \lbrack \p_x,\p_y \rbrack = i \Delta_{xy},\\
3.\;\; & \tag{\ref{eq:ZeroEigenvalueCondition}} i\Delta w = 0 \implies \sum_{x'=1}^{p} w_{x'} \p_{x'} = 0,
\end{align}
can be expressed in the form:
\be \tag{\ref{eq:FieldDef}} \p_x = \sum_{i=1}^s \sqrt{\lambda_i}(u_i)_x \a_i + \sqrt{\lambda_i}(v_i)_x \adag_i, \ee

This means we have found one way to represent the $\p_x$ operators (namely as operators on the Fock space associated to the $\a_i$ and $\adag_i$ operators. Further details of this Fock space will be given in \secref{sec:FockSpace}). A natural question is whether this representation of the $\p_x$ operators is \emph{unique}---do there exist other ways of representing the field operators that are essentially different to this representation?

Through \eqref{eq:FieldDef} the uniqueness of the $\p_x$ operators is equivalent to the uniqueness of the $\a_i$ and $\adag_i$ operators satisfying the commutation relations \eqref{eq:AAC}, \eqref{eq:CCC}, \eqref{eq:ACC}. The Stone-von Neumann theorem \citep[Cor 5.2.15, p34]{Bratteli:1981} ensures that these commutation relations possess a unique (up to unitary equivalence) representation as operators on a Fock space. This means that any other collection of $2s$ operators satisfying the same commutation relations is related to the $\a_i$ and $\adag_i$ by a unitary transformation.

A crucial part of this theorem is the fact that there are a \emph{finite} number of $\a_i$ and $\adag_i$ operators. If there are an infinite number (i.e. if $i$ ranged from $1$ to infinity) then there exist unitarily \emph{in}equivalent representations. It is therefore interesting to wonder what modifications may be necessary to extend the field operator model to infinite causal sets (see \secref{sec:InfiniteCausalSets}).

\subsubsection{Condition 3}

Of the three conditions the first two are natural analogues of the continuum conditions for field operators. It is the third:
\be \tag{\ref{eq:ZeroEigenvalueCondition}} i\Delta w = 0 \implies \sum_{x'=1}^{p} w_{x'} \p_{x'} = 0, \ee
however, that appears the most ad hoc.
It is motivated by the fact that in the continuum $(\Box + m^2)\p(x)$ commutes with all the $\p(y)$ operators (see \eqref{eq:CommutatorKleinGordon}). In the causal set model we have, from \eqref{eq:CommutationCondition},
\be i\Delta w = 0 \implies \left[ \sum_{x'=1}^{p} w_{x'} \p_{x'}, \p_y \right] = (w^T i \Delta)_y = 0.\ee 
Imposing condition 3 therefore ensures that \emph{any} linear combination of the $\p_x$ that commutes with all the $\p_y$ should be set to zero. This is similar to setting $(\Box+m^2)\p(x) = 0$ in the continuum.

The effect of condition 3 is to make the algebra of field operators as small as possible, so long as it's consistent with conditions 1 and 2.
If we had not imposed condition 3 then the most general expression for the $\p_x$ operators (which satisfy conditions 1 and 2) would be:
\def\c{\hat{c}}
\be \p_x = \sum_{i=1}^s \sqrt{\lambda_i}(u_i)_x \a_i + \sqrt{\lambda_i}(v_i)_x \adag_i + \sum_{k=1}^{p-2s} (w_k)_x\c_k, \ee
where the $w_k$ are the $p-2s$ linearly independent real eigenvectors of $i\Delta$ with 0 eigenvalue: $i\Delta w_k = 0$. If all we know is that the $\p_x$ operators satisfy conditions 1 and 2 then the $\c_k$ are arbitrary but satisfy
\be \c_k = \sum_{x=1}^p (w_k)_x \p_x,\ee
and have the following commutation relations:
\be \lbrack \c_k , \a_i \rbrack = \lbrack \c_k , \adag_i \rbrack = 0, \quad \lbrack \c_k , \c_l \rbrack =0, \ee
for $i=1,\ldots,s$ and $k,l=1,\ldots,p-2s$. Imposing condition 3 is equivalent to imposing that $\c_k := 0$ for all $k$. Without this condition we would need to specify what the $\c_k$ operators were if we wanted to completely define the $\p_x$ operators. Thus condition 3 ensures that the $\p_x$ operators only depend on the $\a_i$ and $\adag_i$ operators.

As it turns out, condition 3 also enforces some desirable properties in the operator algebra. 
Suppose $v_x, v_y \in \CS$ are a non-Hegelian pair of causal set elements (meaning they share identical strict causal relations). Condition 3 ensures that the field operators assigned to these elements are equal: $\p_x = \p_y$.

We can see this because the $x$ and $y$ rows and columns in $i \Delta$ are identical (because $v_x$ and $v_y$ share identical strict causal relations). This means that the $p$-component vector $w = \ve_x - \ve_y$ (where the $\ve_i$ are the standard basis vectors only non-zero in the $i\th$ position) satisfies $i\Delta w = 0$. Applying condition 3 to this implies $\p_x - \p_y = 0$ or $\p_x = \p_y$. The same argument applies to any collection of non-Hegelian elements. This is pleasing because non-Hegelian elements are identical as far as the causal set is concerned so it is sensible that they are assigned identical field operators.

In addition, if $v_x \in \CS$ is a single element, unrelated to all others, then $\p_x = 0$. This is because the $x\th$ row and column in $i\Delta$ is equal to zero which ensures that $w = \ve_x$ satisfies $i \Delta w = 0$. Applying condition 3 ensures $\p_x = 0$.

\section{Feynman propagator on a causal set} \label{sec:FeynmanPropCausalSet}

Directly from the equation
\be \tag{\ref{eq:FieldDef}} \p_x = \sum_{i=1}^s \sqrt{\lambda_i}(u_i)_x \a_i + \sqrt{\lambda_i}(v_i)_x \adag_i, \ee
the two-point function can be evaluated as
\begin{align}
\lv \p_x \p_y \rv &= \sum_{i=1}^s \sum_{j=1}^s \sqrt{\lambda_i \lambda_j} (u_i)_x (v_j)_y \lv  \a_i
\adag_j \rv & \\ \nonumber &= \sum_{i=1}^s \sum_{j=1}^s \sqrt{\lambda_i \lambda_j} (u_i)_x (v_j)_y
\delta_{ij} = Q_{xy}.&\end{align}
where we have used $\lv \adag_i = 0$, $\a_i \rv = 0$ and $Q$ is the matrix defined by \eqref{eq:Qdef}.

To define the Feynman propagator we need a notion of time-ordering. On a causal set this is provided by a \emph{linear extension}.

Recall from \secref{sec:Definitions} that a linear extension of a causal set $(\CS,\preceq)$ is a total order $(\CS, \leq)$ which is consistent with the partial order. This means $u \preceq v \implies u \leq v$ for all $u,v \in \CS$.
A linear extension assigns an ordering to all pairs of elements in $\CS$ in a manner entirely analogous to the time-ordering operation in $\mink^d$.

To define the candidate Feynman propagator we fix a linear extension $(\CS,\leq)$. By analogy with \eqref{eq:GEDef} and with time increasing from right to left we then define the Feynman propagator as:
\be (K_F)_{xy} := i \lv T \p_x \p_y \rv := \left\{ \begin{array}{ll} i Q_{yx} & \textrm{if $v_x \leq v_y$} \\ 
i Q_{xy} & \textrm{if $v_y \leq v_x$}. \end{array} \right.\ee 
or
\be \label{eq:KFDef} (K_F)_{xy}= i\left(\Ale_{xy} Q_{yx} + \Ale_{yx}Q_{xy} + \delta_{xy} Q_{xy}\right),\ee
where $\Ale$ denotes the causal matrix of the linear extension and $\delta_{xy}$ is the Kronecker delta.

In general there are multiple different linear extensions which assign an arbitrary order to pairs of unrelated elements. This arbitrariness does not affect $K_F$ because the field operators commute for pairs of unrelated elements.

Observing that $\Ale_{xy} (i\Delta_{xy}) = (iK_R)_{xy}$ and using $Q -Q^T = i\Delta$ we have
\be \Ale_{xy} Q_{yx} = \Ale_{xy}(Q_{xy}-i \Delta_{xy}) = \Ale_{xy} Q_{xy} -i (K_R)_{xy}.\ee
Substituting this into \eqref{eq:KFDef} gives an alternative form
\be \label{eq:KFAlt} K_F = K_R + i Q,\ee
since $\Ale_{xy} + \Ale_{yx} + \delta_{xy} = 1$ for all $x, y=1,\ldots,p$.

Since $i \Delta = Q - Q^*$ the imaginary part of $Q$ is $\Im[Q] = \Delta/2$. Combining this with \eqref{eq:KFAlt} and looking at the real and imaginary parts of $K_F$ gives
\begin{eqnarray}
\label{eq:KFReal} \Re[K_F] &=& K_R - \frac{\Delta}{2} = \frac{K_R + K_A}{2},\\
\label{eq:KFImag} \Im[K_F] &=& \Re[Q],
\end{eqnarray}
(compare \eqref{eq:RealFeynman} and \eqref{eq:KFReal}).

\section{Comparison with the continuum} \label{eq:QFTComparison}

The causal set propagators depend on the particular random causal set that is sprinkled. By calculating their average value for different sprinklings we can compare the causal set and continuum propagators. To do this, first fix a finite causal interval $\cint \subset \mink^d$. Pick two points $\X, \Y \in \cint$. Sprinkle a finite causal set into $\cint$ with density $\rho$. Almost surely this will not contain $\X$ and $\Y$ so add $\X$ and $\Y$ to it to obtain a finite causal set $(\CS,\preceq)$. For definiteness label the causal set element $\X$ as $v_1$ and $\Y$ as $v_2$.

We now calculate $K_R$ and $K_F$ for $(\CS,\preceq)$ and look at $(K_R)_{12}$ and $(K_F)_{12}$, i.e. the propagator values for the pair $(\X, \Y)$. Let $\mathbb{E}(K_R|\X, \Y, \mink^d, \rho)$ denote the expected value of $(K_R)_{1 2}$ (and $\mathbb{E}(K_F|\X, \Y, \mink^d, \rho)$ the expected value of $(K_F)_{1 2}$) averaged over all causal sets sprinkled into $\cint \subset \mink^d$ (with $\X$ and $\Y$ added in the manner described and for a fixed density $\rho$). It was shown in \secref{sec:1+1pathintegral} and \secref{sec:3+1pathintegral} that
\begin{align}
\label{eq:1+1expected} \mathbb{E}(K_R|\X, \Y,\mathbb{M}^2,\rho) &= \GR^{(2)}_m(\Y-\X),\\ 
\label{eq:3+1expected} \lim_{\rho\to\infty}\mathbb{E}(K_R|\X,\Y,\mathbb{M}^4,\rho) &= \GR^{(4)}_m(\Y-\X).
\end{align}
Using these, \eqref{eq:RealFeynman} and \eqref{eq:KFReal} we have
\begin{align}
\label{eq:KF1+1} \Re[\mathbb{E}(K_F|\X,\Y,\mathbb{M}^2,\rho)] &= \Re[\GF^{(2)}_m(\Y-\X)],\\
\label{eq:KF3+1} \lim_{\rho\to\infty} \Re[\mathbb{E}(K_F|\X,\Y,\mathbb{M}^4,\rho)]&= \Re[\GF^{(4)}_m(\Y-\X)].
\end{align}
That is, the real part of the expected value of $K_F$ is correct for $\mink^2$ and correct in the infinite density limit for $\mink^4$.

We can also compare $K_F$ to $\GF^{(d)}_m$ through numerical simulations, this being one way to investigate the behaviour of the imaginary part of $K_F$. To do this we follow similar steps to \secref{sec:PathIntegralComparison} but this time there is more to compare because the Feynman propagator is a complex function non-zero inside and outside the lightcone (in contrast to the real retarded propagator non-zero only inside the lightcone):
\begin{enumerate}
 \item Fix a sprinkling density $\rho$ and a particle mass $m$.
 \item Sprinkle a causal set $\CS$ with density $\rho$ into a causal interval in $\mink^d$.
 \item Compute $C$ and $L$, the causal and link matrices for $\CS$.
 \item Compute $K_F$, the Feynman propagator matrix (this requires computing the $K_R, \ID$ and $Q$ matrices).
 \item For each pair of sprinkled points $v_x$ and $v_y$ compute their proper time separation $\tau_{xy}$ using their coordinates in $\mink^d$. For spacelike separated points this is imaginary.
 \item For causally related pairs of points:
 \begin{enumerate}
   \item Plot $\Re[(K_F)_{xy}]$ against $\tau_{xy}$.
   \item Plot $\Im[(K_F)_{xy}]$ against $\tau_{xy}$.
 \end{enumerate}
 \item For spacelike separated pairs of points:
 \begin{enumerate}
   \item Plot $\Re[(K_F)_{xy}]$ against $|\tau_{xy}|$, the absolute value of the proper time (or rather proper \emph{distance}).
   \item Plot $\Im[(K_F)_{xy}]$ against $|\tau_{xy}|$.
 \end{enumerate}
 \item Compare the plots to the appropriate plots of $\GF^{(d)}_m$ as a function of proper time.
\end{enumerate}
By virtue of \eqref{eq:KFReal}, plots for the real part of $K_F$ for causally related points reproduce the same results as the retarded propagator discussed in \secref{sec:1+1Comparison} and \secref{sec:3+1Comparison} (the real part of $K_F$ is zero for spacelike points). We shall therefore only show the plots for the imaginary part of $K_F$.

\subsection{1+1 dimensions}

\begin{figure}[!hp]
\begin{center}
\includegraphics[width = \textwidth]{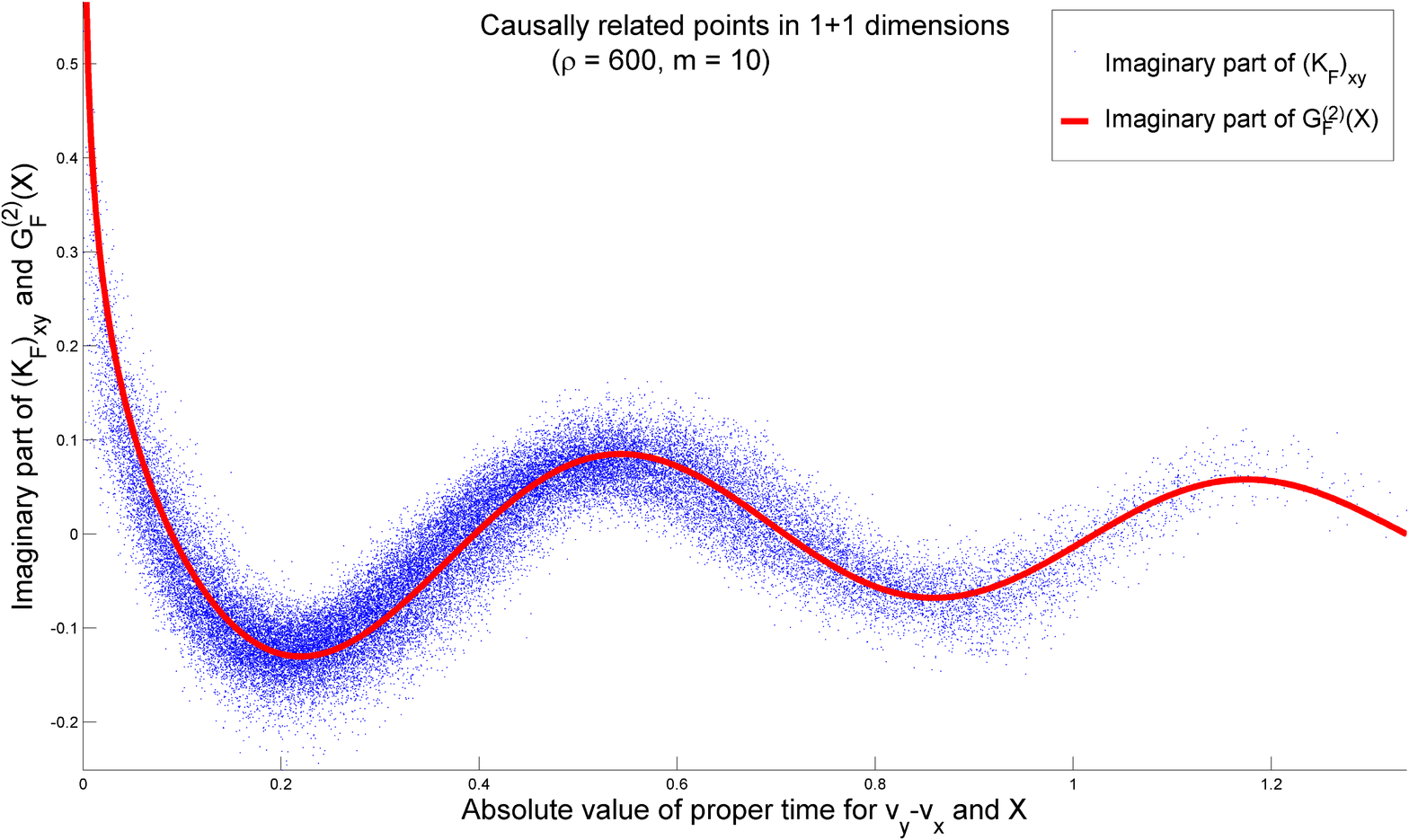}
\\
\caption[Feynman propagator for 1+1 dimensional sprinkling]{The imaginary part of the causal set Feynman propagator for causally related points for a sprinkling in 1+1 dimensions.}
\label{fig:1+1FeynmanTimelike}
\vspace{1cm}
\includegraphics[width = \textwidth]{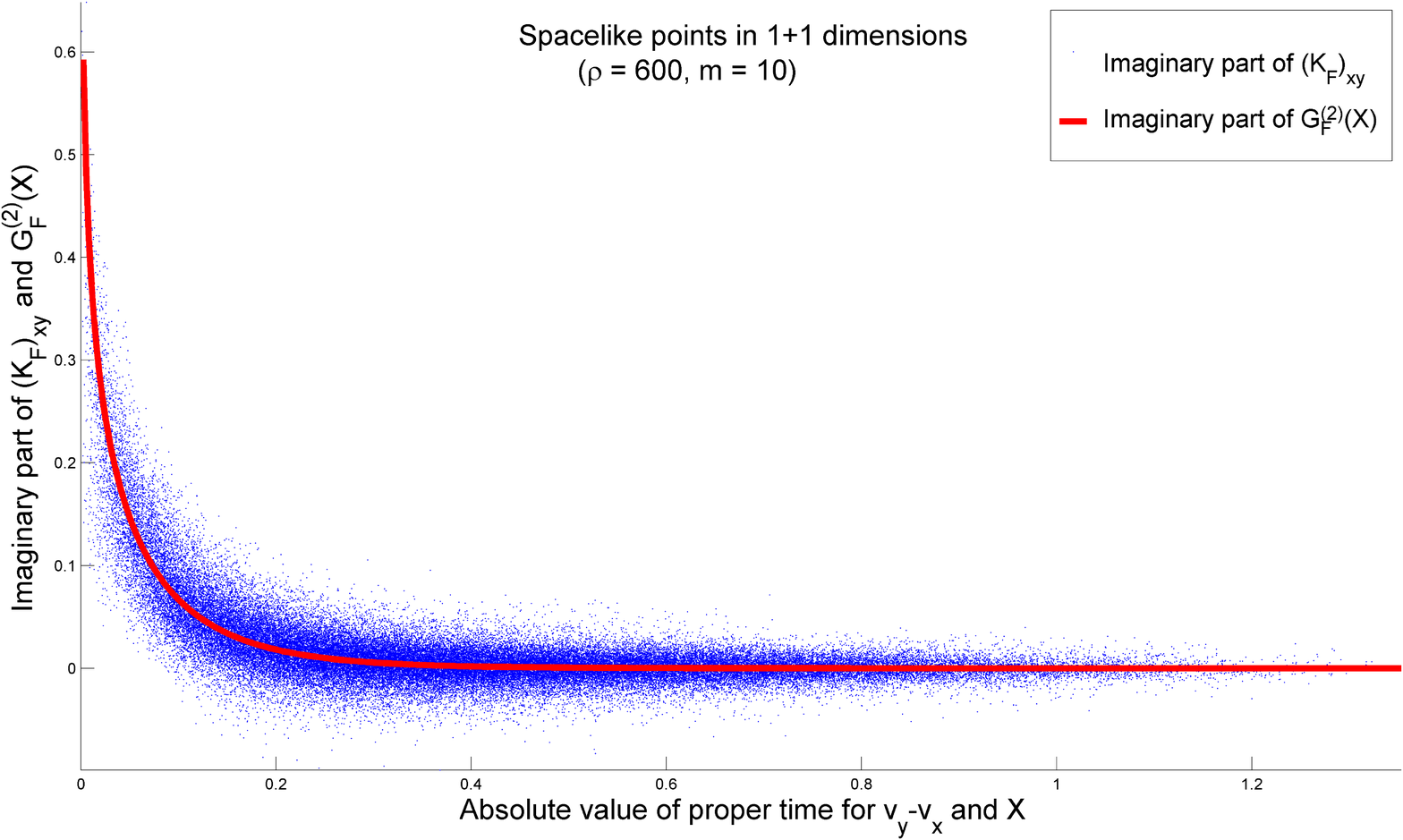}
\caption[Feynman propagator for 1+1 dimensional sprinkling]{The imaginary part of the causal set Feynman propagator for spacelike points for a sprinkling in 1+1 dimensions.}
\label{fig:1+1FeynmanSpacelike}
\end{center}
\end{figure}

In \figref{fig:1+1FeynmanTimelike} and \figref{fig:1+1FeynmanSpacelike} we show the results for a sprinkling in $\mink^2$ with $\rho = 600$ and $m = 10$. We see that the agreement is very good. This holds provided $0 \ll m \ll \sqrt{\rho}$. Note that, although the continuum propagator diverges for small proper times, the causal set propagator remains finite.

\begin{figure}[!hp]
\begin{center}
\includegraphics[width = \textwidth]{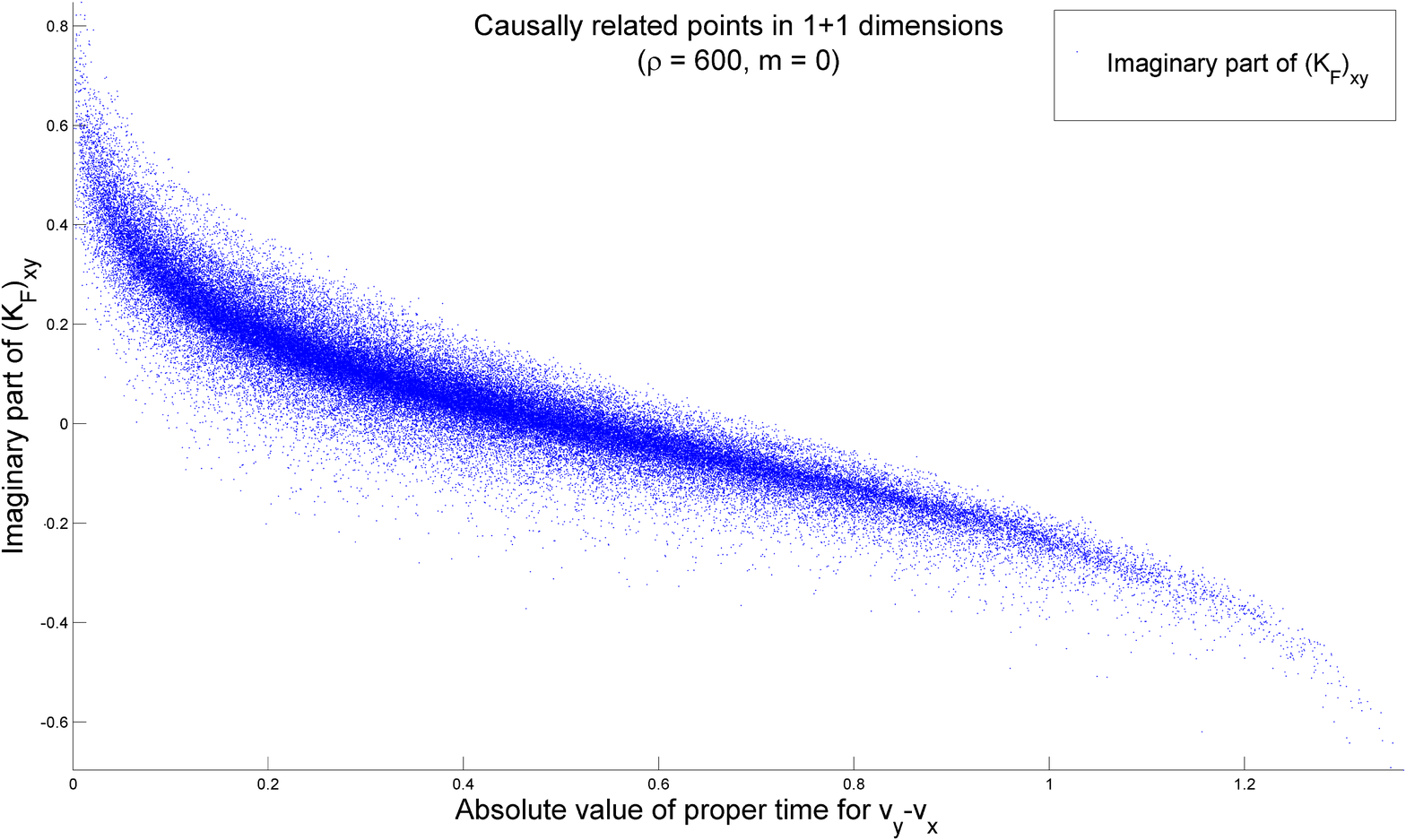}
\\
\caption[Massless Feynman propagator for 1+1 dimensional sprinkling]{The imaginary part of the massless causal set Feynman propagator for causally related points for a sprinkling in 1+1 dimensions.}
\label{fig:1+1FeynmanTimelikem=0}
\vspace{2cm}
\includegraphics[width = \textwidth]{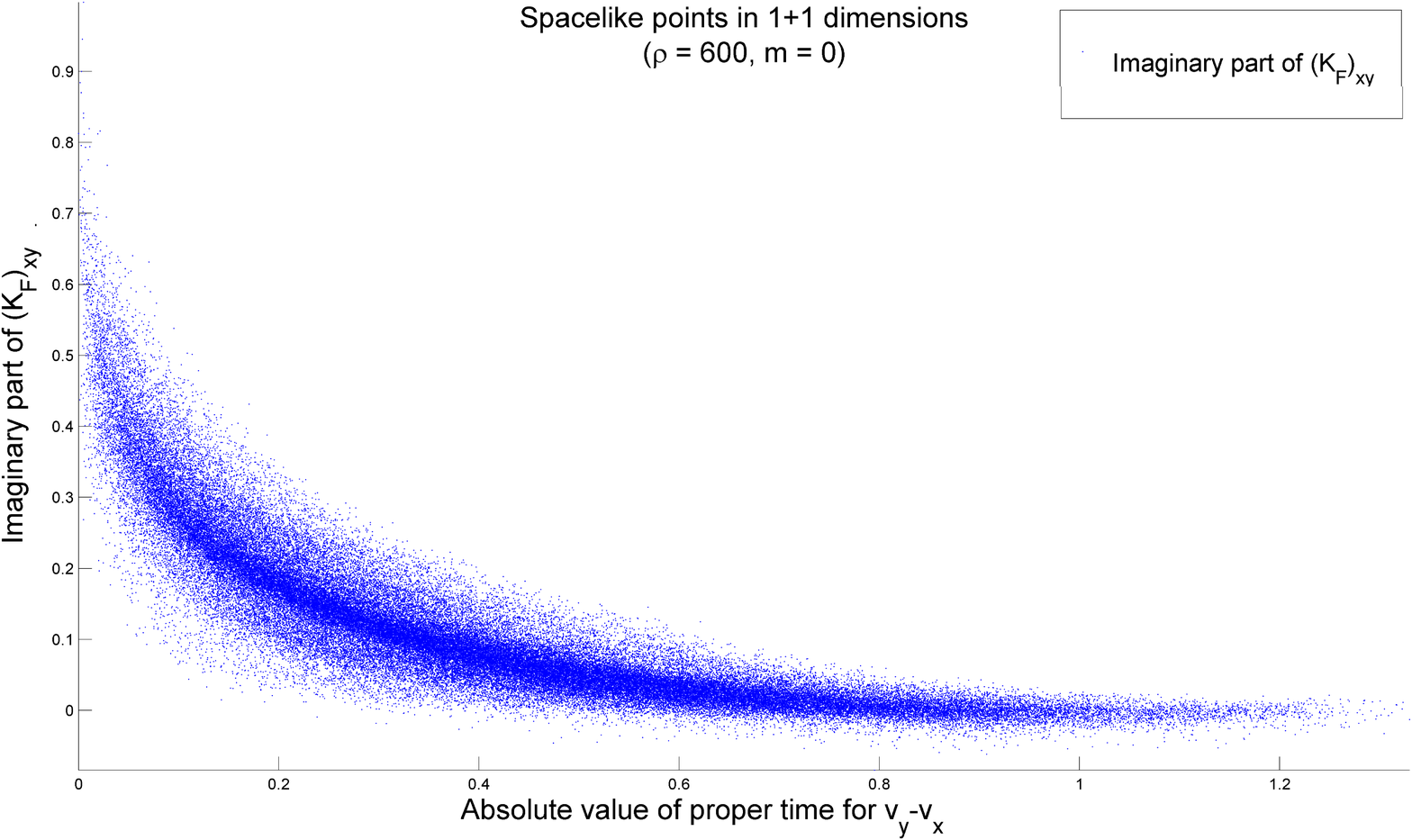}
\caption[Massless Feynman propagator for 1+1 dimensional sprinkling]{The imaginary part of the massless causal set Feynman propagator for spacelike points for a sprinkling in 1+1 dimensions.}
\label{fig:1+1FeynmanSpacelikem=0}
\end{center}
\end{figure}

There is disagreement between the imaginary parts of $K_F$ and $G_F^{(2)}$ as we take the field mass to zero but this is due to the lack of a massless limit of $\Im[G_F^{(2)}]$---indeed it is not even clear what the imaginary part of $G_F^{(2)}$ should be in the massless limit. In \figref{fig:1+1FeynmanTimelikem=0} and \figref{fig:1+1FeynmanSpacelikem=0} we show the imaginary parts of $K_F$ for $\rho = 600$ and $m=0$. It certainly appears to approximate a continuum function but, as yet, we don't know what that function is.

As with the causal set retarded propagator there are deviations from the continuum propagator. These have two causes: (i) the fluctuations due to the random sprinkling being used, (ii) edge effects due to the causal set being sprinkled into a \emph{finite} region of Minkowski spacetime.

The fluctuations appear to be the same order of magnitude as in the 1+1 dimensional retarded case of \secref{sec:1+1Comparison}. The ``edge effects'' get smaller if the results are only plotted for pairs of points away from the edges of the region. By this we mean that we calculate the Feynman propagator for a large causal interval $\cint_1$ but then only plot the values for pairs of points in a smaller causal interval $\cint_2$ in the middle of $\cint_1$. If we do this the spread of the points is less than if we had plotted the values for all pairs of points in $\cint_1$.
\clearpage
\subsection{3+1 dimensions} \label{sec:KF3+1Comparison}

Unfortunately, as with the retarded propagator, in 3+1 dimensions the simulations are less clear. We only expect good agreement with the continuum for large $\rho$ since the infinite density limit is needed in \eqref{eq:3+1expected}. The only way to get a large density simulation without a large causal set is to sprinkle a moderate number of points into a small spacetime volume. The behaviour is then only investigated for small proper times and comparisons are difficult because the small proper-time behaviour of $G_F^{(4)}$ is singular, being dominated by delta-functions and divergent Bessel functions.

\begin{figure}[!hp]
\begin{center}
\includegraphics[width = \textwidth]{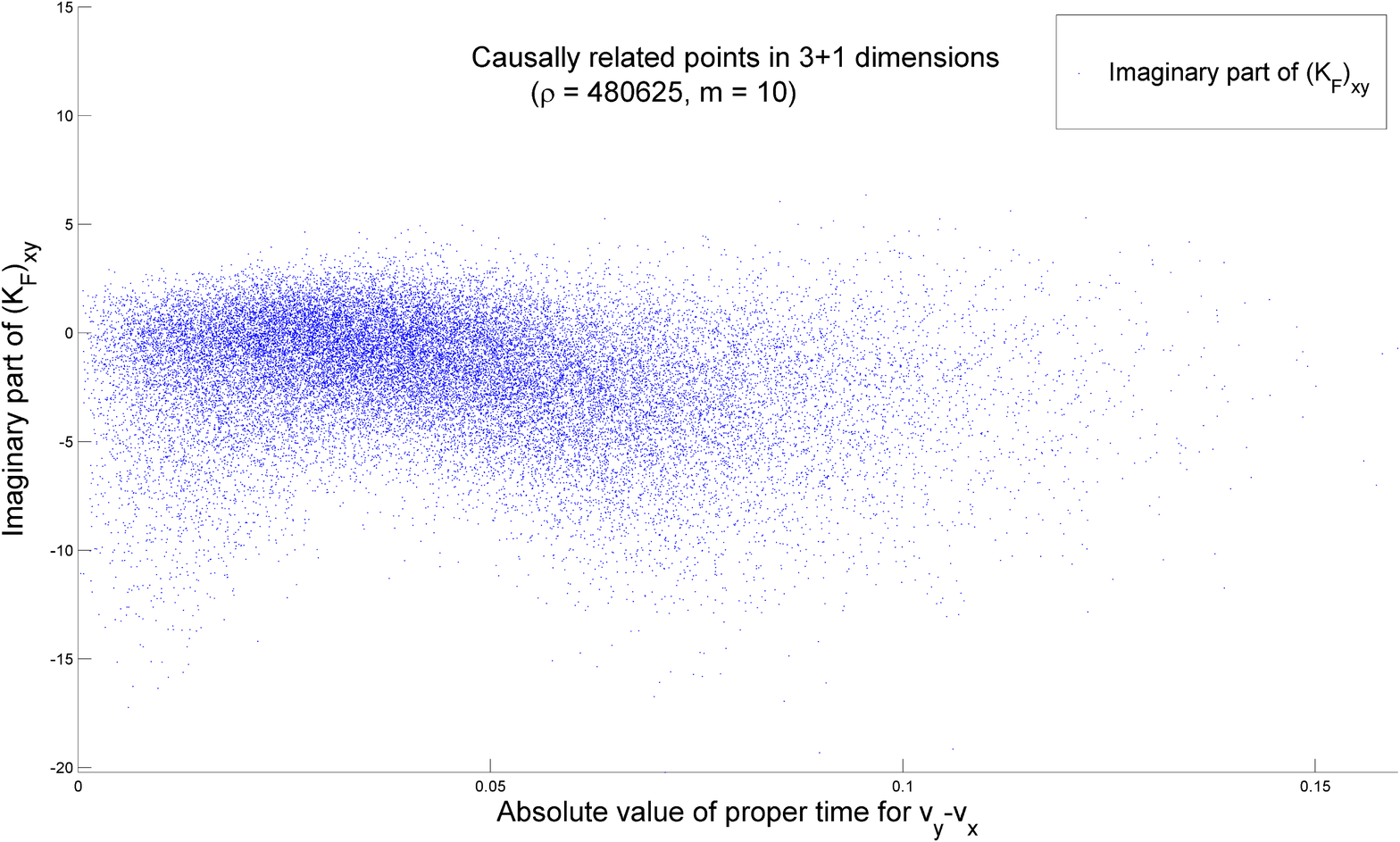}
\caption[Feynman propagator for 3+1 dimensional sprinkling]{The imaginary part of the causal set Feynman propagator for a causally related points for a sprinkling in 3+1 dimensions.}
\label{fig:3+1FeynmanTimelike}
\vspace{2cm} 
\includegraphics[width = \textwidth]{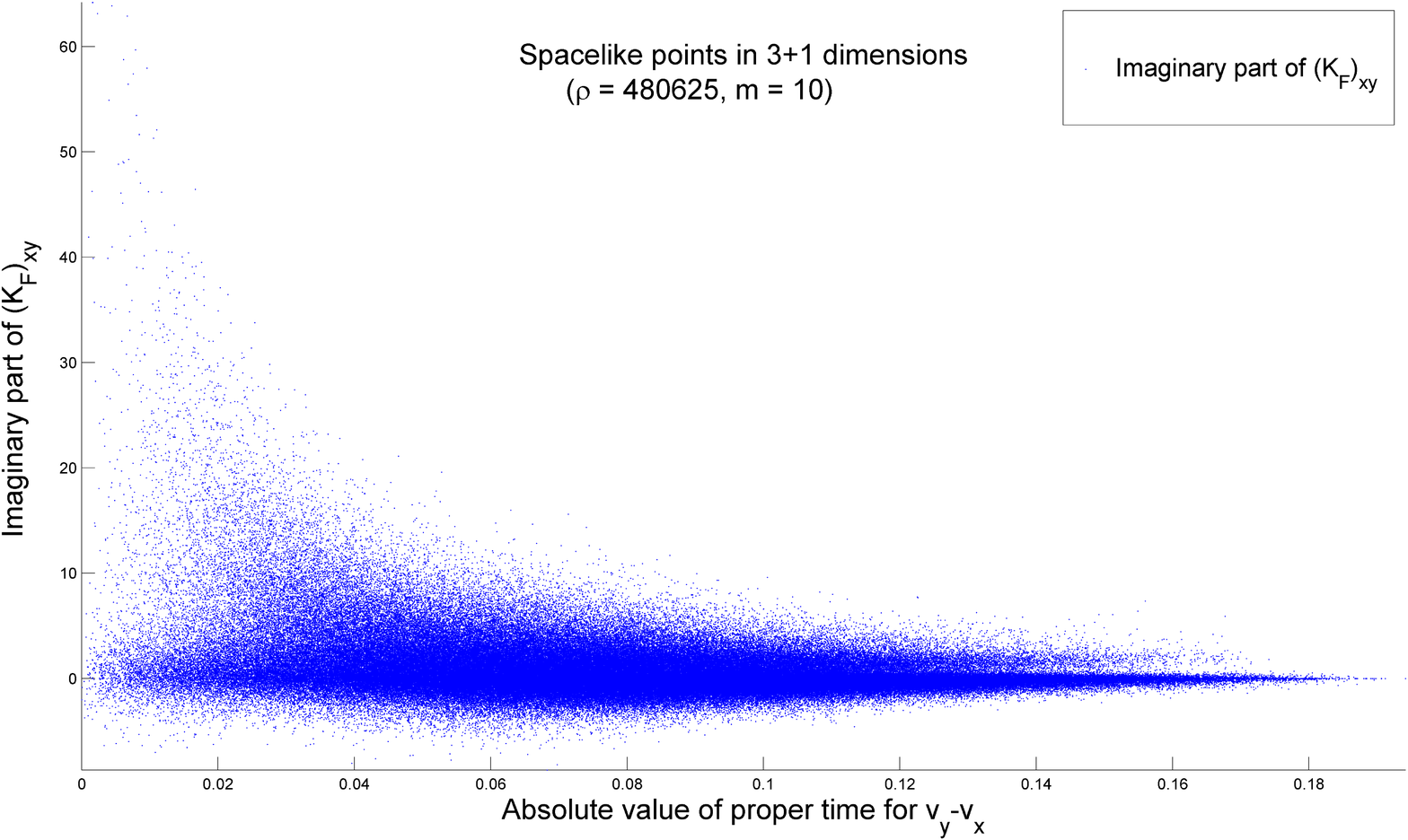}
\caption[Feynman propagator for 3+1 dimensional sprinkling]{The imaginary part of the causal set Feynman propagator for a spacelike points for a sprinkling in 3+1 dimensions.}
\label{fig:3+1FeynmanSpacelike}
\end{center}
\end{figure}

Larger sprinklings are needed to investigate further any agreement between $K_F$ and $G_F^{(4)}$ for a range of proper times. For what it's worth we present the plots for $\rho = 480625$ $m=10$ in \figref{fig:3+1FeynmanTimelike} and \figref{fig:3+1FeynmanSpacelike}. Encouraged by the 1+1 dimensional results we expect to see agreement when it is possible to simulate large densities over a large volume.

\section{Extensions of the model}

The model we've defined so far deals only with a real scalar field on a finite causal set. We now discuss extending the model to complex scalar fields and infinite causal sets.

\subsection{Complex scalar fields}
\def\pa{\hat{\phi}_a}
\def\pb{\hat{\phi}_b}
\def\ch{\hat{\Phi}}

The formalism described so far extends without difficulty to complex scalar fields. The simplest way to achieve this extension is to use two real scalar fields $\pa$ and $\pb$ (with the same mass $m$) which, in addition to conditions 1-3, satisfy
\be \lbrack (\pa)_x , (\pb)_y\rbrack = 0, \ee
for all $x,y=1,\ldots,p$.

We then define 
\be \ch_x:= \frac{1}{\sqrt{2}}\left( (\pa)_x + i(\pb)_x \right),\ee
to be our complex scalar field.

This satisfies
\begin{align}
1.\;\;&\lbrack \ch_x^\dagger, \ch_y \rbrack = i \Delta_{xy}, \qquad \lbrack \ch_x, \ch_y \rbrack = \lbrack \ch_x^\dagger, \ch_y^\dagger \rbrack = 0 \\
2.\;\;& i\Delta w \implies \sum_{x'=1}^p w_{x'} \ch_{x'} = 0
\end{align}

These two conditions could in fact have been used to define the complex scalar field theory, proceeding along the same lines as in the real case---the steps are entirely analogous.

\subsection{Non-finite causal sets} \label{sec:InfiniteCausalSets}

The formalism presented so far is only defined for finite causal sets. For finite $p$-element causal sets the matrix $i \Delta$ is a finite $p \times p$ Hermitian matrix. This is guaranteed to possess $p$ eigenvalues and $p$ eigenvectors and, using these, the field operators $\p_x$ ($x=1,\ldots,p$) can be written as
\be \tag{\ref{eq:FieldDef}} \p_x = \sum_{i=1}^s \sqrt{\lambda_i}(u_i)_x \a_i + \sqrt{\lambda_i}(v_i)_x \adag_i. \ee
for $s$ pairs of creation and annihilation operators $\adag_i$ and $\a_i$.

When the causal set is \emph{infinite} there are two difficulties which are encountered when trying to define the field operators in this way.

\subsubsection{Eigenvectors of Pauli-Jordan function}

Firstly, on the positive side, we note that since a causal set is locally finite $i \Delta$ is well-defined for \emph{any} causal set. On the negative side, when the causal set is infinite, $i\Delta$ takes the form of an infinite matrix and many of the operations that can be performed on finite matrices do not carry over to infinite ones\footnote{See \citet[\chap III, \S 1]{Stone:1932} for a full discussion.}. Instead, the correct way to treat $i\Delta$ is as an operator on a Hilbert space, say (if the causal set is countable) the space of infinite square-summable sequences of complex numbers, denoted $\ell^2$.

If $\psi = (\psi_1,\psi_2,\ldots) \in \ell^2$ is an infinite sequence of complex numbers (such that $\sum_{x=1}^\infty |\psi_x|^2 < \infty$) then the action of $i\Delta$ on $\psi$ is the infinite sequence
\be (i\Delta \psi)_x:=\sum_{y=1}^\infty i\Delta_{xy} \psi_y, \ee
where $x=1,2,\ldots,\infty$.
For a general causal set\footnote{The convergence of these sums depends on the particular causal set as well as the choice of model for the retarded propagator.} however, these infinite sums may not converge for particular values of $x$ and, even if the sums do converge for all $x$, the sequence $(i\Delta \psi)_x$ may not be a vector in $\ell^2$ (i.e. we may not have $\sum_{x=1}^\infty |(i\Delta\psi)_x|^2 < \infty$).

Nevertheless, even though, for a particular causal set, $i\Delta$ may not define an operator valid for all $\psi \in \ell^2$ there will exist $\psi \in \ell^2$ such that the sequence $(i\Delta \psi)_x$ \emph{is} a vector in $\ell^2$. As an example $i\Delta$ acting on the zero vector $0=(0,0,\ldots)$ is always in $\ell^2$, i.e. $i\Delta 0 = 0$. We therefore expect that the action of $i\Delta$ will be only well-defined on a \emph{domain} $D(i\Delta) \subseteq \ell^2$ (in some special cases $D(i\Delta)$ may equal $\ell^2$, i.e. when $\sum_{x=1}^\infty |\sum_{y=1}^\infty i\Delta_{xy} \psi_y|^2 < \infty$ for all $\psi_y \in \ell^2$). The full machinery of operators defined on domains will be covered in the next chapter (see \secref{sec:Operators} for a full explanation of the terms about to be employed).

If we are lucky enough that $i\Delta$ defines a self-adjoint operator then we can try to repeat the procedure we followed for a finite causal set to define the field operators. If $i\Delta$ possesses an orthonormal basis of eigenvectors (i.e. if it has a countable point spectrum \citep[p51]{Jordan:1969}) then we can use these eigenvectors and eigenvalues to define $\p_x$ operators analogously to \eqref{eq:FieldDef}, the summation presumably now being an infinite one. We could then attempt to use this expression for the field operators to develop the theory, although we would expect to encounter issues of convergence relating to the sums involved.

Unfortunately, even if $i\Delta$ is self-adjoint, it may not possess any eigenvalues or eigenvectors \citep[p44]{Jordan:1969}. In this case we would be unable to use eigenvectors of $i\Delta$ to define the field operators $\p_x$ (simply because there would be no eigenvectors!). Nevertheless we could use spectral theory to define an analogue of the two-point function $Q$ when $i\Delta$ is self-adjoint. If the spectral family of $i\Delta$ is $\E_x$ (see \secref{sec:SpectralTheory} for definitions) then we have
\be (\psi, i\Delta \phi) = \int_{-\infty}^\infty x d(\psi,\E_x\phi), \ee
for $\psi, \phi \in D(i\Delta)$. We can define the two-point function $Q$ to be the operator defined by taking the ``positive part'' of this:
\be (\psi, Q \phi) := \int_0^\infty x d(\psi,\E_x\phi). \ee
It would then be possible to define the Feynman propagator as $K_F = K_R + i Q$. We could then try to proceed by just using these ``infinite matrices'' without ever mentioning field operators (for example, trying to sum Feynman diagrams on the causal set without mentioning operators).

\subsubsection{Creation and annihilation operators}

Another difficulty in extending the model to infinite causal sets is that, even if we were able to find appropriate $u_i$ and $v_i$ eigenvectors and wrote down $\p_x$ in the form \eqref{eq:FieldDef} then, presumably, we would have an infinite number of creation and annihilation operators $\adag_i$ and $\a_i$, $i=1,2,\ldots,\infty$ satisfying
\be \lbrack \adag_i, \a_j \rbrack = \delta_{ij}, \quad \lbrack \adag_i, \adag_j \rbrack = \lbrack \a_i, \a_j \rbrack = 0. \ee
Since there are infinitely many pairs of these operators the Stone-von-Neumann theorem does not apply---there will exist unitarily-\emph{inequivalent} representations of these commutation relations as operators on a Hilbert space. The question arises, therefore, which representation should we choose? We make no attempt to answer this question but point out that its resolution may require us to introduce another condition on the $\a_i$ and $\adag_i$ which would restore the  uniqueness of their representation.

\subsubsection{Discussion}

It is clear from the two problems just described that there are mathematical difficulties in generalising the definition of field operators \eqref{eq:FieldDef} from a finite causal set to an infinite one. It appears that these difficulties, however, depend heavily on the particular causal set that is being used. For some infinite causal sets (e.g. the infinite antichain where $i\Delta_{xy} = 0$ for all $x,y=1,2,\ldots$) the theory is simple (i.e. $\p_x = \hat{0}$ for all $x$) whereas for others one can imagine that the domain of $i\Delta$ and a determination of its eigenvectors (if they exist at all) is much more complicated. It's possible that the current model only extends to a certain class of infinite causal sets and that substantial modification will be needed if it is to extend to all infinite causal sets.

An obvious way out of these difficulties is to only use finite (extremely large) causal sets as a model for spacetime. The field operators are well-defined for any finite causal set and for all practical purposes a very large but finite causal set would serve perfectly well as a model for spacetime (consider, for example, a Planckian density sprinkling into a Lorentzian manifold modelling the entire observable universe!).

\section{0+1 dimensional calculation}

It is hard to calculate the eigen-decomposition of $i\Delta$ for general causal sets analytically. For a totally ordered causal set, however, it is possible.

Recall from \secref{sec:0+1dim} that the retarded propagator for a $p$-element sprinkling into 0+1 dimensions is given by \eqref{eq:0+1prop} (here we again use the natural labelling $v_1 \prec v_2 \prec \ldots \prec v_p$):
\everymath{\displaystyle}
\be (K_R)_{xy} = \left\{\begin{array}{cc} \frac{2 \sin \left(2 (y-x) \arcsin\left(\frac{m}{2 \rho}\right)\right)}{m \sqrt{4-\frac{m^2}{\rho^2}}} & \textrm{if $x < y$} \\ \\
0 & \textrm{if $x \geq y$.} \end{array} \right. \ee\everymath{}
for $x,y=1,\ldots,p$.

Using this we see that $i\Delta = i(K_R-K_A)$ takes the form 
\be i\Delta_{xy} = \frac{2i \sin \left(2 (y-x) \arcsin\left(\frac{m}{2 \rho}\right)\right)}{m \sqrt{4-\frac{m^2}{\rho^2}}}, \ee
for $x,y=1,\ldots,p$.

There are $p-2$ eigenvectors which have 0 as an eigenvalue. These are $p$-component column vectors of the (unnormalised) form
\be \label{eq:0+1ZeroEigen} w_k := \rho^2(\ve_{k-1} - 2 \ve_{k}+ \ve_{k+1}) + m^2 \ve_{k},\ee
for $k=2,\ldots,p-1$ (here $\ve_i$ are the standard basis vectors non-zero only in the $i\th$ component). For example 
\be w_2 = \left(\begin{array}{c} \rho^2 \\ -2\rho^2 + m^2 \\ \rho^2 \\0 \\0\\ \vdots \\0\end{array} \right), \qquad w_3 = \left(\begin{array}{c} 0\\ \rho^2 \\ -2\rho^2 + m^2 \\ \rho^2 \\0 \\ \vdots \\0\end{array} \right).\ee

We can see that these are eigenvectors because the $x$-component of the vector $i\Delta w_k$ is equal to 
\be (i\Delta w_k)_x = \rho^2 i\Delta_{x(k-1)} - \left(2\rho^2 - m^2\right) i\Delta_{xk} + \rho^2 i\Delta_{x(k+1)}\ee
\be = \frac{2i\rho^2}{m \sqrt{4-\frac{m^2}{\rho^2}}}\left(\sin \left((k\!-\!x\!-\!1) \theta\right)\!-\!\left(2\!-\!\frac{m^2}{\rho^2}\right) \sin \left((k\!-\!x) \theta\right)\!+\!\sin \left((k\!-\!x\!+\!1)\theta\right)\right),\ee
where $\theta:= 2\arcsin\left(\frac{m}{2 \rho}\right)$.
Trigonometric identities can be used to simplify this and we are left with
\be (i\Delta w_k)_x = 0, \ee
for $k=2,\ldots,p-1$ and $x=1,\ldots,p$.

This means that we have found $p-2$ eigenvectors which have 0 eigenvalue, namely the vectors $w_k$ defined in \eqref{eq:0+1ZeroEigen} for $k=2,\ldots,p-1$. 
Imposing condition 3 immediately gives constraints on the field operators of the form: $\rho^2(\p_{x-1} - 2\p_x + \p_{x+1}) + m^2 \p_x = 0$ for $x=2,\ldots,p-1$. This is a discretisation of the 0+1 dimensional Klein-Gordon equation $(\Box + m^2) \p(x) = 0$.
Thus, in the 0+1 case, condition 3 serves to enforce an appropriate version of the Klein-Gordon equation.
The 2 remaining non-zero eigenvalues are harder to obtain analytically.

\section{Rank of the Pauli-Jordan matrix}

In the 0+1 dimensional example just studied we saw that the vast majority of the eigenvalues were 0. This is peculiar to sprinklings into $\mink^1$. For sprinklings into higher dimensions the number of 0 eigenvalues is much less.

The number of 0 eigenvalues for a $p$-element sprinkling is $p-2s$ where $2s$ is the rank of $\ID$ ($2s$ is also the rank of $\Delta$, the factor of $i$ makes no difference). Here we investigate how this quantity changes as we increase the sprinkling density. We will look at sprinklings into a fixed causal interval in $\mink^2$ at different densities.

\subsection{Mass dependence} \label{sec:MassIndependent}

Firstly we note that the rank of the Pauli-Jordan matrix is mass-independent. In \secref{sec:Nonconstant} we found the general form of the retarded propagator $K_R$ (obtained by summing over trajectories that only go forward in time) to be \eqref{eq:GeneralRetarded}:
\be \label{eq:K_RSeries} K_R = \Phi(I - \Psi \Phi)^{-1} = \Phi + \Phi \Psi \Phi + \Phi \Psi \Phi \Psi \Phi + \ldots, \ee
where $\Phi$ is the matrix of ``hop'' amplitudes and $\Psi$ is a diagonal matrix of ``stop'' amplitudes. Inverting \eqref{eq:K_RSeries} gives:
\be \label{eq:PhiSeries} \Phi = K_R (I+\Psi K_R)^{-1} = K_R - K_R \Psi K_R + K_R \Psi K_R \Psi K_R - \ldots. \ee
In keeping with the mass scattering ideas of \secref{sec:MassScatterings} we assume that when the mass is zero we have $K_R = \Phi$. In this case the massive Pauli-Jordan matrix is $\Delta_m := K_R - K_R^T$ and the massless Pauli-Jordan matrix is $\Delta_0 := \Phi - \Phi^T$. We can show that these two matrices have the same null space\footnote{The null space of a matrix is the vector space spanned its 0-eigenvalue eigenvectors.}, i.e. $\Delta_m w = 0 \iff \Delta_0 w = 0$. To see this we use \eqref{eq:K_RSeries} and \eqref{eq:PhiSeries} to give:
\be \Delta_0 w = 0 \iff \Phi w = \Phi^T w \iff K_R w = K_R^T w \iff \Delta_m w = 0, \ee
where we have used that $\Psi = \Psi^T$.

Since the null spaces of the matrices are the same this shows that the rank of the massive Pauli-Jordan matrix equals the rank of the massless Pauli-Jordan matrix. Therefore in our investigation it is sufficient to look at $p - 2s$ for the massless Pauli-Jordan function.

\subsection{Density dependence}

\begin{figure}[!h]
\begin{center}
\includegraphics[angle = 270,width = 0.9\textwidth]{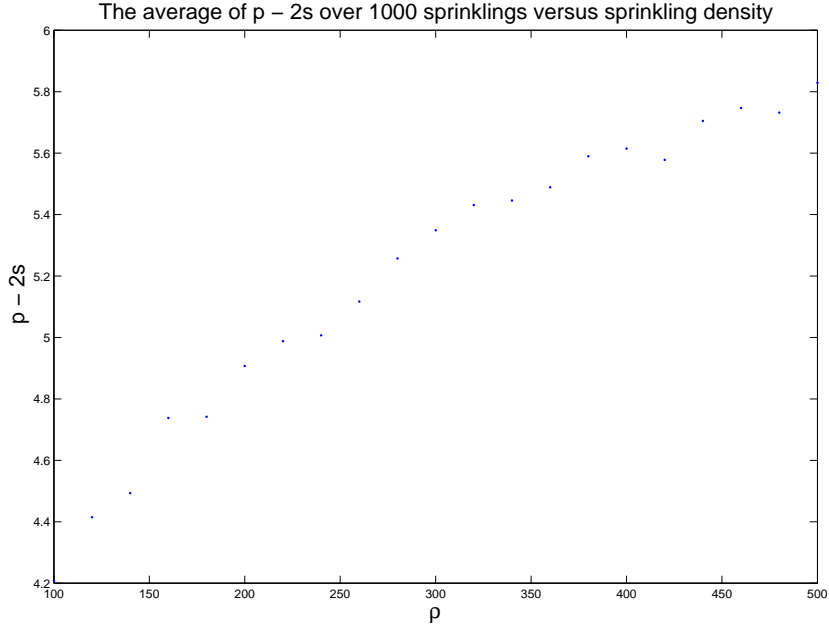}
\caption[The rank of the Pauli-Jordan function versus sprinkling density]{A plot of the average of $p-2s$ for 1000 sprinklings at different densities.}
\label{fig:PJPlot}
\end{center}
\end{figure}

We can investigate the dependence of $p-2s$ on the sprinkling density by performing the following steps:
\begin{enumerate}
 \item Let $\rho$ take values from $\rho_{\textrm{min}}$ to $\rho_{\textrm{max}}$ in increments of $\rho_{\textrm{inc}}$.
 \item For each $\rho$ sprinkle a causal set into a causal interval in $\mink^d$ of unit volume.
 \item Calculate $p - 2s$ (i.e. the number of 0 eigenvalues of $i\Delta$) for the causal set.
 \item Repeat steps 2 and 3 for $N$ sprinklings and take the average of these $N$ numbers.
 \item Plot this average for each density $\rho$.
\end{enumerate}
The result of this is shown in \figref{fig:PJPlot} for $\rho_{\textrm{min}} = 100, \rho_{\textrm{max}} = 500, \rho_{\textrm{inc}} = 20, d=2, N = 1000$. As we can see there is a steady increase in the number of 0 eigenvalues as the density increases.

The number of creation and annihilation operators needed to define the field operators $\p_x$ is $2s$. The results of \figref{fig:PJPlot} show that $2s$ increases only \emph{slightly} slower than the number of causal set elements $p$ (for the sprinklings used in the Figure the expected value of $p$ is $\rho$). This is in contrast to the continuum where the number of creation and annihilation operators is infinite (usually indexed by spatial-momentum), regardless of the size of the region.

\section{Calculation in the continuum} \label{sec:ContinuumCalc}

The idea of using the eigenvectors and eigenvalues of the Pauli-Jordan function as a way to derive the Feynman propagator seems to be a new one. We can try to perform the same procedure in the continuum.

\subsection{Preliminaries}

Suppose we know the Pauli-Jordan function $\ID(x,y)$ for a spacetime region $R$. We shall just work in flat spacetimes for which this is a translation-invariant function depending only on $x-y$. For curved spacetimes this would depend on $x$ and $y$ individually.

Associated to $R$ is the Hilbert space of square-integrable functions $L^2(R)$. We can use $\ID$ to define an integral operator on functions in $L^2(R)$ by:
\be (\ID \psi)(x) = \int_R dy \;i\Delta(x,y) \psi(y), \ee
for $\psi \in L^2(R)$. The convergence of these integrals will depend on $\ID$, $R$ and $\psi$ but for suitable functions $\ID$ and suitable regions $R$ this integral will be well-defined for all $\psi \in L^2(R)$.
An important special case of this is when
\be \int_R dx \int_R dy \;|\ID(x,y)|^2 < \infty. \ee
In this case we say $\ID$ is a Hilbert-Schmidt integral kernel. The operator on $L^2(R)$ it defines is bounded and, since $\ID(x,y) = (\ID(y,x))^*$ it is also self-adjoint \citep[Thm 3.8, p101]{Stone:1932}. The operator is called a Hilbert-Schmidt operator.

This operator is guaranteed to possess either a finite or countable collection of eigenvectors $\lambda_n$ which satisfy the useful identity \citep[Thm 3.8 and Thm 5.14]{Stone:1932}:
\be \label{eq:HilbertSchmidtId} \int_R dx \int_R dy \;|\ID(x,y)|^2 = \sum_n \lambda_n^2, \ee

If $f \in L^2(R)$ is an eigenvector of $\ID$ then it satisfies
\be \int_R dy \;\ID(x,y) f(y) = \lambda f(x). \ee
Applying $\Box_x + m^2$ to both sides and using that $(\Box_x + m^2)\ID(x,y) = 0$ we have that if $f(x)$ is an eigenvector with non-zero eigenvalue then it is a solution of the Klein-Gordon equation: $(\Box+m^2)f(x) = 0$. This is a helpful aid when trying to determine the eigenvectors.
The eigenvectors are defined up to a normalisation and a phase factor.

If we succeed in determining all the eigenvectors with non-zero eigenvalues then we can define the two-point function to be
\be Q(x,y) = \sum_{\lambda > 0} \lambda f(x) f(y)^*, \ee
where we sum over all eigenvalue-eigenvector pairs $(\lambda, f)$ with $\lambda > 0$ and the $f$ are normalised: $||f|| = 1$ where
$||f||^2 = \int_R dx \; |f(x)|^2 $.
We shall concentrate on the two-point function because if we know that then we can define the Feynman propagator (as the sum of the retarded propagator and $iQ(x,y)$, see \eqref{eq:KFAlt}).

\subsection{Causal interval in 0+1 dimensional Minkowski spacetime}

As a warm-up exercise we examine the eigen-decomposition of $\ID$ for a causal interval in $\mink^1$. 
We suppose our interval extends from $x=-L$ to $x = L$ so our region is $R = [-L,L]$. In $\mink^1$ the d'Alembertian and the massive Pauli-Jordan function are (see \eqref{eq:KR1d}):
\be \Box = \frac{\partial^2}{\partial x^2},\qquad \Delta(x) = \frac{\sin(m x)}{m}. \ee
We have
\be \int_{-L}^L dx \int_{-L}^L dy |\ID(x-y)|^2 = \frac{8 L^2 m^2+\cos (4 L m)-1}{4 m^4},\ee
which shows that $\ID$ defines a Hilbert-Schmidt integral operator on $L^2([-L,L])$.

The eigenvectors of $\ID$ with non-zero eigenvalue satisfy $(\Box + m^2)f(x) = 0$. They must therefore take the form
\be f(x) = A \cos(m x) + B \sin(m x), \ee
for some constants $A$ and $B$.

We have
\begin{center}
\vspace{-1cm}
\everymath{\displaystyle}
\[\begin{array}{c|l}f & \ID f \\
\hline \\
\sin(m x)& \left(\frac{i \sin (2L m)}{2m^2}-\frac{i L }{m}\right)\cos(mx) \\ \\
\cos(m x) & \left(\frac{i \sin (2 L m) }{2 m^2}+\frac{i L }{m}\right)\sin (m x) 
\end{array}\]
\everymath{}
\end{center}

Comparing the coefficients of $\sin(mx)$ and $\cos(m x)$ and normalising the eigenvectors we find that
\be f_+(x) := \frac{\sqrt{m}\cos (m x)}{\sqrt{2 L m+\sin (2 L m)}}+\frac{i \sqrt{m}\sin (m x)}{\sqrt{2 L m-\sin (2 L m)}},\ee
\be f_-(x) := \frac{\sqrt{m}\cos (m x)}{\sqrt{2 L m+\sin (2 L m)}}-\frac{i \sqrt{m}\sin (m x)}{\sqrt{2 L m-\sin (2 L m)}},\ee
satisfy
\be \ID f_+ = \lambda f_+, \qquad \ID f_- = -\lambda f_- ,\ee
where \be \lambda = \frac{\sqrt{2 L m-\sin (2 L m)} \sqrt{2 L m+\sin (2 L m)}}{2 m^2}. \ee
We see that $\pm \lambda$ are the only non-zero eigenvalues because 
\be \lambda^2 + (-\lambda)^2 = \frac{8 L^2 m^2+\cos (4 L m)-1}{4 m^4} = \int_{-L}^L dx \int_{-L}^L dy |\ID(x-y)|^2, \ee
(compare with \eqref{eq:HilbertSchmidtId}).

\subsubsection{The two-point function}

Usually we would define the two-point function to be the sum over all the positive eigenvectors. This time, however, we only have one non-zero eigenvalue. This gives:
\be Q(x,y) := \lambda f_+(x) f_+(y)^*. \ee
The real and imaginary parts of this function are
\begin{align} \Re[Q(x,y)] &= \frac{1}{2m C}\cos (m x) \cos (m y) + \frac{C}{2m} \sin (m x) \sin (m y)&\\
 \Im[Q(x,y)] &= \frac{\sin (m x) \cos (m y)}{2 m}-\frac{\cos (m x) \sin (m y)}{2 m} = \frac{\Delta(x-y)}{2}\end{align}
where \be C = \frac{\sqrt{2 L m+\sin (2 L m)}}{\sqrt{2 L m-\sin (2 L m)}}.\ee
Noticing that $\lim_{L \to \infty} C = 1$ therefore gives
\be \label{eq:0+1QLim}\lim_{L \to \infty} Q(x,y) = \frac{\exp(i m(x-y))}{2m}, \ee
which is precisely the continuum two-point function.

For finite $L$ the real part $\Re[Q(x,y)]$ differs from the real part of the continuum two-point function due to the calculation being done in a finite sized interval.

\subsection{Causal interval in 1+1 dimensional Minkowski spacetime} \label{sec:1+1ContinuumCalc}

A more interesting example is that of a massless scalar field in a finite causal interval in $\mink^2$.
We shall work in light-cone coordinates 
\be u = \frac{t+x}{\sqrt{2}}\qquad v=\frac{t-x}{\sqrt{2}}.\ee
For a fixed finite length $L$ the range of these coordinates will be restricted to $u,v \in[-L, L]$. Our attention is therefore restricted to the region $R = [-L,L] \times [-L,L]$.

In these coordinates the 2-dimensional d'Alembertian is simply 
\be \Box = 2\frac{\partial^2}{\partial u \partial v},\ee
and the Pauli-Jordan function is (see \eqref{eq:KR2d})
\be \Delta(u,v) = \frac{1}{2} (\theta(u) \theta(v) - \theta(-u) \theta(-v)) = \frac{1}{2} (\theta(u) + \theta(v) - 1).\ee
We have
\be \int_{-L}^L d u \int_{-L}^L du' \int_{-L}^L d v \int_{-L}^L dv' | \ID(u-u',v-v')|^2 = 2 L^4, \ee
which shows that $\ID$ defines a Hilbert-Schmidt operator.

\subsubsection{Eigenvectors and Eigenvalues}

The eigenvectors of $\ID$ with non-zero eigenvalue satisfy $\Box f = 0$. The most general form of such solutions is $f(u,v) = f_1(u) + f_2(v)$ for arbitrary functions $f_1, f_2 \in L^2([-L,L])$.

By considering how $\ID$ acts on particular functions we will be able to identify the eigenvectors. The action of $i\Delta$ is evaluated by simply performing the integration.
\begin{center}
\vspace{-1cm}
\everymath{\displaystyle}
\[ \begin{array}{c|l}
f & \ID f \\
\hline \\
e^{i k u} & \frac{L}{k} e^{i k u} - \frac{L}{k} \cos(k L) + i \frac{v}{k} \sin(k L)  \\ \\
e^{i k v} & \frac{L}{k} e^{i k v} - \frac{L}{k} \cos(k L) + i \frac{u}{k} \sin(k L)  \\ \\
1 & i L (u+v)\\ \\
\end{array}\]
\end{center}\everymath{}

Using these results we can obtain two families of eigenvectors $f_k$ and $g_k$ given by\footnote{The $g_k$ family were found by Rafael Sorkin.}:
\begin{align} 
\label{eq:feigenvector} f_k(u,v) &:= e^{iku} - e^{i k v}, & &\textrm{with } k = \frac{n \pi}{L}, \; n = \pm 1, \pm 2, \ldots\\
\label{eq:geigenvector} g_k(u,v) &:= e^{iku} + e^{i k v} - 2 \cos(k L), & &\textrm{with } \tan(kL)= 2kL,\; k \neq 0 \end{align}
These satisfy
\be (\ID f_k)(u,v) = \frac{L}{k} f_k(u,v), \qquad (\ID g_k)(u,v) = \frac{L}{k} g_k(u,v). \ee
As written these eigenvectors are unnormalised. By direct computation the square of their $L^2(R)$ norms are $||f_k||^2 = 8 L^2$ and
$||g_k||^2 = 8 L^2 - 16 L^2 \cos^2(k L)$.

\begin{figure}[!h]
\begin{center}
\includegraphics[width = \textwidth]{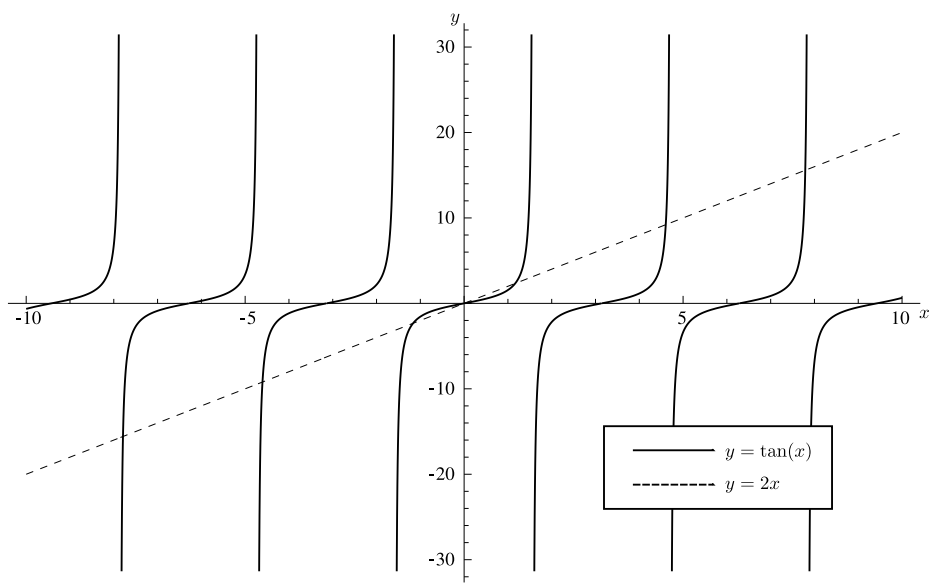}
\caption{A plot of $y=\tan(x)$ and $y = 2x$.}
\label{fig:Tanx}
\end{center}
\end{figure}

For the $g_k$ family of eigenvectors the $k$ parameter satisfies the transcendental equation $\tan(kL) = 2kL$. This has a countably infinite number of real solutions (see \figref{fig:Tanx}).

We now show that the $f_k$ and $g_k$ families are the only eigenvectors with non-zero eigenvalues. We do this, as in the 0+1 case, by summing the eigenvalues we have found so far:
\be \sum_{\begin{subarray}{c}n=-\infty \\ n \neq 0 \end{subarray}}^\infty \left(\frac{L^2}{\pi n}\right)^2 + \sum_{\begin{subarray}{c}\tan(x) = 2x \\ x \neq 0 \end{subarray}} \left(\frac{L^2}{x}\right)^2 \ee
\be = 2L^4\left(\sum_{n=1}^\infty \frac{1}{(\pi n)^2} + \sum_{\begin{subarray}{c}\tan(x) = 2x \\ x > 0 \end{subarray}} \frac{1}{x^2} \right).\ee
The first sum is given by
\be \sum_{n=1}^\infty \frac{1}{(\pi n)^2} = \frac{1}{6}, \ee
and the second sum can be evaluated, using methods described in \citet[Example 3]{Speigel:1953}, to give
\be \sum_{\begin{subarray}{c}\tan(x) = 2x \\ x > 0 \end{subarray}} \frac{1}{x^2} = \frac{5}{6}. \ee
We therefore have that
\begin{align} 2L^4 &= \sum_{\begin{subarray}{c}n=-\infty \\ n \neq 0 \end{subarray}}^\infty \left(\frac{L^2}{\pi n}\right)^2 + \sum_{\begin{subarray}{c}\tan(x) = 2x \\ x \neq 0 \end{subarray}} \left(\frac{L^2}{x}\right)^2 \\ &= \int_{-L}^L d u \int_{-L}^L du' \int_{-L}^L d v \int_{-L}^L dv' | \ID(u-u',v-v')|^2, \end{align}
which shows that we have found all the non-zero eigenvalues (compare with \eqref{eq:HilbertSchmidtId}).

\subsubsection{The two-point function}

Having found all the non-zero eigenvalues and corresponding eigenvectors we are interested in obtaining the analogue of the two-point function by summing over the positive eigenvalues.

We define
\begin{align}
 \nonumber Q(u,v,u',v') := &\sum_{n=1}^\infty \frac{L^2}{\pi n} \frac{1}{||f_k||^2} f_k(u,v) f_k(u',v')^*\\ &\qquad+ \sum_{\begin{subarray}{c}\tan(x) = 2x \\ x > 0 \end{subarray}} \frac{L^2}{x} \frac{1}{||g_k||^2}  g_k(u,v) g_k(u',v')^*,
\end{align}
where $||f_k||^2 = 8 L^2$ and $||g_k||^2 = 8 L^2 - 16 L^2 \cos^2(k L)$ are included to ensure we're summing over products of normalised eigenvectors.

It seems formidable to evaluate this sum in closed-form. We can, however, approximate it by computing the sum over a finite number of eigenvalues numerically. As with the 0+1 case of the previous section the function is not translation invariant (i.e. it depends on $u,v,u',v'$ separately, not just on their differences $u-u',v-v'$).

\begin{figure}[!hp]
\begin{center}
\includegraphics[width = \textwidth]{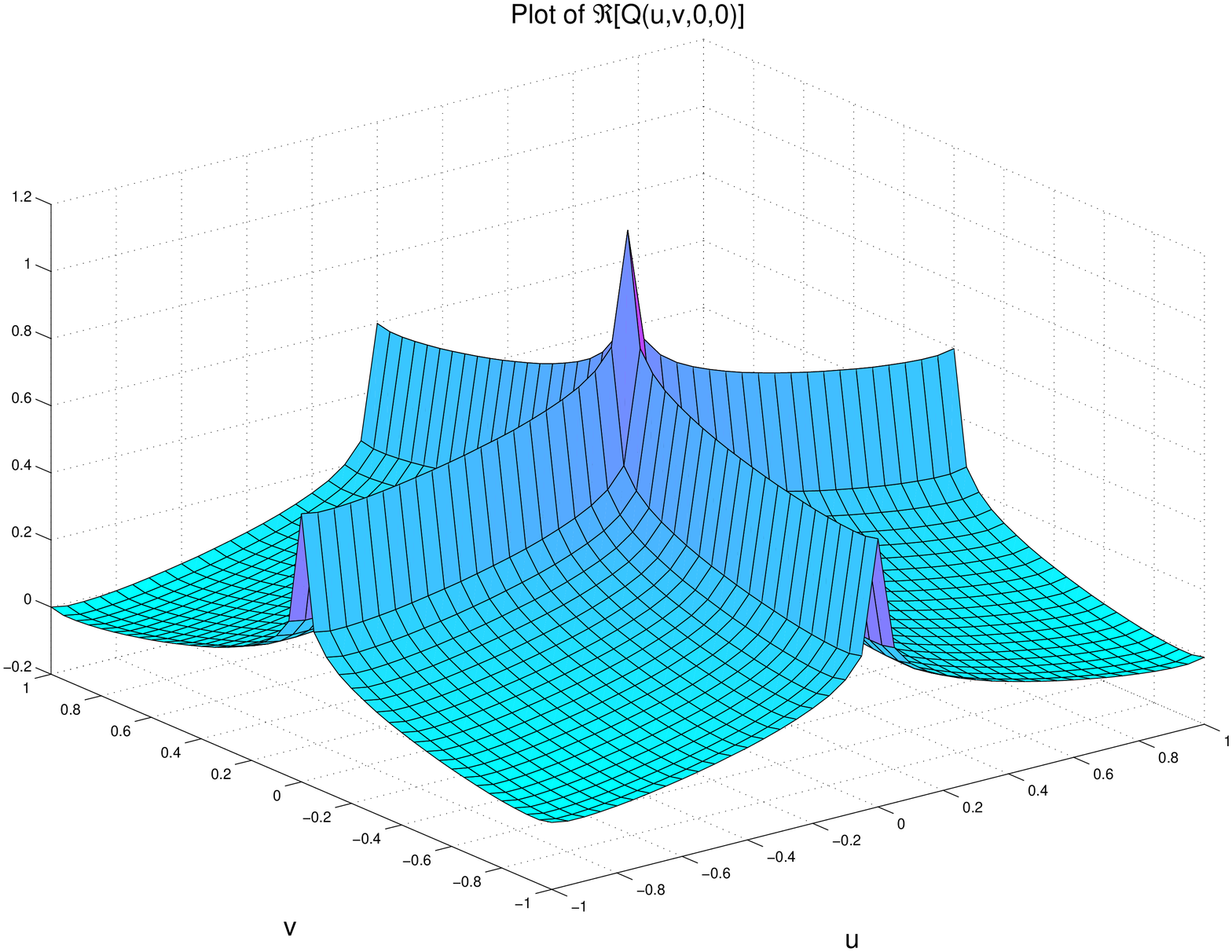}
\caption{The real part of an approximation to $Q(u,v,0,0)$ for $L=1$.}
\label{fig:QRealPlot}
\vspace{1cm} 
\includegraphics[width = \textwidth]{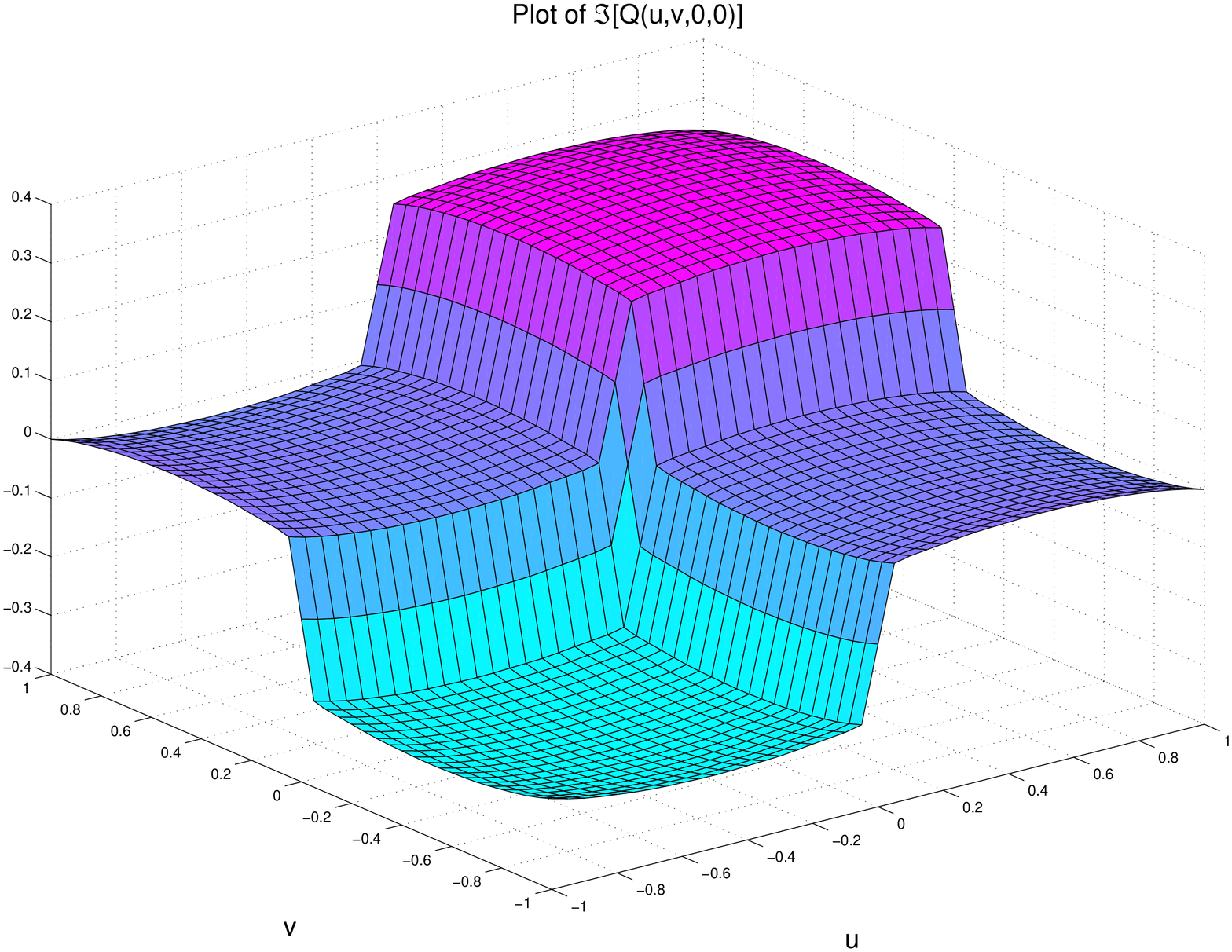}
\caption{The imaginary part of an approximation to $Q(u,v,0,0)$ for $L=1$.}
\label{fig:QImagPlot}
\end{center}
\end{figure}

We plot the real and imaginary parts of $Q(u,v,0,0)$ for $L=1$ in \figref{fig:QRealPlot} and \figref{fig:QImagPlot}. To make the plots we summed over roots of $\tan(x)=2x$ between 0 and 1000.
It is clear from \figref{fig:QImagPlot} that the imaginary part of $Q(u,v,0,0)$ is approaching $\Delta(u,v)/2$. The real part of $Q(u,v,0,0)$ certainly seems to approach a well-defined function but the analytic form of it is not yet known.

\subsubsection{The massive case}

The massive Pauli-Jordan function in $\mink^2$ is given by a Bessel function which decays inside the future and past lightcones. This is bounded above by the massless Pauli-Jordan function and, as such, defines a Hilbert-Schmidt operator. We would like to be able to compute the eigendecomposition for the massive case but have been unable to do due to the complicated integrals involved. Nevertheless since it is a Hilbert-Schmidt operator it does have an eigendecomposition and, in principle, the two-point function could be computed.

It remains possible that the eigendecomposition or the two-point function could be computed by another, less direct, method (e.g. using the Fourier transform). It may be possible, for example, to compute the two-point function directly, without recourse to the eigen-decomposition of $\ID$ (e.g. taking inspiration from \eqref{eq:QSqrt}).

\subsection{Unbounded regions}

The two examples just considered restrict attention to bounded regions $R \subset \mink^d$. If we try to apply the same ideas to unbounded regions, say the whole of $\mink^d$, we encounter divergent expressions. If we ignore these divergences and treat the objects formally we can, with a little care, still obtain the correct two-point function.

\subsubsection{0+1 dimensional Minkowski spacetime}

As a warm-up we consider the calculation in $\mink^1$. As before 
\be i\Delta(x) = \frac{i\sin(m x)}{m} = \frac{1}{2m}\left(e^{imx} - e^{-imx} \right). \ee
This defines an integral operator on functions on $\mink^1$ by
\be (i\Delta f)(x) = \int dy \ID(x-y) f(y), \ee
where we now integrate over all of $\mink^1 = \mathbb{R}$ (not just an interval from $-L$ to $L$).

For $f_c(x) := e^{i c x}$ this gives
\be (\ID f_c)(x) = \frac{1}{2m} \left(\delta(c-m) e^{imx} - \delta(c+m) e^{-imx} \right) \ee
From this we find that as $c$ varies we obtain a family of eigenvectors:
\begin{itemize}
\item If $c \neq \pm m$ then $\ID f_c = \lambda_c f_c = 0$.
\item If $c = m$ then $\ID f_m = \lambda_m f_m$.
\item If $c = -m$ then $\ID f_{-m} = -\lambda_{-m} f_{-m}$. 
\end{itemize}
where we have defined 
\be \label{eq:lambdadef} \lambda_c := \epsilon(c)\frac{\delta(c-m)+\delta(c+m)}{2m} = \epsilon(c)\delta(c^2 -  m^2).\ee
where 
\be 
\epsilon(\alpha) := \left\{ 
\begin{array}{cc} 1 & \textrm{if $\alpha > 0$,} \\ 
0 & \textrm{if $\alpha = 0$,}\\
-1 & \textrm{if $\alpha < 0$.} \end{array} \right. \ee
Note that $\lambda_m$ and $\lambda_{-m}$ are divergent, being proportional to $\delta(0)$.

These eigenvectors are orthogonal and normalised as 
\be \int dx f^*_c(x) f_{c'}(x) = \delta(c-c')\ee

From \eqref{eq:lambdadef} we see the sign of any non-zero eigenvalue $\lambda_c$ is equal to $\epsilon(c)$ (treating $\delta(0)$ as a positive number!).

We define the two-point function by integrating over all eigenvectors but setting the contribution from the negative eigenvalues to zero (this is achieved by a factor of $\theta(\epsilon(c))$ in the integral). This gives:
\begin{align} Q(x,y) :=& \int_{-\infty}^\infty dc \; \theta(\epsilon(c)) \lambda_c f_c(x) f_c(y)^* \\=& \int_0^\infty dc \; \frac{\delta(c-m)}{2m} e^{i c x} e^{-i c y}  = \frac{1}{2m} e^{i m(x-y)} \end{align}
which is precisely the continuum two-point function!

\subsubsection{3+1 dimensional Minkowski spacetime}\label{sec:3+1ContinuumCalc}

The same arguments apply to the calculation in $\mink^d$ for general $d$. Here we look at the $d=4$ case where we have \citep[Appx I, \S B]{Bogoliubov:1959}:
\be \ID(x) = \frac{1}{(2\pi)^3} \int d^4p \epsilon(p_0) \delta(p^2-m^2) e^{i p x} = \frac{1}{(2\pi)^3} \int d^3 \vp \frac{1}{2 \omega_{\vp}} \left(e^{i \tilde{p} x} - e^{- i \tilde{p} x} \right) \ee
where $p := (p_0,\vp)$ is a general momentum 4-vector and $\tilde{p} := (\omega_{\vp},\vp)$ is an on-shell momentum 4-vector with $\omega_{\vp} := \sqrt{\vp^2 + m^2}$.

Treating this as an integral operator acting on functions on $\mink^4$ we have:
\be (\ID f)(x):=\int d^4 y \ID(x-y) f(y) \ee
where we integrate over all of $\mink^4$.

For $f_k(x):=e^{i k x}$ where $k = (k_0,\vk)$ is a general momentum 4-vector (not necessarily on shell) we have that
\begin{align} (\ID f_k) (x) &= \frac{1}{(2\pi)^3} \int d^4 y \int d^3 \vp \frac{1}{2 \omega_{\vp}} \left(e^{i \tilde{p} (x-y)} - e^{- i \tilde{p} (x-y)} \right) e^{i k y} \\
& =  \frac{1}{(2\pi)^3} \int d^3 \vp \frac{1}{2 \omega_{\vp}} \left(e^{i \tilde{p} x} \delta^4(k-\tilde{p}) - e^{- i \tilde{p} x} \delta^4(k+\tilde{p}) \right)\\
& =  \frac{1}{(2\pi)^3} \frac{1}{2 \omega_{\vk}} \left(e^{i \tilde{k}x} \delta(k_0-\omega_{\vk}) - e^{- i \tilde{k}x} \delta(k_0+\omega_{\vk}) \right)
\end{align}
where $\tilde{k} := (\omega_{\vk},\vk)$ is on-shell.

We thus have the following family of eigenvectors:
\begin{itemize}
\item If $k_0 \neq \pm\omega_{\vk}$ then $\ID f_k = \lambda_k f_k = 0$.
\item If $k_0 = \omega_{\vk}$ then $\ID f_k = \lambda_k f_k$.
\item If $k_0 = -\omega_{\vk}$ then $\ID f_{-k} = -\lambda_k f_{-k}$. 
\end{itemize}
where 
\be \lambda_k:= \frac{1}{(2\pi)^3} \epsilon(k_0)\frac{\delta(k_0 - \omega_{\vk}) + \delta(k_0 + \omega_{\vk})}{2 \omega_{\vk}}  = \frac{1}{(2\pi)^3} \epsilon(k_0) \delta(k^2 - m^2)\ee

Note that, unlike the 0+1 calculation, here the non-zero eigenvalues appear with multiplicity---there are multiple eigenvectors for each eigenvalue.

The eigenvectors are orthogonal and normalised as
\be \int d^4 x f_k(x)^* f_{k'}(x) = \delta^4(k - k'). \ee

We see the sign of any non-zero eigenvalue $\lambda_k$ is equal to $\epsilon(k_0)$. To define the two-point function we therefore sum over all eigenvectors but set the contribution from the negative eigenvalues to zero:
\begin{align} Q(x,y) :=& \int d^4k\; \theta(\epsilon(k_0))  \lambda_k f_k(x) f_k(y)^* \\
=& \frac{1}{(2\pi)^3} \int d^4k \;\theta(k_0) \delta(k^2-m^2) e^{i k(x-y)} 
\end{align}
which is exactly the continuum two-point function \citep[Appx I, \S B]{Bogoliubov:1959}!

\subsubsection{Discussion}

The calculations in an unbounded region require working with divergent formal expressions (e.g. eigenvalues proportional to $\delta(0)$). Nevertheless formal manipulations of such expressions do lead to the correct two-point function. It may be possible, therefore, to redo the calculations in a rigorous manner. This could involve (i) using the theory of distributions or (ii) working in a bounded region and taking the limit as the region tends to the unbounded region (compare \eqref{eq:0+1QLim}).

It is also interesting to investigate what we get when this method for obtaining the two-point function is applied to other spacetimes. This has been investigated by \citet{Hustler:2010} for Rindler spacetime.

\section{Mode expansions} \label{sec:NormalModes}

In continuum scalar quantum field theory in $\mink^4$ the field $\phi(x)$ can be expanded in terms of creation and annihilation operators in the form
\be \p(x) = \int d^3\vec{p} \frac{1}{2\omega_{\vec{p}}} \left( \a_{\vec{p}} e^{ipx} + \adag_{\vec{p}} e^{-ipx} \right), \ee
where $\omega_{\vec{p}} := \sqrt{\vec{p}^2+m^2}$ and $\adag_{\vec{p}}$ and $\a_{\vec{p}}$ are the creation and annihilation operators for a particle of on-shell 4-momentum $(\omega_{\vec{p}},\vec{p})$. Similar decompositions hold in other dimensions.

When we compare this to the expansion for our field operators:
\be \tag{\ref{eq:FieldDef}} \p_x = \sum_{i=1}^s \sqrt{\lambda_i}(u_i)_x \a_i + \sqrt{\lambda_i}(v_i)_x \adag_i, \ee
it is tempting to identify the combination $\sqrt{\lambda_i}(u_i)_x$ as playing the role of the plane-waves $\frac{1}{2\omega_{\vec{p}}} e^{i p x}$. We would then be able to identify particular eigenvectors $u_i$ as corresponding to particular on-shell momenta.

To better understand the behaviour of the eigenvectors we look back at the continuum calculation in 1+1 dimensions from \secref{sec:1+1ContinuumCalc}. There the eigenvectors were  of the form 
\begin{align} 
\tag{\ref{eq:feigenvector}} f_k(u,v) &:= e^{iku} - e^{i k v}, & &\textrm{with } k = \frac{n \pi}{L}, \; n = \pm 1, \pm 2, \ldots\\
\tag{\ref{eq:geigenvector}} g_k(u,v) &:= e^{iku} + e^{i k v} - 2 \cos(k L), & &\textrm{with } \tan(kL)= 2kL,\; k \neq 0 \end{align}
We see that each $f_k$ and $g_k$ is a linear combination of plane-waves with on-shell null 2-momenta of energy $k$ (with an additional constant term appearing in $g_k$). In particular there is no multiplicity here---each eigenvalue appears once only.

On the other hand, if we look at the continuum calculation in $\mink^4$ from \secref{sec:3+1ContinuumCalc} we see the eigenvectors are plane-waves (albeit with divergent eigenvalues). In addition the eigenvalues $\lambda_k$ appear with multiplicity---there are multiple plane-waves with the same eigenvalue.

In light of this we suggest the following behaviour. It seems that for sprinklings into finite regions of $\mink^d$ the eigenvectors of the causal set Pauli-Jordan function do not correspond to plane-waves but rather to superpositions of plane-waves (at a fixed energy). As the sprinkling region gets larger and larger the eigenvalues get bunched closer together, e.g. for large causal sets we might have an eigenvalue as $\lambda = 50$ and another at $\lambda' = 50.1$ etc. The corresponding eigenvectors \emph{almost} share the same eigenvalue. Linear combinations of these eigenvectors may then approximate a plane-wave with an energy related to the eigenvalue. In the limit when the region is the whole of $\mink^d$ the eigenvalues appear with multiplicity and the eigenvectors correspond to plane-waves.

This speculative description would bridge the gap between the continuum calculations in bounded and unbounded regions. If it's really what's going on then for large causal sets sprinkled into $\mink^d$ we should be able to obtain eigenvectors (or linear combinations of eigenvectors) to serve as plane-waves.

We can study the behaviour of the eigenvectors for particular sprinklings into finite regions of $\mink^d$ on a computer. We regard the $x$-component of an eigenvector as assigning a complex number to the $x\th$ causal set element. In two-dimensions we can display this as a 3-dimensional plot where two axes are the element's position in $\mink^2$ and the third axis is the real or imaginary part of the eigenvector component.

\begin{figure}[!hp]
\begin{center}
\includegraphics[width = \textwidth]{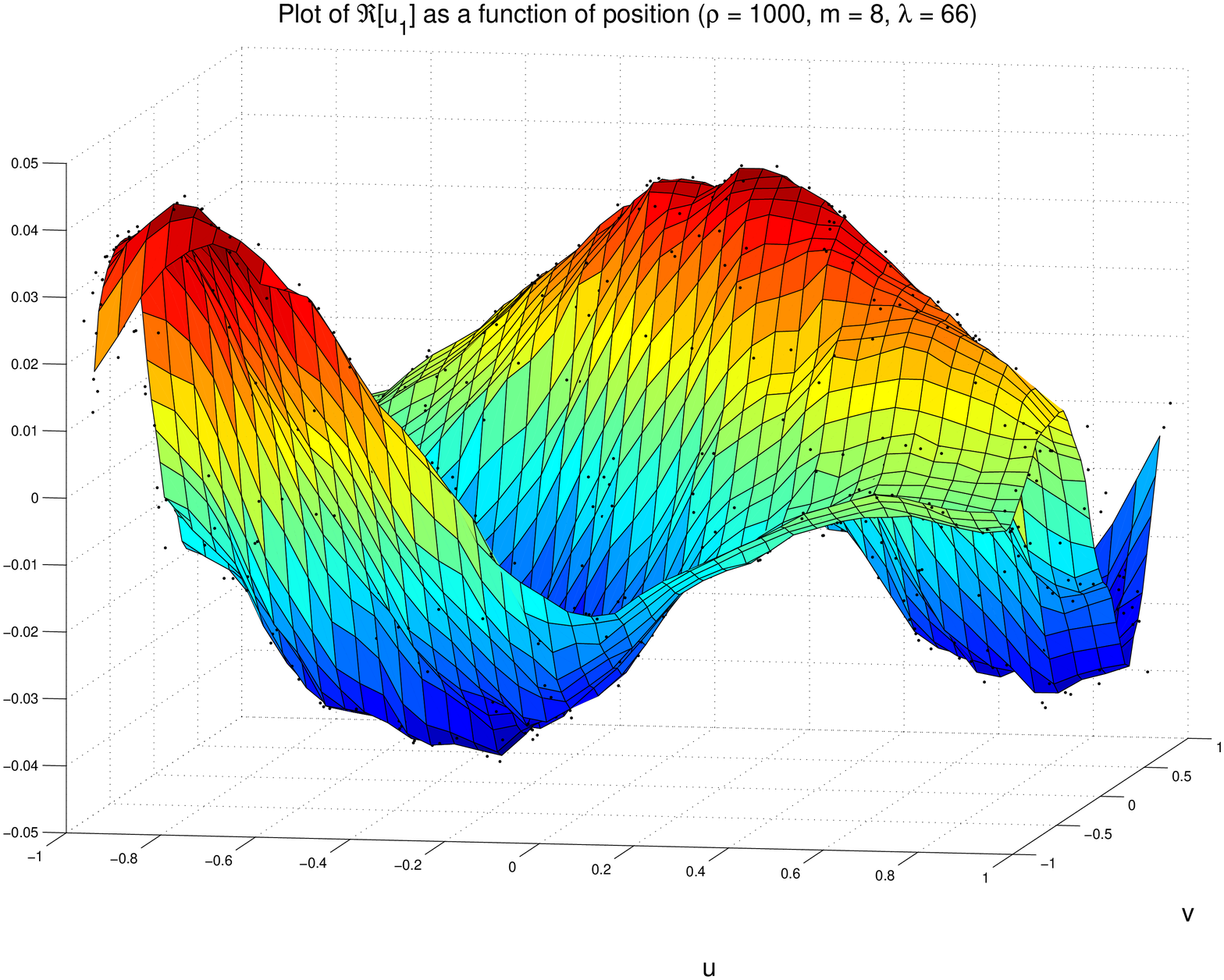}
\caption[The real part of an large-eigenvalue eigenvector]{The real part of $u_1$ for $\rho = 1000, m = 8$ with eigenvalue $\lambda_1 = 66$.}
\label{fig:NMRealPlot}
\vspace{1cm} 
\includegraphics[width = \textwidth]{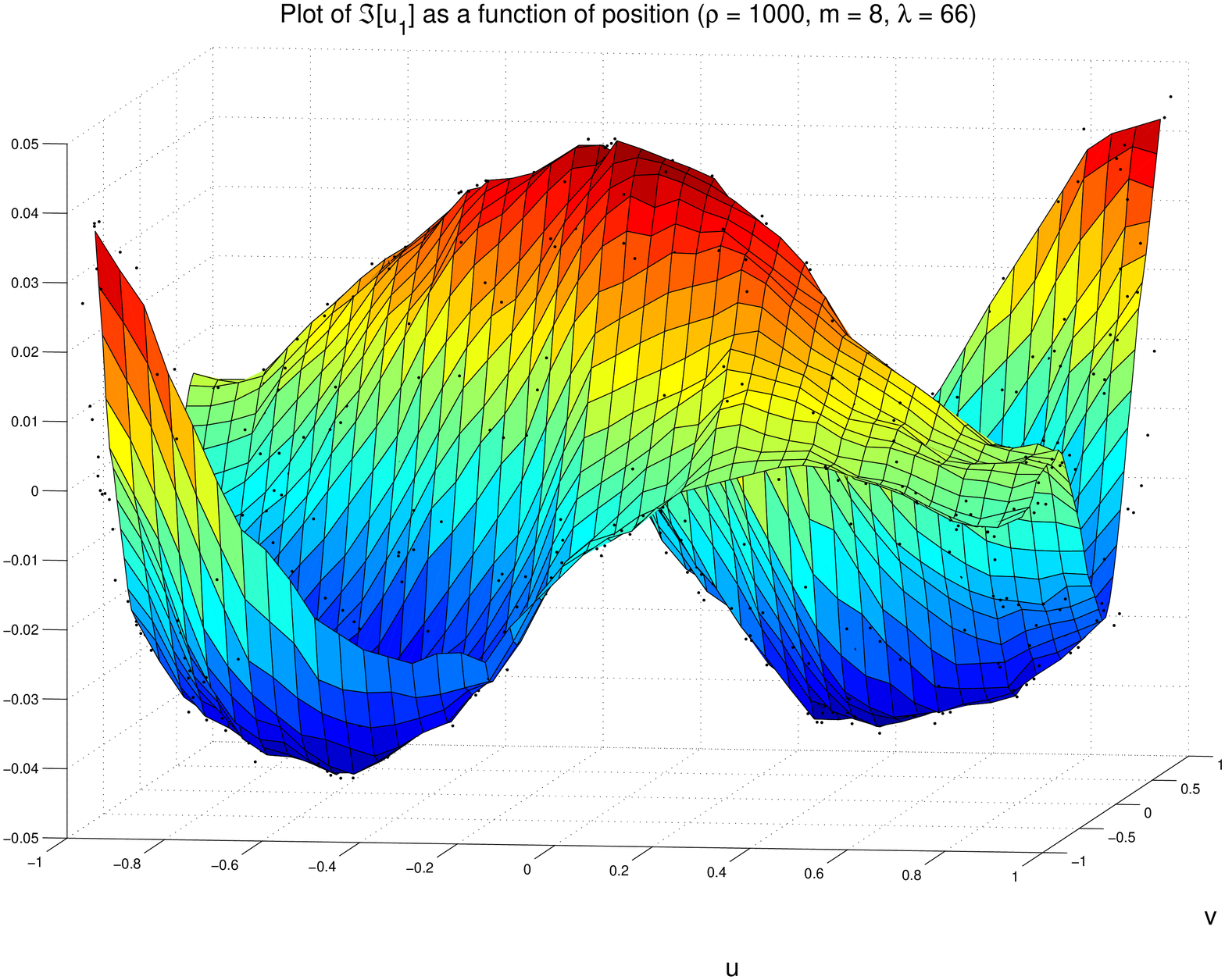}
\caption[The imaginary part of an large-eigenvalue eigenvector]{The imaginary part of $u_1$ for $\rho = 1000, m = 8$ with eigenvalue $\lambda_1 = 66$.}
\label{fig:NMImagPlot}
\end{center}
\end{figure}

In \figref{fig:NMRealPlot} and \figref{fig:NMImagPlot} we plot the real and imaginary parts of the eigenvector (which we denote $u_1$) associated with the largest eigenvalue of $\ID$, for a sprinkling into a causal interval in $\mink^2$. In the plot the  dots denote the values of $(u_1)_x$ at the $x\th$ sprinkled point. The surface is an interpolation between them (we hope there is no confusion between the coordinate $u$ and the eigenvector $u_1$). Interestingly the eigenvectors associated with the largest eigenvalues show relatively smooth oscillating behaviour. There are oscillations in more than one direction which suggests that the eigenvectors best correspond to a sum of plane-waves---if they corresponded to a single plane-wave then the peaks and troughs would proceed in a single spacetime direction.

\begin{figure}[!hp]
\begin{center}
\includegraphics[width = \textwidth]{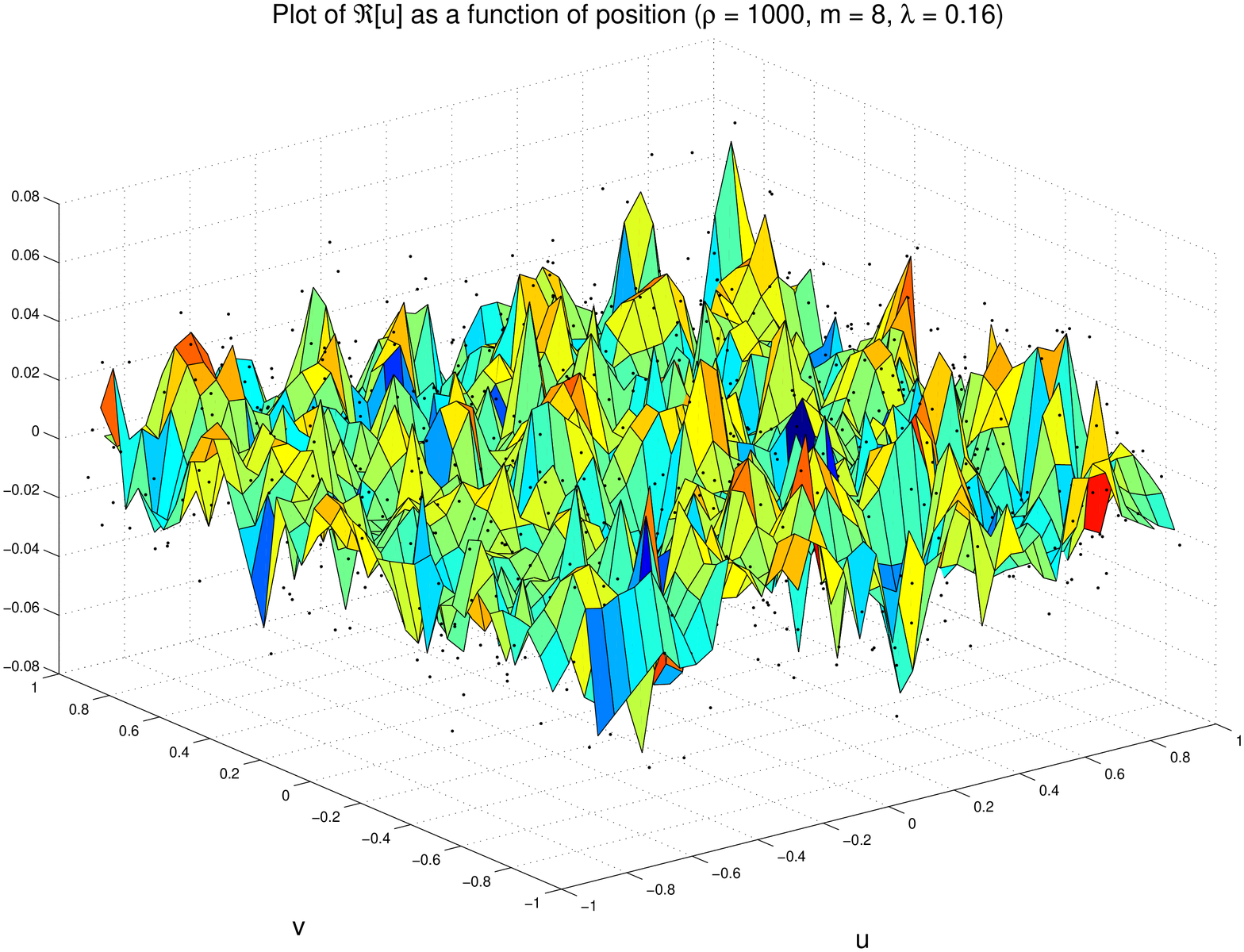}
\caption[The real part of an small-eigenvalue eigenvector]{The real part of an eigenvector $u$ for $\rho = 1000, m = 8$ with eigenvalue $\lambda = 0.16$.}
\label{fig:NMRealPlotJagged}
\vspace{1cm} 
\includegraphics[width = \textwidth]{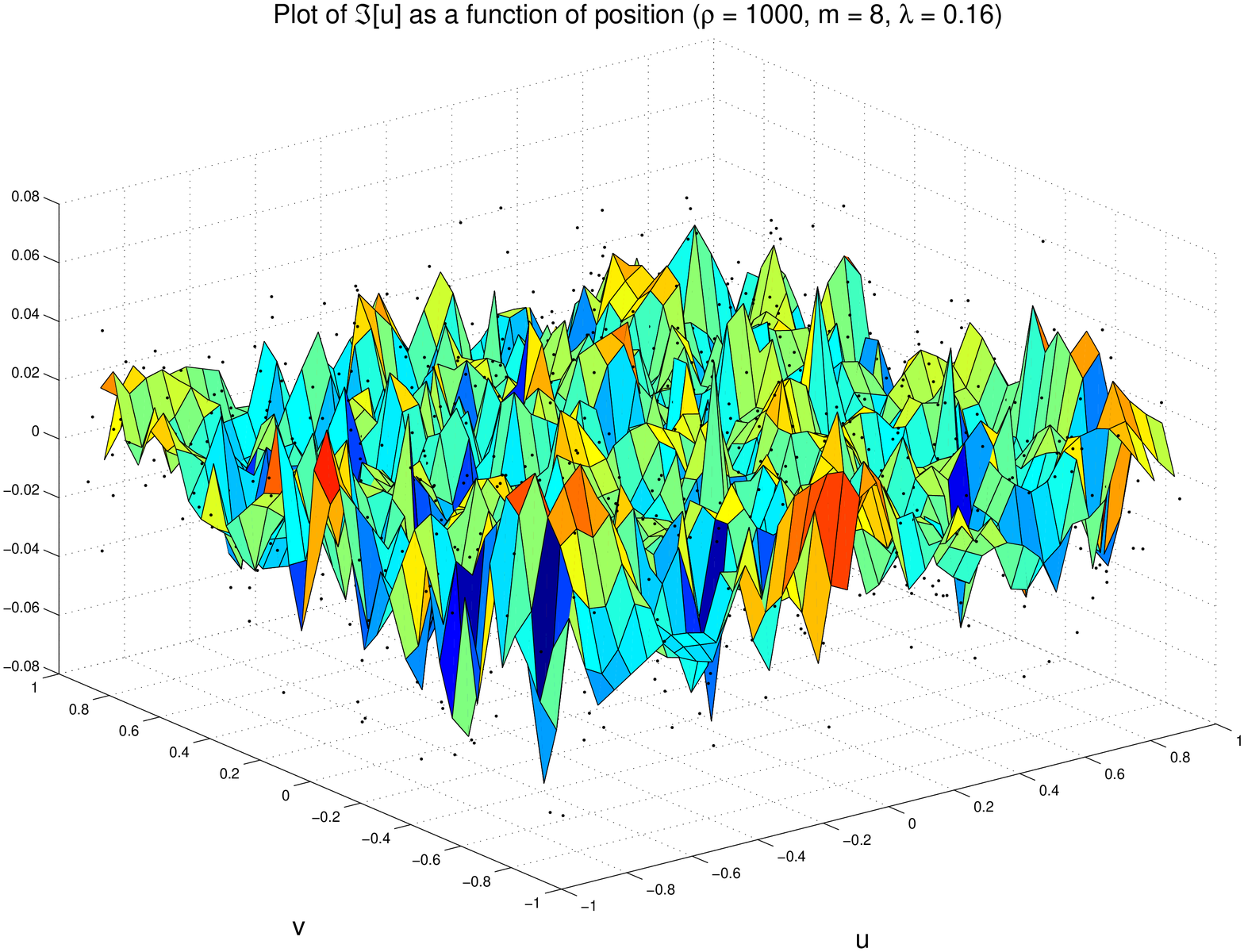}
\caption[The imaginary part of an small-eigenvalue eigenvector]{The imaginary part of an eigenvector $u$ for $\rho = 1000, m = 8$ with eigenvalue $\lambda = 0.16$.}
\label{fig:NMImagPlotJagged}
\end{center}
\end{figure}

The eigenvectors associated with the smaller eigenvalues show less smooth, more jagged behaviour (as shown in \figref{fig:NMRealPlotJagged} and \figref{fig:NMImagPlotJagged}).

In light of \secref{sec:3+1ContinuumCalc} it appears the eigenvalue can be related to the energy of the plane-wave. Since the mass-dimensions of the causal set eigenvalues are $[\lambda_i] = M^{d-2}$ the quantity
\be E_i:=\sqrt{\frac{\rho}{\lambda_i}}, \ee
has dimensions of energy: $[E_i] = M$. This suggests that $E_i$ could be interpreted (up to a dimensionless proportionality constant) as the energy associated with the eigenvector $u_i$. This would agree with the smooth behaviour of ``low energy'' (large eigenvalue) eigenvectors and the jagged behaviour of ``high energy'' (small eigenvalue) eigenvectors. 

In addition the sign of the energy could also be identified with the sign of the eigenvalue: i.e. the $u_i$ would have positive energy and the $v_i$ would have negative energy. This would agree with the the continuum two-point function which is a sum over plane-wave modes of positive energy \citep[Appx I, \S B]{Bogoliubov:1959} (whereas the causal set $Q$ matrix is a sum over the  eigenspace spanned by the $u_i$ eigenvectors).

We should mention that since energy is a frame-dependent quantity we have to decide what frame these statements refer to. For sprinklings into a finite causal interval the causal interval picks out a preferred time-axis (i.e. the axis through the end-points of the interval). Presumably the energy assigned to the eigenvectors is the energy of the plane-waves in a frame with this time-axis.

\subsection{Preferred set of modes}

An important aspect in conventional quantum field theory in curved spacetimes is the lack of a preferred choice of modes with which to expand the field \citep[p45]{Birrell:1982}. Taking inspiration from the causal set model it is possible that the eigenvectors of the Pauli-Jordan function could be used to expand the field in an arbitrary curved spacetime (or possibly a finite sub-region of such a spacetime).

As mentioned in \secref{sec:CurvedSpacetimes} the Klein-Gordon equation in a globally hyperbolic spacetime is guaranteed to possess unique retarded and advanced Green's functions. This ensures that the Pauli-Jordan function exists and is unique. If its eigenvectors could be determined (along the lines of \secref{sec:ContinuumCalc}) then these could be used to either (i) define a field operator expansion or (ii) define a Feynman propagator directly without mentioning operators. This approach could provide a unique way to define the Feynman propagator in curved spacetimes---something which, if possible, would have important ramifications for the usual theory.

We can phrase this suggestion in another way. The usual theory involves generalising the field's Fourier mode-decomposition from $\mink^d$ to a curved Lorentzian manifold. In most cases there is no unique way to perform this generalisation. The present formalism, however, suggests that the Fourier modes and the eigenvectors of $\ID$ coincide in $\mink^d$. From this point of view it is the eigenvectors of $\ID$, \emph{not} the Fourier modes, which should be generalised to curved spacetimes. These eigenvectors (if they exist!) would then provide a preferred set of modes with which to expand the field.

\def\SM{S-matrix }
\def\SO{$\hat{S}$-operator }
\def\SS{\hat{S}}
\def\p{\hat{\phi}}
\def\h{\hat{h}}
\def\A{\hat{A}}
\def\H{\hat{H}}
\def\G{\hat{G}}
\def\I{\mathbb{I}}
\def\U{\hat{U}}
\def\lv{\langle 0 |}
\def\rv{| 0 \rangle}
\def\adag{\hat{a}^\dagger}
\def\a{\hat{a}}
\def\Ale{\bar{A}}
\def\HH{\mathcal{H}}
\def\FF{\mathcal{F}_s(\HH)}
\def\f{f}
\def\g{g}

\chapter{Interacting Quantum Field Theory} \label{chap:InteractingQFT}

\begin{quote}
All my efforts up to that point had been directed toward building a complete convergent theory. Finding out that after all the series diverged convinced me that was as far as one could go\ldots. That was of course a terrible blow to all my hopes. It really meant that this whole program made no sense.

\flushright{Freeman Dyson in \citet[p565]{Schweber:1994}}
\end{quote}

In the previous chapter we described a model for free quantum scalar field theory on a causal set. Here we extend that model to include interactions between the scalar fields. This takes us one step closer to a realistic theory since all known matter fields are involved in interactions. Unfortunately there are, as yet, no known scalar fields in nature. In the next chapter we discuss possible models for particles with non-zero spin.

Ultimately our aim is to calculate scattering amplitudes and interaction cross-sections for particular interacting theories. If a physically realistic theory could be modelled on a causal set (this would require modelling spin-half and spin-one particles) then these scattering amplitudes would differ from their continuum counterparts because of the causal set discreteness. Comparing the causal set calculations to experiments would, at the very least, place bounds on the allowed discreteness scale. A phenomenological approach to this is described in \secref{sec:Phenomenology}.

\section{Operators on Hilbert space} \label{sec:Operators}

\def\A{\hat{A}}
\def\Adag{\hat{A}^\dagger}
\def\B{\hat{B}}

We start by reviewing the mathematical tools we shall need in this chapter. These will be needed to put the free field theory of the previous chapter on a rigorous foundation from which we can then define the interacting theory.

We assume the reader is already familiar with the basics of Hilbert space theory (if not, \citet[\chap 48-54]{Geroch:1985} and \citep[\chap II]{Reed:1972} provide good introductions). Our definitions follow \citet[\chap 55-56]{Geroch:1985} and \citet[\chap VIII]{Reed:1972}.

\subsection{Definitions}

Let $\HH$ be a Hilbert space. An \emph{operator} on $\HH$ consists of a vector subspace of $D(\A)$ (called the \emph{domain} of $\A$) of $\HH$ together with a linear mapping $\A: D(\A) \to \HH$ (the domain may be equal to $\HH$ or it may be strictly contained within it). An operator whose domain is a proper subset of $\HH$ is called \emph{unbounded}.
Two operators $D(\A), \A$ and $D(\B), \B$ are \emph{equal} if $D(\A) = D(\B)$ and $\A = \B$ on this common domain.

An operator $D(\A),\A$ is an \emph{extension} an operator $D(\B), \B$ if $D(\B) \subseteq D(\A)$ and $\A=\B$ whenever both are defined (i.e. on $D(\B)$).

The domain $D(\A)$ of an operator is \emph{dense} in $\HH$ if, for every $v \in \HH$ and every $\epsilon > 0$ there exists a vector $v' \in D(\A)$ such that $||h - h'|| < \epsilon$ (where $||\cdot||$ denotes the norm in $\HH$). An operator with a dense domain is called \emph{densely defined}.

The sum of operators $D(\A), \A$ and $D(\B),\B$ is the operator with domain $D(\A+\B):=D(\A) \cap D(\B)$ defined as $\A + \B$ on this domain.

The product of operators $D(\A), \A$ and $D(\B),\B$ is the operator with domain $D(\A \B):=\left\{ v \in D(\B) : \B v \in D(\A)\right\}$ and defined as $\A\B$ on this domain.

An operator $D(\A),\A$ is \emph{closed} if whenever a sequence of vectors $v_n \in D(\A)$ converges to a limit vector $v$ and the sequence of vectors $\A v_n$ converges to a limit vector $u$, then $v \in D(\A)$ and $\A v = u$.

An operator $D(\A), \A$ is \emph{symmetric} if $(\A u, v) = (u, \A v)$ for all $u, v \in D(\A)$ (here $(\cdot,\cdot)$ denotes the inner-product in $\HH$).

\subsection{Adjoint of an operator}

Let $D(\A), \A$ be a densely defined operator on $\HH$. We can define another operator $D(\Adag), \Adag$ called the \emph{adjoint} of $\A$ as follows \citep[p252]{Reed:1972}, \citep[\chap 56]{Geroch:1985}.

We let $D(\Adag)$ be the set of all $v \in \HH$ such that there exists a $u \in \HH$ with
\be \left( \A w, v\right) = (w, u), \quad \textrm{ for all $w \in D(\A)$}. \ee
For each such $v \in D(\Adag)$ we define $\Adag v := u$. If $D(\A)$ is dense in $\HH$ then this $u$ is unique.

An operator $D(\A), \A$ is \emph{self-adjoint} if it is equal to its adjoint: $D(\A) = D(\Adag)$ and $\A = \Adag$ on this common domain.

An operator $D(\U), \U$ is \emph{unitary} if $D(\U) = \HH$ and $\U \U^\dagger = \U^\dagger \U = \I$, the identity operator on $\HH$.

\subsubsection{Example}

Let $\HH = L^2(\mathbb{R})$ be the Hilbert space of square-integrable functions $\psi : \mathbb{R} \to \mathbb{C}$. The \emph{position operator} $D(\q),\q$ is the operator defined by
\be D(\q) := \left\{ \psi \in L^2(\mathbb{R}) : \int_{-\infty}^\infty dq \; |q \psi(q)|^2 < \infty \right\},\ee
and $ \q \psi = q \psi $ for $\psi \in D(\q)$. It can be shown that this operator is self-adjoint \citep[p331-332]{Geroch:1985}.

\subsection{Spectral theorem} \label{sec:SpectralTheory}

An important result concerning self-adjoint operators is the spectral theory. This has a number of equivalent definitions and we follow \citet[\S14]{Jordan:1969}. We start with a few definitions.

A \emph{projection operator} $D(\E), \E$ is a self-adjoint operator defined on all of $\HH$ (i.e. $D(\E) = \HH)$ such that $\E^2 = \E$.

A family of projection operators $\E_x$ depending on a real parameter $x$ is a \emph{spectral family} if it satisfies:
\begin{enumerate}
\item If $x \leq y$ then $\E_x \leq \E_y$ (meaning $\E_x \E_y = \E_y \E_x = \E_x$),
\item For $\epsilon > 0$ we have $\E_{x+\epsilon} v \to \E_x v$ as $\epsilon \to 0$ for any $v \in \HH$ and any $x \in \mathbb{R}$,
\item $\E_x v \to 0$ as $x \to -\infty$ and $\E_x v \to v$ as $x \to +\infty$ for any vector $v \in \HH$.
\end{enumerate}

For any self-adjoint operator $D(\A),\A$ there exists a unique spectral family of projection operators $\E_x$ such that
\be (u, \A v) = \int_{-\infty}^\infty x d(u, \E_x v), \ee
for all $u, v \in D(\A)$.
where the integrals are Riemann-Stieltjes integrals\footnote{Meaning that $\int_a^b g(x) dF(x)$ is the limit of \be \sum_{k=1}^n g(x_k)\lbrack F(x_k) - F(x_{k-1})\rbrack,\ee
as $n\to \infty$ and $a < x_1 < x_2 < \ldots < x_n \leq b$ divides the range of integration into smaller and smaller pieces.}.

\subsubsection{Example}

For the position operator $D(\q),\q$ on $L^2(\mathbb{R})$ its spectral family $\E_x$ are the operators defined by \citep[p43]{Jordan:1969}:
\be (\E_x \psi)(q) :=\left\{ \begin{array}{ll} \psi(q) & \textrm{ if } x \leq q \\ 0 & \textrm{ if } x > q. \end{array} \right. \ee

We then have, for $\psi, \phi \in D(\q)$ that
\begin{align} \int_{-\infty}^\infty x d(\psi, \E_x \phi) &= \int_{-\infty}^\infty x d \int_{-\infty}^\infty dq \;\psi^*(q)(\E_x \phi)(q) &\\
& = \int_{-\infty}^\infty x d \int_{-\infty}^x dq \;\psi^*(q) \phi(q) = \int_{-\infty}^\infty x \psi^*(x) \phi(x) &\\&= (\psi,\q \phi)&\end{align}

\subsection{Functional calculus} \label{sec:FunctionalCalculus}

The spectral decomposition of a self-adjoint operator can be used to define functions of the operator \citep[\S 15]{Jordan:1969}. For a function $f : \mathbb{R} \to \mathbb{C}$ we have
\be (u, f(\A) v) := \int_{-\infty}^\infty f(x) d(u, \E_x v). \ee
If the function $f$ is bounded for all $x \in \mathbb{R}$ then this formula is valid for all $u, v \in \HH$. If $f$ is unbounded then the $u$ and $v$ are restricted to appropriate domains \citep[p49, footnote 16]{Jordan:1969}.

\section{Fock space} \label{sec:FockSpace}
The appropriate arena for discussing interacting scalar quantum field theory on a causal set is the Fock space obtained in the previous chapter (see \secref{sec:FieldsOnCausalSet}). Here we look at this space and the field operators that act on it in a more rigorous manner. A clear account of the construction of a Fock space and its associated creation and annihilation operators is given in \citet[Example 2, Sec II.4]{Reed:1972}, \citet[Sec X.7]{Reed:1975}, \citet[\chap 21]{Geroch:1985}.

The bosonic Fock space for real scalar fields on a causal set is based on the ``one particle'' Hilbert space $\HH := \mathbb{C}^s$ (where $2s$ is the rank of the matrix $i\Delta$). The Fock space is then
\be \FF := \bigoplus_{n=0}^\infty S_n (\HH^{(n)}), \ee
where $S_n$ is the symmetrising operator, $\HH^{(0)} := \mathbb{C}$ and $\HH^{(n)} = \otimes_{k=1}^n \HH$ for $n \geq 1$.

The symmetrising operator is a linear mapping from $\HH^{(n)}$ to itself which ensures that states in $S_n(\HH^{(n)})$ are invariant under a permutation of the one particle Hilbert spaces in the tensor product.

As an example, $S_2$ is the map from $\HH^{(2)}$ to itself defined by
\be S_2(u \otimes v):=\frac{1}{2}\left(v \otimes u + u \otimes v\right).\ee Similarly for $\HH^{(3)}$ we have\be S_3(u \otimes v \otimes w) := \frac{1}{6}(u \otimes v \otimes w + u \otimes w \otimes v+v \otimes u \otimes w+v \otimes w \otimes u+w\otimes v \otimes u+w \otimes u \otimes v),\ee
and similarly for general $\HH^{(n)}$ (see \citet[p115]{Geroch:1985} for full details). The extra factors of $1/n!$ are included to ensure that $S_n$ is a projection operator (i.e. $S_n^2 = S_n$ and $S_n^\dagger = S_n$, where $\dagger$ denotes the operator adjoint).

Vectors in $\FF$ are sequences of vectors $(\psi_0,\psi_1,\psi_2,\ldots)$ with $\psi_n \in S_n (\HH^{(n)})$ for $n = 0,1,2,\ldots$ which satisfy
\be \label{eq:FockConverge} \sum_{n=0}^\infty ||\psi_n||_n < \infty,\ee
where $|| \cdot ||_n$ denotes the norm in $\HH^{(n)}$.

The inner-product of vectors $\psi = (\psi_0,\psi_1,\psi_2,\ldots)$ and $\phi = (\phi_0,\phi_1,\phi_2,\ldots)$ in $\FF$ is defined to be
\be (\psi,\phi) := \sum_{n=0}^\infty (\psi_n, \phi_n)_n, \ee
where $(\cdot,\cdot)_n$ denotes the inner-product in $\HH^{(n)}$. For all $\psi,\phi \in \FF$ this sum converges.

Vectors for which only a finite number of the $\psi_n$ are non-zero are called \emph{finite particle vectors}. The set of all such vectors is denoted $F_0$ and is dense in $\FF$.

The vector $\rv := (1,0,0,0,\ldots) \in \FF$ is the \emph{vacuum state}.

Vectors in $\HH$ are taken to represent the possible states of a single particle. Vectors in $S_n(\HH^{(n)})$ represent the states of $n$ identical particles. The particles are bosons which is why their states are symmetric under permutations of the single-particle Hilbert spaces. For further discussion of $\HH$ see \secref{sec:InterpretationofFock}.

\subsection{Creation and annihilation operators}

For any vector $\f \in \HH$ we can define a pair of mutually adjoint operators: the annihilation operator $\a(\f)$ and the creation operator $\adag(\f)$. These are unbounded operators defined on dense subspaces of $\FF$.

Suppose $\psi = (\psi_0,\psi_1,\psi_2,\psi_3,\ldots) \in \FF$ then the creation operator acts as:
\be\label{eq:CreationDef} \adag(\f) \psi := (0, \sqrt{1}\f \psi_0, \sqrt{2} S_2(\f \otimes \psi_1), \sqrt{3} S_3(\f \otimes \psi_2), \cdots ).\ee

It's simplest to define the annihilation operator for unsymmetrised ``finite particle states'':
\begin{align}
\a(\f) (\psi_0,0,0,\ldots) &:= (0,0,0,0,\ldots) \\
\a(\f)(0,\psi_1,0,0,\ldots) &:= (\sqrt{1}(f,\psi_1),0,0,0,\ldots) \\
\a(\f) (0,0,\psi_1 \otimes \phi_1,0,\ldots) &:= (0,\sqrt{2}(f,\psi_1)\phi_1,0,\ldots) \\
\a(\f) (0,0,0,\psi_1 \otimes \phi_1 \otimes \chi_1,0,\ldots) &:= (0,0,\sqrt{3}(f,\psi_1)\phi_1 \otimes \chi_1,0,\ldots) \\
\nonumber& \vdots
\end{align}
where $\psi_0 \in \mathbb{C}$, $\psi_1,\phi_1,\chi_1 \in \HH$ and $(u,v)$ denotes the inner-product of $u,v \in \HH$. These definitions are then extended by linearity to vectors $\psi \in \FF$.

Both $\a(\f)$ and $\adag(\f)$ are not defined on all of $\FF$. This is because there are vectors in $\FF$ such that the right-hand-side of, for example \eqref{eq:CreationDef}, is not a vector in $\FF$ (because the sum analogous to \eqref{eq:FockConverge} does not converge for that particular right hand side of \eqref{eq:CreationDef}). The domains of $\a(\f)$ (resp. $\adag(\f)$) are those vectors $\psi \in \FF$ such that $\a(\f) \psi \in \FF$ (resp. $\adag(\f)\psi \in \FF$). In particular, both $\a(\f)$ and $\adag(\f)$ are well-defined on the finite particle vectors, i.e. the domains of $\a(\f)$ and $\adag(\f)$ contain $F_0$.

It can be shown that $\a(\f)$ and $\adag(\f)$ are mutually adjoint operators on $\FF$ and that, on the appropriate domains, they satisfy the following commutation rules:
\begin{align}
\label{eq:CommutationRelations} \nonumber \left[ \a(\f), \a(\g) \right] &= \left[ \adag(\f), \adag(\g) \right]  = 0, \\
\left[ \a(\f), \adag(\g) \right] &= (\f,\g)\I,
\end{align}
where $(\f,\g)$ is the inner-product of $\f, \g \in \HH$ and $\I$ is the identity operator on $\FF$.

We note that the map $f \to \adag(f)$ is linear whereas the map $f \to \a(f)$ is anti-linear.
The $n$-particle states in $\FF$ are spanned by states of the form $\adag(f_1) \adag(f_2) \ldots \adag(f_n) \rv$ for $f_1,\ldots,f_n \in \HH$.

\subsection{Schr\"odinger representation} \label{sec:SchrodingerRep}

The Fock space based on $\mathbb{C}^s$ can be represented in a different way known as the Schr\"odinger representation \citep[Example 5.2.16, p36]{Bratteli:1981}. We start with the $L^2(\mathbb{R}^s)$ Hilbert space and define
\be \label{eq:SchroRep} \adag(f) = \frac{1}{\sqrt{2}} \sum_{i=1}^s f_i \left(q_i - \frac{\partial}{\partial q_i}\right), \qquad \a(f) = \frac{1}{\sqrt{2}} \sum_{i=1}^s f^*_i \left(q_i + \frac{\partial}{\partial q_i}\right). \ee
In this representation the vacuum state (defined by $\a(f)\rv = 0$ for all $f \in \mathbb{C}^s$ and $\lv 0 \rangle$ = 1) is a multi-dimensional Gaussian:
\be \rv = \pi^{-s/2} \exp\left(-\frac{q_1^2 + q_2^2 + \ldots q_s^2}{2}\right).\ee

\section{Field Operators}

Having laid the groundwork we can now return to defining scalar quantum field theory on a causal set.

To represent a free scalar field on a causal set $(\CS,\preceq)$ we assign, to every element $v_x \in \CS$, an operator $\p_x$ defined by
\be \label{eq:FieldDefinition} \p_x := \a(\f_x) + \adag(\f_x), \textrm{ with } \f_x:= \sum_{i=1}^s \sqrt{\lambda_i} (v_i)_x \vec{e}_i,\ee
a vector in $\HH$. Here $x=1,\ldots,p$ and $\vec{e}_i$ is an orthonormal set of basis vectors in $\mathbb{C}^s$.
Note that these $\f_x \in \mathbb{C}^s$ vectors have inner-products $(\f_x, \f_y) = Q_{xy}$.

We can use the linearity of $\adag$ and anti-linearity of $\a$ to expand this expression as:
\be \p_x = \sum_{i=1}^s \sqrt{\lambda_i} (u_i)_x \a(\vec{e}_i) + \sqrt{\lambda_i} (v_i)_x \adag(\vec{e}_i), \ee
which is of the form \eqref{eq:FieldDef} if we identify $\a_i := \a(\vec{e}_i)$ and $\adag_i := \adag(\vec{e}_i)$.

In turn we can express the creation and annihilation operators in terms of the field operators. If $g \in \mathbb{C}^s$ then
\be \a(g) := \sum_{i=1}^s \sum_{x=1}^{p} g_i^* \frac{1}{\sqrt{\lambda_i}} (v_i)_x \p_x, \qquad \adag(g) := \sum_{i=1}^s \sum_{x=1}^{p} g_i\frac{1}{\sqrt{\lambda_i}} (u_i)_x \p_x.\ee
where we use the orthogonality of the eigenvectors \eqref{eq:EigenOrtho}.

We mention that $\p_x$ is related to the \emph{Segal field operator} \citep[p209-210]{Reed:1975} $\Phi_S(\f)$ by $\p_x = \Phi_S(\sqrt{2} \f_x)$. Each $\p_x$ is essentially self-adjoint\footnote{The domain of $\p_x$ contains $F_0$ which contains a dense set of analytic vectors. See \citet[Thm X.41]{Reed:1975}.}. This means it has a unique self-adjoint extension (which we also denote $\p_x$).

These field operators $\p_x$ therefore satisfy:
\begin{enumerate}
\item $\p_x$ is self-adjoint.
\item For $\psi \in F_0$, $\lbrack \p_x,\p_y\rbrack\psi  =i \Delta_{xy} \psi$. This follows from \eqref{eq:CommutationRelations} and \eqref{eq:FieldDefinition}.
\item $i\Delta w = 0 \implies \sum_{x'=1}^{p} w_{x'} \p_{x'} = 0$. This follows because $i\Delta w = 0 \implies w^\dag v_i = w^\dag u_i = 0$ for $i=1,\ldots,s$ and from the linearity of $\f \mapsto \adag(\f)$ and anti-linearity of $\f \mapsto \a(\f)$.
\end{enumerate}
These are the three conditions for the $\p_x$ operators given in \secref{sec:FieldsOnCausalSet}, but with more attention paid to operator domains in condition 2.

\subsection{Field operators in the Schr\"odinger representation} \label{sec:FieldSchrodinger}

Using the Schr\"odinger representation for the creation and annihilation operators \eqref{eq:SchroRep} the field operators become:
\be \p_x = \frac{1}{\sqrt{2}} \sum_{i=1}^s (f_x)_i \left(q_i - \frac{\partial}{\partial q_i}\right) + (f_x)^*_i \left(q_i + \frac{\partial}{\partial q_i}\right). \ee
Denoting the familiar position and momentum operators by $\hat{\mathbf{q}} := (q_1,q_2,\ldots,q_s)$ and $\hat{\mathbf{p}} := -i(\frac{\partial}{\partial q_1}, \frac{\partial}{\partial q_2}, \ldots, \frac{\partial}{\partial q_s})$
we have
\be \label{eq:FieldSchro} \p_x = A_x \cdot \hat{\mathbf{q}} + B_x \cdot \hat{\mathbf{p}}, \ee
where $A_x = \sqrt{2} \Re[f_x]$ and $B_x = \sqrt{2} \Im[f_x]$ are the real and imaginary part of the $f_x$ vectors. This alternative representation may be useful for calculations.

\section{Scattering amplitudes in the continuum}

One of the main aims of any quantum field theory is the calculation of scattering amplitudes. These are amplitudes for particular interactions to occur and can be used to calculate experimentally measurable quantities such as scattering cross-sections. See \citet{Bogoliubov:1959,Schweber:1961,Weinberg:1995} for full details.

To model a scattering experiment using quantum field theory in $\mink^4$ we represent the incoming particles by an initial in-state $|\psi_{\textrm{in}}\rangle \in F$. This is a vector in the Fock space associated with the free field operators (which we denote by $F$) and represents the state of the system in the infinite past. We can ask what is the amplitude that the final state of the system in the infinite future is a free field out-state $|\psi_{\textrm{out}} \rangle \in F$. The answer is that the amplitude equals the matrix-element $\langle \psi_\textrm{out} | \SS | \psi_\textrm{in} \rangle$ of a unitary ``scattering operator'' $\SS$ (which we shall refer to as the $\SS$-operator). The collection of these amplitudes for all incoming and outgoing states is called the S-matrix. To calculate scattering amplitudes we thus have to calculate the $\SS$-operator.

\subsection{Calculating the S-matrix}  \label{sec:Smatrixcontinuum}

To define a particular interacting scalar field theory in $\mink^d$ it is necessary to specify an \emph{interaction Hamiltonian density} $\H(x)$. This is a polynomial in the free field operators which make up the interacting theory, for example $\H(x) = \lambda \p^4(x)$ for a self-interacting scalar field.

Once the interaction Hamiltonian is known the \SO can be computed perturbatively by the well-known Dyson series:
\be \label{eq:DysonSeries} \SS := \sum_{n=0}^\infty \SS^{(n)}, \ee
where $\SS^{(0)}:= \I$ is the identity operator on the Fock space for the field operators and
\be \SS^{(n)} := \frac{(-i)^n}{n!} \int \!\!\ldots\!\! \int d^dx_1 \ldots d^dx_n T(\H(x_1) \ldots \H(x_n)),\ee
where the time-ordering $T$ has time increasing from right to left.

Without discussing the convergence of the series, the Dyson series provides only a formal definition of $\SS$. Unfortunately, it appears that the series for $\SS$ may be divergent \citep[p207]{Bogoliubov:1959}. This mathematical difficulty has not stopped the formalism being used to good effect---calculations are performed using formal manipulations and renormalisation. Nevertheless, attempts to go beyond formal manipulations and define the \SM rigorously run into difficulties \citep[p223]{Reed:1975}.

The scattering amplitude between an in-state $|\psi_\textrm{in}\rangle$ and an out-state $|\psi_\textrm{out}\rangle$ is given by the matrix element $\langle \psi_\textrm{out} | \SS | \psi_\textrm{in} \rangle$. The matrix elements $\langle \psi_\textrm{out} | \SS^{(n)} | \psi_\textrm{in} \rangle$ are then the $n\th$ order contributions to this scattering amplitude. These $n\th$ order contributions can be evaluated by summing appropriate Feynman diagrams with $n$ vertices.

\subsection{Feynman diagrams}

The matrix elements of $\SS^{(n)}$ between states which contain a finite number of particles can be calculated by using a Feynman diagram expansion. This is based on Wick's theorem\footnote{Clear descriptions can be found in \citet{Wick:1950},  \citet[\S19.2, p233]{Bogoliubov:1959} and \citet[\S13c, p435]{Schweber:1961}.} and provides a way to express scattering amplitudes as sums of products of Feynman propagators together with factors associated with the incoming and outgoing states. Each product then corresponds to a diagram in the familiar way (see \citet[\chap 6]{Weinberg:1995} for details).

There are essentially two common views one can take about Feynman diagrams:
\begin{itemize}
\item they provide a convenient device to keep track of terms in a mathematical expression, or
\item they are a graphical depiction of different particle trajectories occurring in a superposition, i.e. they represent different ``histories'' in a sum-over-histories theory.
\end{itemize}
(an interesting account of the various interpretations of Feynman diagrams is given by \citet{Kaiser:2000}).

\subsubsection{Feynman diagrams on a causal set}

When trying to define the interacting field theory on a causal set it is physically appealing to base it on a Feynman diagram expansion\footnote{We mention that Feynman diagrams on causal sets have been considered before by \citet{Meyer:1997} but were not followed up.}. Following their sum-over-histories interpretation, the amplitude for a particular process is simply the sum of the amplitudes for all the ways the process can occur. In the continuum this sum diverges but if it was computed on a causal set one could hope that the sum-over-diagrams would be finite. This would be similar to the path integral models of \chapref{chap:PathIntegrals} in which a finite number of trajectories were summed over. The sum would be finite because, in essence, ``you could only squeeze a finite number of diagrams onto a finite causal set''.

Unfortunately, it seems difficult to start from scratch with a sum-over-diagrams scheme on a causal set. While, on the positive side, the Feynman propagator of \chapref{chap:FreeQFT} is available to be assigned to the legs of the diagrams, the difficulty is deciding \emph{which} diagrams to sum over!

As an example, suppose we consider $\phi^4$-theory. Here we sum over diagrams with vertices which join four lines. When we sum over the contribution from all two-vertex diagrams, say, we are really summing the amplitudes for each diagram as the two vertices range over all possible spacetime positions. When doing this sum on a causal set, do we include the contribution when the two vertices happen to coincide at the same causal set element (i.e. when they na\"ively appear as \emph{one} vertex joining \emph{eight} lines)? Similar questions follow---Do we allow loop-diagrams starting from one element and returning to itself? Do we sum over \emph{all} diagrams or only \emph{connected} diagrams? etc.

To define the theory on a causal set we would like to follow some definite guidance to help answer these questions (rather than making ad hoc choices---\emph{do} sum over coincident vertices, \emph{don't} sum disconnected diagrams etc). For this reason we  concentrate on defining a causal set analogue of the \SO which, as we shall see, gives a Feynman diagram expansion automatically.

\section{Scattering amplitudes on a causal set}

We now attempt to extend the free field theory defined in \chapref{chap:FreeQFT} to include interacting fields. We shall focus on obtaining a causal set analogue of the continuum $\SS$-operator and treat its matrix elements as scattering amplitudes.
The approach will concentrate on the mathematical apparatus of the formalism, delaying its interpretation until \secref{sec:SMatrixDiscussion}.

Our first definition of a causal set \SO will be as a perturbative series \eqref{eq:SMatrixSeries} similar to the Dyson series. This is then used to motivate a \emph{non-perturbative} unitary \SM operator \eqref{eq:SMatrix}.

For definiteness we shall consider self-interacting $\p^4$-theory. We take the interaction Hamiltonian density\footnote{The term ``density'' is used only to emphasise that $\H_x$ is defined at individual elements.} at a causal set element $v_x$ to be the operator
\be \H_x:=\frac{\lambda}{\rho} \p_x^4,\ee
where the coupling constant $\lambda$ is a dimensionless real number. We include a factor of $1/\rho$ to ensure that $\H_x$ is dimensionless (for sprinklings into $\mink^4$ we have the following mass-dimensions: $[\p_x] = M$, $[\rho]= M^4$). This is to ensure that products of $\H_x$ operators all have the same mass-dimension and so can be added together.

We note that these operators satisfy
\be \label{eq:Microcausality} \textrm{$v_x$ and $v_y$ unrelated} \implies \lbrack \H_x, \H_y \rbrack = 0. \ee

This $\H_x$ is a self-adjoint operator on $\FF$ (as can be seen by applying \citet[Thm X.25, p180]{Reed:1975} twice). The formalism we define will apply to any dimensionless self-adjoint Hamiltonian $\H_x$ defined on $\FF$ satisfying \eqref{eq:Microcausality}.

\subsection{Perturbative Dyson series}

As discussed in \secref{sec:FeynmanPropCausalSet} the notion of time ordering on a causal set $(\CS,\preceq)$ is provided by a \emph{linear extension} $(\CS,\leq)$ of the partial order. If we fix one such linear extension the time-ordered product of $n$ operators $\H_x$ (indexed by causal set elements $v_x \in \CS$) is just the product of the operators taken in the order of the linear extension (with the earlier elements to the right and the later elements to the left). As an example: if the linear extension for a 4 element causal set is $v_1 < v_2 < v_3 < v_4$ then $T(\H_2 \H_1 \H_3 \H_4 \H_2) := \H_4 \H_3 \H_2 \H_2 \H_1$.
The time-ordered product does not depend on the particular linear extension that is used provided the $\H_x$ operators satisfy \eqref{eq:Microcausality}.

Motivated by \eqref{eq:DysonSeries} we define the \SO Dyson series on a causal set as the series
\be \label{eq:SMatrixSeries} \SS:=\sum_{n=0}^\infty \SS^{(n)},\ee
with
\be\SS^{(0)} := \I,\quad \SS^{(n)}:=\frac{(-i)^n}{n!} \sum_{x_1=1}^p \ldots \sum_{{x_n}=1}^p T(\H_{x_1} \cdots \H_{x_n}),\ee
where $n \geq 1$ and $\mathbb{I}$ is the identity operator on $\FF$.

Each $\SS^{(n)}$ operator is well-defined on any finite causal set. The convergence of the series \eqref{eq:SMatrixSeries}, on the other hand, depends on the structure of the causal set and the particular choice of $\H_x$. It seems that, in general, the series does not converge, although establishing this is a task for future work.
Formal manipulations of the series (similar to those in \citet[Sec 18.2]{Bogoliubov:1959}) can be done which show that $\SS \SS^\dag = \SS^\dag \SS = \I$, i.e. that $\SS$ is a (formally) unitary operator.

\subsection{Feynman diagram expansion} \label{sec:FeynmanDiagramsCausalSet}

The matrix elements $\langle \psi_{\textrm{out}} |\SS^{(n)} |\psi_\textrm{in} \rangle$ between finite-particle states $|\psi_{\textrm{in}}\rangle, |\psi_\textrm{out}\rangle \in \FF$ can be evaluated using Wick's theorem just as in the continuum.

To start, let's just consider vacuum expectation values of time-ordered products of field operators. In this case, when applied to our causal set field operators Wick's theorem states that if $n$ is a positive even integer then:
\be \lv T \p_{x_1} \ldots \p_{x_n} \rv = \sum \F_{y_1 y_2} \F_{y_3 y_4} \cdots \F_{y_{n-1} y_{n}}, \ee
where $\F_{xy} = -i (K_F)_{xy}$ and $K_F$ is the matrix \eqref{eq:KFAlt} for the Feynman propagator on the causal set\footnote{The factor of $-i$ appears because of the $i$ in $(K_F)_{xy} = i \lv T\p_x\p_y \rv$.}. The sum is over all pairings of the indices $x_1,\ldots,x_n$. If $n$ is odd then the $n$-point function is zero.

As an example we have:
\begin{align}
\lv T \p_x \rv &= 0\\
\lv T \p_x \p_y \rv &= \F_{xy}\\
\lv T \p_x \p_y \p_z \rv &= 0\\
\label{eq:4ptfunction} \lv T \p_x \p_y \p_z \p_w\rv &= \F_{xy} \F_{zw} + \F_{xz} \F_{yw} + \F_{xw} \F_{yz}
\end{align}
The calculation of such $n$-point functions is therefore reduced to the combinatorial problem of summing over all such pairings of their indices. This can be achieved using a generating function \citep{Polyak:2005}.

\begin{figure}[h!]
\begin{center}
\includegraphics[width = \textwidth]{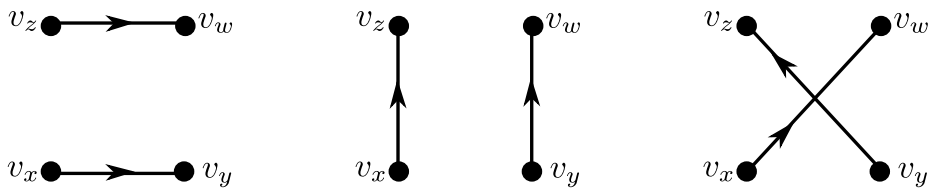}
\caption[3 simple Feynman diagrams]{Vacuum Feynman diagrams for \eqref{eq:4ptfunction}.}
\label{fig:FeynmanDiag}
\end{center}
\end{figure}

Each sum-over-pairings can be represented by a Feynman diagram. Each pairing $F_{xy}$ is drawn as a directed line from causal set element $v_x$ to $v_y$. As an example \eqref{eq:4ptfunction} can be represented by three vacuum-diagrams in \figref{fig:FeynmanDiag}. This is all just as in the usual continuum theory.

If we were to evaluate the expectation values between states that have a non-zero number of particles there would be additional pairings and extra factors related to the incoming and outgoing states.
For example if the incoming state is a two-particle state $\adag(g) \adag(h) \rv$ and the outgoing state is a two-particle state $\adag(f) \rv$ (for $f,g,h \in \mathbb{C}^s$) then we have
\begin{align} \lv \a(f) \p_x \adag(g) \adag(h) \rv &= \lv \a(f) \left(\a(f_x) + \adag(f_x)\right) \adag(g) \adag(h) \rv \\
&= \lv \a(f) \a(f_x) \adag(g) \adag(h) \rv \\ &= (f,g) (f_x,h) + (f,h) (f_x,g)
\end{align}
where $(\cdot,\cdot)$ denotes the inner-product in $\mathbb{C}^s$ and we have applied Wick's theorem to the creation and annihilation operators directly (rather than at the level of the field operators).

\subsection{Non-perturbative operator}

The apparent non-convergence of the Dyson series \eqref{eq:SMatrixSeries} blocks us from using the series to define a non-perturbative $\SS$-operator.
We can still make progress, however, by another route which takes \eqref{eq:SMatrixSeries} as its inspiration. We express \eqref{eq:SMatrixSeries} as the familiar time-ordered exponential:
\be \SS = T\left(\exp\left(-i\sum_{x=1}^p H_x\right)\right),\ee
(which is just another \emph{notation} for \eqref{eq:SMatrixSeries}).

This form for $\SS$ suggests a non-perturbative definition of the $\SS$-operator. The idea is to split $\exp(-i \sum_{x=1}^p \H_x)$ into a product of exponentials. In general, however, for non-commuting operators $\hat{A}$ and $\hat{B}$ this splitting is ambiguous: $\exp(\hat{A} +\hat{B}) \neq \exp(\hat{A}) \exp(\hat{B})$. From this point of view therefore it is helpful that we have the  time-ordering operator to define the order in which to factorise the exponential. We therefore define the non-perturbative \SM as a single time-ordered product:
\be \label{eq:SMatrix} \SS := T\left( \prod_{x=1}^p \exp\left(-i H_x\right)\right).\ee
Since each $\H_x$ operator is self-adjoint there exists a unique unitary operator $\U_x:=\exp(-i H_x)$ defined on all of $\FF$. Unfortunately, since $\H_x$ is in general an unbounded operator, the definition of $\U_x$ requires the use of a suitable functional calculus (see \secref{sec:FunctionalCalculus} and \citet[Sec VIII.4]{Reed:1972}).

If the causal set is labelled using a natural labelling then we have the simple expression
\be \label{eq:SMatrixUnitary} \SS = \U_p \U_{p-1} \cdots \U_3 \U_2 \U_1. \ee
Writing $\SS$ as a finite product of unitary operators (taken in a particular order) provides a manifestly unitary expression for the $\SS$-operator. It does not depend on the linear extension used because if $v_x$ and $v_y$ are unrelated then, using \eqref{eq:Microcausality} and \citet[Thm VIII.13, p271]{Reed:1972}, $\lbrack \U_x,\U_y\rbrack = 0$.

\def\R{\mathcal{R}}

The formalism still holds for models in which the interaction is only turned on within a particular subregion $\R \subset \CS$. We simply have $\H_x = 0$ (so that $\U_x = \I$) for all $v_x \not\in \R$.
In the simplest case, if the interaction is only turned on at a single element $v_x$, we have that $\SS = \U_x$. This gives us a better understanding of \eqref{eq:SMatrixUnitary}: the \SO for the whole causal set is the time-ordered product of individual $\SS$-operators defined for each causal set element.

\subsection[The U operators]{The $\U_x$ operators}

Clearly a thorough understanding of the $\U_x$ operators is crucial if we are to work with the non-perturbative operator $\SS$ defined in \eqref{eq:SMatrixUnitary}. Unfortunately this is a non-trivial task.

\subsubsection{Functional calculus}

One way to define the $\U_x$ operators is through the functional calculus described in \secref{sec:FunctionalCalculus}. If the spectral decomposition of $\p_x$ is given by
\be \langle \psi_1 |\p_x | \psi_2 \rangle = \int_{-\infty}^\infty z\, d\langle \psi_1 |\E_z | \psi_2 \rangle, \ee
then, for the $\phi^4$-theory, we have\footnote{For other interacting theories the $z^4$ exponent would be replaced by a different polynomial.}
\be \langle \psi_1 |\U_x | \psi_2 \rangle = \int_{-\infty}^\infty e^{-i\frac{\lambda}{\rho}z^4} \, d\langle \psi_1 |\E_z | \psi_2 \rangle. \ee
where $|\psi_1\rangle, |\psi_2\rangle \in \FF$. The difficulty here is determining the $\E_z$ projection operators. This remains a task for future work.

\subsubsection{Weyl correspondence}

\def\tha{\vec{\tau}}

An alternative approach which side-steps the functional calculus is to work in the Schr\"odinger representation. Here the field operator is an operator on $L^2(\mathbb{R}^s)$ given by (see \secref{sec:FieldSchrodinger})
\be \tag{\ref{eq:FieldSchro}} \p_x = A_x \cdot \hat{\mathbf{q}} + B_x \cdot \hat{\mathbf{p}}, \ee
where $\hat{\mathbf{q}} := (q_1,q_2,\ldots,q_s)$ and $\hat{\mathbf{p}} := -i(\frac{\partial}{\partial q_1}, \frac{\partial}{\partial q_2}, \ldots, \frac{\partial}{\partial q_s})$ and $A_x = \sqrt{2} \Re[f_x]$ and $B_x = \sqrt{2} \Im[f_x]$ are $s$-component vectors of real numbers.

For $\phi^4$-theory we are interested in the operator
\be \U_x := \exp\left(-i\frac{\lambda}{\rho} \p_x^4\right) = \exp\left(-i\frac{\lambda}{\rho} \left(A_x \cdot \hat{\mathbf{q}} + B_x \cdot \hat{\mathbf{p}}\right)^4\right). \ee
One way to define this is to use the \emph{Weyl correspondence} (see \citet[p274-275]{Weyl:1931} and \citet[\chap 1, \S1]{Follard:1989}). This correspondence is a way to assign operators (which act on $L^2(\mathbb{R}^s)$) to functions on the ``phase space'' $\mathbb{R}^s \times \mathbb{R}^s$.

Before defining the correspondence we consider the unitary operator on $L^2(\mathbb{R}^s)$ defined by
\be \hat{F}(\sh,\tha):=\exp\left(i \sh \cdot \hat{\mathbf{q}} + i \tha \cdot \hat{\mathbf{p}}\right), \ee
for $\sh, \tha \in \mathbb{R}^s$. By using the Baker-Campbell-Hausdorff formula\footnote{A simplified version of this states that if $\A$ and $\B$ are two self-adjoint operators such that $[\A,\B]$ commutes with $\A$ and $\B$ then \[ \exp( \A + \B ) = \exp(\A) \exp(\B) \exp([\A,\B]/2)\]} and
\begin{align}
\exp(i \sh\cdot \hat{\mathbf{q}}) \psi(\vq) &= \exp(i \sh \cdot \vq) \psi(\vq),\\
\exp(i \tha\cdot \hat{\mathbf{p}}) \psi(\vq) &= \psi(\vq + \tha) \\
[i \sh \cdot \hat{\mathbf{q}},i \tha \cdot \hat{\mathbf{p}}] &= i \sh \cdot \tha
\end{align}
this can be more explicitly expressed as \citep[eq 1.23, p22]{Follard:1989}
\be \hat{F}(\sh,\tha)\psi(\vq) = \exp(i \sh\cdot (\vq + \tha/2)) \psi(\vq + \tha), \ee
for $\psi \in L^2(\mathbb{R}^s)$.

To define the Weyl correspondence we now suppose we have a general function $U(\vq,\vp)$ on $\mathbb{R}^s \times \mathbb{R}^s$. This is assigned an operator by first expressing it in terms of its Fourier transform
\be U(\vq,\vp) = \int d^s\sh\int   d^s \tha \;\; \widetilde{U}(\sh,\tha) \exp(i \sh \cdot \vq + i \tha \cdot \vp), \ee
and then replacing $\vq$ and $\vp$ by $\hat{\mathbf{q}}$ and $\hat{\mathbf{p}}$, i.e. we assign $U(\vq,\vp)$ the operator:
\be \hat{U} = \int d^s\sh \int d^s\tha \;\;\widetilde{U}(\sh,\tha) \hat{F}(\sh,\tha). \ee
For $\psi \in L^2(\mathbb{R}^s)$ this acts as
\be \hat{U}\psi(\vq) = \int d^s\sh \int d^s\tha \;\widetilde{U}(\sh,\tha) \exp(i\sh\cdot (\vq + \tha/2)) \psi(\vq +\tha). \ee
This way to assign an operator $\U$ to a function $U(\vq,\vp)$ is the Weyl correspondence.

Using this correspondence the operator $\U_x$ could be defined using the Fourier transform of
\be \exp\left(- i\frac{\lambda}{\rho} (A_x \cdot \vq + B_x \cdot \vp)^4\right). \ee
Further exploration of this idea is a task for future work.

\section{CT symmetry} \label{sec:CT}

The scattering amplitudes calculated using \eqref{eq:SMatrixUnitary} are invariant under the causal set analogs of ``charge conjugation'' and ``time reversal'', i.e. a $C T$-transformation.

The $C$ transformation (``charge conjugation'') involves flipping the sign of the coupling constant: i.e. $\lambda \to -\lambda$. The $T$ transformation (``time reversal'') reverses the time-direction: i.e. we take the anti-time-ordered product in our expressions. If we simultaneously perform both these operators then we have
\be \SS \to \SS^\dag,\ee
or
\be \langle \psi_1 | \SS | \psi_2 \rangle \to \langle \psi_2 | \SS^\dag | \psi_1 \rangle = \langle \psi_1 | \SS | \psi_2\rangle^*, \ee
(where, due to the $T$ transformation, we have also swapped the roles of the incoming and outgoing states in the matrix element).

The result is that the probabilities $ |\langle \psi_1 | \SS | \psi_2 \rangle |^2 = |\langle \psi_1 | \SS | \psi_2 \rangle^* |^2$ are unchanged under a $CT$-transformation.

In continuum quantum field theories an important symmetry is the $CPT$-transformation where $P$ denotes reversing spatial parity. In the causal set framework it seems there is no obvious analogue for such a $P$-transformation. See \secref{sec:Parity} for a discussion of this point.

\section{Multiple fields}

The formalism developed so far deals only with a single self-interacting scalar field. It can be extended to deal with $N$ mutually interacting scalar fields. For simplicity we shall describe the extension to $N=2$.

Suppose we have two scalar fields $\p_a$ and $\p_b$ (which may have different masses). The free field Fock space for each of these is $\FF$ where $\HH = \mathbb{C}^s$. Note that even though $\p_a$ and $\p_b$ may have different masses their single-particle Hilbert spaces are the same since the rank of $i\Delta$ is mass-independent (see \secref{sec:MassIndependent}).

The Hilbert space to describe the interacting theory of the two fields is then $\FF \otimes \FF$. The interaction Hamiltonian densities are now of the form
\be \H_x = \sum_{n=0}^\infty \sum_{m=0}^\infty \lambda_{nm} (\p_a)_x^n \otimes (\p_b)_x^m,\ee
for suitable coupling constants $\lambda_{nm}$. For example we could have
\be \H_x = \lambda \p_a^2 \otimes \p_b, \ee
(which would correspond to Feynman diagrams with vertices involving two $a$-particles and one $b$-particle).
The \SO would then be the unitary operator on $\FF \otimes \FF$ defined using this interaction Hamiltonian density in \eqref{eq:SMatrixUnitary}.

In theories with multiple fields there is the possibility of symmetries between the fields---i.e. the theory is invariant under some transformation of the fields amongst themselves. These ideas (of internal symmetry groups acting on field multiplets) carry over from the continuum to the causal set framework without change.

In continuum gauge theories promoting a global symmetry to a local (i.e. position-dependent) symmetry helps to determine the allowed field interactions. It is interesting to wonder what would result if this idea was applied to the causal set model.

\section{Discussion} \label{sec:SMatrixDiscussion}

We have developed a mathematical framework to define scattering amplitudes for interacting scalar fields on a causal set. Here we discuss how it could be used.

\subsection{Interpretation of the Fock space} \label{sec:InterpretationofFock}

The Fock space $\FF$ that the field operators act on is based on the ``single-particle'' Hilbert space $\HH = \mathbb{C}^s$. A natural question is ``in what sense do vectors in $\HH$ represent states of the particles?''.

We address this question by again looking at the field operators expanded in terms of creation and annihilation operators:
\be \tag{\ref{eq:FieldDef}} \p_x = \sum_{i=1}^s \sqrt{\lambda_i}(u_i)_x \a_i + \sqrt{\lambda_i}(v_i)_x \adag_i, \ee
Following on from the discussion of mode expansions in \secref{sec:NormalModes} we suggest the following interpretation: the operators $\adag_i = \adag(\vec{e}_i)$ act to create a particle with a particular momentum. The momentum of the particle is related to the eigenvector $u_i$ (or $v_i$) and its energy is related to the eigenvalue $\lambda_i$. For large sprinkled causal sets these eigenvectors (or linear combinations of them) should approximate plane-waves which correspond to a definite momentum. It is not clear, however, how to determine \emph{which} momentum each eigenvector would correspond to.
We therefore suggest that the vectors in $\HH$ represent (possibly superpositions of) states with different on-shell momenta.

This interpretation could allow the calculation of scattering amplitudes between states with definite momentum. Such scattering amplitudes are most suited to describing the results of particle accelerator experiments.

An alternative approach is to use wave packets. These would represent particles without a precisely defined position or momentum. It's possible that the definition of the wave packets could use the spacetime geometry. We could chose a particular subset of the causal set (e.g. a chain or family of chains, say) and use that to help define a wave-packet localised there with momentum in a particular direction.

\subsection{Poincar\'e invariance}

When we sprinkle a causal set into a finite region of $R \subset \mink^d$ we break the Poincar\'e invariance of the spacetime. The region $R$ picks out a preferred frame in which the region takes a certain shape. If $R$ is a causal interval, for example, a time-axis is chosen by the end-points of the interval.

The only way to remove this is to perform the calculations with infinite causal sets generated by sprinkling into all of $\mink^d$. As already discussed in \secref{sec:InfiniteCausalSets}, however, there are difficulties in extending the free (let alone the interacting) field theory to infinite causal sets.

\subsubsection{Representations of the Poincar\'e group}

An important difference between the causal set and continuum theories is that there is no notion of a Poincar\'e transformation on a causal set.

This may prove to be a problem because the Poincar\'e group plays an important role in particle physics. Particles are classified according to how they transform under the Poincar\'e group \citep[\S2.5, p62]{Weinberg:1995} so without the group we have to find another way to define particles with non-zero spin (since these have non-trivial spacetime transformation properties).

\subsection{Asymptotic regions}

Another difference between the causal set and continuum theories is that a finite causal set cannot accommodate regions in the infinite past and infinite future. These regions are available in $\mink^d$ and are used to define the incoming and outgoing states in scattering theory.

The lack of asymptotic regions like these is not a major difficulty --- presumably on a large causal set one could select regions in the ``far past'' and ``far future'' to serve as the incoming and outgoing regions.

\subsection{Gravitational interaction}

All the work presented so far has made no mention of the gravitational effects of the matter fields---here we present a few remarks on this topic.

Firstly we observe that the causal set framework for interacting scalar fields is well-defined for any finite causal set. If the finite causal set is generated by sprinkling into a curved Lorentzian manifold then we have the causal set equivalent of quantum field theory in curved spacetime: the causal set is fixed but we work with non-back-reacting quantum matter on it.

Correctly incorporating the gravitational effects of the matter is likely to be very difficult.
In \secref{sec:Dynamics} two approaches were discussed for defining (gravitation) dynamics of causal sets. These currently only deal with pure gravity without matter. It is possible that future developments could combine those gravitational dynamics with the current work.

We can already see hints that the causal set framework for interacting fields is incomplete. Since we are dealing with bosons we can have states of the field which contain arbitrarily large numbers of particles\footnote{This is not the case with fermions---eventually a finite causal set would ``get filled up'' due to the Pauli exclusion principle.}. Similarly the Feynman diagram expansion would contain diagrams with ever greater numbers of vertices. Physically such large numbers of (virtual) particles would have a gravitational effect on spacetime which has not included in the framework presented here. Including this effect could mean that Feynman diagrams with large numbers of particles would be suppressed---possibly ensuring that, if gravitation is included, the Dyson series converges.

\subsection{Parity} \label{sec:Parity}

As mentioned in \secref{sec:CT} there is no natural notion of a parity transformation on a causal set. This $P$ operation, which sends $(x^0,\vec{x})$ to $(x^0,-\vec{x})$, plays an important role in particle physics (for example in assigning intrinsic parity to elementary particles) and is an ingredient of the celebrated $CPT$-theorem.

If $x$ and $y$ are two causally related points in $\mink^d$ we see that $P$ preserves the causal relations: $Px \preceq Py \iff x \preceq y$. In contrast a time-reversal transformation reverses the causal relations: $Tx \preceq Ty \iff y \preceq x$.
This suggests that the causal structure of $\mink^d$ is blind to spatial parity.

As discussed in \secref{sec:WhatIsACausalSet} we expect the causal relations as well as information about the spacetime volume to be sufficient to define a Lorentzian manifold. If the causal relations are blind to parity then perhaps the volume information could be the key? Certainly the volume of a manifold is an oriented quantity (depending on the manifold's volume form). By changing the way causal set theory defines spacetime volumes it's possible that parity could be defined. We make no attempt to follow up this suggestion---perhaps we need to modify the definition of a causal set, or perhaps there is \emph{already} a potential definition of parity within causal set theory as it stands.

A notion of parity on a causal set would give clues toward defining field theories for particles with non-zero spin (since such fields have non-trivial transformation properties under a partity transformation).

\subsection{Use of the continuum}

We conclude this discussion on a more philosophical note. In \chapref{chap:Intro} we presented arguments against the use of the continuum in physics. Modelling spacetime by a causal set is certainly one way of rejecting the continuum but, as our formalism has developed we see that the causal set quantum field theory relies heavily on continuum concepts.

Starting in \chapref{chap:PathIntegrals} the $a$ and $b$ amplitudes in the path integral model were allowed to be real numbers.
In \chapref{chap:FreeQFT} a scalar field was modelled by operators acting on a Hilbert space. The functional analysis this requires is based on a host of continuum concepts (e.g. complex Hilbert spaces, Cauchy-sequence convergence etc). Even the eigendecomposition of $\ID$ requires that we work with complex numbers (since the eigenvectors of a matrix may be complex).
The current chapter has also made heavy use of continuum-based tools, such as the spectral theory and functional calculus.

It seems that we have built-up a theory that goes against one of the main motivations for considering causal sets in the first place. Part of the reason for this is that we have sought to emulate the continuum theories for matter. The continuum-concepts that these theories are based on have therefore cropped up again in our model.

It is possible that the causal set quantum field theory could be reformulated and based on discrete combinatorial concepts. The sum-over-trajectories of \chapref{chap:PathIntegrals} and the Feynman diagram expansion in \secref{sec:FeynmanDiagramsCausalSet}, for example, are both combinatorial in nature. If we could discover a simpler combinatorial basis for a theory of interacting matter fields on a causal set then this would be more in harmony with the philosophical ideas that motivated the use of causal sets in the first place.

\section{Phenomenology} \label{sec:Phenomenology}

Here we discuss what phenomenology could be done based on the ideas presented in the current work.

The simplest approach would be to use the usual continuum quantum field theory but replace the continuum propagators by the expected values of the causal set propagators. That is, we replace the continuum propagator $G(x)$ (be it retarded, advanced or Feynman) with the expectation value of the causal set equivalent $\mathbb{E}(K|0,x,\mathbb{M}^d,\rho)$ (for a suitable matrix $K$). See \secref{eq:QFTComparison} for the definition of the expected value.

The physically interesting case is in $\mink^4$ but unfortunately we only know the analytic expectation value
for the massless retarded propagator (the other expectation values have proved too hard to evaluate for finite
$\rho$). As an example of the general replacements, therefore, the massless retarded propagator in $\mink^4$ \be
\tag{\ref{eq:KR4dMassless}} \GR_0^{(4)}(x)= \theta(x^0)\theta(\tau^2) \frac{1}{2 \pi} \delta(\tau^2), \ee would
be replaced by \be \mathbb{E}(K_R|0,x,\mathbb{M}^4,\rho) = \theta(x^0) \theta(\tau^2) \frac{\sqrt{\rho}}{2\pi
\sqrt{6}} \exp\left(-\rho \frac{\pi}{24} \tau^4\right).\ee This replacement smears out a delta-function on the
future lightcone to a highly-peaked Gaussian. In fact this particular substitution is very similar to a
suggestion by \citet{Feynman:1948:RelCutOff, Feynman:1948:RelCutOffQuantum} for a relativistic cut-off to tame
the divergences in classical and quantum electrodynamics.

For a finite sprinkling density $\rho$ the $d=4$ propagators would differ from their continuum counterparts. Nevertheless one could still use them in quantum field theory calculations, e.g. one could use $\mathbb{E}(K_F|0,x,\mathbb{M}^4,\rho)$ to replace the Feynman propagator in Feynman diagrams. One advantage to this is that, for finite $\rho$, the expectation values are not as singular as the continuum propagators. We could therefore hope that divergent expressions in the continuum could be regularised by this substitution.

By working in the continuum but using the expectation values of the causal set propagators we have, in a sense, the best of both worlds. The expectation values are less divergent \emph{and} we can use all the usual framework of the continuum spacetime (Poincar\'e transformations, parity transformations, momentum space, spinors etc).

One could also apply the substitution to fields with non-zero spin. Propagators for the Dirac equation, for example, can be obtained from the Klein-Gordon propagators by\footnote{See \citep[Appx I, \S C]{Bogoliubov:1959}. We have added a minus sign, however, to agree with the \chapref{chap:Spinors}.}
\be G_{D}(x) := -(i \gamma^\mu \partial_\mu + m) G_{KG}(x)\ee
where $G_D$ is the Dirac propagator, $G_{KG}$ the Klein-Gordon propagator and $(i \gamma^\mu \partial_\mu + m)$ is a differential operator we'll meet in the next chapter. By applying this differential operator to the expectation values of the causal set propagators we would obtain causal-set-inspired propagators for particles with non-zero spin.

In this way we could define a phenomenological model, inspired by causal set theory, for realistic matter. Any calculations performed in this theory would have a dependence on the sprinkling density $\rho$. This, in turn, defines the scale of any spacetime discreteness. By comparing the results of real-world experiments with the phenomenological model one may be able to fit $\rho$ to the data and determine the best-fit for the scale of spacetime discreteness!

\renewcommand\slash[1]{\not\!#1}
\def\hf{$\frac{1}{2}$ }

\chapter{Spin-half Particles} \label{chap:Spinors}

\begin{quote}
I was playing around with the three components $\sigma_1, \sigma_2, \sigma_3,$ which I had used to describe the spin of an electron, and I noticed that if you formed the expression $\sigma_1 p_1 + \sigma_2 p_2 + \sigma_3 p_3$ and squared it, $p_1, p_2$ and $p_3$ being the three components of momentum, you got just $p_1^2 + p_2^2 + p_3^2$, the square of the momentum. This was a pretty mathematical result. I was quite excited over it. It seemed that it must be of some importance.

\flushright{Paul Dirac in \citet[p295]{Mehra:2000}}
\end{quote}

The causal set formalism presented so far deals with scalar particles. While this is the simplest type of quantum field there are, as yet,\footnote{If the Higgs boson is discovered then this would be a fundamental scalar field.} no fundamental scalar fields in nature. If we wish to build a model for realistic particle physics on a causal set we have to be able to model elementary particles with spin half (e.g. the electron) and spin one (e.g. the photon). If we are only interested in a phenomenological model we could follow the ideas in \secref{sec:Phenomenology}.

Here we discuss some approaches to the problem of modelling spin-\hf particles on a causal set.

\section{Spin-half particles in the continuum}

A spin-\hf particle in $\mink^d$ is represented by a spinor-valued function of position $\psi(x)$ satisfying the Dirac equation: 
\be (i \gamma^\mu \partial_\mu - m) \psi(x) = 0, \ee
where $\gamma^\mu$ are the Dirac matrices which satisfy $\{ \gamma^\mu, \gamma^\nu \} = 2\eta^{\mu \nu}$, the Minkowski metric. 
In $d=4$ the field $\psi(x)$ is a 4-component object that transforms under Lorentz transformations as a spinor (in other spacetime dimensions $\psi$ has different numbers of components). The components of $\psi$ describe the spin degrees of freedom of the field---they describe the particle's superposition between  spin-up and spin-down states. Green's functions for the Dirac equation are matrix-valued functions of position $G(x)$ satisfying
\be (i \gamma^\mu \partial_\mu - m) G(x) = \delta^d(x) I, \ee
where $I$ is an appropriately-sized identity matrix.
For full details of the Dirac theory see any quantum field theory textbook (e.g. \citet{Bogoliubov:1959}, \citet{Weinberg:1995}).

The formalism we have briefly outlined is the called ``4-spinor'' theory but there also exists a ``2-spinor'' theory in which the 4-spinor field $\psi$ corresponds to a pair of 2-spinor fields $(\xi^A, \eta_{A'})$ (for more details see \citet[\S18-20]{Geroch:1973}, \citet{Penrose:1984}). When $m=0$ these two fields don't interact but when $m \neq 0$ they become coupled, each serving as the source for the other. This way of representing the particle leads to a ``mass scattering'' series similar to that in \secref{sec:MassScatterings}, see \citet[p412-417]{Penrose:1984} for details.

One way to obtain the Dirac equation is to ``take the square-root'' of the Klein-Gordon equation. This relies on the factorisation of the Klein-Gordon operator as:
\be \label{eq:KGSqrt} \Box + m^2 = (-i \gamma^\mu \partial_\mu - m) (i \gamma^\nu\partial_\nu - m), \ee
where we use that $\gamma^\mu a_\mu \gamma^\nu b_\nu = a^\mu b_\mu = a b$. We shall make use of this in \secref{sec:SquareRootProp}.

\subsection{Spinors in curved spacetime}

It is non-trivial to extend the notion of spinors to general Lorentzian manifolds. To even be able to define a spinor field on a curved spacetime it must possess \emph{spinor structure} (See \citet[p48-56]{Penrose:1984}). This ensures that the global topology of the manifold is suitable to define spinor fields.

\section{Spin-half on a causal set}

The most direct way to model spin-\hf particles on a causal set is to try to define a causal set analogue of the usual continuum theory. This would require defining spinors on a causal set. One approach to this has been considered by \citet{Sverdlov:2008c}. In that work an attempt is made to re-write the Lagrangian for the Dirac field in terms that can be re-expressed on a causal set. The resulting expressions, however, are very complicated and it is not clear in what way the Lorentz-transformation properties of a spinor are included.

Ultimately, however, we do not necessarily want a model for \emph{spinors} on a causal set\footnote{Indeed, since spinors cannot be defined for an arbitrary Lorentzian manifold, it may be over-optimistic to try to define them for an arbitrary causal set.}, but rather a model for \emph{spin-half particles}. This could require completely rephrasing the usual continuum spinor-based theory and replacing it with something that serves just as well but could be generalised to a causal set.
This is clearly a formidable task.

For now we outline two ideas which could lead to models for spin-\hf particles on the causal set.

\section{The Feynman checkerboard} \label{sec:Checkerboard}
\begin{figure}[!h]
\begin{center}
\includegraphics[width = 0.5\textwidth]{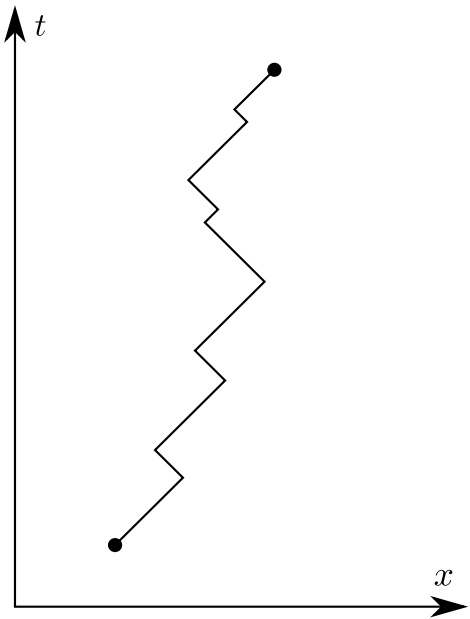}
\caption{An example path in the Feynman checkerboard.}
\label{fig:Checkerboard}
\end{center}
\end{figure}

In \citet[p35-36]{FeynmanHibbs} a path-integral model is presented for the retarded propagator of the Dirac equation in 2 dimensional Minkowski spacetime. This model has since become known as the \emph{Feynman checkerboard}. Good references include \citet{Jacobson:1985, Schweber:1986}.

The paths that are summed over in the model are zig-zags in spacetime made from sequences of null geodesics (see \figref{fig:Checkerboard}). Such paths correspond to a particle shuttling forward and backward (along the one spatial dimension) at the speed of light.

The sum-over-paths is calculated by discretising the time variable. If time is discretised to progress in steps of size $\epsilon$ we can restrict the sum to the finite number of paths which have zig-zag corners at times $n \epsilon$ where $n$ is an integer (this restricts the class of paths to those that lie on a lightcone-lattice). After calculating this finite sum we can then take the continuum limit as $\epsilon$ tends to $0$. Each corner in the zig-zag path corresponds to a reversal of the direction the particle is travelling in (for example the path in \figref{fig:Checkerboard} has 10 corners).

The amplitude assigned to a path with $R$ corners (i.e. $R$ reversals of direction) is then $(i m \epsilon)^R$ where $m$ is the mass of the particle.

\citeauthor{FeynmanHibbs} left it to the reader to show that this path integral model correctly reproduces the retarded Dirac propagator in $\mink^2$ (detailed solutions showing this appear in \citet{Jacobson:1984} and \citet{Kull:1999}).

The method of solution involves representing the Dirac propagator in $\mink^2$ as a $2 \times 2$ matrix depending on position. If we label its elements as
\be K(x) = \left(\begin{array}{cc} K_{++}(x) & K_{+-}(x) \\ K_{-+}(x) & K_{--}(x) \end{array}\right),\ee
then in the checkerboard model $K_{++}(x-y)$ is the amplitude that the particle started at $x$ with positive velocity and arrived at $y$ with positive velocity, $K_{+-}(x-y)$ is the amplitude that the particle started at $x$ with positive velocity and arrived at $y$ with negative velocity, with similar expressions for $K_{-+}(x-y)$ and $K_{--}(x-y)$. In particular the amplitudes $K_{++}$ and $K_{--}$ (resp. $K_{+-}$ and $K_{-+}$) involve summing over paths with an even (resp. odd) number of corners.

Generalising the model to 3+1 dimensions was attempted by Feynman but not published \citep{Schweber:1986} and has been attempted by subsequent authors (including \citet{Jacobson:1984, Jacobson:1984a} and others).

\subsection{The checkerboard on a causal set}

It is a natural step to attempt to formulate a checkerboard-type model for spin-\hf particles on a causal set. Indeed, the zig-zag trajectories which are summed over in the checkerboard model are reminiscent of the hop-and-stop path integral model of \chapref{chap:PathIntegrals}.

The first task is to characterise the appropriate zig-zag trajectories on a causal set. Since each straight segment in the original model is a null ray the most na\"ive choice for the straight parts of the causal set trajectories would be links. Each link is certainly a good candidate for a null segment of the trajectory but the difficulty comes when one tries to identify the number of corners for a trajectory (or indeed where the corners are).

\begin{figure}[!h]
\begin{center}
\includegraphics[width = 0.5\textwidth]{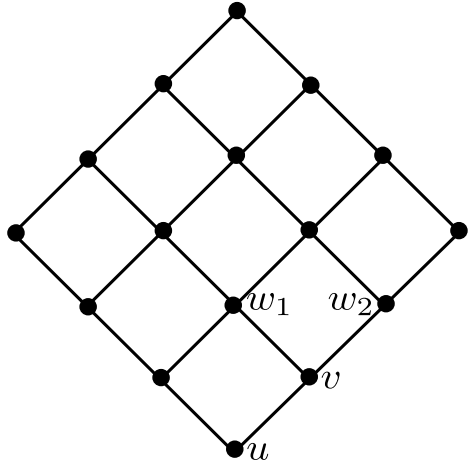}
\caption{A segment of the light-cone lattice drawn as a Hasse diagram.}
\label{fig:CheckerboardCorner}
\end{center}
\end{figure}

To see the difficulty it helps to look at the light-cone lattice that the continuum checkerboard-model is formulated on. Treating this as a causal set we let $u \link v$ be a link in the light-cone lattice (see \figref{fig:CheckerboardCorner}). To continue the path with another link we could choose the next element to be either $w_1$ or $w_2$.

We would want the path $u \link v \link w_1$ to count as having 1 corner but the path $u \link v \link w_2$ to count as having 0 corners. To distinguish between these two trajectories we need more information than is provided by just knowing that the elements are linked.

We can make this distinction, for the light-cone lattice at least, by using the idea of a \emph{chain interval}. This is a subset of the causal set which is both a chain and an interval, i.e. a causal interval which happens to be totally ordered. In the continuum\footnote{See \citet[Lemma 2]{Levichev:1987}.} and on the light-cone lattice these are precisely the null-geodesics.

Every chain interval in a causal set is a path and, to be more explicit, a chain interval of length $n$ is a sequence of elements $v_0 \link v_1 \link v_2 \ldots \link v_n$ such that the causal interval $[v_0,v_n]$ is totally ordered. We will denote such a chain interval by its interval: $[v_0,v_n]$.

In translating the checkerboard model to the causal set we take the straight parts of the trajectories to be chain intervals and the trajectories themselves to be concatenations of chain intervals---that is, each trajectory is a sequence of chain intervals joining two elements $v_x$ and $v_y$ related as $[v_x,v_1]$, $[v_1,v_2]$, $[v_2,v_3], \ldots, [v_n,v_y]$. Such a sequence is not unique and can always be lengthened by adding in extra copies of single-element chain intervals $[v_i,v_i]$.

Each of these concatenations of chain intervals is a path. The sum-over-trajectories, therefore, is really a sum over all paths joining two elements. This is similar to the hop-and-stop model of \chapref{chap:PathIntegrals},  the main difference being in the amplitude that is assigned to each path.

\subsubsection{Corners}

We can assign a non-negative integer $R$ to a path in a causal set by using the fact that every path can be re-expressed as a sequence of chain intervals. To see this we simply realise that every link in the path is a chain interval but acknowledge that there may be larger chain intervals in the path which contain more than two elements.

We define the number of corners $R$ of a path to be the \emph{minimum} number of chain intervals into which it can be resolved, minus one.

To illustrate this definition we return to \figref{fig:CheckerboardCorner}. Here $[u,v]$ and $[v, w_1]$ are both chain intervals but $[u, w_1]$ is not a chain interval (since it is not totally ordered). Thus the path $u \link v \link w_1$ can be resolved into a minimum of two chain intervals so is assigned 1 corner.

For the other path we have that $[u,v]$ and $[v,w_2]$ are both chain intervals but so is $[u, w_2]$. Therefore $u \link v \link w_2$ can be resolved into one chain interval so is assigned 0 corners.

Our definition of ``number of corners'' agrees with usual notion of corner when applied to paths in the light-cone lattice. A slightly unsatisfactory aspect to our definition, however, is that for a general causal set (in contrast to the light-cone lattice) we may be unable to assign the corners to particular elements in the path.

\begin{figure}[!h]
\begin{center}
\includegraphics[height = 0.3\textheight]{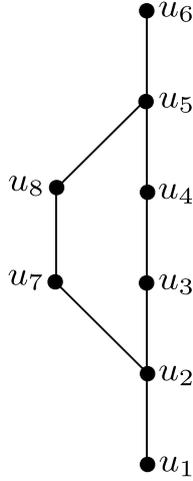}
\caption[Checkerboard corners]{Corners applied to a particular example.}
\label{fig:CornerDefEx}
\end{center}
\end{figure}

Consider, for example, \figref{fig:CornerDefEx}. The path $P=u_1 \link u_2 \link u_3 \link u_4 \link u_5 \link u_6$ is not a chain interval. It can be resolved, however, into two chain intervals: $[u_1,u_3],[u_3,u_6]$. We therefore assign $P$ one corner and, looking at the resolution into chain intervals, we'd be tempted to assign the corner to $u_3$.

This is wrong, however, because $P$ can also be resolved as $[u_1,u_4], [u_4,u_6]$. From this resolution we would be tempted to assign the corner to $u_4$. It is thus clear that while our definition assigns one corner to $P$ it does not assign it to any particular element. In other words, we should not think of the corners as being assigned to a particular element but rather that the ``number of corners'' is a property of the path as a whole.

\subsection{The path integral}

Having established which trajectories we shall sum over we now consider what amplitudes we should assign to them.

The continuum Dirac propagator in $\mink^2$ has mass-dimension $M$. We therefore want the amplitudes we assign to paths to have dimension $M$. Inspired by the usual continuum checkerboard model as well as the hop-and-stop path integral of \chapref{chap:PathIntegrals} we assume the amplitude assigned to a path with $n$ corners is $a^{n+1} b^n$ where $a$ and $b$ are appropriate constants (we think of each corner having a ``stop amplitude'' $b$).

On dimensional grounds we have
\be [a] = M, \quad [b] = M^{-1}, \quad [ab] = 1, \ee
and if our causal set is generated by a sprinkling into $\mink^2$ with density $\rho$ the only dimensionful quantities available are $\rho$ and $m$:
\be [\rho] = M^2, \quad [m] = M. \ee
Therefore, in a similar spirit to the 1+1 dimensional scalar path integral of \secref{sec:1+1pathintegral}, we take
\be \label{eq:CheckerboardAmps} a = A\sqrt{\rho}, \qquad b = B \frac{m}{\rho}, \ee
where $A$ and $B$ are dimensionless (possibly complex) constants independent of $\rho$ or $m$.

To compute the sum-over-paths on a $p$-element causal set we can define a family of $p\times p$ matrices $T^{(n)}$ ($n=0,1,2,\ldots$) as:
\be T^{(n)}_{xy} := \textrm{The number of paths from $v_x$ to $v_y$ with $n$ corners.}\ee
Calculating these matrices explicitly is difficult but we mention that 
\be T^{(0)}_{xy} := \left\{ \begin{array}{ll} 1 & \textrm{ if $[v_x, v_y]$ is a chain interval} \\ 0 & \textrm{otherwise.} \end{array} \right. \ee 
can be computed from the causal matrix $C$ as:
\be T^{(0)}_{xy} = \left(I + C^{\left(1+(C^2)_{xy}\right)}\right)_{xy},\ee
where $I$ is the $p\times p$ identity matrix. This follows because if $[v_x,v_y]$ is a chain interval and $v_x \prec v_y$ then there is exactly 1 chain of length $1+(C^2)_{xy} = |[v_x,v_y]|-1$ from $v_x$ to $v_y$. If $[v_x,v_y]$ is not a chain interval then $\left(C^{\left(1+(C^2)_{xy}\right)}\right)_{xy} = 0$ since there are no chains of length $1+(C^2)_{xy}$ from $v_x$ to $v_y$.

The 4-elements of the propagator $K_{++},K_{+-},K_{-+},K_{--}$ would then be calculated as
\begin{align} K_{++} = K_{--} &= a T^{(0)}+a^3 2 T^{(2)} + a^5 b^4 T^{(4)} + \ldots = \sum_{n \textrm{ even}} a^{n+1} b^{n} T^{(n)}, \\ K_{+-} = K_{-+} &= a^2 b T^{(1)} + a^4 b^3 T^{(3)} + a^6 b^5 T^{(5)} + \ldots = \sum_{n \textrm{ odd}} a^{n+1} b^{n} T^{(n)},
 \end{align}
where we sum over odd or even numbers of corners in the same way as the usual checkerboard model.

By further analysing this model for causal sets generated by sprinkling into $\mink^2$ it is possible that we could choose the $A$ and $B$ constants to give a good model for the Dirac propagator.

\section{Square root of the propagator} \label{sec:SquareRootProp}

An alternative approach to modelling spin-\hf particles attempts to replicate Dirac's insight that by ``taking a suitable square-root'' of the Klein-Gordon equation we can obtain the Dirac equation.

The insight is that the Klein-Gordon operator can be factorised as:
\be \tag{\ref{eq:KGSqrt}}\Box + m^2 = (-i \gamma^\mu \partial_\mu - m) (i \gamma^\nu\partial_\nu - m). \ee
To use this we now consider the retarded propagator for the Dirac equation, which we'll denote by $R_m(x)$, which satisfies:
\be (i \gamma^\mu\partial_\mu - m) R_m(x) = \delta(x) I.\ee
Here $R_m(x)$ is a matrix-valued function of position and $I$ is an identity matrix the same size as the Dirac matrices.

If we now define the matrix-valued function
\be \label{eq:SqrtSMDef} S_m(x-y) := -\int d^dz R_m(x-z) R_{-m}(z-y), \ee
then we have that
\begin{align}
(\Box + m^2) S_m(x-y) &= (i \gamma^\mu \partial_\mu + m) (i \gamma^\nu\partial_\nu - m) \int d^dz R_m(x-z) R_{-m}(z-y),\\
&= (i \gamma^\mu\partial_\mu  + m) \int d^dz \left((i \gamma^\nu\partial_\nu - m) R_m(x-z)\right) R_{-m}(z-y), \\
&= (i \gamma^\mu\partial_\mu + m) R_{-m}(x-y) = \delta(x-y) I. \end{align}
This means that the diagonal of the matrix-valued function $S_m(x)$ is a propagator for the Klein-Gordon equation. Since $R_m(x)$ is retarded so is $S_m(x)$ which means that the diagonal of $S_m(x)$ must be the retarded Klein-Gordon propagator (which is the unique retarded propagator for the Klein-Gordon equation). The off-diagonal entries of $S_m(x)$ are retarded \emph{solutions} of the Klein-Gordon equation (i.e. they are zero).

This observation suggests that we can obtain the retarded Dirac propagator as the square-root of the retarded Klein-Gordon 
propagator.

To implement this on a causal set with $p$ elements we seek a square matrix $R_m$ such that
\be -\frac{1}{\rho} R_m R_{-m} = K_R \otimes I, \ee
or
\be \label{eq:RmDef} R_m R_{-m} = -\rho K_R \otimes I, \ee
where $K_R$ is the retarded propagator on the causal set obtained from \chapref{chap:PathIntegrals} and we have included a factor of $\rho$ to mirror the dimensionful integration in \eqref{eq:SqrtSMDef}. This ensures that the $R_m$ has the correct mass dimension of $M^{d-1}$.

The size of the identity matrix $I$ corresponds to the size of the ``spin-space'' for the particles. In turn this depends on which dimension we are working in and whether we are using the 4-spinor or 2-spinor theory (e.g. for the 4-spinor theory in $d=4$, $I$ would be a $4\times 4$ matrix). Ideally, we would like the size of this spin-space to be determined by the act of taking the square root of $K_R$, i.e. we'd like it to only work for $I$ of a particular size. This would be similar to the way in which \eqref{eq:KGSqrt} only holds for Dirac matrices of a particular size.

\subsection{Mass scattering}

The square-root of the propagator is consistent with the mass scattering ideas from \secref{sec:MassScatterings}. To see this we
follow the results from \chapref{chap:PathIntegrals} and suppose that the retarded Klein-Gordon propagator takes the form
\be K_R = \Phi\left(I + \frac{m^2}{\rho} \Phi\right)^{-1},\ee
(where $\Phi$ is the matrix of ``hop'' amplitudes).

For the $m=0$ case of \eqref{eq:RmDef} we then have
\be (R_0)^2 = -\rho \Phi \otimes I.\ee
If we now define a ``mass scattering series''
\be \label{eq:RMassiveDef} R_m = R_0 + \bb (R_0)^2 + \bb^2 (R_0)^3 + \ldots = R_0(I - \bb R_0)^{-1},\ee 
(where we suppose $\bb$ is a mass-dependent amplitude such that if $m$ is replaced by $-m$ then $\bb$ becomes $-\bb$) then the product $R_m R_{-m}$ is:
\begin{align} R_m R_{-m} &= R_0 (I - \bb R_0)^{-1} (I + \bb R_0)^{-1} R_0 = R_0^2 (I - \bb^2 (R_0)^2)^{-1}\\
&= -\rho (\Phi \otimes I) \left(I + \bb^2 \rho \Phi \otimes I \right)^{-1} = -\rho \left(\Phi (I + \bb^2 \rho \Phi)^{-1}\right) \otimes I.\end{align}
This is equal to $-\rho \left(K_R \otimes I\right)$ if $\bb^2 \rho = m^2/\rho$, i.e. $\bb = \pm \frac{m}{\rho}$.

We see, therefore, that if we determine $R_0$ we can obtain $R_m$ through the series \eqref{eq:RMassiveDef}. The $\bb$ amplitude is the same form as the $b$ amplitude in \eqref{eq:CheckerboardAmps}. This may suggest that the checkerboard and square-root models take the same form in $\mink^2$.

The square roots of a matrix (if they exist) are in general not unique. It is possible that further conditions must be imposed to define the correct square root of $-\rho K_R \otimes I$ to serve as the retarded Dirac propagator. This remains a task for future work.

\section{General approach}

We have only been able to present a few suggestions for obtaining the retarded propagator for the Dirac equation on a causal set. Nevertheless here we outline how one would proceed if we had obtained the propagator.

To model a spin-\hf particle on a causal set we would expect to require a multi-component field $\hat{\psi}_{\alpha x}$ (where $\alpha = 1,\ldots, S$ denotes the spin degrees of freedom and $x = 1,\ldots, p$ indexes the causal set elements $v_x$).

Now suppose the retarded propagator is $(K_R)_{\alpha x, \beta y}$ (which, if we treat $\alpha x$ and $\beta y$ as single indices, we can think of as a $pS \times pS$ matrix). This represents the amplitude that a particle at $v_x$ propagates to $v_y$ (with spin degrees-of-freedom described by the $\alpha$ and $\beta$ indices).

The Pauli-Jordan function would then be
\be \Delta_{\alpha x, \beta y} = (K_R)_{\alpha x, \beta y} - (K_R)_{\beta y, \alpha x}.\ee

Since these would represent fermionic fields we would expect to quantize them with anti-commutation relations along the lines of\footnote{We use the adjoint $\psi^\dagger$ but this may need to be replaced by an analogue of the Dirac adjoint $\bar{\psi} = \psi^\dagger \gamma^0$ used in the continuum theory.}:
\be \{ \hat{\psi}_{\alpha x}, \hat{\psi}^\dagger_{\beta y} \} = \ID_{\alpha x, \beta y}, \ee
where $\{\hat{A},\hat{B} \}:= \hat{A} \hat{B} + \hat{B} \hat{A}$ is the anticommutator.

The eigendecomposition of $\ID$ could then be used to express the $\hat{\psi}$ operators as linear combinations of anticommuting creation and annihilation operators acting on a \emph{fermionic} Fock space (see \citet[Chapter 21]{Geroch:1985}). Together with a suitable analogue of condition 3 \eqref{eq:ZeroEigenvalueCondition} this might be sufficient to define a free field theory for spin-\hf particles on a causal set.

\chapter{Conclusions}

We conclude by summarising the work presented and discussing directions for future work.

\section{Summary}

\chapref{chap:Intro} gave an overview of the motivations for causal set theory, including the problem of quantum gravity as well as difficulties with continuum models for spacetime. \chapref{chap:CausalSets} introduced causal set theory in detail, including definitions and an overview of current developments. 

In \chapref{chap:PathIntegrals} we presented a model for discrete path integrals on a causal set. For a suitable choice of parameters (and when evaluated on causal sets generated by sprinkling into $d$-dimensional Minkowski spacetime) these give agreement with the continuum retarded propagator for the Klein-Gordon equation in $d$-dimensional Minkowski spacetime (for $d=1,2,3,4$). Variations on the models as well as possible extensions were discussed.

The path integrals of \chapref{chap:PathIntegrals} were used as the starting-point for \chapref{chap:FreeQFT}. Here we defined a model for free quantum scalar field theory on a causal set. This was done by assigning each causal set element a field operator, the collection of which were subject to three simple conditions. These operators were then used to define a Feynman propagator on a causal set (as the vacuum expectation value of a time-ordered product of field operators). The formal properties of this model were discussed and evidence for the agreement between the Feynman propagator on the causal set and the Feynman propagator in Minkowski spacetime was presented. The method for obtaining the Feynman propagator was also applied to calculations in the continuum.

In \chapref{chap:InteractingQFT} the free quantum scalar field theory was extended to include interacting scalar fields. This included defining a causal set analogue of the Dyson series for the S-matrix. This was then used to motivate a non-perturbative definition of the S-matrix on a causal set. Discussion of how this framework could be used was also given.

In \chapref{chap:Spinors} two speculative suggestions were given for how to model spin-half particles on a causal set. One of these involved translating the Feynman checkerboard model to a causal set.

\section{Directions for future work}

The current work opens up a number of areas for future study. Firstly we would like to better understand the models already presented, including:
\begin{itemize}
\item Calculating the expectation values of the retarded and Feynman propagators for finite $\rho$ in $d$-dimensional Minkowski spacetime (including, at least, the physical $d=4$ case).
\item Calculating the variance of the causal set propagators and investigating how the variance behaves in the $\rho \to \infty$ limit.
\item Investigating the behaviour of the propagators for sprinklings into curved spacetimes.
\item Investigating the convergence of the perturbative causal set Dyson series.
\item Better understanding the non-perturbative S-matrix (for example, calculating its matrix elements).
\item Better understanding how the framework can be extended from finite to infinite causal sets.
\item Applying the interacting theory to massless scalar fields with an interaction term $\H_x = \frac{m^2}{\rho} \p_x^2$ to see if we recover free massive scalar field theory.
\end{itemize}
If we better understood the expectation values of the causal set propagators then we could follow the phenomenological approach described in \secref{sec:Phenomenology} to get a causal-set inspired model for a realistic field theory.

Extending the work presented here from scalar fields to realistic interacting matter fields would represent the conclusion of the ideas developed so far. If we had, for example, the Standard Model on a causal set then we could calculate experimentally measurable quantities within the causal set framework. These calculations would have a dependence on the fundamental spacetime volume for each causal set element. Comparing the calculations with experimental results would provide evidence on whether spacetime is discrete and, if it is, what the discreteness scale is. To achieve this goal there are (at least) two key tasks to do:
\begin{itemize}
\item Developing a theory of spin-half and spin-one particles on a causal set.
\item Better understanding physical quantities such as energy, momentum and angular momentum on a causal set.
\end{itemize}
If these tasks can be completed then we would have a realistic theory of matter on a causal set spacetime background!

\appendix

\chapter{Expected Number of Paths} \label{chap:ExpectedNumberOfPaths}

In \secref{sec:SummingOverPaths} we discussed calculating the expected number of paths of length $n$ between two points in a sprinkling into $d$-dimensional Minkowski spacetime. No analytic expression is yet known for this quantity for general values of $d$ and $n$ but here we present some approaches to the problem.

\section{Preliminaries}

We first describe the quantity we are trying to calculate. We start by fixing two points $x, y \in \mink^d$. We then sprinkle a causal set with density $\rho$ into $\mink^d$. There is zero probability that the sprinkled causal set will contain $x$ and $y$ so we add them to the causal set. We now count the number of paths within the causal set of length $n$ from $x$ to $y$ (now regarded as elements of the causal set). The number we get is then averaged over all sprinklings with the same density $\rho$ to give us a density-dependent function $P_n(y-x)$. Our goal is to determine this function as fully as possible.

Before we begin we fix some notation. For two functions $f, g$ on $\mink^d$ their convolution is defined as:
\be (f*g)(x):= \int d^dy f(x-y) g(y). \ee
We define two useful functions (which we met earlier in \secref{sec:SummingOverChains} and \secref{sec:SummingOverPaths}):
\be C(y-x) :=\left\{ \begin{array}{ll} 1 & \textrm{ if } x \preceq y \\ 0 & \textrm{otherwise,}
\end{array} \right.\ee
\be P(y-x) := C(y-x) e^{-\rho V(x-y)} = \left\{\begin{array}{ll} e^{- \rho V(x-y)} & \textrm{ if } x \preceq y \\ 0 & \textrm{otherwise,} \end{array} \right.\ee 
where $V(x-y)$ is the $d$-dimensional Minkowski spacetime volume of the causal interval between $x$ and $y$ (see \eqref{eq:CausalVolume}).

As discussed in \secref{sec:SummingOverChains}, the expected number of chains of length one from $x$ to $y$ is $C_1(y-x) := C(y-x)$. The expected
number of chains of length $n > 1$ from $x$ to $y$ is given by
\be \label{eq:ChainDef} C_n(y-x) := \rho (C*C_{n-1})(y-x) = \rho^{n-1}(C*C*\ldots *C)(y-x),\ee
where there are $n$ copies of $C$ and $n-1$ convolutions in the final expression.

From the definition we have:
\be \label{eq:ChainConvolve} \rho C_n * C_m = C_{n+m}.\ee

As discussed in \ref{sec:SummingOverPaths}, the expected number of paths of length one from $x$ to $y$ is given by $P_1(y-x) := P(y-x)$. The expected number of paths of length $n > 1$ from $x$ to $y$ is given:
\be P_n(y-x) := \rho(P * P_{n-1})(y-x) = \rho^{n-1}(P * P * \ldots * P)(y-x),\ee 
where there are $n$ copies of $P$ and $n-1$ convolutions.

The expression for $C_n(x-y)$ has been calculated explicitly by \citet[Theorem III.2, p50]{Meyer:1988}. For $n \geq 1$ we have:
\be \label{eq:ChainForm} C_n(y-x) = C(y-x) (\rho V(x-y))^{n-1} D_n, \ee
where $D_1:=1$ and
\be D_n:=\frac{1}{n-1} \left( \frac{\Gamma(d+1)}{2}\right)^{(n-2)} \frac{\Gamma(\omega) \Gamma(2 \omega)}
{\Gamma((n-1) \omega) \Gamma(n \omega)}, \ee
for $n > 1$ is a real dimensionless constant with $\omega = d/2$. We have $D_2 = 1$ in all dimensions. 

Using these expressions the expected total number of chains is $C_T(y-x)$:
\begin{align} C_T(y-x):=&\sum_{n=1}^\infty C_n(y-x)\\ =& C(y-x)\!\phantom{1}_0F_{d-1} \left(\left. \begin{array}{c} (\;) \\ 2 \frac{k}{d}, k=1,\ldots,d-1 \end{array} \right\vert \frac{\Gamma(d)}{\left(\frac{d}{2}\right)^{d-1}} \rho V \right),\end{align}
where $\!\!\phantom{1}_0F_{d-1}$ is a generalised hypergeometric function.

For particular values of $d$ this simplifies. For example in $d=2$ we have:
\be C_T(y-x) = C(y-x)\!\phantom{1}_0F_1 \left(\left. \begin{array}{c} (\;) \\ 1 \end{array} \right\vert \rho V \right) = I_0(2 \sqrt{\rho V}), \ee
and in $d=4$ we have 
\be \label{eq:TotalChains} C_T(y-x) = C(y-x)\phantom{1}_0F_3 \left(\left. \begin{array}{c} (\;) \\ \frac{1}{2}, \frac{2}{2}, \frac{3}{2} \end{array} \right\vert \frac{3}{4} \rho V \right) = \frac{1}{z}\left(I_1(z)+J_1(z)\right), \ee
where $z = 2 (12 \rho V)^\frac{1}{4}$ (compare to \eqref{eq:3+1SumOverChains} in \secref{sec:3+1pathintegral}).

\section{Convolution approach}

One approach to determining the $P_n$ functions involves expanding $P(x)$ in terms of the functions $C_n(x)$ to give:
\be \label{eq:PintermsofC} P(x) = C(x) e^{-\rho V(x)} = C(x) \sum_{n=0}^\infty \frac{(-\rho V(x))^n}{n!} = \sum_{n=0}^\infty \frac{(-1)^n C_{n+1}(x)}{n! D_{n+1}},\ee
where we have used \eqref{eq:ChainForm}.

In this form we can use \eqref{eq:ChainConvolve} to compute expressions for $P_n$
As an example we have
\be P_2 := \rho P * P = \sum_{n=0}^\infty \sum_{m=0}^\infty \frac{(-1)^{n+m} \rho C_{n+1}*C_{m+1}}{n! m! D_{n+1} D_{m+1}} = \sum_{n=0}^\infty \sum_{m=0}^\infty \frac{(-1)^{n+m} C_{n+m+2}}{n! m! D_{n+1} D_{m+1}}\ee
\be = \sum_{n=0}^\infty \sum_{m=0}^\infty \frac{(-1)^{n+m} C (\rho V)^{n+m+1}D_{n+m+2}}{n! m! D_{n+1} D_{m+1}}, \ee
where we have used \eqref{eq:ChainConvolve} and \eqref{eq:ChainForm}.

By resumming the terms in this series we can replace the double infinite summation by one infinite and one finite summation to give
\be P_2 = \sum_{n=0}^\infty \sum_{k=0}^n \frac{(-1)^n C_{n+2}}{k! (n-k)! D_{k+1} D_{n-k+1}}= \sum_{n=0}^\infty \sum_{k=0}^n \frac{(-1)^n C (\rho V)^{n+1} D_{n+2}}{k! (n-k)! D_{k+1} D_{n-k+1}}.\ee 

In fact this procedure can be performed for general $P_n$. In effect we treat $P$ as a formal power series in a variable $C$ where the $n\th$ ``power'' of the variable is $C_n$ (i.e. the ``product'' of two functions $f$ and $g$ is $\rho f * g$).

We can use the following result \citep[0.314, p14]{Gradshteyn:1965} that:
\be \left(\sum_{k=0}^\infty a_k x^k\right)^{\!\!n} = \sum_{k=0}^\infty c_k x^k, \ee
with
\be c_0 = a_0^n, \quad c_m = \frac{1}{m a_0} \sum_{k=1}^m (kn - m +k)a_k c_{n-k}. \ee
Applying this to \eqref{eq:PintermsofC} we have, for $n \geq 1$,
\be P_n = \sum_{m=0}^\infty g_m C_{m+n} = \sum_{m=0}^\infty g_m D_{m+n} C (\rho V)^{n+m-1}, \ee
where the $g_m$ coefficients satisfy the recurrence relation
\be g_0 = 1 \qquad g_m = \frac{1}{m} \sum_{k=1}^m (k(n+1)-m) \frac{(-1)^k}{k! D_{k+1}} g_{m-k}. \ee
As an example, for $d=2$ we have (for causally related points)
\begin{align}
P_1 & = 1-z+\frac{z^2}{2}-\frac{z^3}{6}+\frac{z^4}{24}-\frac{z^5}{120}+\frac{z^6}{720}+O\left(z^7\right)&\\
P_2 & = z-\frac{z^2}{2}+\frac{5 z^3}{36}-\frac{z^4}{36}+\frac{z^5}{225}-\frac{13 z^6}{21600}+\frac{151 z^7}{2116800}+O\left(z^8\right)&\\
P_3 &= \frac{z^2}{4}-\frac{z^3}{12}+\frac{z^4}{64}-\frac{31 z^5}{14400}+\frac{7 z^6}{28800}-\frac{101 z^7}{4233600}+\frac{857 z^8}{406425600}+O\left(z^9\right)&\\
P_4  &= \frac{z^3}{36} - \frac{z^4}{144} + \frac{7 z^5}{7200} - \frac{13 z^6}{129600} + \frac{31 z^7}{3628800} - \frac{13 z^8}{20321280} + O\left(z^{9} \right)& 
\end{align}
where $z = \rho V$.

\subsection{Total number of paths}

We are also interested in the expected total number of paths from $x$ to $y$. Calling this $P_T$ we have
\be P_T(y-x) := \sum_{n=1}^\infty P_n(y-x). \ee
Treating this as a formal power series in $P$ (again, with the $n^\textrm{th}$ ``power'' of the $P$ equal to $P_n$) we have
\be P_T = \frac{P}{1-P}. \ee

We can now use the expression for $P$ as a formal power series in $C$ and use the following result \citep[0.313, p14]{Gradshteyn:1965}:
\be \frac{\sum_{k=0}^\infty b_k x^k}{\sum_{k=0}^\infty a_k x^k} = \frac{1}{a_0}\sum_{k=0}^\infty c_k x^k, \ee
with
\be c_n = b_n - \frac{1}{a_0} \sum_{k=1}^n a_k c_{n-k}, \ee
to give
\be P_T = \sum_{n=0}^\infty a_{n+1} C_{n+1} =\sum_{n=0}^\infty a_{n+1} D_{n+1} C (\rho V)^n, \ee
with
\be a_0 = 1, \quad a_n = \sum_{k=1}^n a_{n-k} \frac{(-1)^{k-1}}{(k-1)! D_k}. \ee
It turns out that $a_1 = 1, a_2 = 0$ in all dimensions. 

As an example, for $d=2$ we have (for causally related points):
\be P_T = 1+\frac{z^2}{4}-\frac{z^3}{12}+\frac{7 z^4}{288}-\frac{77 z^5}{14400}+\frac{497 z^6}{518400}-\frac{919 z^7}{6350400}+O(z^8),\ee
and in $d=4$ we have
\be P_T = 1+\frac{9 z^2}{20}-\frac{331 z^3}{2100}+\frac{319 z^4}{7840}-\frac{435679 z^5}{52920000}+\frac{459235787 z^6}{332972640000}+O(z^7),\ee
where $z = \rho V$.

In principal these series expansions for $P_n$ or $P_T$ can be use to compute the functions to high accuracy.

\subsection{Comparison to previous work}

In \citet{Bombelli:1987:PhD} the expected total number of paths was approximated for sprinklings into $\mink^2$. They used a Laplace transform and a saddle-point approximation to obtain \citep[2.5.15]{Bombelli:1987:PhD}:
\be \label{eq:BombelliApprox} P_T(y-x) \approx \frac{e}{2\pi}\frac{\gamma^2}{2-\gamma^2} \frac{e^{\gamma \sqrt{\rho} l}}{\sqrt{\gamma l \sqrt{\rho} }}, \ee
where $d=2$ and $\gamma \approx 0.93254288$ and $l$ is the proper time from $x$ to $y$.

\begin{figure}[!h]
\begin{center}
\includegraphics[width = 0.8\textwidth]{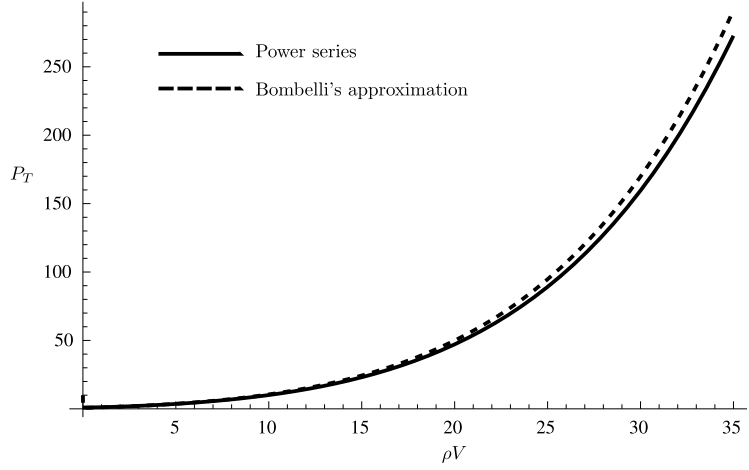}
\caption[Approximations for expected total number of paths]{Comparison of the power series for $P_T$ and Bombelli's approximation \eqref{eq:BombelliApprox}.}
\label{fig:PTPlotCompare}
\end{center}
\end{figure}

We can compare this approximation by plotting it against our power series. In \figref{fig:PTPlotCompare} we plot the two functions as a function of $\rho V$ (the power series has been summed to the first 500 terms).
As we can see the approximation is very good.

\subsection{Comparison to total number of chains}

Since every path is a chain the total number of paths is always less than the total number of chains.
Here we plot the power series for $P_T$ against the total number of chains \eqref{eq:TotalChains} for $d=2$ and $d=4$ dimensions.

\subsubsection{Two dimensions}

\begin{figure}[!h]
\begin{center}
\includegraphics[width = 0.8\textwidth]{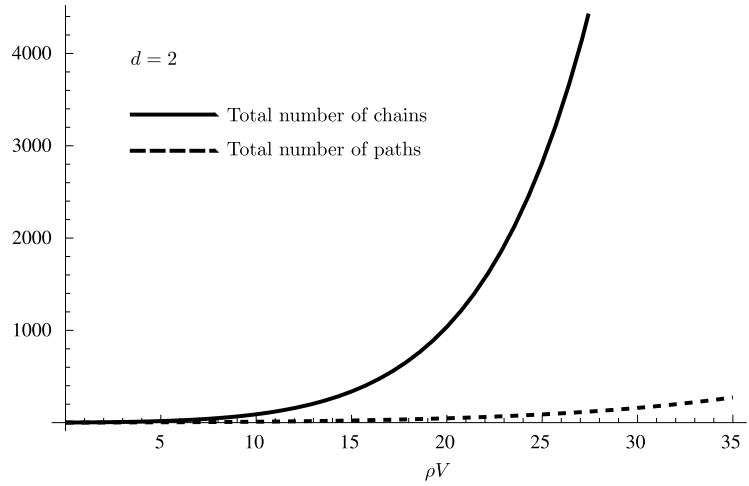}
\caption[Total number of paths vs chains in 1+1 dimensional sprinkling]{Comparison of the expected total number of chains and paths for sprinklings into two dimensional Minkowski spacetime.}
\label{fig:PTvsCT2d}
\end{center}
\end{figure}
From \figref{fig:PTvsCT2d} we see that in the $d=2$ case the vast majority of chains are not paths.

\subsubsection{Four dimensions}

\begin{figure}[!h]
\begin{center}
\includegraphics[width = 0.8\textwidth]{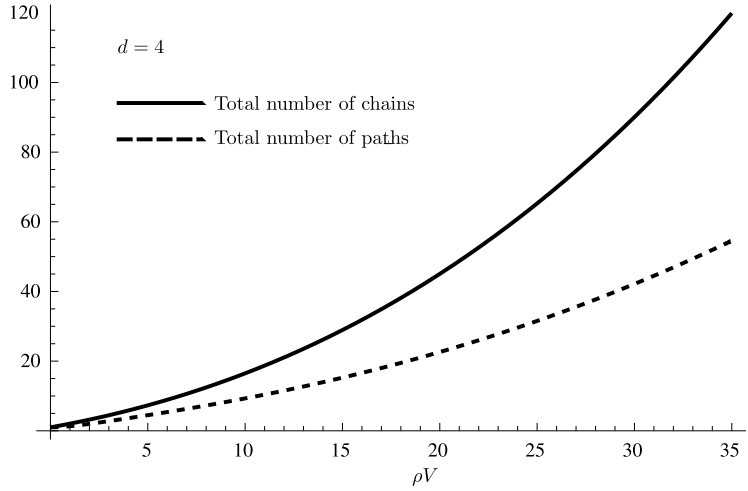}
\caption[Total number of paths vs chains in 3+1 dimensional sprinkling]{Comparison of the expected total number of chains and paths for sprinklings into four dimensional Minkowski spacetime.}
\label{fig:PTvsCT4d}
\end{center}
\end{figure}
From \figref{fig:PTvsCT4d} we see that in the $d=4$ case around half of all chains are paths.

\section{Green's function approach}

Another approach to determining $P_n$ is to try to find a differential equation satisfied by the $P_n$ and $P_T$ functions.

We start with the observation that\footnote{Fractional powers of $\Box$ are defined through the Fourier transform.} \citep[eq 22 with $\alpha = d/2$]{Bollini:1993}
\be \Box^{d/2} C(x) = A \delta(x), \ee
where $\delta$ is the $d$-dimensional delta-function, $\Box = \partial^2_{x^0} - \partial_{\vec{x}}^2$ is the d'Alembertian and $A$ is a constant equal to
\be A = 2^{d-1} \pi^{d/2-1} \Gamma(d/2).\ee
From this and \eqref{eq:ChainDef} we have
\be \Box^{d/2} C_n = \rho \left(\Box^{d/2} C\right) * C_{n-1} = A \rho \delta * C_{n-1} =  A \rho C_{n-1}. \ee
Recalling that 
\be \tag{\ref{eq:PintermsofC}} 
P(x) = C(x) e^{-\rho V(x)} = C(x) \sum_{n=0}^\infty \frac{(-\rho V(x))^n}{n!} = \sum_{n=0}^\infty \frac{(-1)^n C_{n+1}(x)}{n! D_{n+1}}, \ee
it immediately follows that 
\be \Box^{d/2} P = A \delta + A \rho \sum_{n=1}^\infty \frac{(-1)^n C_{n}}{n! D_{n+1}}  =  A \delta + A \rho C \sum_{n=1}^\infty \frac{(-1)^n (\rho V)^{n-1} D_{n}}{n! D_{n+1}}. \ee

This infinite sum can be evaluated and is equal to
\be C \sum_{n=1}^\infty \frac{(-1)^n (\rho V)^{n-1} D_{n}}{n! D_{n+1}} \!= -\frac{1}{(d-1)!} \left(\frac{d}{2}H\!+\! 1\right)\!\left(\frac{d}{2}H\!+\!2\right)\cdots\left(\frac{d}{2}H\!+\!d\!-\!1\right)P, \ee
where $H = \rho \frac{\partial}{\partial \rho}$ is a differential operator. We can see this holds by expanding both sides in powers of $\rho V$.

In 1+1 dimensions we have:
\be C \sum_{n=1}^\infty \frac{(-1)^n (\rho V)^{n-1} D_{n}}{n! D_{n+1}} = (-1 + \rho V) P = -(H+1)P.\ee

In 3+1 dimensions we have
\begin{align} C \sum_{n=1}^\infty \frac{(-1)^n (\rho V)^{n-1} D_{n}}{n! D_{n+1}} &= (-1 + 9 \rho V - 8(\rho V)^2 + \frac{4}{3}(\rho V)^3) P \\ &=   -\frac{1}{6}(2H+1)(2H+2)(2H+3)P.\end{align}

We now specialise to the two and four-dimensional cases.

\subsection{1+1 dimensions}
In 1+1 dimensions we have
\be \Box C = 2 \delta, \ee
which leads to
\be \Box P = 2 \delta + 2 \rho  (-1 + \rho V) P, \ee
or
\be \label{eq:PPDE} \left(\Box +2\rho(1 - \rho V)\right) P = \left(\Box +2\rho\left(H + 1\right)\right) P = 2 \delta. \ee

Now for $n > 1$ we have $P_n = \rho^{n-1} P * P * \cdots * P$ (with $n$ copies of $P$). This means that\footnote{$ H(  c(f * g)) = (Hc) (f*g) + c(H f) * g + c f*(Hg)$}
\begin{align} H P_n &= (n-1) P_n + \rho^{n-1} \lbrack(HP)* P * \cdots *P \\ & \qquad + P*(HP)* P * \cdots *P + \ldots +  P*\cdots *P*(HP)\rbrack\\
&= (n-1) P_n + n \rho (HP)* P_{n-1},\end{align}
or
\be (H+1)P_n = n \rho ((H+1)P) * P_{n-1}.\ee
Therefore
\begin{align} P_n = \rho (\Box P) * P_{n-1} &= \rho(2\delta + 2 \rho (H+1) P) * P_{n-1}\\
\label{eq:PnPDE} &= 2 \rho P_{n-1} - \frac{2}{n}  \rho(H+1)P_n, \end{align}
or
\be \left(\Box +\frac{2}{n}\rho\left(1 + \rho\frac{\partial}{\partial \rho}\right)\right) P_n = 2 \rho P_{n-1}. \ee
This partial differential equation for $P_n$ could, in principle, be solved one $n$ at a time starting with $n=2$.

\subsubsection{Total number of paths}

It's helpful to define the generating function for the expected number of paths:
\be P_T(z):=\sum_{n=1}^\infty z^{n-1} P_n = P_1 + z P_2 + z^2 P_3 +\ldots,\ee
for some real number $z$. Note that $P_T(1)$ is the expected total number of paths.

Using \eqref{eq:PPDE} and \eqref{eq:PnPDE} we have
\be (H+1)P_1 = \frac{\delta}{\rho} - \frac{\Box}{2\rho}P_1, \qquad (H+1)P_n = n\left(P_{n-1} - \frac{\Box}{2\rho}P_n\right). \ee
Using these and applying $H+1$ term-by-term to $P_T$ gives: 
\be (H+1) P_T = \left(\frac{\delta}{\rho} - \frac{\Box}{2\rho}P_1\right) + 2z \left(P_1 - \frac{\Box}{2\rho}P_2 \right)+3z^2 \left(P_2 - \frac{\Box}{2\rho}P_3 \right)  +\ldots, \ee
or
\begin{align} (H+1)P_T &= \frac{\delta}{\rho} + \left(2 z P_1 + 3z^2 P_2 + \ldots \right) - \frac{\Box}{2\rho}\left(P_1 + 2z P_2 + 3z^2 P_3 + \ldots \right) \\
&= \frac{\delta}{\rho} + \frac{\partial}{\partial z} (z^2 P_T) - \frac{\Box}{2\rho} \frac{\partial}{\partial z} (z P_T).\end{align}
We have therefore obtained a partial differential equation satisfied by $P_T(z)$, namely:

\be \rho \frac{\partial P_T}{\partial \rho} +P_T- \frac{\partial}{\partial z} (z^2P_T) + \frac{\Box}{2\rho} \frac{\partial}{\partial z} (z P_T) = \frac{\delta}{\rho},\ee
with the boundary condition that $P_T(0)= P_1$. If this could be solved we would be able to determine $P_T(z)$.

\subsection{3+1 dimensions} \label{sec:3+1PathPDE}
In four dimensions we have
\be \Box^2 C = 8 \pi \delta. \ee
This gives
\begin{align} \Box^2 P &= 8 \pi \delta + 8 \pi \rho P \left(-1 + 8(\rho V) - 9(\rho V)^2 + \frac{4}{3} (\rho V)^3\right), \\
&=  8 \pi \delta - 8 \pi \rho \frac{1}{6}(2H+1)(2H+2)(2H+3) P. \end{align}
Due to the complexity of this differential equation we will not attempt to derive differential equations satisfied by general $P_n$ or $P_T$.

Nevertheless, we mention that this equation provides a surprising link with the causal set d'Alembertian work of \citet{Benincasa:2010}. In that work they show that the expected value of the causal set d'Alembertian (applied to a field $\phi$ and evaluated at a point $x \in \mink^4$) is $(\bar{B} * \phi)(x)$ for a suitable integral kernel $\bar{B}(x)$. As it turns out we have
\be \Box^2 P = \frac{2 \pi \sqrt{6}}{\sqrt{\rho}} \bar{B},\ee
(compare with \citet[eq. (4)]{Benincasa:2010}). Defining $K_R:=\frac{\sqrt{\rho}}{2 \pi \sqrt{6}} P$ then we have
\be \Box^2 K_R = \bar{B}. \ee
But $K_R$ is the expected value of the 4 dimensional massless retarded propagator defined in \secref{sec:3+1pathintegral}!
This surprising connection shows that the current work and the work of \citeauthor{Benincasa:2010} are mutually consistent. We can see this by evaluating $\bar{B} * \phi$ and taking the infinite density limit:
\be \bar{B} * \phi = (\Box^2 K_R) * \phi = (\Box K_R) * (\Box \phi). \ee
The work of \secref{sec:3+1pathintegral} showed that in the infinite density limit we have
\be \lim_{\rho\to\infty} \Box K_R = \delta,\ee
which means that 
\be \lim_{\rho\to\infty} \bar{B} * \phi = \lim_{\rho\to\infty} (\Box K_R) * (\Box \phi) = \delta * (\Box \phi) = \Box \phi, \ee
as was shown by other means by \citet{Benincasa:2010}.

\setlength{\bibsep}{2pt}
\bibliographystyle{kluwer}
\clearpage
\pagestyle{plain}
\phantomsection
\addcontentsline{toc}{chapter}{Bibliography}
\small{\bibliography{thesisreferences}}

\end{document}